\documentclass{pasj01}

\RequirePackage{mathpazo}
\RequirePackage[T1]{fontenc}

\usepackage[OT2,T1]{fontenc}

\def\commenta{$^*$}
\def\commentb{$^\dagger$}
\def\commentc{$^\ddagger$}
\def\commentd{$^\S$}

\newcounter{author}
\def\authorcount#1#2{\refstepcounter{author}\label{#1}
                     \altaffiltext{\ref{#1}}{#2}}

\def\Ohtprep{T. Ohshima et al. in preparation}

\begin{document}
\SetRunningHead{T. Kato et al.}{Period Variations in SU UMa-Type Dwarf Novae X}

\Received{201X/XX/XX}
\Accepted{201X/XX/XX}

\title{Survey of Period Variations of Superhumps in SU UMa-Type Dwarf Novae.
    X: The Tenth Year (2017)}

\author{Taichi~\textsc{Kato},\altaffilmark{\ref{affil:Kyoto}*}
        Keisuke~\textsc{Isogai},\altaffilmark{\ref{affil:Kyoto}}$^,$\altaffilmark{\ref{affil:KyotoOkayama}}
        Yasuyuki~\textsc{Wakamatsu},\altaffilmark{\ref{affil:Kyoto}}
        Franz-Josef~\textsc{Hambsch},\altaffilmark{\ref{affil:GEOS}}$^,$\altaffilmark{\ref{affil:BAV}}$^,$\altaffilmark{\ref{affil:Hambsch}}
        Hiroshi~\textsc{Itoh},\altaffilmark{\ref{affil:Ioh}}
        Tam\'as~\textsc{Tordai},\altaffilmark{\ref{affil:Polaris}}
        Tonny~\textsc{Vanmunster},\altaffilmark{\ref{affil:Vanmunster}}
        Pavol~A.~\textsc{Dubovsky},\altaffilmark{\ref{affil:Dubovsky}}
        Igor~\textsc{Kudzej},\altaffilmark{\ref{affil:Dubovsky}}
        Tom\'a\v{s}~\textsc{Medulka},\altaffilmark{\ref{affil:Dubovsky}}
        Mariko~\textsc{Kimura},\altaffilmark{\ref{affil:Kyoto}}
        Ryuhei~\textsc{Ohnishi},\altaffilmark{\ref{affil:Kyoto}}
        Berto~\textsc{Monard},\altaffilmark{\ref{affil:Monard}}$^,$\altaffilmark
{\ref{affil:Monard2}}
        Elena~P.~\textsc{Pavlenko},\altaffilmark{\ref{affil:CrAO}}
        Kirill~A.~\textsc{Antonyuk},\altaffilmark{\ref{affil:CrAO}}
        Nikolaj~V.~\textsc{Pit},\altaffilmark{\ref{affil:CrAO}}
        Oksana~I.~\textsc{Antonyuk},\altaffilmark{\ref{affil:CrAO}}
        Julia~V.~\textsc{Babina},\altaffilmark{\ref{affil:CrAO}}
        Aleksei~V.~\textsc{Baklanov},\altaffilmark{\ref{affil:CrAO}}
        Aleksei~A.~\textsc{Sosnovskij},\altaffilmark{\ref{affil:CrAO}}
        Roger~D.~\textsc{Pickard},\altaffilmark{\ref{affil:BAAVSS}}$^,$\altaffilmark{\ref{affil:Pickard}}
        Ian~\textsc{Miller},\altaffilmark{\ref{affil:Miller}}
        Yutaka~\textsc{Maeda},\altaffilmark{\ref{affil:Mdy}}
        Enrique~\textsc{de Miguel},\altaffilmark{\ref{affil:Miguel}}$^,$\altaffilmark{\ref{affil:Miguel2}}
        Stephen~M.~\textsc{Brincat},\altaffilmark{\ref{affil:Brincat}}
        Domenico~\textsc{Licchelli},\altaffilmark{\ref{affil:Lic}}$^,$\altaffilmark{\ref{affil:Lic2}}
        Lewis~M.~\textsc{Cook},\altaffilmark{\ref{affil:LewCook}}
        Sergey~Yu.~\textsc{Shugarov},\altaffilmark{\ref{affil:Sternberg}}$^,$\altaffilmark{\ref{affil:Slovak}}
        Anna~M.~\textsc{Zaostrojnykh},\altaffilmark{\ref{affil:Kazan}}
        Drahomir~\textsc{Chochol},\altaffilmark{\ref{affil:Slovak}}
        Polina~\textsc{Golysheva},\altaffilmark{\ref{affil:Sternberg}}
        Natalia~\textsc{Katysheva},\altaffilmark{\ref{affil:Sternberg}}
        Alexandra~M.~\textsc{Zubareva},\altaffilmark{\ref{affil:RASInstAst}}$^,$
\altaffilmark{\ref{affil:Sternberg}}
        Geoff~\textsc{Stone},\altaffilmark{\ref{affil:AAVSO}}
        Kiyoshi~\textsc{Kasai},\altaffilmark{\ref{affil:Kai}}
        Peter~\textsc{Starr},\altaffilmark{\ref{affil:Starr}}
        Colin~\textsc{Littlefield},\altaffilmark{\ref{affil:LCO}}
        Seiichiro~\textsc{Kiyota},\altaffilmark{\ref{affil:Kis}}
        Maksim~V.~\textsc{Andreev},\altaffilmark{\ref{affil:Terskol}}$^,$\altaffilmark{\ref{affil:ICUkraine}}
        Alexandr~V.~\textsc{Sergeev},\altaffilmark{\ref{affil:Terskol}}$^,$\altaffilmark{\ref{affil:ICUkraine}}
        Javier~\textsc{Ruiz},\altaffilmark{\ref{affil:Ruiz1}}$^,$\altaffilmark{\ref{affil:Ruiz2}}$^,$\altaffilmark{\ref{affil:Ruiz3}}
        Gordon~\textsc{Myers},\altaffilmark{\ref{affil:Myers}}
        Andrii~O.~\textsc{Simon},\altaffilmark{\ref{affil:TarasShevshenko}}
        Volodymyr~V.~\textsc{Vasylenko},\altaffilmark{\ref{affil:TarasShevshenko}}
        Francisco~\textsc{Sold\'an},\altaffilmark{\ref{affil:Soldan}}
        Yenal~\textsc{\"Ogmen},\altaffilmark{\ref{affil:Ogmen}}
        Kazuhiro~\textsc{Nakajima},\altaffilmark{\ref{affil:Njh}}
        Peter~\textsc{Nelson},\altaffilmark{\ref{affil:Nelson}}
        Gianluca~\textsc{Masi},\altaffilmark{\ref{affil:Masi}}
        Kenneth~\textsc{Menzies},\altaffilmark{\ref{affil:Menzies}}
        Richard~\textsc{Sabo},\altaffilmark{\ref{affil:Sabo}}
        Greg~\textsc{Bolt},\altaffilmark{\ref{affil:Bolt}}
        Shawn~\textsc{Dvorak},\altaffilmark{\ref{affil:Dvorak}}
        Krzysztof~Z.~\textsc{Stanek},\altaffilmark{\ref{affil:Ohio}}
        Joseph~V.~\textsc{Shields},\altaffilmark{\ref{affil:Ohio}}
        Christopher~S.~\textsc{Kochanek},\altaffilmark{\ref{affil:Ohio}}
        Thomas~W.-S.~\textsc{Holoien},\altaffilmark{\ref{affil:Ohio}}
        Benjamin~\textsc{Shappee},\altaffilmark{\ref{affil:Carnegie}}
        Jos\'e~L.~\textsc{Prieto},\altaffilmark{\ref{affil:DiegoPortales}}$^,$\altaffilmark{\ref{affil:Princeton}}
        Tadashi~\textsc{Kojima},\altaffilmark{\ref{affil:Kojima}}
        Hideo~\textsc{Nishimura},\altaffilmark{\ref{affil:Nishimura}}
        Shizuo~\textsc{Kaneko},\altaffilmark{\ref{affil:Kaneko}}
        Shigehisa~\textsc{Fujikawa},\altaffilmark{\ref{affil:Fujikawa}}
        Rod~\textsc{Stubbings},\altaffilmark{\ref{affil:Stubbings}}
        Eddy~\textsc{Muyllaert},\altaffilmark{\ref{affil:VVSBelgium}}
        Gary~\textsc{Poyner},\altaffilmark{\ref{affil:Poyner}}
        Masayuki~\textsc{Moriyama},\altaffilmark{\ref{affil:Myy}}
        Hiroyuki~\textsc{Maehara},\altaffilmark{\ref{affil:OAO}}
        Patrick~\textsc{Schmeer},\altaffilmark{\ref{affil:Schmeer}}
        Denis~\textsc{Denisenko},\altaffilmark{\ref{affil:Sternberg}}
}

\authorcount{affil:Kyoto}{
     Department of Astronomy, Kyoto University, Kyoto 606-8502, Japan}
\email{$^*$tkato@kusastro.kyoto-u.ac.jp}

\authorcount{affil:KyotoOkayama}{
     Okayama Observatory, Kyoto University, 3037-5 Honjo, Kamogatacho,
     Asakuchi, Okayama 719-0232, Japan}

\authorcount{affil:GEOS}{
     Groupe Europ\'een d'Observations Stellaires (GEOS),
     23 Parc de Levesville, 28300 Bailleau l'Ev\^eque, France}

\authorcount{affil:BAV}{
     Bundesdeutsche Arbeitsgemeinschaft f\"ur Ver\"anderliche Sterne
     (BAV), Munsterdamm 90, 12169 Berlin, Germany}

\authorcount{affil:Hambsch}{
     Vereniging Voor Sterrenkunde (VVS), Oude Bleken 12, 2400 Mol, Belgium}

\authorcount{affil:Ioh}{
     Variable Star Observers League in Japan (VSOLJ),
     1001-105 Nishiterakata, Hachioji, Tokyo 192-0153, Japan}

\authorcount{affil:Polaris}{
     Polaris Observatory, Hungarian Astronomical Association,
     Laborc utca 2/c, 1037 Budapest, Hungary}

\authorcount{affil:Vanmunster}{
     Center for Backyard Astrophysics Belgium, Walhostraat 1A,
     B-3401 Landen, Belgium}

\authorcount{affil:Dubovsky}{
     Vihorlat Observatory, Mierova 4, 06601 Humenne, Slovakia}

\authorcount{affil:Monard}{
     Bronberg Observatory, Center for Backyard Astrophysics Pretoria,
     PO Box 11426, Tiegerpoort 0056, South Africa}

\authorcount{affil:Monard2}{
     Kleinkaroo Observatory, Center for Backyard Astrophysics Kleinkaroo,
     Sint Helena 1B, PO Box 281, Calitzdorp 6660, South Africa}

\authorcount{affil:CrAO}{
     Federal State Budget Scientific Institution ``Crimean Astrophysical
     Observatory of RAS'', Nauchny, 298409, Republic of Crimea}

\authorcount{affil:BAAVSS}{
     The British Astronomical Association, Variable Star Section (BAA VSS),
     Burlington House, Piccadilly, London, W1J 0DU, UK}

\authorcount{affil:Pickard}{
     3 The Birches, Shobdon, Leominster, Herefordshire, HR6 9NG, UK}

\authorcount{affil:Miller}{
     Furzehill House, Ilston, Swansea, SA2 7LE, UK}

\authorcount{affil:Mdy}{
     Kaminishiyamamachi 12-14, Nagasaki, Nagasaki 850-0006, Japan}

\authorcount{affil:Miguel}{
     Departamento de Ciencias Integradas, Facultad de Ciencias
     Experimentales, Universidad de Huelva,
     21071 Huelva, Spain}

\authorcount{affil:Miguel2}{
     Center for Backyard Astrophysics, Observatorio del CIECEM,
     Parque Dunar, Matalasca\~nas, 21760 Almonte, Huelva, Spain}

\authorcount{affil:Brincat}{
     Flarestar Observatory, San Gwann SGN 3160, Malta}

\authorcount{affil:Lic}{
     R. P. Feynman Observatory, Gagliano del Capo, 73034, Italy}

\authorcount{affil:Lic2}{
     Center for Backyard Astrophysics -- Gagliano del Capo, 73034, Italy}

\authorcount{affil:LewCook}{
     Center for Backyard Astrophysics Concord, 1730 Helix Ct. Concord,
     California 94518, USA}

\authorcount{affil:Sternberg}{
     Sternberg Astronomical Institute, Lomonosov Moscow State University, 
     Universitetsky Ave., 13, Moscow 119992, Russia}

\authorcount{affil:Slovak}{
     Astronomical Institute of the Slovak Academy of Sciences,
     05960 Tatranska Lomnica, Slovakia}

\authorcount{affil:Kazan}{
     Institute of Physics, Kazan Federal University,
     Ulitsa Kremlevskaya 16a, Kazan 420008, Russia}

\authorcount{affil:RASInstAst}{
     Institute of Astronomy, Russian Academy of Sciences,
     Moscow 119017, Russia}

\authorcount{affil:AAVSO}{
     American Association of Variable Star Observers, 49 Bay State Rd.,
     Cambridge, MA 02138, USA}

\authorcount{affil:Kai}{
     Baselstrasse 133D, CH-4132 Muttenz, Switzerland}

\authorcount{affil:Starr}{
     Warrumbungle Observatory, Tenby, 841 Timor Rd,
     Coonabarabran NSW 2357, Australia}

\authorcount{affil:LCO}{
     Department of Physics, University of Notre Dame, 
     225 Nieuwland Science Hall, Notre Dame, Indiana 46556, USA}

\authorcount{affil:Kis}{
     VSOLJ, 7-1 Kitahatsutomi, Kamagaya, Chiba 273-0126, Japan}

\authorcount{affil:Terskol}{
     Terskol Branch of Institute of Astronomy, Russian Academy of Sciences,
     361605, Peak Terskol, Kabardino-Balkaria Republic, Russia}

\authorcount{affil:ICUkraine}{
     International Center for Astronomical, Medical and Ecological Research
     of NASU, Ukraine 27 Akademika Zabolotnoho Str. 03680 Kyiv,
     Ukraine}

\authorcount{affil:Ruiz1}{
     Observatorio de C\'antabria, Ctra. de Rocamundo s/n, Valderredible, 
     39220 Cantabria, Spain}

\authorcount{affil:Ruiz2}{
     Instituto de F\'{\i}sica de Cantabria (CSIC-UC), Avenida Los Castros s/n, 
     E-39005 Santander, Cantabria, Spain}

\authorcount{affil:Ruiz3}{
     Agrupaci\'on Astron\'omica C\'antabria, Apartado 573,
     39080, Santander, Spain}

\authorcount{affil:Myers}{
     Center for Backyard Astrophysics San Mateo, 5 inverness Way,
     Hillsborough, CA 94010, USA}

\authorcount{affil:TarasShevshenko}{
     Astronomy and Space Physics Department, Taras Shevshenko National
     University of Kyiv, Volodymyrska str. 60, Kyiv, 01601, Ukraine}

\authorcount{affil:Soldan}{
     Observatorio Amanecer de Arrakis, Alcal\'a de Guada\'{\i}ra,
     Petrarca 6, 1A 41006. Seville, Spain}

\authorcount{affil:Ogmen}{
     Green Island Observatory, Ge\c{c}itkale, Magosa, via Mersin, North Cyprus}

\authorcount{affil:Njh}{
     Variable Star Observers League in Japan (VSOLJ),
     124 Isatotyo, Teradani, Kumano, Mie 519-4673}

\authorcount{affil:Nelson}{
     1105 Hazeldean Rd, Ellinbank 3820, Australia}

\authorcount{affil:Masi}{
     The Virtual Telescope Project, Via Madonna del Loco 47, 03023
     Ceccano (FR), Italy}

\authorcount{affil:Menzies}{
     Center for Backyard Astrophysics (Framingham), 
     318A Potter Road, Framingham, MA 01701, USA}

\authorcount{affil:Sabo}{
     2336 Trailcrest Dr., Bozeman, Montana 59718, USA}

\authorcount{affil:Bolt}{
     Camberwarra Drive, Craigie, Western Australia 6025, Australia}

\authorcount{affil:Dvorak}{
     Rolling Hills Observatory, 1643 Nightfall Drive,
     Clermont, Florida 34711, USA}

\authorcount{affil:Ohio}{
     Department of Astronomy, the Ohio State University, Columbia,
     OH 43210, USA}

\authorcount{affil:Carnegie}{
     Carnegie Observatories, 813 Santa Barbara Street, Pasadena,
     CA 91101, USA}

\authorcount{affil:DiegoPortales}{
     N\'ucleo de Astronom\'ia de la Facultad de Ingenier\'ia, Universidad
     Diego Portales, Av. Ej\'ercito 441, Santiago, Chile}

\authorcount{affil:Princeton}{
     Department of Astrophysical Sciences, Princeton University,
     NJ 08544, USA}

\authorcount{affil:Kojima}{
     Kanbara, Tsumagoi-mura, Agatsuma-gun, Gunma 377-1524, Japan}

\authorcount{affil:Nishimura}{
     Miyawaki 302-6, Kakegawa, Shizuoka 436-0086, Japan}

\authorcount{affil:Kaneko}{
     14-7 Kami-Yashiki, Kakegawa, Shizuoka 436-0049, Japan}

\authorcount{affil:Fujikawa}{
     107 Iseki, Oonohara-cho, Kannonji, Kagawa 769-1621, Japan}

\authorcount{affil:Stubbings}{
     Tetoora Observatory, 2643 Warragul-Korumburra Road, Tetoora Road,
     Victoria 3821, Australia}

\authorcount{affil:VVSBelgium}{
     Vereniging Voor Sterrenkunde (VVS), Moffelstraat 13 3370
     Boutersem, Belgium}

\authorcount{affil:Poyner}{
     BAA Variable Star Section, 67 Ellerton Road, Kingstanding,
     Birmingham B44 0QE, UK}

\authorcount{affil:Myy}{
     290-383, Ogata-cho, Sasebo, Nagasaki 858-0926, Japan}

\authorcount{affil:OAO}{
     Subaru Telescope Okayama Branch Office, National Astronomical Observatory
     of Japan, NINS, Asakuchi, Okayama 719-0232, Japan}

\authorcount{affil:Schmeer}{
     Bischmisheim, Am Probstbaum 10, 66132 Saarbr\"{u}cken, Germany}



\maketitle

\begin{abstract}
Continuing the project described by Kato et al. (2009,
PASJ, 61, S395), we collected times of superhump maxima for
102 SU UMa-type dwarf novae observed mainly during
the 2017 season and characterized these objects.
WZ Sge-type stars identified in this study
are PT And, ASASSN-17ei, ASASSN-17el, ASASSN-17es,
ASASSN-17fn, ASASSN-17fz, ASASSN-17hw, ASASSN-17kd,
ASASSN-17la, PNV J20205397$+$2508145 and TCP J00332502$-$3518565.
We obtained new mass ratios for 7 objects using
growing superhumps (stage A).
ASASSN-17gf is an EI Psc-type object below the period minimum.
CRTS J080941.3$+$171528 and DDE 51 are objects in the period
gap and both showed long-lasting phase of stage A superhumps.
We also summarized the recent advances in understanding
of SU UMa-type and WZ Sge-type dwarf novae.
\end{abstract}

\section{Introduction}

   This is a continuation of series of papers \citet{Pdot},
\citet{Pdot2}, \citet{Pdot3}, \citet{Pdot4}, \citet{Pdot5},
\citet{Pdot6}, \citet{Pdot7}, \citet{Pdot8} and \citet{Pdot9}
reporting new observations of superhumps in
SU UMa-type dwarf novae.
[see e.g. \citet{war95book} for SU UMa-type dwarf novae
and CVs in general].

   Upon recommendation from the previous reviewer
and the PASJ office, we provide the result in a concise
form: presenting the results in the Supporting Information (SI)
and only the list the objects (table \ref{tab:outobs}),
the obtained parameters (table \ref{tab:perlist})
and the references (section \ref{sec:list})
are given in the main paper.
For the details of the analysis, terminology and definitions
see \citet{Pdot} and for the initial and current
aims of this survey, see \citet{Pdot} and \citet{Pdot9},
respectively.  For superhump stages, see \citet{Pdot}
and a concise version in e-section 1 in SI.
A short description of the data analysis is given
in e-section 2 in SI.
In table \ref{tab:perlist}, $P_1$ and $P_2$
represent periods in stage B and C, respectively
($P_1$ is averaged during the entire course of the observed
segment of stage B),
and $E_1$ and $E_2$ represent intervals (in cycle numbers)
to determine $P_1$ and $P_2$, respectively.

\section{Data source}\label{sec:data}

   The CCD time-series observations were were obtained
under campaigns led by  the VSNET Collaboration \citep{VSNET}.
We also used the public data from
the AAVSO International Database\footnote{
   $<$http://www.aavso.org/data-download$>$.
}.

   Outburst detections of many new and known objects
relied on the ASAS-SN CV patrol \citep{dav15ASASSNCVAAS}\footnote{
   $<$http://cv.asassn.astronomy.ohio-state.edu/$>$.
}, the MASTER network \citep{MASTER}, and
Catalina Real-time Transient Survey
(CRTS; \cite{CRTS})\footnote{
   $<$http://nesssi.cacr.caltech.edu/catalina/$>$.
   For the information of the individual Catalina CVs, see
   $<$http://nesssi.cacr.caltech.edu/catalina/AllCV.html$>$.
}.  There were also outburst detections reported to
VSNET, AAVSO\footnote{
  $<$https://www.aavso.org/$>$.
}, BAAVSS alert\footnote{
  $<$https://groups.yahoo.com/neo/groups/baavss-alert/$>$.
} and cvnet-outburst.\footnote{
  $<$https://groups.yahoo.com/neo/groups/cvnet-outburst/$>$.
}

\begin{table*}
\caption{List of Superoutbursts.}\label{tab:outobs}
\begin{center}
\begin{tabular}{cccl}
\hline
Object & Year & Observers or references\commenta & \multicolumn{1}{c}{ID\commentb} \\
\hline
PT And        & 2017 & DPV, Ioh & \\
DH Aql        & 2017 & Ioh & \\
V1047 Aql     & 2017 & SGE, deM, Trt & \\
NN Cam        & 2017 & DPV & \\
V391 Cam      & 2017 & \citet{Pdot9} & \\
KP Cas        & 2017 & Ioh & \\
VW CrB        & 2017 & COO, Kis & \\
V503 Cyg      & 2017 & IMi & \\
V632 Cyg      & 2017 & deM & \\
GP CVn        & 2017 & Trt & \\
GQ CVn        & 2017 & deM, Mdy, COO & \\
HO Del        & 2017 & BSM & \\
MN Dra        & 2017 & COO & \\
OV Dra        & 2015 & Trt & \\
              & 2017 & DPV, Lis, Ioh, Trt & \\
V1454 Cyg     & 2006 & \citet{Pdot} & \\
BE Oct        & 2017 & HaC & \\
V521 Peg      & 2017 & Ioh, BSM, RPc & \\
V368 Per      & 2017 & CRI, Ter, Ioh, Trt, IMi, RPc, DPV & \\
XY Psc        & 2017 & KU, HaC, Ioh & \\
V701 Tau      & 2017 & BSM & \\
V1208 Tau     & 2017 & MZK, CRI, Trt & \\
TU Tri        & 2017 & Trt, Ioh, RPc, DPV & \\
SU UMa        & 2017 & DPV & \\
\hline
  \multicolumn{4}{l}{\parbox{500pt}{\commenta Key to observers:
BSM\commentc (S. Brincat),
CRI (Crimean Astrophys. Obs.),
deM (E. de Miguel),
DKS\commentc (S. Dvorak),
DPV (P. Dubovsky),
GBo (G. Bolt),
HaC (F.-J. Hambsch, remote obs. in Chile),
IMi\commentc (I. Miller),
Ioh (H. Itoh),
KU (Kyoto U., campus obs.),
Kai (K. Kasai),
Kis (S. Kiyota),
LCO (C. Littlefield),
Lic (D. Licchelli),
Lis (Lisnyky Obs.),
NGW\commentc (G. Myers),
MLF (B. Monard),
MZK\commentc (K. Menzies),
Mas (G. Masi),
Mdy (Y. Maeda),
NKa (N. Katysheva and S. Shugarov),
Nel (P. Nelson),
Njh (K. Nakajima),
OYE\commentc (Y. \"Ogmen),
RPc\commentc (R. Pickard),
Rui (J. Ruiz),
SGE\commentc (G. Stone),
SPE\commentc (P. Starr),
SRI\commentc (R. Sabo),
Shu (S. Shugarov team),
Sol (F. Sold\'an),
Ter (Terskol Obs.),
Trt (T. Tordai),
Van (T. Vanmunster),
AAVSO (AAVSO database)
}} \\
  \multicolumn{4}{l}{\commentb Original identifications, discoverers or data sou
rce.} \\
  \multicolumn{4}{l}{\commentc Inclusive of observations from the AAVSO database
.} \\
\end{tabular}
\end{center}
\end{table*}

\addtocounter{table}{-1}
\begin{table*}
\caption{List of Superoutbursts (continued).}\label{tab:outobs}
\begin{center}
\begin{tabular}{cccl}
\hline
Object & Year & Observers or references\commenta & \multicolumn{1}{c}{ID\commentb} \\
\hline
HS Vir        & 2007 & Njh & \\
              & 2017 & HaC, DPV, Mdy & \\
V406 Vir      & 2017 & MLF, MGW, HaC, Nel & \\
NSV 35        & 2017 & MGW, HaC & \\
1RXS J161659  & 2017 & deM, IMi & 1RXS J161659.5$+$620014 \\
ASASSN-13ce   & 2017 & Van, Ioh & \\
ASASSN-13dh   & 2017 & SGE, DPV, IMi, BSM & \\
ASASSN-14ca   & 2017 & Ter, Trt, Lis, DPV & \\
ASASSN-14cr   & 2017 & DPV & \\
ASASSN-14kb   & 2017 & HaC & \\
ASASSN-14lk   & 2017 & MLF & \\
ASASSN-15fu   & 2017 & HaC & \\
ASASSN-15fv   & 2017 & Van & \\
ASASSN-15qu   & 2017 & MLF, HaC & \\
ASASSN-17ei   & 2017 & MLF, HaC, SPE & \\
ASASSN-17el   & 2017 & MLF, HaC & \\
ASASSN-17eq   & 2017 & Van, Ioh & \\
ASASSN-17es   & 2017 & HaC, Van, Ioh & \\
ASASSN-17et   & 2017 & MLF, HaC & \\
ASASSN-17ew   & 2017 & HaC & \\
ASASSN-17ex   & 2017 & HaC & \\
ASASSN-17fh   & 2017 & Van & \\
ASASSN-17fi   & 2017 & Van & \\
ASASSN-17fj   & 2017 & HaC & \\
ASASSN-17fl   & 2017 & HaC & \\
ASASSN-17fn   & 2017 & Van, Ioh, DPV, Trt, Mdy, Shu, Lic, CRI & \\
ASASSN-17fo   & 2017 & Mdy, Kis, HaC, Lic, COO, RPc, Ioh, CRI & \\
ASASSN-17fp   & 2017 & MLF, HaC & \\
ASASSN-17fz   & 2017 & MLF, HaC, SPE & \\
ASASSN-17gf   & 2017 & MLF, HaC & \\
ASASSN-17gh   & 2017 & Ioh, Van & \\
ASASSN-17gv   & 2017 & MLF, HaC & \\
\hline
\end{tabular}
\end{center}
\end{table*}

\addtocounter{table}{-1}
\begin{table*}
\caption{List of Superoutbursts (continued).}\label{tab:outobs}
\begin{center}
\begin{tabular}{cccl}
\hline
Object & Year & Observers or references\commenta & \multicolumn{1}{c}{ID\commentb} \\
\hline
ASASSN-17hm   & 2017 & HaC & \\
ASASSN-17hw   & 2017 & MLF, HaC, BSM, Ioh, SPE, Shu, Van & \\
ASASSN-17hy   & 2017 & HaC & \\
ASASSN-17id   & 2017 & HaC & \\
ASASSN-17if   & 2017 & HaC & \\
ASASSN-17ig   & 2017 & GBo, HaC & \\
ASASSN-17il   & 2017 & Van & \\
ASASSN-17iv   & 2017 & HaC & \\
ASASSN-17iw   & 2017 & HaC & \\
ASASSN-17ix   & 2017 & HaC & \\
ASASSN-17ji   & 2017 & IMi, Trt, RPc & \\
ASASSN-17jr   & 2017 & HaC & \\
ASASSN-17kc   & 2017 & HaC & \\
ASASSN-17kd   & 2017 & HaC & \\
ASASSN-17kg   & 2017 & HaC, RPc, Van, Trt & \\
ASASSN-17kp   & 2017 & Trt, Van, RPc & \\
ASASSN-17la   & 2017 & COO, Van, DPV, IMi, Trt, NKa, KU & \\
ASASSN-17lr   & 2017 & IMi & \\
ASASSN-17me   & 2017 & LCO, CRI & \\
ASASSN-17np   & 2017 & MLF, HaC & \\
ASASSN-17nr   & 2017 & HaC & \\
ASASSN-17of   & 2017 & Van, Ioh, KU, IMi, CRI & \\
ASASSN-17oo   & 2017 & KU, HaC & \\
ASASSN-17ou   & 2017 & Shu, KU, HaC, Trt & \\
ASASSN-17pb   & 2017 & Van, CRI, IMi, KU & \\
CRTS J044027  & 2017 & HaC, Van & CRTS J044027.1$+$023301 \\
CRTS J080941  & 2017 & Van, HaC, CRI, Trt & CRTS J080941.3$+$171528 \\
CRTS J120052  & 2017 & Mdy & CRTS J120052.9$-$152620 \\
CRTS J122221  & 2017 & \citet{neu17j1222} & CRTS J122221.6$-$311524 \\
CRTS J162806  & 2017 & Trt & CRTS J162806.2$+$065316 \\
CRTS J214934  & 2017 & HaC, Ioh & CRTS J214934.1$-$121908 \\
\hline
\end{tabular}
\end{center}
\end{table*}

\addtocounter{table}{-1}
\begin{table*}
\caption{List of Superoutbursts (continued).}\label{tab:outobs}
\begin{center}
\begin{tabular}{cccl}
\hline
Object & Year & Observers or references\commenta & \multicolumn{1}{c}{ID\commentb} \\
\hline
CRTS J223235  & 2017 & IMi, Van & CRTS J223235.4$+$304105 \\
CTCV J1940    & 2017 & HaC & CTCV J1940$-$4724 \\
DDE 51        & 2017 & Mdy, Trt, RPc, Rui, CRI, IMi & \\
MASTER J132501 & 2017 & Kai, Lic, deM, Van & MASTER OT J132501.00$+$431846.1 \\
MASTER J174305 & 2017 & Mdy, Kai, DPV, Lic, Trt & MASTER OT J174305.70$+$231107.8 \\
MASTER J192757 & 2017 & Van & MASTER OT J192757.03$+$404042.8 \\
MASTER J200904 & 2017 & KU, deM, Lic & MASTER OT J200904.69$+$825153.6 \\
MASTER J205110 & 2017 & LCO, Lic, Ioh, KU, Trt & MASTER OT J205110.36$+$044842.2 \\
MASTER J212624 & 2017 & Shu, DPV, BSM, Trt, RPc, Ioh & MASTER OT J212624.16+253827.2 \\
OT J182142     & 2017 & DPV, Ioh & OT J182142.8$+$212154 \\
OT J204222    & 2017 & Ioh, LCO, Mas, RPc, Trt, Mdy & OT J204222.3$+$271211 \\
PNV J202053   & 2017 & deM, CRI, AAVSO, COO, SGE, Lic, Ioh, Trt, & PNV J20205397$+$2508145 \\
              &      & Van, OYE, Sol, DPV, Rui, RPc, Kis & \\
SDSS J152857  & 2017 & Van & SDSS J152857.86$+$034911.7 \\
SDSS 153015   & 2017b & KU, Trt, CRI & SDSS J153015.04$+$094946.3 \\
SDSS J204817  & 2017 & BSM, Ioh & SDSS J204817.85$-$061044.8 \\
TCP J003325   & 2017 & MLF, HaC & TCP J00332502$-$3518565 \\
TCP J201005   & 2017 & Van, Kai, deM, HaC, SRI, SGE, Trt, DKS, & TCP J20100517$+$1303006 \\
              &      & Ioh, Kai, BSM, DPV & \\
\hline
\end{tabular}
\end{center}
\end{table*}

\begin{table*}
\caption{Superhump Periods and Period Derivatives}\label{tab:perlist}
\begin{center}
\begin{tabular}{c@{\hspace{7pt}}c@{\hspace{7pt}}c@{\hspace{7pt}}c@{\hspace{7pt}}c@{\hspace{7pt}}c@{\hspace{7pt}}c@{\hspace{7pt}}c@{\hspace{7pt}}c@{\hspace{7pt}}c@{\hspace{7pt}}c@{\hspace{7pt}}c@{\hspace{7pt}}c@{\hspace{7pt}}c}
\hline
Object & Year & $P_1$ (d) & err & \multicolumn{2}{c}{$E_1$\commenta} & $P_{\rm dot}$\commentb & err\commentb & $P_2$ (d) & err & \multicolumn{2}{c}{$E_2$\commenta} & $P_{\rm orb}$ (d)\commentc & Q\commentd \\
\hline
V1047 Aql & 2017 & 0.073914 & 0.000098 & 0 & 19 & -- & -- & -- & -- & -- & -- & -- & C \\
V391 Cam & 2017 & -- & -- & -- & -- & -- & -- & 0.056728 & 0.000012 & 209 & 263 & 0.05620 & C \\
KP Cas & 2017 & -- & -- & -- & -- & -- & -- & 0.085143 & 0.000242 & 0 & 13 & -- & C \\
VW CrB & 2017 & 0.071985 & 0.000528 & 0 & 11 & -- & -- & -- & -- & -- & -- & -- & C \\
V632 Cyg & 2017 & 0.0655 & 0.0003 & 0 & 2 & -- & -- & -- & -- & -- & -- & -- & C \\
OV Dra & 2017 & 0.060398 & 0.000033 & 0 & 98 & 14.5 & 2.4 & 0.060032 & 0.000057 & 94 & 150 & 0.058736 & B \\
GQ CVn & 2017 & 0.089476 & 0.000091 & 0 & 37 & -- & -- & -- & -- & -- & -- & -- & C \\
BE Oct & 2017 & 0.077115 & 0.000132 & 0 & 40 & -- & -- & -- & -- & -- & -- & -- & C \\
V521 Peg & 2017 & 0.061646 & 0.000065 & 0 & 29 & -- & -- & -- & -- & -- & -- & -- & C \\
V368 Per & 2017 & 0.079224 & 0.000028 & 0 & 41 & -- & -- & 0.078602 & 0.000166 & 63 & 79 & -- & B \\
XY Psc & 2017 & 0.060675 & 0.000045 & 0 & 83 & 13.7 & 2.3 & 0.060230 & 0.000053 & 82 & 99 & -- & C \\
V701 Tau & 2017 & 0.069026 & 0.000037 & 0 & 31 & -- & -- & -- & -- & -- & -- & -- & C \\
V1208 Tau & 2017 & 0.0698 & 0.0040 & 0 & 3 & -- & -- & -- & -- & -- & -- & 0.0681 & C \\
TU Tri & 2017 & 0.076246 & 0.000080 & 0 & 20 & -- & -- & -- & -- & -- & -- & -- & C \\
SU UMa & 2017b & 0.078924 & 0.000123 & 0 & 64 & -- & -- & -- & -- & -- & -- & 0.07635 & C \\
HS Vir & 2017 & 0.080313 & 0.000063 & 0 & 103 & 3.7 & 4.9 & -- & -- & -- & -- & 0.0769 & CG \\
V406 Vir & 2017 & 0.056960 & 0.000016 & 0 & 88 & 8.1 & 1.5 & -- & -- & -- & -- & 0.05592 & B \\
1RXS J161659 & 2017 & 0.071028 & 0.000032 & 0 & 70 & $-$10.6 & 3.3 & -- & -- & -- & -- & -- & CG \\
ASASSN-13dh & 2017 & -- & -- & -- & -- & -- & -- & 0.091322 & 0.000056 & 38 & 100 & -- & B \\
\hline
  \multicolumn{14}{l}{\commenta Interval used for calculating the period.} \\
  \multicolumn{14}{l}{\commentb Unit $10^{-5}$.} \\
  \multicolumn{14}{l}{\parbox{440pt}{\commentc References: \\
  V391 Cam \citep{kap06j0532},
V1208 Tau \citep{pat05SH},
SU UMa \citep{tho86suuma},
HS Vir \citep{men99hsvir},
V406 Vir \citep{zha06j1238},
ASASSN-14kb \citep{wyr14asassn14kbatel6690},
PT And, OV Dra, ASASSN-17ei, ASASSN-17el,
ASASSN-17es, ASASSN-17fn, ASASSN-17fo,
ASASSN-17hw, ASASSN-17la, PNV J202053,
TCP J003325 (this work)
  }} \\
  \multicolumn{14}{l}{\parbox{440pt}{\commentd Data quality and comments. A: excellent, B: partial coverage or slightly low quality, C: insufficient coverage or observations with large scatter, G: $P_{\rm dot}$ denotes global $P_{\rm dot}$, M: observational gap in middle stage, U: uncertainty in alias selection, 2: late-stage coverage, the listed period may refer to $P_2$, a: early-stage coverage, the listed period may be contaminated by stage A superhumps, E: $P_{\rm orb}$ refers to the period of early superhumps.}} \\
\end{tabular}
\end{center}
\end{table*}

\addtocounter{table}{-1}
\begin{table*}
\caption{Superhump Periods and Period Derivatives (continued)}
\begin{center}
\begin{tabular}{c@{\hspace{7pt}}c@{\hspace{7pt}}c@{\hspace{7pt}}c@{\hspace{7pt}}c@{\hspace{7pt}}c@{\hspace{7pt}}c@{\hspace{7pt}}c@{\hspace{7pt}}c@{\hspace{7pt}}c@{\hspace{7pt}}c@{\hspace{7pt}}c@{\hspace{7pt}}c@{\hspace{7pt}}c}
\hline
Object & Year & $P_1$ & err & \multicolumn{2}{c}{$E_1$} & $P_{\rm dot}$ & err & $P_2$ & err & \multicolumn{2}{c}{$E_2$} & $P_{\rm orb}$ & Q \\
\hline
ASASSN-14ca & 2017 & 0.067036 & 0.000014 & 0 & 45 & $-$4.7 & 3.0 & -- & -- & -- & -- & -- & C \\
ASASSN-14cr & 2017 & -- & -- & -- & -- & -- & -- & 0.068698 & 0.000055 & 0 & 45 & -- & C \\
ASASSN-14kb & 2017 & 0.070420 & 0.000030 & 0 & 86 & 3.0 & 3.9 & -- & -- & -- & -- & 0.068106 & CG \\
ASASSN-14lk & 2017 & -- & -- & -- & -- & -- & -- & 0.061054 & 0.000094 & 0 & 34 & -- & C \\
ASASSN-15fu & 2017 & 0.074592 & 0.000071 & 0 & 28 & -- & -- & -- & -- & -- & -- & -- & CG \\
ASASSN-15fv & 2017 & 0.0682 & 0.0040 & 0 & 1 & -- & -- & -- & -- & -- & -- & -- & C \\
ASASSN-15qu & 2017 & 0.080449 & 0.000038 & 0 & 78 & $-$7.5 & 3.8 & -- & -- & -- & -- & -- & CG \\
ASASSN-17ei & 2017 & 0.057257 & 0.000011 & 34 & 247 & 3.4 & 0.4 & -- & -- & -- & -- & 0.05646 & BE \\
ASASSN-17el & 2017 & 0.055183 & 0.000013 & 48 & 213 & 5.1 & 0.3 & 0.054911 & 0.000184 & 230 & 271 & 0.05434 & BE \\
ASASSN-17eq & 2017 & 0.072197 & 0.000069 & 0 & 28 & -- & -- & -- & -- & -- & -- & -- & C \\
ASASSN-17es & 2017 & 0.057858 & 0.000023 & 33 & 105 & 0.6 & 4.4 & -- & -- & -- & -- & 0.05719 & BE \\
ASASSN-17et & 2017 & -- & -- & -- & -- & -- & -- & 0.095636 & 0.000060 & 0 & 63 & -- & C \\
ASASSN-17ew & 2017 & -- & -- & -- & -- & -- & -- & 0.078497 & 0.000027 & 0 & 65 & -- & C \\
ASASSN-17ex & 2017 & -- & -- & -- & -- & -- & -- & 0.068306 & 0.000096 & 0 & 31 & -- & C \\
ASASSN-17fh & 2017 & 0.064 & 0.001 & 0 & 1 & -- & -- & -- & -- & -- & -- & -- & C \\
ASASSN-17fi & 2017 & 0.058833 & 0.000011 & 0 & 52 & -- & -- & -- & -- & -- & -- & -- & C \\
ASASSN-17fj & 2017 & 0.066266 & 0.000021 & 0 & 77 & 8.4 & 2.3 & 0.065950 & 0.000044 & 75 & 135 & -- & B \\
ASASSN-17fl & 2017 & 0.062632 & 0.000123 & 0 & 18 & -- & -- & -- & -- & -- & -- & -- & C \\
ASASSN-17fn & 2017 & 0.061584 & 0.000014 & 37 & 169 & $-$2.8 & 1.4 & -- & -- & -- & -- & 0.06096 & BE \\
ASASSN-17fo & 2017 & 0.063240 & 0.000028 & 8 & 80 & 7.3 & 3.8 & -- & -- & -- & -- & 0.061548 & B \\
ASASSN-17fz & 2017 & 0.054404 & 0.000025 & 41 & 152 & 7.0 & 2.0 & -- & -- & -- & -- & -- & B \\
ASASSN-17gf & 2017 & 0.052551 & 0.000010 & 31 & 129 & 5.2 & 1.0 & -- & -- & -- & -- & -- & B \\
ASASSN-17gh & 2017 & 0.061394 & 0.000348 & 0 & 9 & -- & -- & -- & -- & -- & -- & -- & C \\
ASASSN-17gv & 2017 & 0.060897 & 0.000039 & 0 & 88 & -- & -- & -- & -- & -- & -- & -- & CG \\
ASASSN-17hm & 2017 & 0.088586 & 0.000073 & 0 & 37 & -- & -- & 0.088140 & 0.000059 & 34 & 59 & -- & C \\
ASASSN-17hw & 2017 & 0.059717 & 0.000013 & 29 & 218 & 0.3 & 0.9 & -- & -- & -- & -- & 0.05886 & BE \\
ASASSN-17hy & 2017 & 0.071475 & 0.000048 & 0 & 72 & 16.3 & 4.3 & -- & -- & -- & -- & -- & C \\
ASASSN-17id & 2017 & 0.078613 & 0.000074 & 0 & 39 & -- & -- & -- & -- & -- & -- & -- & C2 \\
ASASSN-17if & 2017 & 0.058827 & 0.000031 & 0 & 154 & 8.2 & 0.8 & 0.058568 & 0.000041 & 153 & 223 & -- & B \\
\hline
\end{tabular}
\end{center}
\end{table*}

\addtocounter{table}{-1}
\begin{table*}
\caption{Superhump Periods and Period Derivatives (continued)}
\begin{center}
\begin{tabular}{c@{\hspace{7pt}}c@{\hspace{7pt}}c@{\hspace{7pt}}c@{\hspace{7pt}}c@{\hspace{7pt}}c@{\hspace{7pt}}c@{\hspace{7pt}}c@{\hspace{7pt}}c@{\hspace{7pt}}c@{\hspace{7pt}}c@{\hspace{7pt}}c@{\hspace{7pt}}c@{\hspace{7pt}}c}
\hline
Object & Year & $P_1$ & err & \multicolumn{2}{c}{$E_1$} & $P_{\rm dot}$ & err & $P_2$ & err & \multicolumn{2}{c}{$E_2$} & $P_{\rm orb}$ & Q \\
\hline
ASASSN-17ig & 2017 & 0.094947 & 0.000084 & 0 & 25 & -- & -- & 0.094393 & 0.000024 & 25 & 96 & -- & C \\
ASASSN-17iv & 2017 & -- & -- & -- & -- & -- & -- & 0.070237 & 0.000044 & 15 & 87 & -- & C \\
ASASSN-17iw & 2017 & 0.055906 & 0.000047 & 0 & 90 & 11.3 & 5.9 & -- & -- & -- & -- & -- & C \\
ASASSN-17ix & 2017 & 0.062449 & 0.000048 & 0 & 82 & 18.2 & 4.0 & -- & -- & -- & -- & -- & C \\
ASASSN-17ji & 2017 & 0.0589 & 0.0001 & 0 & 18 & -- & -- & -- & -- & -- & -- & -- & C \\
ASASSN-17jr & 2017 & 0.061706 & 0.000038 & 0 & 98 & 8.0 & 3.0 & -- & -- & -- & -- & -- & C \\
ASASSN-17kc & 2017 & 0.063764 & 0.000028 & 0 & 81 & 12.8 & 1.4 & 0.063320 & 0.000024 & 80 & 160 & -- & B \\
ASASSN-17kd & 2017 & 0.060919 & 0.000016 & 33 & 213 & 2.7 & 0.8 & -- & -- & -- & -- & -- & B \\
ASASSN-17kg & 2017 & 0.057620 & 0.000017 & 36 & 228 & 5.4 & 0.5 & 0.057427 & 0.000025 & 242 & 297 & -- & A \\
ASASSN-17kp & 2017 & 0.057957 & 0.000030 & 0 & 51 & 9.3 & 5.7 & -- & -- & -- & -- & -- & C \\
ASASSN-17la & 2017 & 0.061571 & 0.000021 & 27 & 175 & 7.9 & 0.5 & -- & -- & -- & -- & 0.06039 & BE \\
ASASSN-17lr & 2017 & 0.058635 & 0.000057 & 0 & 102 & $-$8.8 & 3.2 & -- & -- & -- & -- & -- & CG \\
ASASSN-17me & 2017 & 0.0614 & 0.0004 & 0 & 1 & -- & -- & -- & -- & -- & -- & -- & C \\
ASASSN-17np & 2017 & 0.089227 & 0.000047 & 0 & 26 & -- & -- & 0.088730 & 0.000032 & 25 & 82 & -- & C \\
ASASSN-17nr & 2017 & 0.056376 & 0.000027 & 0 & 107 & 5.8 & 1.6 & -- & -- & -- & -- & -- & CU \\
ASASSN-17of & 2017 & 0.064175 & 0.000067 & 0 & 74 & -- & -- & 0.063567 & 0.000030 & 74 & 109 & -- & C \\
ASASSN-17oo & 2017 & 0.06781 & 0.00005 & -- & -- & -- & -- & -- & -- & -- & -- & -- & C2 \\
ASASSN-17ou & 2017 & 0.057128 & 0.000045 & 0 & 70 & -- & -- & -- & -- & -- & -- & -- & C \\
ASASSN-17pb & 2017 & 0.076092 & 0.000049 & 47 & 101 & $-$1.0 & 8.6 & -- & -- & -- & -- & -- & C \\
CRTS J044027 & 2017 & -- & -- & -- & -- & -- & -- & 0.064361 & 0.000034 & 49 & 97 & -- & C \\
CRTS J080941 & 2017 & 0.100467 & 0.000122 & 20 & 62 & -- & -- & -- & -- & -- & -- & -- & B \\
CRTS J214934 & 2017 & 0.071482 & 0.000005 & 0 & 65 & -- & -- & 0.071222 & 0.000041 & 64 & 107 & -- & C \\
CRTS J223235 & 2017 & 0.062994 & 0.000136 & 0 & 32 & -- & -- & -- & -- & -- & -- & -- & C \\
CTCV J1940 & 2017 & 0.076668 & 0.000027 & 0 & 79 & $-$3.7 & 3.2 & -- & -- & -- & -- & -- & CU \\
DDE 51 & 2017 & 0.100277 & 0.000020 & 49 & 108 & $-$0.5 & 2.1 & -- & -- & -- & -- & -- & B \\
MASTER J174305 & 2017 & 0.069949 & 0.000079 & 0 & 14 & -- & -- & 0.069425 & 0.000074 & 27 & 44 & -- & C \\
MASTER J192757 & 2017 & 0.08161 & 0.00005 & 0 & 12 & -- & -- & -- & -- & -- & -- & -- & C \\
MASTER J200904 & 2017 & 0.073646 & 0.000115 & 0 & 20 & -- & -- & -- & -- & -- & -- & -- & C \\
MASTER J205110 & 2017 & 0.080710 & 0.000044 & 0 & 59 & 7.6 & 4.7 & -- & -- & -- & -- & -- & C \\
MASTER J212624 & 2017 & 0.090888 & 0.000074 & 43 & 75 & -- & -- & -- & -- & -- & -- & -- & B \\
NSV 35 & 2017 & 0.081034 & 0.000039 & 0 & 112 & $-$1.1 & 2.4 & -- & -- & -- & -- & -- & BG \\
OT J182142 & 2017 & 0.082140 & 0.000095 & 0 & 40 & -- & -- & -- & -- & -- & -- & -- & C2 \\
OT J204222 & 2017 & 0.056152 & 0.000045 & 65 & 167 & $-$1.1 & 6.6 & -- & -- & -- & -- & -- & C \\
PNV J202053 & 2017 & 0.057392 & 0.000010 & 53 & 250 & 4.3 & 0.4 & 0.056443 & 0.000153 & 246 & 263 & 0.056509 & AE \\
SDSS J152857 & 2017 & 0.06319 & 0.00024 & 0 & 3 & -- & -- & -- & -- & -- & -- & -- & C \\
SDSS J153015 & 2017b & 0.075310 & 0.000134 & 0 & 32 & -- & -- & -- & -- & -- & -- & -- & C \\
TCP J003325 & 2017 & 0.055222 & 0.000019 & 91 & 256 & 4.6 & 0.3 & -- & -- & -- & -- & 0.05485 & BE \\
TCP J201005 & 2017 & 0.081030 & 0.000046 & 0 & 44 & -- & -- & -- & -- & -- & -- & -- & B2 \\
\hline
\end{tabular}
\end{center}
\end{table*}

\section{Major findings in objects in this paper}\label{sec:findings}

   We list major findings in this paper.

\begin{itemize}

\item Suspected WZ Sge-type dwarf novae XY Psc and
V406 Vir underwent long-awaited superoutbursts,
but neither of them showed WZ Sge-type characteristics.

\item ASASSN-17fo is a deeply eclipsing SU UMa-type
dwarf nova.

\item ASASSN-17gf is an EI Psc-type object below
the period minimum.

\item ASASSN-17kg showed a dip before the termination
of the superoutburst.

\item ASASSN-17la is a WZ Sge-type dwarf nova with
an intermediate mass ratio [0.084(5)] and
a medium long orbital period [0.06039(3)~d].

\item CRTS J080941 and DDE 51 are in the period gap
and had a long-lasting stage A.

\item MASTER J212624 is a long-period system with
a long-lasting stage A.

\item WZ Sge-type stars identified in this study
are PT And, ASASSN-17ei, ASASSN-17el, ASASSN-17es,
ASASSN-17fn, ASASSN-17fz, ASASSN-17hw, ASASSN-17kd,
ASASSN-17la, PNV J202053 and TCP J003325.

\item New mass ratios from stage A superhumps
(using \cite{kat13qfromstageA})
are: ASASSN-17ei 0.074(3),
ASASSN-17el 0.071(3),
ASASSN-17es 0.095(9),
ASASSN-17fn 0.097(1),
ASASSN-17hw 0.078(1),
CRTS J122221 0.032(2),
PNV J202053 0.090(3)

\end{itemize}

\section{Summary of recent progress in understanding of
SU UMa-type dwarf novae}\label{sec:progress}

   In this section, we provide brief descriptions of
recent progress in understanding of SU UMa-type dwarf novae
based on this series of papers and other published papers
upon the request from the reviewer.

\subsection{SU UMa-type dwarf novae and superhump stages}

   For SU UMa-type dwarf novae in general, we have verified
that the relation between the period derivative
($P_{\rm dot}$) for stage B
versus the oribital period ($P_{\rm orb}$)
that we found in \citet{Pdot} essentially
applies to most of ordinary superoutbursts.  The refined
relation was shown in \citet{Pdot8} and \citet{Pdot9}
[we consider that \citet{Pdot9} to be the final regular
summary of the statistics].  The stage A, B and C are
now well-established and used in many publications
by various authors:
\citet{kat14wzsgestarsproc}, \citet{bak14czev404},
\citet{skl16asassn14cv}, \citet{neu17j1222},
\citet{bak17mndra}, \citet{neu18j1222gwlib},
\citet{skl18nyser}, \citet{lit18j0359},
\citet{pal18qzlib}, \citet{pal19j1238},
\citet{pav19asassn18fk}, \citet{mca19DNeclipse},
\citet{cou19zcha} (this work also illustrates the difficulty
in determining superhump times in a deeply eclipsing system).
The rapid growth of papers referring to our superhump
stages indicates that this concept and application are
now widely accepted in this field.

\subsection{SU UMa-type/WZ Sge-type relation and period bouncers}

   The SU UMa-type/WZ Sge-type relation and the nature
of period bouncers would be one of the most intriguing
subjects for many readers.  We have already give
a conclusion to this subject as a review \citep{kat15wzsge}.
The distinction between SU UMa-type and WZ Sge-type
dwarf novae is the manifestaion of the 2:1 resonance
in the latter, and this classification is now widely
accepted (such as in AAVSO VSX\footnote{
  $<$https://www.aavso.org/vsx/$>$.
}).  After the release of
\citet{kat15wzsge}, there have been increasing number
of WZ Sge-type dwarf novae mainly thanks to the ASAS-SN
survey.  The major advance since then has been
the increase of examples of type-E outbursts.
The objects with type-E outbursts have an initial
superoutburst corresponding to the 2:1 resonance
(high-inclination systems show early superhumps) and
the second superoutburst showing the development
of ordinary superhumps.  They are considered to be
the best candidates for the still elusive population
of period bouncers.  The papers dealing with
type-E outbursts are
\citet{kim16asassn15jd} (ASASSN-15jd),
\citet{kim18asassn16dtasassn16hg} (ASASSN-16dt
and ASASSN-16hg),
\citet{iso19nsv1440} (NSV 1440, AM CVn star).
Among them ASASSN-15jd and ASASSN-16hg showed
a transitional feature between single superoutburst
and the type-E outburst.  These observations suggest
that type-E outbursts can be understood as a smooth
extension of WZ Sge-type dwarf novae toward a lower
mass ratio (i.e. period bouncers).  The examples
are still increasing and the results are pending
publication.

\subsection{Systems near stability border of 3:1 resonance}

   The major recent advance in SU UMa-type dwarf novae
is around the stability borderline of the 3:1 resonance.
When \citet{Pdot} was published, it was a mystery
why some long-$P_{\rm orb}$ systems
show a strong decrease of the superhump
periods [cf. MN Dra and UV Gem, see subsection 4.10
in \citet{Pdot}].  An idea to solve this issue required
five years to appear and \citet{Pdot6} gave a working hypothesis
that the 3:1 resonance grows slowly in systems near
the stability border of the 3:1 resonance.
This idea has been reinforced by subsequent observations
\citep{kat16v1006cyg}.  \citet{Pdot8} and \citet{Pdot9}
increased the number of candidate systems showing
this feature.  Some of these objects are known to show
post-superoutburst rebrightenings, which had been usually
considered to be a feature unique to WZ Sge-type
dwarf novae (cf. \cite{kat15wzsge}).  With the increasing
number of long-$P_{\rm orb}$ object showing rebrightenings
[V1006 Cyg, \citet{kat16v1006cyg}; ASASSN-14ho
\citet{kat19asassn14ho}], it is now considered
that the weak 3:1 resonance could cause the decoupling
of the tidal and thermal instabilities, leading to
premature quenching of the superoutburst.
This idea was originally proposed for extremely low
mass-ratio systems such as WZ Sge-type dwarf novae
\citep{hel01eruma}.  Recent findings suggest that
the same mechanism could work in systems near
the stability border of the 3:1 resonance and that
such systems can mimic WZ Sge-type outbursts.
A long precursor followed by a dip and an ordinary
superoutburst in CS Ind \citep{kat19csind} also
strengthens this interpretation.  Theoretical supports
are still lacking and a futher advance would be expected
in this regime.

\subsection{SU UMa-type dwarf nova showing standstills}

   Currently there is only one known SU UMa-type dwarf nova
(NY Ser) which showed standstills in 2018 \citep{kat19nyser}.
This is a single known bona-fide a hybrid SU UMa + Z Cam-type
dwarf nova.  It was shown that superoutbursts arose from
standstills in NY Ser, and the disk should grow in radius
to reach the 3:1 resonance during standstills.

\section{List of references}\label{sec:list}

The references cited in SI are:
\citet{alk00ptand}, \citet{ant96vwcrb}, \citet{ant02var73dra}, \citet{aug10CTCVCV2}, \citet{avi10j1238}, \citet{bal13j1616}, \citet{bal14j1927atel6024}, \citet{bal12j1743atel4022}, \citet{bal17j1325atel10470}, \citet{bal14j2009atel5974}, \citet{boy10kpcas}, \citet{can25dhaql}, \citet{car17asassn17fpatel10334}, \citet{LOWESS}, \citet{dav14asassn14caatel6211}, \citet{den17j1325atel10480}, \citet{den13j2126atel5111}, \citet{dil08SDSSCV}, \citet{dra14CRTSCVs}, \citet{era73v701tau}, \citet{fer89error}, \citet{gre82PGsurveyCV}, \citet{PGsurvey}, \citet{gru58ptand}, \citet{har95v503cyg}, \citet{hen01xypsc}, \citet{hof49newvar}, \citet{hof49newvar1}, \citet{hof57VSchart}, \citet{hof57MVS245}, \citet{hof63VSS61}, \citet{hof64an28849}, \citet{hof67an29043}, \citet{ima17qzvir}, \citet{kat02v503cyg}, \citet{kat15wzsge}, \citet{kat16j0333}, \citet{Pdot6}, \citet{Pdot7}, \citet{Pdot4}, \citet{Pdot5}, \citet{Pdot8}, \citet{Pdot}, \citet{Pdot9}, \citet{Pdot3}, \citet{kat12DNSDSS}, \citet{Pdot2}, \citet{kat13j1222}, \citet{kat98hsvir}, \citet{kat95hsvir}, \citet{kat16v1006cyg}, \citet{kat01hvvir}, \citet{kat01hsvir}, \citet{kat17j0026}, \citet{khr05nsv1485}, \citet{kim16alcom}, \citet{kin00VSID4896}, \citet{lit13sbs1108}, \citet{liu00CVspec3}, \citet{liu99CVspec1}, \citet{luy38propermotion2}, \citet{mar17asassn17fpatel10354}, \citet{mas03faintCV}, \citet{men99hsvir}, \citet{mot96CVROSAT}, \citet{mro15OGLEDNe}, \citet{nak13j2112j2037}, \citet{nam17asassn15po}, \citet{neu17j1222}, \citet{nog95dhaql}, \citet{nog03var73dra}, \citet{nov97vwcrb}, \citet{osa13v1504cygKepler}, \citet{osa13v344lyrv1504cyg}, \citet{ohn19ovboo}, \citet{ohs12eruma}, \citet{osm85hsvir}, \citet{pat05SH}, \citet{pat03suumas}, \citet{pat08j1507}, \citet{pav12v503cyg}, \citet{ASAS3}, \citet{pri14asassn14kbatel6688}, \citet{ric69v1233aql}, \citet{rin93thesis}, \citet{rod05hs2219}, \citet{rom78tutri}, \citet{ros72xypsciauc1}, \citet{ros72xypsciauc2}, \citet{sha91tutri}, \citet{sha89ptand}, \citet{sha92tutri}, \citet{she07v701tau}, \citet{she08j1227}, \citet{she07CVspec}, \citet{shu17j2051atel10790}, \citet{sta13asassn13aoatel5118}, \citet{PDM}, \citet{szk09SDSSCV7}, \citet{szk03SDSSCV2}, \citet{szk06SDSSCV5}, \citet{tho02j2329}, \citet{uem02j2329letter}, \citet{waa17asassn17fpaan580}, \citet{wak17asassn16eg}, \citet{wen89v632cygv630cyg}, \citet{wil10j1924}, \citet{woo11v344lyr}, \citet{wou10CVperiod}, \citet{wou12SDSSCRTSCVs}, \citet{wyr14asassn14kbatel6690}, \citet{zem13eruma}, \citet{zha06j1238}, \citet{zhe10ptandcbet2574}, \citet{zlo04tutri}.

\section*{Acknowledgements}
This work was also partially supported by
Grant VEGA 2/0008/17 (by Shugarov, Chochol) and APVV-15-0458 
(by Shugarov, Chochol, Dubovsky, Kudzej, Medulka), 
RSF-14-12-00146 (Golysheva for processing observation data 
from Slovak Observatory).
ASAS-SN is supported by the Gordon and Betty Moore
Foundation through grant GBMF5490 to the Ohio State
University and NSF grant AST-1515927.
The authors are grateful to observers of VSNET Collaboration and
VSOLJ observers who supplied vital data.
We acknowledge with thanks the variable star
observations from the AAVSO International Database contributed by
observers worldwide and used in this research.
We are also grateful to the VSOLJ database.
This work is helped by outburst detections and announcement
by a number of variable star observers worldwide,
including participants of CVNET and BAA VSS alert.
The CCD operation of the Bronberg Observatory is partly sponsored by
the Center for Backyard Astrophysics.
We are grateful to the Catalina Real-time Transient Survey
team for making their real-time detection of transient objects
and the past photometric database available to the public.
We are also grateful to the ASAS-3 team for making
the past photometric database available to the public.
This research has made use of the SIMBAD database,
operated at CDS, Strasbourg, France.
This research has made use of the International Variable Star Index 
(VSX) database, operated at AAVSO, Cambridge, Massachusetts, USA.

\section*{Supporting information}

For reader's convenience, supporting information
(sections, figures and tables starting with E-)
is combined in this arXiv version.

In the final form of PASJ publication, supporting
information is separated in the online version.

\def\thesection{E-section \arabic{section}}
\renewcommand{\tablename}{E-table}
\renewcommand{\figurename}{E-figure}
\makeatletter
    \renewcommand{\theequation}{%
    E\arabic{equation}}
    \@addtoreset{equation}{section}
\makeatother

\setcounter{section}{0}
\setcounter{table}{0}
\setcounter{figure}{0}

\section{Superhump Stages}

\begin{figure}
  \begin{center}
    \FigureFile(80mm,110mm){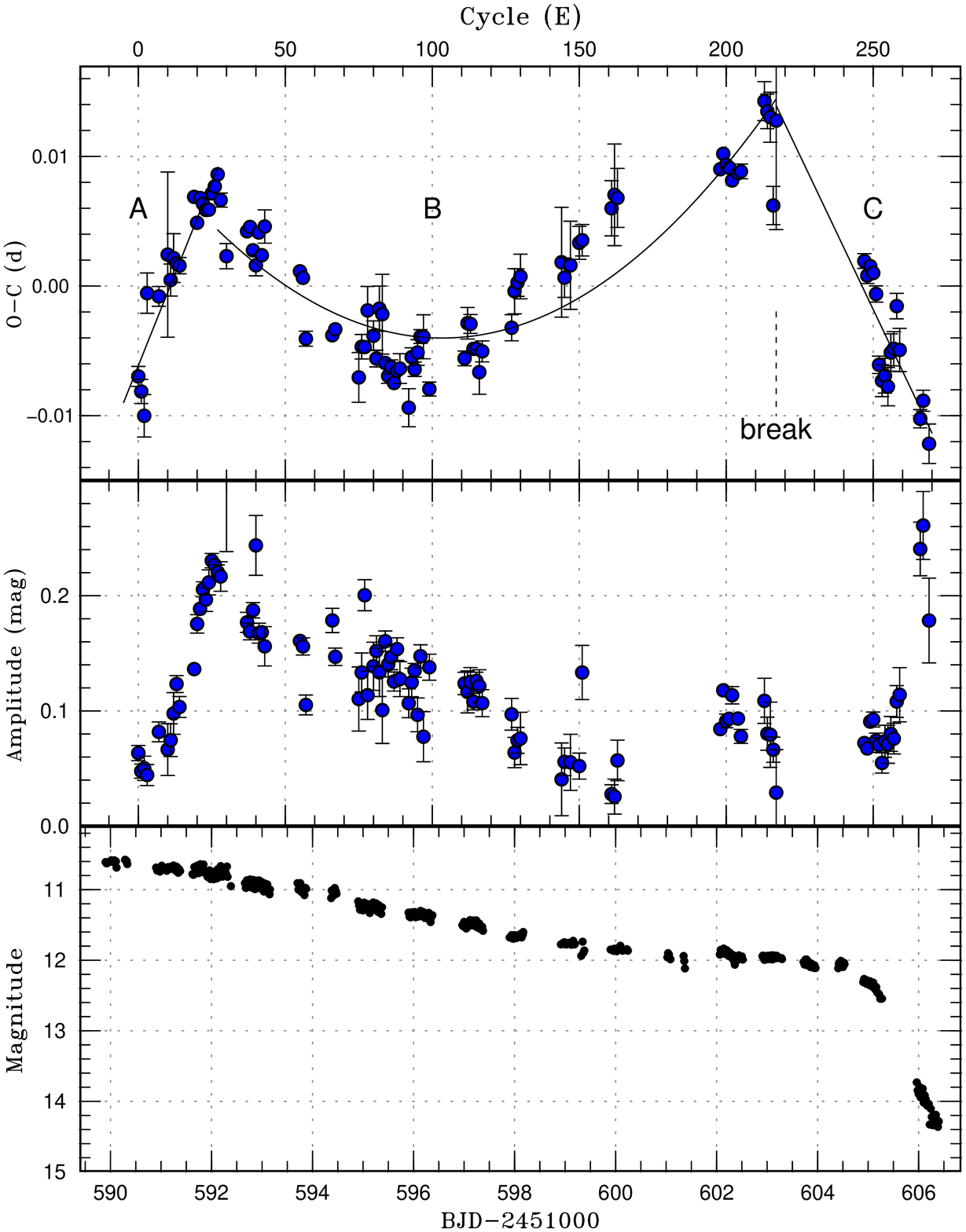}
  \end{center}
  \caption{Representative $O-C$ diagram showing three stages (A--C)
  of $O-C$ variation.  The data were taken from the 2000 superoutburst
  of SW UMa.  (Upper:) $O-C$ diagram.  Three distinct stages
  (A -- evolutionary stage with a longer superhump period, 
  B -- middle stage, and C -- stage after
  transition to a shorter period) and the location of the period break
  between stages B and C are shown. (Middle): Amplitude of superhumps.
  During stage A, the amplitude of the superhumps grew.
  (Lower:) Light curve.
  (Reproduction of figure 1 in \cite{kat13qfromstageA})}
  \label{fig:stagerev}
\end{figure}

It has become evident since \citet{Pdot} that
the superhump periods systematically vary in a way
common to many objects.  \citet{Pdot} introduced
superhump stages (stages A, B and C):
initial growing stage with a long period (stage A) and
fully developed stage with a systematically
varying period (stage B) and later stage C with a shorter,
almost constant period (see figure \ref{fig:stagerev}).
[This part is an excerpt from \citet{Pdot9}].

\section{Data Analysis}

   This part includes an excerpt from \citet{Pdot9}.

   The data analysis was performed in the same way described
in \citet{Pdot} and \citet{Pdot6} and we mainly used
R software\footnote{
   The R Foundation for Statistical Computing:\\
   $<$http://cran.r-project.org/$>$.
} for data analysis.

   In de-trending the data, we mainly used locally-weighted
polynomial regression (LOWESS: \cite{LOWESS})
and sometimes lower (1--3rd order) polynomial fitting
when the observation baseline was short.
The times of superhumps maxima were determined by
the template fitting method as described in \citet{Pdot}.
The times of all observations are expressed in 
barycentric Julian days (BJD).

   We used phase dispersion minimization (PDM; \cite{PDM})
for period analysis and 1$\sigma$ errors for the PDM analysis
was estimated by the methods of \citet{fer89error} and \citet{Pdot2}.
We have used a variety of bootstrapping in
estimating the robustness of the result of the PDM analysis
since \citet{Pdot3}.

   We used \citet{kat13qfromstageA} to determine the
mass ratio ($q$) from the superhump period ($P_{\rm SH}$)
and the orbital period ($P_{\rm orb}$,
when early superhumps were observed, we assumed
them to have the same period as the orbital one).
The fractional superhump excess (in frequency)
$\epsilon^* \equiv 1-P_{\rm orb}/P_{\rm SH}$
is equal to the dynamical precession rate
when the pressure effect can be neglected
as in stage A superhumps \citep{kat13qfromstageA}.

\section{Individual Objects}\label{sec:individual}

\subsection{PT Andromedae}\label{obj:ptand}

   PT And was originally discovered as a nova in M31
(R15 = M31N 1957-10b in \cite{gru58ptand}).  \citet{sha89ptand}
reported a short outburst in 1983 and a long one
in 1986.  \citet{sha89ptand} suggested this object
to be an SU UMa-type dwarf nova as judged from
the outburst behavior.  \citet{alk00ptand} studied
the 1998 outburst and past ones and suggested that
this object is more likely a recurrent nova in M31
based on the lack of plateau phase in SU UMa-type
superoutbursts.

   There was a long outburst in 2010 December
(originally reported as M31N 2010-12a on 2010 December 1 by
K. Nishiyama and F. Kabashima, cf. \cite{zhe10ptandcbet2574}).
During this outburst, likely superhumps were detected
(vsnet-alert 12484, 12497, 12527).\footnote{
   These vsnet-alert messages can be seen at
   $<$http://ooruri.kusastro.kyoto-u.ac.jp/pipermail/vsnet-alert/$>$.
}
Due to the short observational runs and interference
by the moonlight,
it was difficult to determine the superhump period.
A spectrum by A. Arai showed no prominent lines,
confirming the dwarf nova-type nature of this object
(vsnet-alert 12528).
The outburst showed a plateau phase followed by
rapid fading on 2010 December 24--25 (vsnet-alert 12530).
This feature also supported the SU UMa-type interpretation
contrary to what was stated in \citet{alk00ptand}.

\begin{figure}
  \begin{center}
    \FigureFile(85mm,110mm){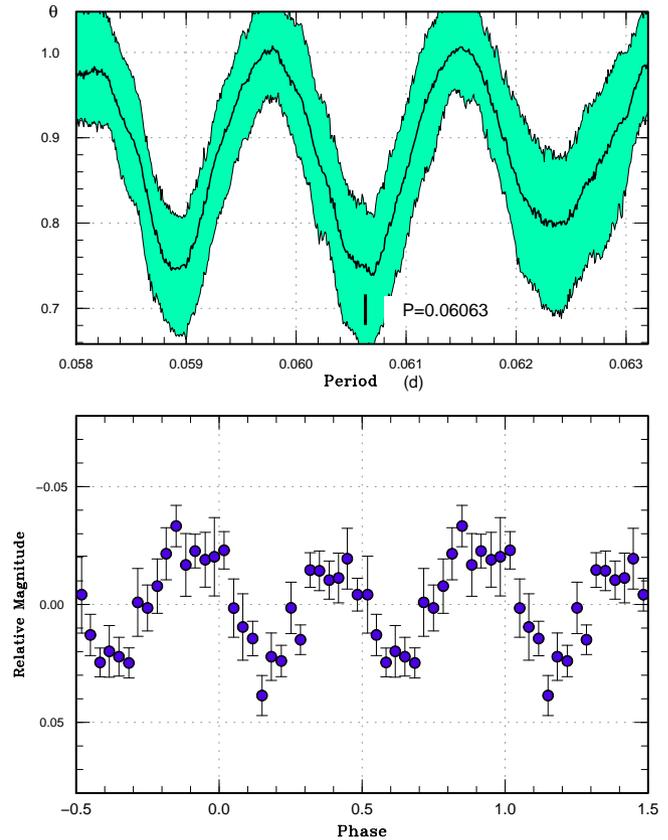}
  \end{center}
  \caption{Early superhumps in PT And (2017), drawn with
     a period of 0.06063~d.
     (Upper): PDM analysis.
     (Lower): Phase-averaged profile.}
  \label{fig:ptandeshpdm}
\end{figure}

\begin{figure}
  \begin{center}
    \FigureFile(85mm,110mm){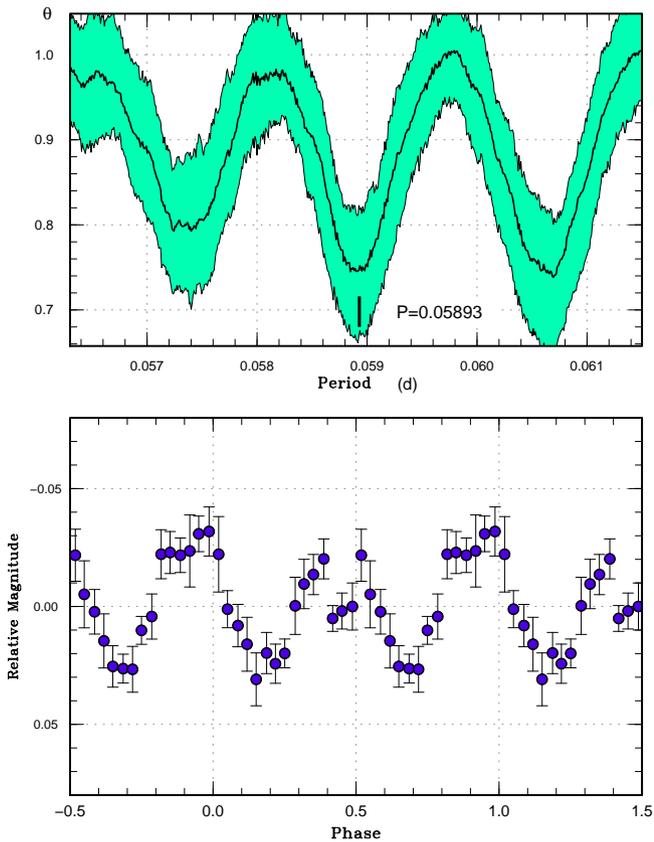}
  \end{center}
  \caption{Early superhumps in PT And (2017), drawn with
     a period of 0.05893~d.
     (Upper): PDM analysis.
     (Lower): Phase-averaged profile.}
  \label{fig:ptandeshpdm2}
\end{figure}

   The 2017 outburst was detected on 2017 August 15
at an unfiltered CCD magnitude of 16.02 by E. Muyllaert
(cvnet-outburst 7623).  Time-resolved CCD photometry
recorded double-wave modulations (vsnet-alert 21356).
They were most likely early superhumps.
Due to the lack of observations, we could not select
the alias.  The two most likely periods were
0.06063(7)~d or 0.05893(7)~d.  These periods are
equally acceptable and phase-averaged profiles
are in e-figures \ref{fig:ptandeshpdm} and \ref{fig:ptandeshpdm2}.
Although there were observations after these two
nights, the data quality was not sufficient and
we could not determine the superhump period.
The object is thus a WZ Sge-type dwarf nova.
The rapidly fading light curves resembling those of
fast novae were probably caused by viscous decay
at the start of the superoutburst (cf. \cite{kat15wzsge}).

\subsection{DH Aquilae}\label{obj:dhaql}

   DH Aql was discovered as a Mira-type variable
(=HV 3899) with a range of 12.5 to fainter than 16
in photographic range \citep{can25dhaql}.  The SU UMa-type
nature of this object was clarified by \citet{nog95dhaql}.
Refer to \citet{Pdot6} for more history.

   The 2017 superoutburst was detected by M. Noriyama
at an unfiltered CCD magnitude of 14.5 while the object
was still rising.  The object was observed to be
at peak magnitude of 12.4 on the next night.
Two superhump maxima were obtained from single-night
observations: BJD 2458036.9216(4) ($N$=106) and
2458037.0013(2) ($N$=152).

\subsection{V1047 Aquilae}\label{obj:v1047aql}

   V1047 Aql was discovered as a dwarf nova (S~8191)
by \citet{hof64an28849}.  The object was identified
to be an SU UMa-type dwarf nova in 2005 by Greg Bolt.
Observations of the 2016 superoutburst were reported
in \citet{Pdot9}, which obtained only two superhump
maxima.

   The 2017 superoutburst was visually detected by 
R. Stubbings at a magnitude of 15.2 on 2017 June 26.
The ASAS-SN data recorded it at $V$=16.2 on 2017 June 20, 
which further brightened to $V$=15.5 on 2017 June 24.
The times of superhump maxima are listed in
e-table \ref{tab:v1047aqloc2017}.  The superhump
stage is unknown.

   As reported in \citet{Pdot9}, this object shows
regular superoutbursts with a short supercycle.
We extracted superoutbursts in the ASAS-SN data
(e-table \ref{tab:v1047aqlout}).  These superoutbursts
can be well expressed by a supercycle of 89.1(3)~d
with maximum $|O-C|$ of 4~d.  There were also
apparently frequent normal outbursts expected for
this short supercycle.

\begin{table}
\caption{Superhump maxima of V1047 Aql (2017)}\label{tab:v1047aqloc2017}
\begin{center}
\begin{tabular}{rp{55pt}p{40pt}r@{.}lr}
\hline
\multicolumn{1}{c}{$E$} & \multicolumn{1}{c}{max\commenta} & \multicolumn{1}{c}{error} & \multicolumn{2}{c}{$O-C$\commentb} & \multicolumn{1}{c}{$N$\commentc} \\
\hline
0 & 57931.4690 & 0.0026 & $-$0&0030 & 29 \\
1 & 57931.5464 & 0.0006 & 0&0006 & 52 \\
4 & 57931.7697 & 0.0005 & 0&0021 & 90 \\
5 & 57931.8408 & 0.0005 & $-$0&0007 & 95 \\
6 & 57931.9159 & 0.0008 & 0&0004 & 48 \\
13 & 57932.4346 & 0.0005 & 0&0018 & 67 \\
14 & 57932.5088 & 0.0005 & 0&0021 & 72 \\
18 & 57932.8002 & 0.0011 & $-$0&0022 & 85 \\
19 & 57932.8752 & 0.0009 & $-$0&0011 & 92 \\
\hline
  \multicolumn{6}{l}{\commenta BJD$-$2400000.} \\
  \multicolumn{6}{l}{\commentb Against max $= 2457931.4719 + 0.073914 E$.} \\
  \multicolumn{6}{l}{\commentc Number of points used to determine the maximum.} \\
\end{tabular}
\end{center}
\end{table}

\begin{table}
\caption{List of superoutbursts of V1047 Aql in the ASAS-SN data}\label{tab:v1047aqlout}
\begin{center}
\begin{tabular}{ccccc}
\hline
Year & Month & Day & max\commenta & $V$ mag \\
\hline
2015 &  7 & 19 & 57222 & 15.0 \\
2015 & 10 & 13 & 57308 & 15.1 \\
2016 &  4 & 10 & 57489 & 15.2 \\
2016 &  7 &  8 & 57578 & 15.0 \\
2016 & 10 &  9 & 57671 & 15.2 \\
2017 &  4 &  3 & 57847 & 15.1 \\
2017 &  6 & 27 & 57932 & 15.1 \\
2017 &  9 & 26 & 58022 & 15.1 \\
\hline
  \multicolumn{5}{l}{\commenta JD$-$2400000.} \\
\end{tabular}
\end{center}
\end{table}

\subsection{NN Camelopardalis}\label{obj:nncam}

   NN Cam = NSV 1485 was identified as a dwarf nova
by \citet{khr05nsv1485}.  For more history, see \citet{Pdot7}.
The 2017 superoutburst was detected by the ASAS-SN
team at $V$=13.14 on 2017 August 30.  Only single
superhump maximum was recorded at BJD 2458004.4408(9) ($N$=76).

\subsection{V391 Camelopardalis}\label{obj:v391cam}

   The material is the same as in \citet{Pdot9}.
A re-analysis of the data yielded a positive detection
of post-superoutburst superhumps whose period
is consistent with the past observations of
stage C superhumps (e-table \ref{tab:v391camoc2017}).
Note that we used a narrower range than in \citet{Pdot9}
to determine the maximum on BJD 2457829, resulting
a slightly different value.

\begin{figure}
  \begin{center}
    \FigureFile(85mm,70mm){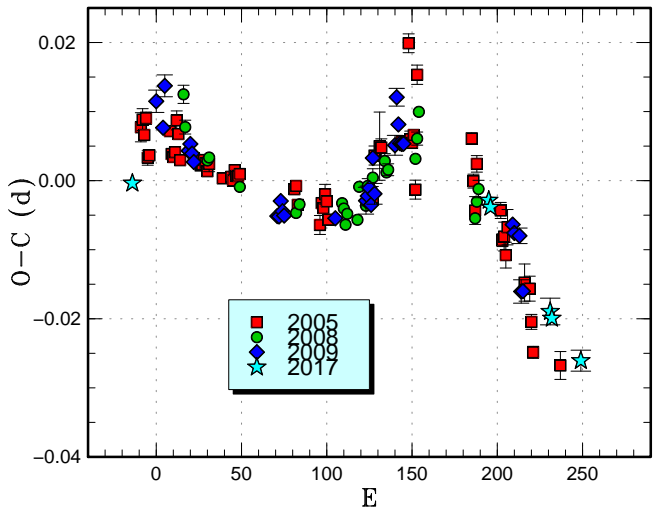}
  \end{center}
  \caption{Comparison of $O-C$ diagrams of V391 Cam between different
  superoutbursts.  A period of 0.05716~d was used to draw this figure.
  Approximate cycle counts ($E$) after the start of the superoutburst
  were used.  For the 2005 and 2008 superoutbursts, we used
  the zero points as defined in \citet{Pdot}.  The 2005 superoutburst
  had a separate precursor and the zero point was shifted by
  9 cycles.  For the 2017 superoutburst, we had to shift by
  40 cycles if we assume the the superoutburst started at the time
  of the ASAS-SN detection.  It was likely that this detection
  referred to a precursor outburst, which is consistent with
  the declining trend on 2017 March 16--17 in AAVSO visual
  observations.}
  \label{fig:v391camcomp2}
\end{figure}

   A comparison of $O-C$ diagrams of V391 Cam between different
superoutbursts (e-figure \ref{fig:v391camcomp2}) suggests
that there was a separate precursor outburst in the 2017
superoutburst and the initial superhump detection
referred to a stage A superhump, when the object
was still rising toward the full maximum of the superoutburst.

\begin{table}
\caption{Superhump maxima of V391 Cam (2017)}\label{tab:v391camoc2017}
\begin{center}
\begin{tabular}{rp{55pt}p{40pt}r@{.}lr}
\hline
\multicolumn{1}{c}{$E$} & \multicolumn{1}{c}{max\commenta} & \multicolumn{1}{c}{error} & \multicolumn{2}{c}{$O-C$\commentb} & \multicolumn{1}{c}{$N$\commentc} \\
\hline
0 & 57829.3181 & 0.0002 & $-$0&0035 & 185 \\
209 & 57841.2621 & 0.0005 & 0&0102 & 41 \\
210 & 57841.3182 & 0.0007 & 0&0093 & 59 \\
245 & 57843.3037 & 0.0019 & $-$0&0031 & 58 \\
246 & 57843.3599 & 0.0011 & $-$0&0040 & 58 \\
263 & 57844.3255 & 0.0015 & $-$0&0088 & 59 \\
\hline
  \multicolumn{6}{l}{\commenta BJD$-$2400000.} \\
  \multicolumn{6}{l}{\commentb Against max $= 2457829.3216 + 0.057083 E$.} \\
  \multicolumn{6}{l}{\commentc Number of points used to determine the maximum.} \\
\end{tabular}
\end{center}
\end{table}

\subsection{KP Cassiopeiae}\label{obj:kpcas}

   This object (S 3865) was discovered by \citet{hof49newvar}.
A finding chart was provided in \citet{hof57MVS245}.
\citet{kin00VSID4896} provided correct identification.
The object, however, has not been regularly monitored
before the chance detection of a bright (13.0 mag) outburst
by Y. Sano on 2008 October 25 (cf. vsnet-alert 10629).
The 2008 outburst was well observed and the object
was confirmed to be an SU UMa-type dwarf nova
(\cite{Pdot}; \cite{boy10kpcas}).
Although several outbursts were recorded since 2008,
all of them were likely normal outbursts.

   The 2017 superoutburst was recorded by Y. Maeda
at an unfiltered CCD magnitude of 13.5 and H. Maehara
at a visual magnitude of 14.2 on 2017 November 11
(cf. vsnet-alert 21581).
Observations starting on 2017 November 18 detected
superhumps.  The times of maxima are listed in e-table
\ref{tab:kpcasoc2017}.  The observations covered the relatively
late phase of the superoutburst, and these superhumps were
likely stage C ones.  The resultant period agrees with
the period of stage C superhumps during the 2008
superoutburst.

\begin{table}
\caption{Superhump maxima of KP Cas (2017)}\label{tab:kpcasoc2017}
\begin{center}
\begin{tabular}{rp{55pt}p{40pt}r@{.}lr}
\hline
\multicolumn{1}{c}{$E$} & \multicolumn{1}{c}{max\commenta} & \multicolumn{1}{c}{error} & \multicolumn{2}{c}{$O-C$\commentb} & \multicolumn{1}{c}{$N$\commentc} \\
\hline
0 & 58076.0720 & 0.0008 & 0&0014 & 81 \\
1 & 58076.1541 & 0.0007 & $-$0&0016 & 92 \\
10 & 58076.9245 & 0.0012 & 0&0025 & 96 \\
11 & 58077.0035 & 0.0012 & $-$0&0036 & 92 \\
13 & 58077.1786 & 0.0010 & 0&0012 & 94 \\
\hline
  \multicolumn{6}{l}{\commenta BJD$-$2400000.} \\
  \multicolumn{6}{l}{\commentb Against max $= 2458076.0705 + 0.085143 E$.} \\
  \multicolumn{6}{l}{\commentc Number of points used to determine the maximum.} \\
\end{tabular}
\end{center}
\end{table}

\subsection{VW Coronae Borealis}\label{obj:vwcrb}

   VW CrB was discovered as a dwarf nova (Antipin Var 21)
by \citet{ant96vwcrb}.  \citet{nov97vwcrb} established
the SU UMa-type nature of this object.  For more
information, see \citet{Pdot8}.
The 2017 superoutburst was detected by the ASAS-SN team
at $V$=14.4 on 2017 May 11 (we later knew that an AAVSO observer
detected a rising phase at $V$=15.69 on May 10).
The outburst was also detected by M. Hiraga at
an unfiltered CCD magnitude of 14.2 on May 13.

\begin{figure}
  \begin{center}
    \FigureFile(85mm,70mm){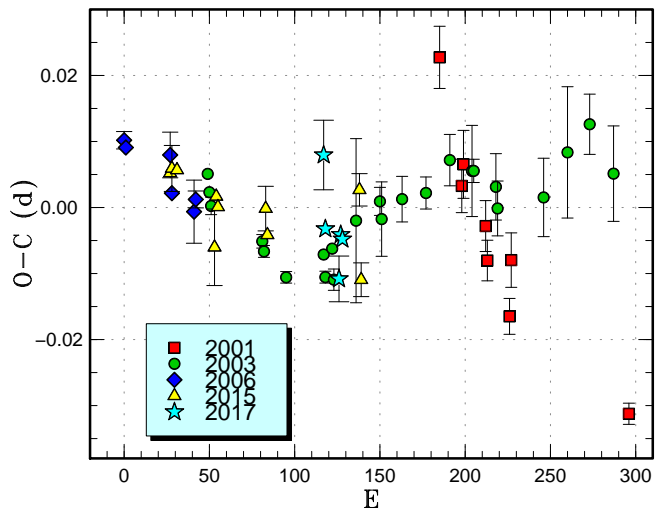}
  \end{center}
  \caption{Comparison of $O-C$ diagrams of VW CrB between different
  superoutbursts.  A period of 0.07290~d was used to draw this figure.
  Approximate cycle counts ($E$) after the start of the superoutburst
  were used.}
  \label{fig:vwcrbcomp3}
\end{figure}

   We observed this superoutburst on two nights and
detected superhumps (e-table \ref{tab:vwcrboc2017}).
Although our observations were carried out relatively
late, they were likely in the middle of stage B
(see e-figure \ref{fig:vwcrbcomp3}) since the duration
of superoutbursts in VW CrB is long.

\begin{table}
\caption{Superhump maxima of VW CrB (2017)}\label{tab:vwcrboc2017}
\begin{center}
\begin{tabular}{rp{55pt}p{40pt}r@{.}lr}
\hline
\multicolumn{1}{c}{$E$} & \multicolumn{1}{c}{max\commenta} & \multicolumn{1}{c}{error} & \multicolumn{2}{c}{$O-C$\commentb} & \multicolumn{1}{c}{$N$\commentc} \\
\hline
0 & 57892.0529 & 0.0053 & 0&0053 & 84 \\
1 & 57892.1146 & 0.0013 & $-$0&0050 & 117 \\
9 & 57892.6903 & 0.0035 & $-$0&0053 & 37 \\
10 & 57892.7699 & 0.0005 & 0&0024 & 106 \\
11 & 57892.8421 & 0.0005 & 0&0026 & 108 \\
\hline
  \multicolumn{6}{l}{\commenta BJD$-$2400000.} \\
  \multicolumn{6}{l}{\commentb Against max $= 2457892.0476 + 0.071985 E$.} \\
  \multicolumn{6}{l}{\commentc Number of points used to determine the maximum.} \\
\end{tabular}
\end{center}
\end{table}

\subsection{GP Canum Venaticorum}\label{obj:gpcvn}

   This object was originally selected as a CV
(SDSS J122740.83$+$513925.0) during the course of
the SDSS \citep{szk06SDSSCV5}.  The object is
an eclipsing SU UMa-type dwarf nova.  The SU UMa-type
nature was confirmed during the 2017  superoutburst
(\cite{she08j1227}; \cite{Pdot}).
For more history, see \citet{Pdot9}.

   The 2017 superoutburst was detected by the ASAS-SN
team at $V$=14.8 on 2017 July 5.
Subsequent single-night observations detected
two superhumps: BJD 2457941.3788(3) ($N$=40)
and 2457941.4457(3) ($N$=50).

\subsection{GQ Canum Venaticorum}\label{obj:gqcvn}

   This object was discovered as ASASSN-13ao by the ASAS-SN
team on 2013 June 8 \citep{sta13asassn13aoatel5118}.
The 2013 superoutburst was studied in \citet{Pdot5}
yielding only two superhump maxima.

\begin{figure}
  \begin{center}
    \FigureFile(85mm,110mm){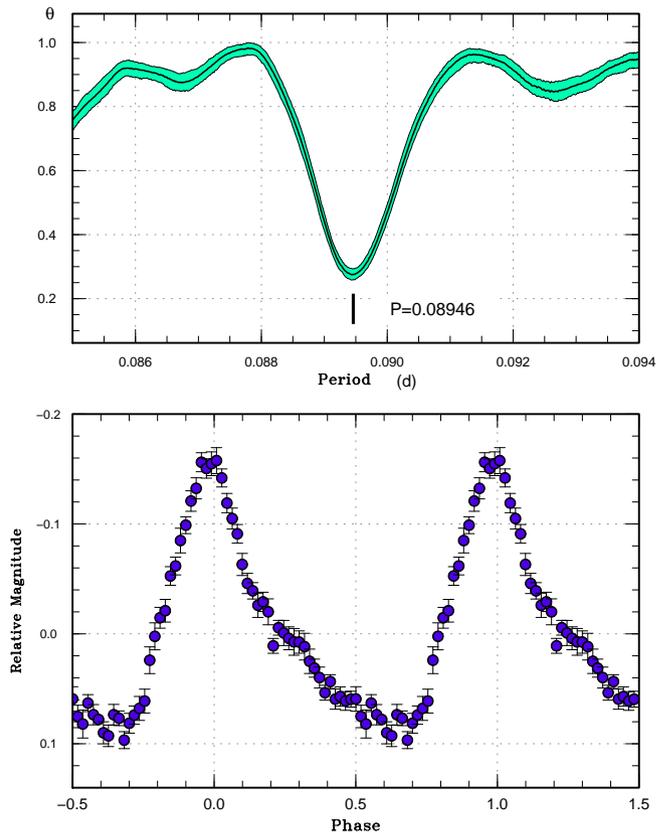}
  \end{center}
  \caption{Superhumps in GQ CVn (2017).
     (Upper): PDM analysis.
     (Lower): Phase-averaged profile.}
  \label{fig:gqcvnshpdm}
\end{figure}

   The 2017 superoutburst was detected by the ASAS-SN team
at $V$=14.7 on 2017 April 8.  Superhumps were better
observed than in the 2013 superoutburst
(vsnet-alert 20899, 20913; e-figure \ref{fig:gqcvnshpdm}).
The times of superhump maxima are listed in
e-table \ref{tab:gqcvnoc2017}.  Although there was some
tendency of a period decrease, we could not determine
the superhump stages.

   Three superoutbursts have been known in the ASAS-SN
data (2013 June 7, $V$=15.0; 2016 January 27, $V$=14.8
and the present one).  The shortest interval between
superoutbursts was 437~d.

\begin{table}
\caption{Superhump maxima of GQ CVn (2017)}\label{tab:gqcvnoc2017}
\begin{center}
\begin{tabular}{rp{55pt}p{40pt}r@{.}lr}
\hline
\multicolumn{1}{c}{$E$} & \multicolumn{1}{c}{max\commenta} & \multicolumn{1}{c}{error} & \multicolumn{2}{c}{$O-C$\commentb} & \multicolumn{1}{c}{$N$\commentc} \\
\hline
0 & 57854.4146 & 0.0005 & $-$0&0022 & 97 \\
1 & 57854.5026 & 0.0009 & $-$0&0037 & 81 \\
11 & 57855.4046 & 0.0006 & 0&0035 & 97 \\
12 & 57855.4919 & 0.0006 & 0&0013 & 94 \\
13 & 57855.5820 & 0.0007 & 0&0020 & 78 \\
14 & 57855.6713 & 0.0010 & 0&0018 & 68 \\
29 & 57857.0064 & 0.0044 & $-$0&0052 & 71 \\
30 & 57857.1038 & 0.0006 & 0&0027 & 181 \\
31 & 57857.1941 & 0.0015 & 0&0035 & 53 \\
37 & 57857.7236 & 0.0004 & $-$0&0038 & 169 \\
\hline
  \multicolumn{6}{l}{\commenta BJD$-$2400000.} \\
  \multicolumn{6}{l}{\commentb Against max $= 2457854.4168 + 0.089476 E$.} \\
  \multicolumn{6}{l}{\commentc Number of points used to determine the maximum.} \\
\end{tabular}
\end{center}
\end{table}

\subsection{V503 Cygni}\label{obj:v503cyg}

   For this famous SU UMa-type dwarf nova with a short
supercycle and negative superhumps \citep{har95v503cyg}.
\citet{kat02v503cyg} reported a dramatic variation in
the number of normal outbursts, and this finding led to
the discovery of the state with negative superhumps
suppressing the number of normal outbursts in other
objects (\cite{ohs12eruma}; \cite{zem13eruma}; 
\cite{osa13v1504cygKepler}; \cite{osa13v344lyrv1504cyg}).
\citet{pav12v503cyg} indeed confirmed temporal
disappearance of negative superhumps in 2010 and
shortening of the outburst cycle.
The 2017 July superoutburst was detected at a visual
magnitude of 14.1 by Alain Klotz on 2017 July 11.
Single-night observations detected two superhumps:
BJD 2457951.4358(28) ($N$=54) and 2457951.5257(3) ($N$=121).

\subsection{V632 Cygni}\label{obj:v632cyg}

   This object was discovered by \citet{hof49newvar1},
who recorded three outbursts on 1938 December 11 (13.9 mag),
1939 November 3 (12.8 mag) and 1940 June 13 (magnitude
unknown).  \citet{hof57VSchart} provided a finding chart.
Although this object had long been introduced in
monitoring programs by the AAVSO and AFOEV, the position
of the object was not correctly marked \citep{wen89v632cygv630cyg}.
Outbursts started to be detected by visual observers
since 1988--1989 when the chart error was corrected
(the old AAVSO chart marked the object at a 16 mag
unrelated star).  Unpublished $I$-band photometry by
one of the authors (T.K) in 1991 suggested a large
outburst amplitude and the object was suspected to be
an SU UMa-type dwarf nova.  Based on this information,
VSOLJ observers monitored the object and obtained
some time-resolved CCD photometry, but it did not
lead to a successful detection of superhumps.
\citet{liu99CVspec1} reported a spectrum and 
suggested that the orbital period is likely short.
\citet{she07CVspec} determined its orbital period
to be 0.06377(8)~d.  The SU UMa-type nature was finally
established during the 2008 superoutburst \citep{Pdot}.

   The 2017 superoutburst was visually detected
by L. Kocsmaros at a magnitude of 13.8 on 2017 June 16
(cvnet-outburst 7520).  Only single-night observations
were obtained.  The times of superhump maxima are
listed in e-table \ref{tab:v632cygoc2017}.
The superhump period of 0.0655(3)~d was determined
by the PDM method.

   We give a list of recent superoutburst in
e-table \ref{tab:v632cygout}.  The supercycle is
around 210~d.

\begin{table}
\caption{Superhump maxima of V632 Cyg (2017)}\label{tab:v632cygoc2017}
\begin{center}
\begin{tabular}{rp{55pt}p{40pt}r@{.}lr}
\hline
\multicolumn{1}{c}{$E$} & \multicolumn{1}{c}{max\commenta} & \multicolumn{1}{c}{error} & \multicolumn{2}{c}{$O-C$\commentb} & \multicolumn{1}{c}{$N$\commentc} \\
\hline
0 & 57926.3774 & 0.0012 & $-$0&0002 & 56 \\
1 & 57926.4438 & 0.0009 & 0&0004 & 68 \\
2 & 57926.5089 & 0.0006 & $-$0&0002 & 57 \\
\hline
  \multicolumn{6}{l}{\commenta BJD$-$2400000.} \\
  \multicolumn{6}{l}{\commentb Against max $= 2457926.3776 + 0.065742 E$.} \\
  \multicolumn{6}{l}{\commentc Number of points used to determine the maximum.} \\
\end{tabular}
\end{center}
\end{table}

\begin{table}
\caption{List of recent superoutbursts V632 Cyg}\label{tab:v632cygout}
\begin{center}
\begin{tabular}{cccccc}
\hline
Year & Month & Day & max\commenta & mag & source \\
\hline
2013 & 10 & 22 & 56588 & 14.0V & AAVSO \\
2016 &  5 &  5 & 57513 & 13.7V & ASAS-SN \\
2016 & 11 & 28 & 57721 & 13.8v & AAVSO \\
2017 &  6 & 16 & 57920 & 14.1V & ASAS-SN \\
\hline
  \multicolumn{5}{l}{\commenta JD$-$2400000.} \\
\end{tabular}
\end{center}
\end{table}

\subsection{V1454 Cygni}\label{obj:v1454cyg}

   Since it turned out the alias selection in \citet{Pdot}
was wrong for the 2006 superoutburst, we list
a corrected $O-C$ table (e-table \ref{tab:v1454cygoc2006}).
The $P_{\rm dot}$ for stage B has been corrected to be
$P_{\rm dot}$ of $+7.4(1.4) \times 10^{-5}$
(120$\le E \le$277).

\begin{table}
\caption{Superhump maxima of V1454 Cyg (2006)}\label{tab:v1454cygoc2006}
\begin{center}
\begin{tabular}{rp{55pt}p{40pt}r@{.}lr}
\hline
\multicolumn{1}{c}{$E$} & \multicolumn{1}{c}{max\commenta} & \multicolumn{1}{c}{error} & \multicolumn{2}{c}{$O-C$\commentb} & \multicolumn{1}{c}{$N$\commentc} \\
\hline
0 & 54063.9540 & 0.0017 & $-$0&0013 & 77 \\
120 & 54070.8827 & 0.0008 & 0&0082 & 84 \\
121 & 54070.9396 & 0.0011 & 0&0075 & 85 \\
173 & 54073.9205 & 0.0046 & $-$0&0099 & 81 \\
190 & 54074.9047 & 0.0015 & $-$0&0059 & 83 \\
207 & 54075.8866 & 0.0025 & $-$0&0042 & 63 \\
208 & 54075.9447 & 0.0032 & $-$0&0038 & 64 \\
248 & 54078.2566 & 0.0022 & 0&0017 & 32 \\
277 & 54079.9323 & 0.0010 & 0&0052 & 70 \\
294 & 54080.9099 & 0.0017 & 0&0026 & 84 \\
\hline
  \multicolumn{6}{l}{\commenta BJD$-$2400000.} \\
  \multicolumn{6}{l}{\commentb Against max $= 2454063.9553 + 0.057660 E$.} \\
  \multicolumn{6}{l}{\commentc Number of points used to determine the maximum.} \\
\end{tabular}
\end{center}
\end{table}

\subsection{HO Delphini}\label{obj:hodel}

   HO Del (=S 10066) was discovered as a dwarf nova by \citet{hof67an29043}.
The SU UMa-type nature was confirmed during the
1994 superoutburst.  See \citet{Pdot8} for more history.
The 2017 superoutburst was detected by the ASAS-SN team
at $V$=14.75 on 2017 July 30.  The outburst was also
visually detected at 14.0 mag on 2017 July 31 by
R. Stubbings.  Single-night observations detected
three superhumps: BJD 2457969.4926(15) ($N$=13),
2457969.5558(10) ($N$=23) and 2457969.6212(17) ($N$=19).

\subsection{MN Draconis}\label{obj:mndra}

   This object was discovered as a dwarf nova \citep{ant02var73dra}.
The object was identified as an SU UMa-type
dwarf nova in the period gap \citep{nog03var73dra}.
The object has both a short supercycle and negative
superhumps in quiescence citep{pav10mndra}.
It was suggested that the large negative $P_{\rm dot}$
for superhumps reflected stage A-B transition \citep{Pdot6}.
For more information, see \citet{Pdot6}.

   The 2017 June superoutburst was detected by G. Poyner
at an unfiltered CCD magnitude of 16.43 on 2017 June 19
(vsnet-alert 21143).
Only one superhump maximum was measured:
BJD 2457926.8051(11) ($N$=208).

\subsection{OV Draconis}\label{obj:ovdra}

   This object (=SDSS J125023.85$+$665525.5) is a CV 
selected during the course of 
the Sloan Digital Sky Survey (SDSS) \citep{szk03SDSSCV2}.
\citet{dil08SDSSCV} confirmed the deeply eclipsing nature.
The 2008 and 2009 superoutbursts were reported in \citet{Pdot2} and
another one in 2011 was reported in \citet{Pdot3}.
The 2013 superoutburst was reported in \citet{Pdot5}.

   The 2015 superoutburst was detected by the ASAS-SN
team at $V$=15.9 on 2015 February 11.  Although
time-resolved observations were reported on two nights,
we could not detect convincing superhumps.  These observations
are used in refining the eclipse ephemeris.

\begin{figure}
  \begin{center}
    \FigureFile(85mm,70mm){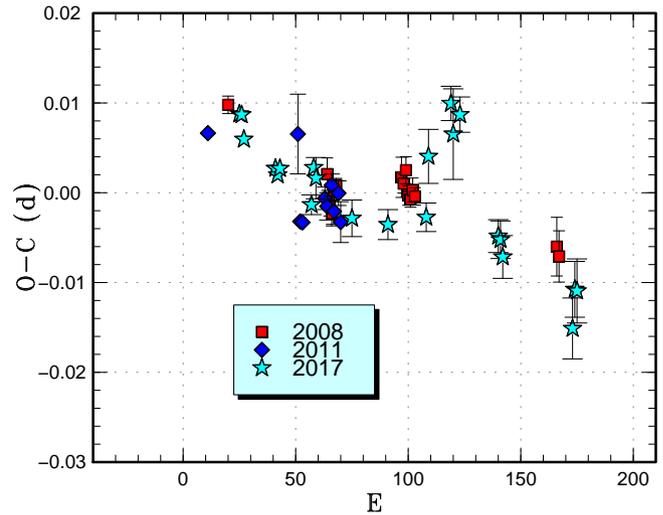}
  \end{center}
  \caption{Comparison of $O-C$ diagrams of OV Dra between different
  superoutbursts.  A period of 0.06040~d was used to draw this figure.
  Approximate cycle counts ($E$) after the start of the superoutburst
  were used.  For the 2008 superoutburst, we shifted by
  20 cycles to best match the others.}
  \label{fig:ovdracomp2}
\end{figure}

   The 2017 superoutburst was detected by the ASAS-SN
team at $V$=15.56 on 2017 May 26.
We updated the eclipse ephemeris using our
2008--2017 observations using the MCMC analysis \citep{Pdot4}:
\begin{equation}
{\rm Min(BJD)} = 2456305.98940(7) + 0.0587356736(13) E
\label{equ:ovdraecl}.
\end{equation}
The epoch corresponds to the center of all the observations.
The times of superhump maxima are listed in
e-table \ref{tab:ovdraoc2017}.  Stages B and C are
clearly seen (e-figure \ref{fig:ovdracomp2}).
It is the first time to show a positive
$P_{\rm dot}$ and transition to stage C so clearly
in a deeply eclipsing system.

\begin{table}
\caption{Superhump maxima of OV Dra (2017)}\label{tab:ovdraoc2017}
\begin{center}
\begin{tabular}{rp{50pt}p{30pt}r@{.}lcr}
\hline
\multicolumn{1}{c}{$E$} & \multicolumn{1}{c}{max\commenta} & \multicolumn{1}{c}{error} & \multicolumn{2}{c}{$O-C$\commentb} & \multicolumn{1}{c}{phase\commentc} & \multicolumn{1}{c}{$N$\commentd} \\
\hline
0 & 57901.4366 & 0.0003 & 0&0024 & 0.17 & 69 \\
1 & 57901.4969 & 0.0004 & 0&0024 & 0.20 & 65 \\
2 & 57901.5546 & 0.0008 & $-$0&0003 & 0.18 & 35 \\
16 & 57902.3969 & 0.0009 & $-$0&0022 & 0.52 & 36 \\
17 & 57902.4566 & 0.0006 & $-$0&0028 & 0.54 & 65 \\
18 & 57902.5177 & 0.0010 & $-$0&0020 & 0.58 & 27 \\
32 & 57903.3593 & 0.0011 & $-$0&0048 & 0.91 & 25 \\
33 & 57903.4238 & 0.0010 & $-$0&0005 & 0.01 & 26 \\
34 & 57903.4830 & 0.0023 & $-$0&0016 & 0.01 & 16 \\
50 & 57904.4450 & 0.0020 & $-$0&0046 & 0.39 & 35 \\
66 & 57905.4107 & 0.0017 & $-$0&0038 & 0.83 & 39 \\
83 & 57906.4383 & 0.0016 & $-$0&0014 & 0.33 & 27 \\
84 & 57906.5055 & 0.0030 & 0&0054 & 0.47 & 21 \\
94 & 57907.1154 & 0.0019 & 0&0123 & 0.86 & 58 \\
95 & 57907.1723 & 0.0050 & 0&0089 & 0.83 & 37 \\
98 & 57907.3557 & 0.0020 & 0&0114 & 0.95 & 27 \\
115 & 57908.3690 & 0.0018 & $-$0&0006 & 0.20 & 28 \\
116 & 57908.4290 & 0.0021 & $-$0&0009 & 0.22 & 27 \\
117 & 57908.4875 & 0.0024 & $-$0&0027 & 0.22 & 25 \\
148 & 57910.3519 & 0.0034 & $-$0&0078 & 0.96 & 21 \\
149 & 57910.4167 & 0.0031 & $-$0&0033 & 0.06 & 28 \\
150 & 57910.4769 & 0.0036 & $-$0&0034 & 0.09 & 27 \\
\hline
  \multicolumn{7}{l}{\commenta BJD$-$2400000.} \\
  \multicolumn{7}{l}{\commentb Against max $= 2457901.4342 + 0.060307 E$.} \\
  \multicolumn{7}{l}{\commentc Orbital phase.} \\
  \multicolumn{7}{l}{\commentd Number of points used to determine the maximum.} \\
\end{tabular}
\end{center}
\end{table}

\subsection{BE Octantis}\label{obj:beoct}

   BE Oct was discovered as a possible dwarf nova (S 6633)
\citep{hof63VSS61}.  J. Kemp and J. Patterson obtained
a superhump period of 0.07712(13)~d from observations on
1996 August 17 and 18 (vsnet-obs 3461).  Although outbursts
have been rather regularly recorded, no further observations
of superhumps were reported.  \citet{mas03faintCV} reported
a typical dwarf nova-type spectrum in quiescence.

\begin{figure}
  \begin{center}
    \FigureFile(85mm,110mm){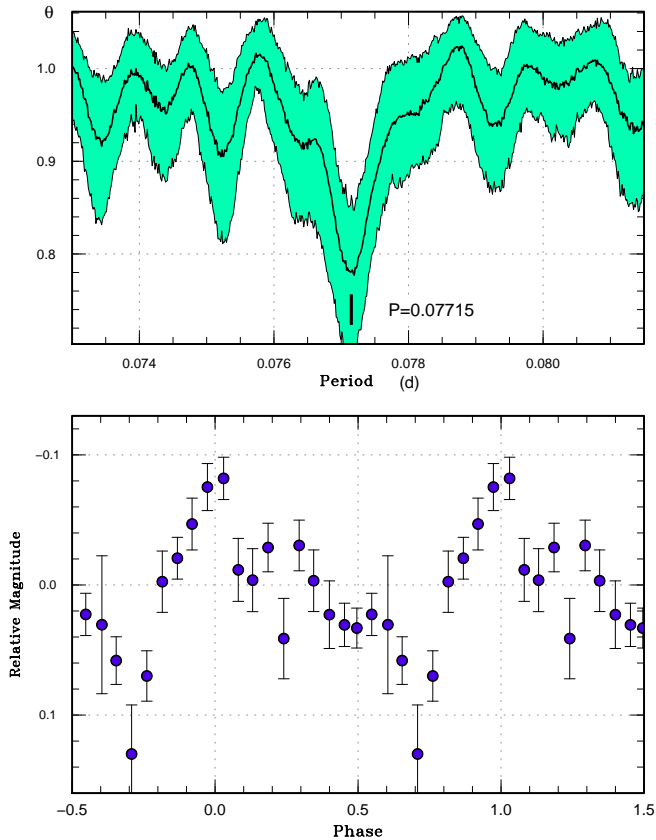}
  \end{center}
  \caption{Superhumps in BE Oct (2017).
     (Upper): PDM analysis.
     (Lower): Phase-averaged profile.}
  \label{fig:beoctshpdm}
\end{figure}

   The 2017 superoutburst was detected at a visual magnitude
of 15.4 by R. Stubbings and at $V$=15.97 by the ASAS-SN
team on 2017 July 1.  Subsequent observations detected
superhumps (figure \ref{fig:beoctshpdm}).
The times of superhump maxima are listed in
e-table \ref{tab:beoctoc2017}.  Although there were observations
after BJD 2457941, the object became too faint
to measure individual superhump maxima.
The light curve indicated brightening on July 7
(BJD 2457941), suggesting that there was stage B-C
transition around here.
The best superhump period based on the first four nights
(figure \ref{fig:beoctshpdm}) was determined to be
0.07715(7)~d by the PDM method.

\begin{table}
\caption{Superhump maxima of BE Oct (2017)}\label{tab:beoctoc2017}
\begin{center}
\begin{tabular}{rp{55pt}p{40pt}r@{.}lr}
\hline
\multicolumn{1}{c}{$E$} & \multicolumn{1}{c}{max\commenta} & \multicolumn{1}{c}{error} & \multicolumn{2}{c}{$O-C$\commentb} & \multicolumn{1}{c}{$N$\commentc} \\
\hline
0 & 57938.7720 & 0.0058 & $-$0&0062 & 12 \\
1 & 57938.8606 & 0.0026 & 0&0053 & 23 \\
14 & 57939.8594 & 0.0020 & 0&0016 & 23 \\
39 & 57941.7878 & 0.0038 & 0&0021 & 19 \\
40 & 57941.8600 & 0.0032 & $-$0&0028 & 23 \\
\hline
  \multicolumn{6}{l}{\commenta BJD$-$2400000.} \\
  \multicolumn{6}{l}{\commentb Against max $= 2457938.7782 + 0.077115 E$.} \\
  \multicolumn{6}{l}{\commentc Number of points used to determine the maximum.} \\
\end{tabular}
\end{center}
\end{table}

\subsection{V521 Pegasi}\label{obj:v521peg}

   This object (=HS 2219$+$1824) is a dwarf nova reported
in \citet{rod05hs2219}.  The SU UMa-type nature was
confirmed by \citet{rod05hs2219}.  For more information,
see \citet{Pdot5} and \citet{Pdot6}.

   The 2017 outburst was detected by the ASAS-SN team at
$V$=12.8 on 2017 August 24 and was found to be fading rapidly
on the same night by K. Wenzel and E. Muyllaert.
This outburst turned out to be a precursor outburst
and the true superoutburst occurred 7~d after
(cf. vsnet-alert 21384: detection by the ASAS-SN team at $V$=12.0
and visually by H. Maehara on 2017 August 31).
The times of superhump maxima are listed in
e-table \ref{tab:v521pegoc2017}.  Although there were
observations during the rapidly fading part,
we could not determine superhump maxima.

   A comparison of $O-C$ diagrams suggests that the 2017
observations recorded the final part of stage B
and superhump started to develop 52 cycles (3.2~d) before
the detection of the superoutburst.  This suggests
that superhumps started to develop several days after
the precursor outburst [the actual growth time may have
been longer, see \citet{kat16j0333}, \citet{ima17qzvir}].

\begin{figure}
  \begin{center}
    \FigureFile(88mm,70mm){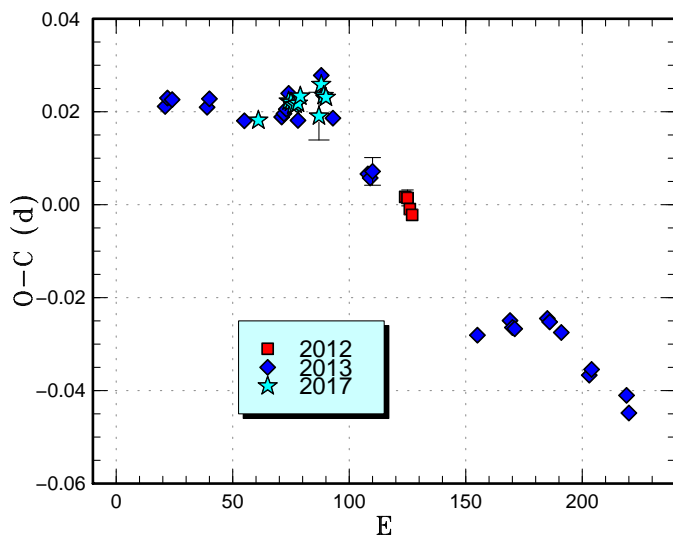}
  \end{center}
  \caption{Comparison of $O-C$ diagrams of V521 Peg between different
  superoutbursts.  A period of 0.06150~d was used to draw this figure.
  Approximate cycle counts ($E$) after the start of the superoutburst
  were used.  The 2017 diagram was shifted by 52 cycles
  (against the start of the main superoutburst) to best match
  the others.
  }
  \label{fig:v521pegcomp2}
\end{figure}

\begin{table}
\caption{Superhump maxima of V521 Peg (2017)}\label{tab:v521pegoc2017}
\begin{center}
\begin{tabular}{rp{55pt}p{40pt}r@{.}lr}
\hline
\multicolumn{1}{c}{$E$} & \multicolumn{1}{c}{max\commenta} & \multicolumn{1}{c}{error} & \multicolumn{2}{c}{$O-C$\commentb} & \multicolumn{1}{c}{$N$\commentc} \\
\hline
0 & 57997.5283 & 0.0004 & $-$0&0011 & 64 \\
13 & 57998.3319 & 0.0003 & 0&0011 & 42 \\
14 & 57998.3931 & 0.0004 & 0&0007 & 45 \\
15 & 57998.4539 & 0.0004 & $-$0&0002 & 41 \\
16 & 57998.5157 & 0.0003 & $-$0&0001 & 47 \\
17 & 57998.5774 & 0.0004 & 0&0000 & 46 \\
18 & 57998.6404 & 0.0018 & 0&0014 & 12 \\
26 & 57999.1281 & 0.0051 & $-$0&0040 & 22 \\
27 & 57999.1964 & 0.0003 & 0&0026 & 128 \\
28 & 57999.2556 & 0.0003 & 0&0001 & 129 \\
29 & 57999.3167 & 0.0009 & $-$0&0005 & 67 \\
\hline
  \multicolumn{6}{l}{\commenta BJD$-$2400000.} \\
  \multicolumn{6}{l}{\commentb Against max $= 2457997.5294 + 0.061646 E$.} \\
  \multicolumn{6}{l}{\commentc Number of points used to determine the maximum.} \\
\end{tabular}
\end{center}
\end{table}

\subsection{V368 Persei}\label{obj:v368per}

   V368 Per was discovered by \citet{ric69v1233aql}.
The SU UMa-type nature was identified by I. Miller
in 2012 (BAAVSS alert 3113).  For more information
see \citet{Pdot5}.

\begin{figure}
  \begin{center}
    \FigureFile(88mm,70mm){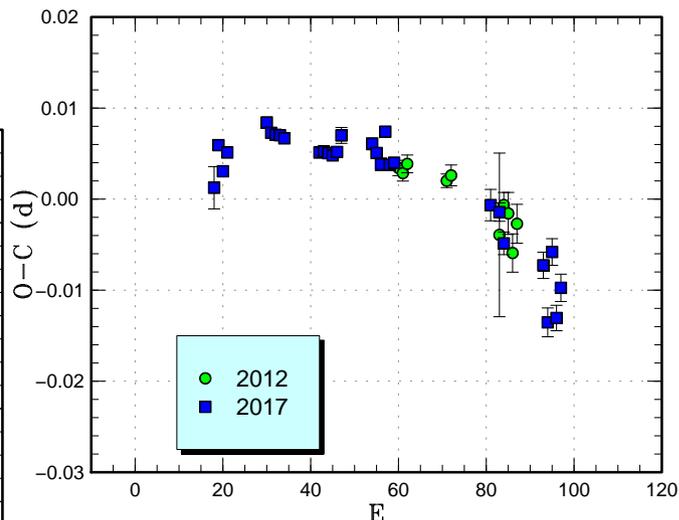}
  \end{center}
  \caption{Comparison of $O-C$ diagrams of V368 Per between different
  superoutbursts.  A period of 0.07922~d was used to draw this figure.
  Approximate cycle counts ($E$) after the start of the superoutburst
  were used.  The start of the 2012 superoutburst was not well
  constrained and we shifted the $O-C$ diagram by 38 cycles
  to best fit the 2017 one.}
  \label{fig:v368percomp}
\end{figure}

   The 2017 superoutburst was detected by the ASAS-SN
team at $V$=15.43 on 2017 September 26.
Subsequent observations detected superhumps
(vsnet-alert 21477, 21485, 21500).
In contrast to the 2012 superoutburst, when only
stage C superhumps were observed, we could observe
both stages B and C (e-figure \ref{fig:v368percomp};
the maxima for $E \le$ may be stage A superhumps).
The $P_{\rm dot}$ for stage B
was not determined due to the shortness of stage B
for this relatively long-$P_{\rm SH}$ object.

   Although this field has been monitored by the ASAS-SN
team since 2012 January, the present outburst was the
first well-recorded superoutburst.  Although the 2012
superoutburst was detected on a single night, it was
impossible to recognize it to be a superoutburst
by the ASAS-SN data only.

\begin{table}
\caption{Superhump maxima of V368 Per (2017)}\label{tab:v368peroc2017}
\begin{center}
\begin{tabular}{rp{55pt}p{40pt}r@{.}lr}
\hline
\multicolumn{1}{c}{$E$} & \multicolumn{1}{c}{max\commenta} & \multicolumn{1}{c}{error} & \multicolumn{2}{c}{$O-C$\commentb} & \multicolumn{1}{c}{$N$\commentc} \\
\hline
0 & 58024.3453 & 0.0023 & $-$0&0081 & 26 \\
1 & 58024.4292 & 0.0003 & $-$0&0032 & 163 \\
2 & 58024.5056 & 0.0004 & $-$0&0059 & 86 \\
3 & 58024.5869 & 0.0002 & $-$0&0036 & 83 \\
12 & 58025.3031 & 0.0004 & 0&0015 & 79 \\
13 & 58025.3812 & 0.0003 & 0&0006 & 86 \\
14 & 58025.4602 & 0.0003 & 0&0006 & 160 \\
15 & 58025.5394 & 0.0002 & 0&0007 & 263 \\
16 & 58025.6183 & 0.0004 & 0&0006 & 84 \\
24 & 58026.2505 & 0.0005 & 0&0007 & 135 \\
25 & 58026.3298 & 0.0004 & 0&0010 & 85 \\
26 & 58026.4088 & 0.0003 & 0&0010 & 151 \\
27 & 58026.4878 & 0.0002 & 0&0010 & 235 \\
28 & 58026.5674 & 0.0003 & 0&0015 & 209 \\
29 & 58026.6485 & 0.0009 & 0&0036 & 37 \\
36 & 58027.2021 & 0.0007 & 0&0041 & 104 \\
37 & 58027.2803 & 0.0005 & 0&0033 & 191 \\
38 & 58027.3583 & 0.0007 & 0&0022 & 95 \\
39 & 58027.4411 & 0.0004 & 0&0060 & 86 \\
40 & 58027.5167 & 0.0003 & 0&0026 & 86 \\
41 & 58027.5961 & 0.0004 & 0&0030 & 74 \\
63 & 58029.3343 & 0.0017 & 0&0028 & 27 \\
65 & 58029.4920 & 0.0010 & 0&0025 & 83 \\
66 & 58029.5677 & 0.0012 & $-$0&0008 & 83 \\
75 & 58030.2783 & 0.0014 & $-$0&0013 & 44 \\
76 & 58030.3513 & 0.0016 & $-$0&0074 & 44 \\
77 & 58030.4382 & 0.0015 & 0&0005 & 44 \\
78 & 58030.5102 & 0.0014 & $-$0&0065 & 29 \\
79 & 58030.5927 & 0.0015 & $-$0&0030 & 33 \\
\hline
  \multicolumn{6}{l}{\commenta BJD$-$2400000.} \\
  \multicolumn{6}{l}{\commentb Against max $= 2458024.3534 + 0.079017 E$.} \\
  \multicolumn{6}{l}{\commentc Number of points used to determine the maximum.} \\
\end{tabular}
\end{center}
\end{table}

\subsection{XY Piscium}\label{obj:xypsc}

   XY Psc was discovered as a transient located close to
the galaxy UGC 729 on 1972 October 5 at a photographic
magnitude of 13.0 and was detected at a magnitude of 15.0
on 1972 October 17 \citep{ros72xypsciauc1}.  Although
the object was suspected to be a supernova of this galaxy,
it was suggested to be a dwarf nova based on
the rapid rise and decline \citep{ros72xypsciauc2}.

   VSOLJ members started monitoring this object in 1984
and S. Fujino recorded a possible outburst at a photographic
magnitude (using a filter adjusted to reproduce the $V$ band)
of 15.0 on 1984 January 21.  There was only one positive
record and, unfortunately, a negative film was lost
and the identity of the object remained unclear.
Since the 1990s, the object started to be monitored more
regularly by observers worldwide.  No outburst,
however, was recorded.

   In the meantime, \citet{hen01xypsc} obtained deep
CCD images of this field and identified a quiescent
blue counterpart of $V$=21.1.  \citet{hen01xypsc}
also provided some more details of historical
observations of this object.
\citet{kat01hvvir} listed this object as a candidate
WZ Sge-type dwarf nova.

\begin{figure}
  \begin{center}
    \FigureFile(85mm,110mm){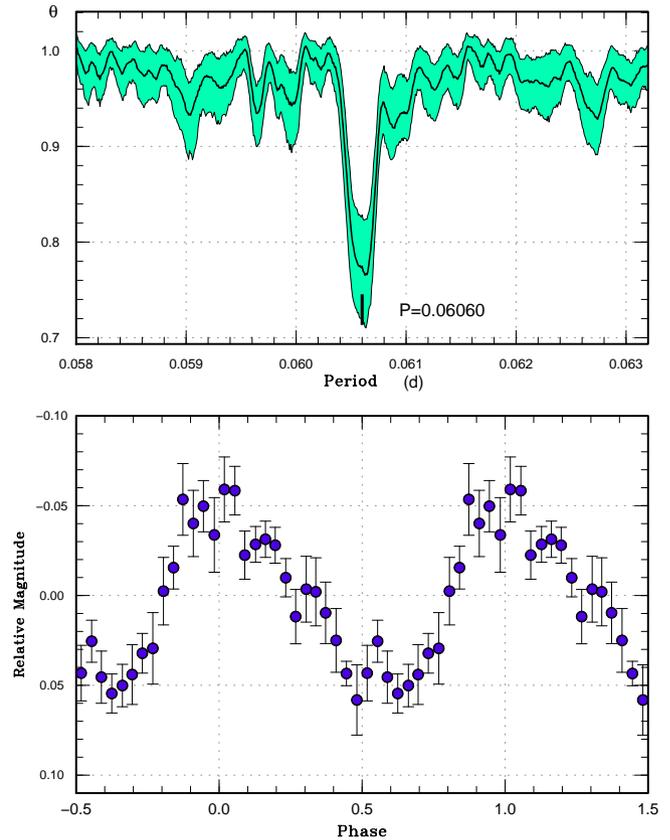}
  \end{center}
  \caption{Superhumps in XY Psc (2017).
     (Upper): PDM analysis.
     (Lower): Phase-averaged profile.}
  \label{fig:xypscshpdm}
\end{figure}

   The 2017 outburst was detected by the ASAS-SN team
at $V$=13.1 on 2017 June 2 (vsnet-alert 21085).
Two night before, the object was fainter than $V$=16.3.
Since the object was still low in the morning sky,
amateur observers had not yet started monitoring.
Due to the short observing windows, it took several
days to detect superhumps (vsnet-alert 21111).
Later observations pinned down the superhump period
as observations accumulated (vsnet-alert 21119, 21131;
e-figure \ref{fig:xypscshpdm}).

   The times of superhump maxima are listed in
e-table \ref{tab:xypscoc2017}.
Despite the limited number of observations,
stage B with a clearly positive $P_{\rm dot}$ and
transition to stage C were recorded.

   The initial sign of superhumps was recorded
on 2017 June 10 (8~d after the outburst detection).
Although there were earlier observations, it was
impossible to detect modulation due to the shortness
of observations.  The relatively early appearance
of ordinary superhumps and a strongly positive
$P_{\rm dot}$ would suggest either an ordinary
SU UMa-type dwarf nova or a WZ Sge/SU UMa-type
borderline object.  Since the object is not suited
for observation around the solar conjunction,
past outbursts near solar conjunctions may have been
easily missed.  We probably need to wait for a superoutburst
occurring in the favorable season of the year to
possibly detect early superhumps.

   The object showed a rebrightening at $V$=15.9
on 2017 June 27 (ASAS-SN detection, vsnet-alert 21171).
According to the ASAS-SN observations, there was no
evidence of multiple rebrightenings.
Regular monitoring by the ASAS-SN team started
in 2012 October without any previous detection.

\begin{table}
\caption{Superhump maxima of XY Psc (2017)}\label{tab:xypscoc2017}
\begin{center}
\begin{tabular}{rp{55pt}p{40pt}r@{.}lr}
\hline
\multicolumn{1}{c}{$E$} & \multicolumn{1}{c}{max\commenta} & \multicolumn{1}{c}{error} & \multicolumn{2}{c}{$O-C$\commentb} & \multicolumn{1}{c}{$N$\commentc} \\
\hline
0 & 57915.8730 & 0.0006 & 0&0043 & 17 \\
17 & 57916.8986 & 0.0011 & $-$0&0013 & 17 \\
33 & 57917.8665 & 0.0010 & $-$0&0040 & 24 \\
50 & 57918.9012 & 0.0012 & $-$0&0004 & 16 \\
56 & 57919.2630 & 0.0006 & $-$0&0026 & 53 \\
66 & 57919.8703 & 0.0013 & $-$0&0018 & 22 \\
82 & 57920.8461 & 0.0011 & 0&0034 & 17 \\
83 & 57920.9077 & 0.0012 & 0&0044 & 11 \\
89 & 57921.2686 & 0.0009 & 0&0014 & 48 \\
99 & 57921.8706 & 0.0013 & $-$0&0032 & 21 \\
\hline
  \multicolumn{6}{l}{\commenta BJD$-$2400000.} \\
  \multicolumn{6}{l}{\commentb Against max $= 2457915.8687 + 0.060658 E$.} \\
  \multicolumn{6}{l}{\commentc Number of points used to determine the maximum.} \\
\end{tabular}
\end{center}
\end{table}

\subsection{V701 Tauri}\label{obj:v701tau}

   V701 Tau was discovered by \citet{era73v701tau} as
an eruptive object.  The SU UMa-type nature was confirmed
in 1995 (reported in \cite{Pdot}).  \citet{she07v701tau}
further reported the 2005 superoutburst.  For more details,
see \citet{Pdot7}.

\begin{figure}
  \begin{center}
    \FigureFile(88mm,70mm){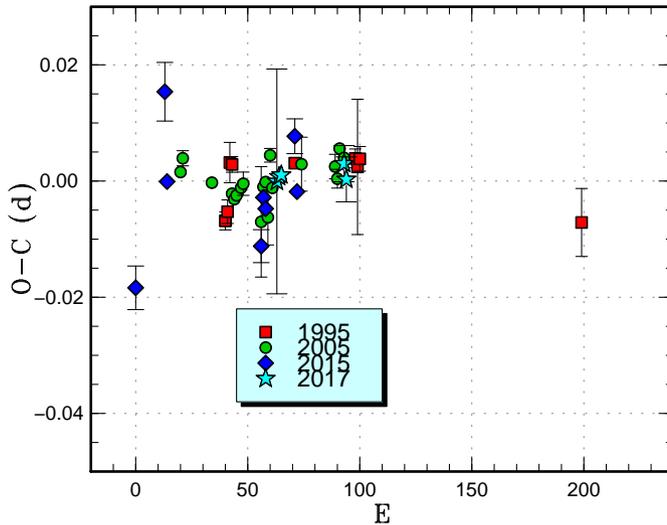}
  \end{center}
  \caption{Comparison of $O-C$ diagrams of V701 Tau between different
  superoutbursts.  A period of 0.06899~d was used to draw this figure.
  Approximate cycle counts ($E$) after the start of the superoutburst
  were used.  We assumed that the 2015 superoutburst was detected
  soon after the maximum and shifted other superoutbursts to
  get the best fit.}
  \label{fig:v701taucomp3}
\end{figure}

   The 2017 superoutburst was detected by the ASAS-SN team
at $V$=14.88 on 2017 August 17.  Time-resolved photometric
observations were carried out on two nights and
the times of superhump maxima are listed in
e-table \ref{tab:v701tauoc2017}.  A comparison of
$O-C$ diagrams suggests that we observed the late
part of stage B superhumps (e-figure \ref{fig:v701taucomp3}).

\begin{table}
\caption{Superhump maxima of V701 Tau (2017)}\label{tab:v701tauoc2017}
\begin{center}
\begin{tabular}{rp{55pt}p{40pt}r@{.}lr}
\hline
\multicolumn{1}{c}{$E$} & \multicolumn{1}{c}{max\commenta} & \multicolumn{1}{c}{error} & \multicolumn{2}{c}{$O-C$\commentb} & \multicolumn{1}{c}{$N$\commentc} \\
\hline
0 & 57987.4742 & 0.0193 & $-$0&0006 & 11 \\
1 & 57987.5440 & 0.0014 & 0&0002 & 31 \\
2 & 57987.6132 & 0.0014 & 0&0004 & 29 \\
30 & 57989.5469 & 0.0031 & 0&0014 & 21 \\
31 & 57989.6132 & 0.0039 & $-$0&0014 & 31 \\
\hline
  \multicolumn{6}{l}{\commenta BJD$-$2400000.} \\
  \multicolumn{6}{l}{\commentb Against max $= 2457987.4748 + 0.069026 E$.} \\
  \multicolumn{6}{l}{\commentc Number of points used to determine the maximum.} \\
\end{tabular}
\end{center}
\end{table}

\subsection{V1208 Tauri}\label{obj:v1208tau}

   V1208 Tau was originally identified as a CV
during the course of identification of ROSAT sources
(=1RXS J045942.9$+$192625, \cite{mot96CVROSAT}).
P. Schmeer detected the first-ever recorded outburst in 2000
(vsnet-alert 4118).  The SU UMa-type nature was confirmed
during this outburst.  Although \citet{pat05SH} gave
an orbital period of 0.0681(2)~d, the source was unclear.
The 2000 and 2002 superoutbursts were reported in \citet{Pdot}.
The 2011 superoutburst was reported in \citet{Pdot4}.
The entire course of the superoutburst, however, has not
yet been well observed.

   The 2017 superoutburst was detected at an unfiltered
CCD magnitude of 14.8 on 2017 November 8 by Y. Maeda.
This superoutburst was also recorded in the ASAS-SN data
($V$=15.90 on 2017 November 9, vsnet-alert 21576).
Superhumps were detected on 2017 November 11 (vsnet-alert 21576).
The times of superhump maxima are listed in
e-table \ref{tab:v1208tauoc2017}.
Although there were observations on three nights after
these observations, no clear superhumps were detected.
It may have been that we only observed the terminal portion
of the superoutburst, and that there could have been
a rebrightening on 2017 November 17.  The data were
too sparse to depict the outburst behavior unambiguously.
This object still need better observations to obtain
precise values of superhump and orbital periods.

\begin{table}
\caption{Superhump maxima of V1208 Tau (2017)}\label{tab:v1208tauoc2017}
\begin{center}
\begin{tabular}{rp{55pt}p{40pt}r@{.}lr}
\hline
\multicolumn{1}{c}{$E$} & \multicolumn{1}{c}{max\commenta} & \multicolumn{1}{c}{error} & \multicolumn{2}{c}{$O-C$\commentb} & \multicolumn{1}{c}{$N$\commentc} \\
\hline
0 & 58068.7076 & 0.0009 & $-$0&0010 & 40 \\
1 & 58068.7797 & 0.0016 & 0&0017 & 39 \\
2 & 58068.8471 & 0.0012 & $-$0&0003 & 40 \\
3 & 58068.9165 & 0.0011 & $-$0&0003 & 39 \\
\hline
  \multicolumn{6}{l}{\commenta BJD$-$2400000.} \\
  \multicolumn{6}{l}{\commentb Against max $= 2458068.7086 + 0.069413 E$.} \\
  \multicolumn{6}{l}{\commentc Number of points used to determine the maximum.} \\
\end{tabular}
\end{center}
\end{table}

\subsection{TU Trianguli}\label{obj:tutri}

   TU Tri was discovered as a dwarf nova (GR 287) with
a photographic range of 14.8 to fainter than 18.0 by
\citet{rom78tutri}.  The coordinates given in this paper,
however, was incorrect \citep{sha92tutri} and
\citet{sha91tutri} independently discovered this dwarf nova.
The observation by \citet{sha91tutri} recorded a long outburst
starting on 1982 November 9 and lasted at least up to
1982 November 14.  \citet{liu00CVspec3} obtained a spectrum
in quiescence without emission lines.

   There was an outburst on 2013 January 1 at a visual
magnitude of 14.6 detected by M. Simonsen.  Although
K. Torii's observations could not detect superhumps
(vsnet-campaign-dn 3237, 3262), \citet{zlo04tutri}
recorded superhumps with a period of 0.0745~d
from single-night observations.

\begin{figure}
  \begin{center}
    \FigureFile(85mm,110mm){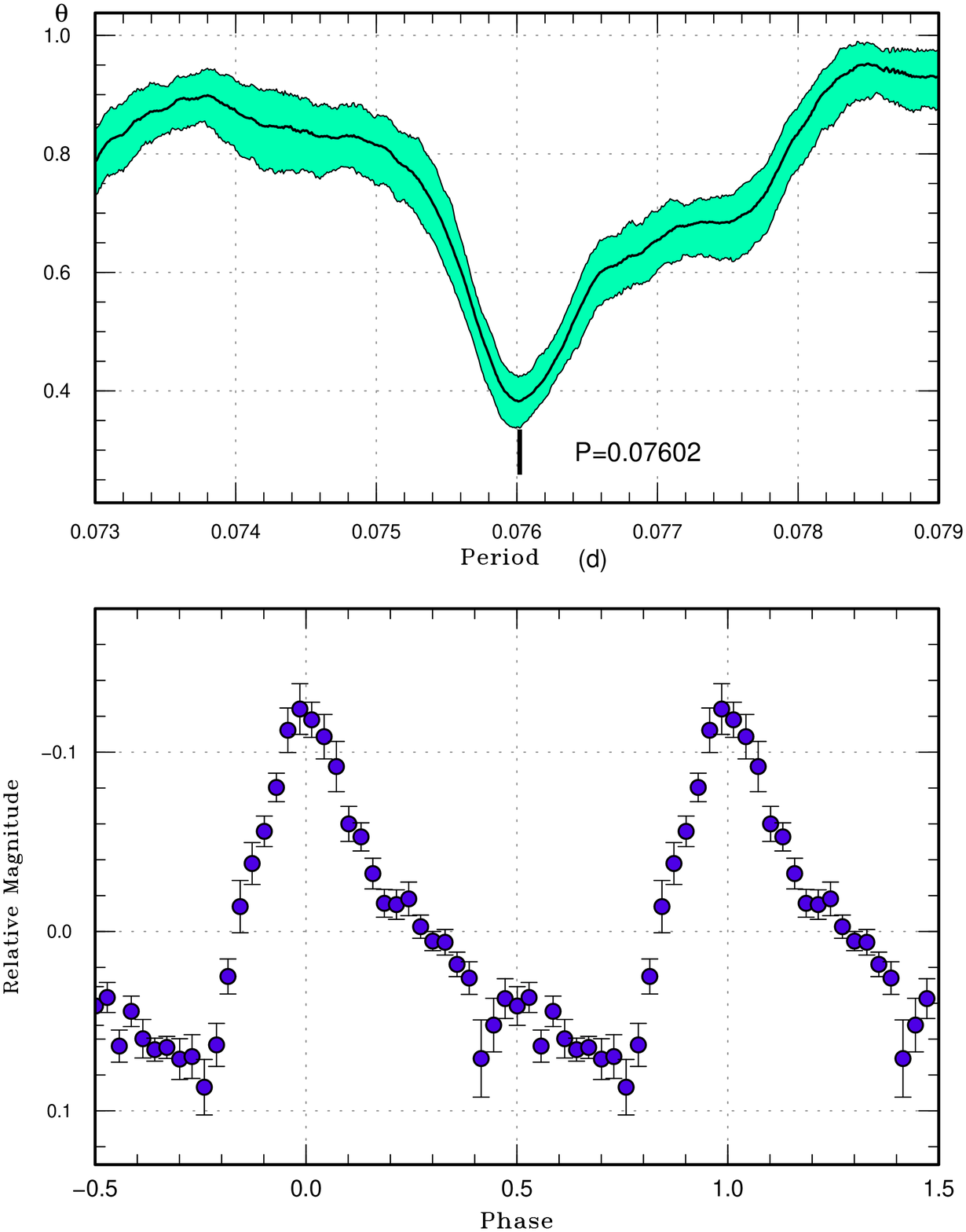}
  \end{center}
  \caption{Superhumps in TU Tri (2017).
     (Upper): PDM analysis.
     (Lower): Phase-averaged profile.}
  \label{fig:tutrishpdm}
\end{figure}

   Although an outburst on 2017 January 29 was observed
(P. Dubovsky), it faded quickly and must have been
a normal outburst.
The 2017 superoutburst was detected by the ASAS-SN team
at $V$=14.98 on 2017 October 19.  Subsequent observations
detected superhumps (vsnet-alert 21540, 21544).
The resultant period of 0.07602(2)~d (e-figure \ref{fig:tutrishpdm})
was significantly longer than the measurement
by \citet{zlo04tutri}.
The times of superhump maxima are listed in
e-table \ref{tab:tutrioc2017}.  There looks like to
have been stage B-C transition between $E$=20
and $E$=59.

\begin{table}
\caption{Superhump maxima of TU Tri (2017)}\label{tab:tutrioc2017}
\begin{center}
\begin{tabular}{rp{55pt}p{40pt}r@{.}lr}
\hline
\multicolumn{1}{c}{$E$} & \multicolumn{1}{c}{max\commenta} & \multicolumn{1}{c}{error} & \multicolumn{2}{c}{$O-C$\commentb} & \multicolumn{1}{c}{$N$\commentc} \\
\hline
0 & 58050.0377 & 0.0023 & $-$0&0038 & 42 \\
1 & 58050.1166 & 0.0009 & $-$0&0009 & 64 \\
16 & 58051.2561 & 0.0018 & $-$0&0017 & 51 \\
17 & 58051.3355 & 0.0005 & 0&0017 & 87 \\
18 & 58051.4131 & 0.0005 & 0&0032 & 105 \\
19 & 58051.4875 & 0.0006 & 0&0017 & 44 \\
20 & 58051.5639 & 0.0006 & 0&0020 & 24 \\
59 & 58054.5242 & 0.0009 & $-$0&0022 & 84 \\
\hline
  \multicolumn{6}{l}{\commenta BJD$-$2400000.} \\
  \multicolumn{6}{l}{\commentb Against max $= 2458050.0416 + 0.076015 E$.} \\
  \multicolumn{6}{l}{\commentc Number of points used to determine the maximum.} \\
\end{tabular}
\end{center}
\end{table}

\subsection{SU Ursae Majoris}\label{obj:suuma}

   This object is the prototype of SU UMa-type
dwarf novae.  See \citet{Pdot7} for the history.
The second superoutburst in 2017 was detected by
P. Schmeer at a visual magnitude of 12.3 on
2017 October 16.  The object further brightened to
11.2 on 2017 October 18 (J. Toone, baavss-alert 4810).
Superhumps were observed on two nights and
the times of superhump maxima are listed in
e-table \ref{tab:suumaoc2017b}.

\begin{table}
\caption{Superhump maxima of SU UMa (2017b)}\label{tab:suumaoc2017b}
\begin{center}
\begin{tabular}{rp{55pt}p{40pt}r@{.}lr}
\hline
\multicolumn{1}{c}{$E$} & \multicolumn{1}{c}{max\commenta} & \multicolumn{1}{c}{error} & \multicolumn{2}{c}{$O-C$\commentb} & \multicolumn{1}{c}{$N$\commentc} \\
\hline
0 & 58046.5183 & 0.0003 & 0&0027 & 111 \\
1 & 58046.5920 & 0.0006 & $-$0&0026 & 140 \\
63 & 58051.4806 & 0.0015 & $-$0&0073 & 149 \\
64 & 58051.5740 & 0.0009 & 0&0072 & 140 \\
\hline
  \multicolumn{6}{l}{\commenta BJD$-$2400000.} \\
  \multicolumn{6}{l}{\commentb Against max $= 2458046.5157 + 0.078924 E$.} \\
  \multicolumn{6}{l}{\commentc Number of points used to determine the maximum.} \\
\end{tabular}
\end{center}
\end{table}

\subsection{HS Virginis}\label{obj:hsvir}

   HS Vir was discovered as an ultraviolet excess object PG 1341$-$079,
and was confirmed by spectroscopy to be a cataclysmic variable
(\cite{gre82PGsurveyCV}; \cite{PGsurvey}).
\citet{osm85hsvir} reported from photographic observations
that this object shows relatively abundant short, faint outbursts,
together with a bright ($\sim$12.8 mag) one.
\citet{rin93thesis} suggested an orbital period of 0.0836 day
(with possible aliasing problems) from radial-velocity
measurements.  Since this observation, HS Vir has been
considered to be a candidate for an SU UMa-type dwarf nova.
\citet{kat95hsvir} reported frequent occurrence of short
outbursts.
Although the bright outburst in \citet{osm85hsvir} was
suggestive of a superoutburst, it was only in 1996 March
when superhumps with a mean period of 0.08059(3)~d were
indeed detected, confirming the SU UMa-type nature \citep{kat98hsvir}.
\citet{pat03suumas} observed the same superoutburst and
reported a superhump period of 0.08045(19)~d.
\citet{men99hsvir} obtained an orbital period of 0.07692(3)~d
from radial-velocity measurements.
\citet{kat01hsvir} suggested a supercycle of 186~d or 371~d.
\citet{Pdot} analyzed the 1996 data again, and yielded
a slightly shorter superhump period of 0.08003(3)~d.

   Although this object was observed several times during
later superoutbursts [2007 March, 2008 June \citep{Pdot},
and 2014 February], these observations were performed
only for a short time and there have been no new
measurement of the superhump period.

   The 2017 superoutburst was detected by H. Maehara
at a visual magnitude of 13.5 on 2017 April 22.
Superhumps were observed (vsnet-alert 20958).
The times of superhump maxima are listed in
e-table \ref{tab:hsviroc2017}.  There was a gap
in the observation following the initial superhump
detection and maxima from later observations were
not of very good quality due to the low sampling rate
and the complex profile of superhumps.
The derived mean superhump period was 0.08031(6)~d.
Since observations were made during the relatively
late phase, this period was probably affected by
stage C superhumps, which were not apparent on
the $O-C$ diagram due to scatter.

\begin{figure}
  \begin{center}
    \FigureFile(85mm,70mm){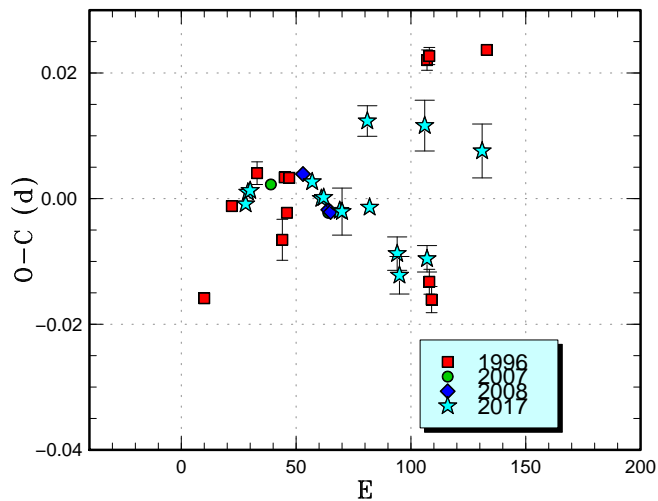}
  \end{center}
  \caption{Comparison of $O-C$ diagrams of HS Vir between different
  superoutbursts.  A period of 0.08031~d was used to draw this figure.
  Approximate cycle counts ($E$) after the start of the superoutburst
  were used.  Since the start of the 1996 superoutburst was
  not well constrained, we shifted to best match the others.
  Growing superhumps were observed during the 1996 superoutburst
  \citep{kat98hsvir}, confirming the validity of this treatment.}
  \label{fig:hsvircomp}
\end{figure}

   We also provide yet unpublished maxima of the 2007
superoutburst: BJD 2454188.0930(5) ($N$=57) and
2454190.0962(6) ($N$=89).
A comparison of the $O-C$ diagram (e-figure \ref{fig:hsvircomp})
cannot tell much other than the presence of stage A
in the 1996 superoutburst.

\begin{table}
\caption{Superhump maxima of HS Vir (2017)}\label{tab:hsviroc2017}
\begin{center}
\begin{tabular}{rp{55pt}p{40pt}r@{.}lr}
\hline
\multicolumn{1}{c}{$E$} & \multicolumn{1}{c}{max\commenta} & \multicolumn{1}{c}{error} & \multicolumn{2}{c}{$O-C$\commentb} & \multicolumn{1}{c}{$N$\commentc} \\
\hline
0 & 57868.3646 & 0.0003 & $-$0&0007 & 75 \\
1 & 57868.4468 & 0.0002 & 0&0012 & 73 \\
2 & 57868.5274 & 0.0006 & 0&0014 & 57 \\
29 & 57870.6971 & 0.0010 & 0&0027 & 14 \\
33 & 57871.0157 & 0.0013 & 0&0001 & 83 \\
34 & 57871.0962 & 0.0008 & 0&0002 & 99 \\
41 & 57871.6563 & 0.0009 & $-$0&0019 & 22 \\
42 & 57871.7364 & 0.0037 & $-$0&0021 & 11 \\
53 & 57872.6343 & 0.0024 & 0&0123 & 25 \\
54 & 57872.7008 & 0.0011 & $-$0&0014 & 22 \\
66 & 57873.6572 & 0.0027 & $-$0&0088 & 19 \\
67 & 57873.7340 & 0.0030 & $-$0&0123 & 15 \\
78 & 57874.6413 & 0.0040 & 0&0115 & 19 \\
79 & 57874.7004 & 0.0021 & $-$0&0097 & 21 \\
103 & 57876.6450 & 0.0043 & 0&0074 & 22 \\
\hline
  \multicolumn{6}{l}{\commenta BJD$-$2400000.} \\
  \multicolumn{6}{l}{\commentb Against max $= 2457868.3653 + 0.080313 E$.} \\
  \multicolumn{6}{l}{\commentc Number of points used to determine the maximum.} \\
\end{tabular}
\end{center}
\end{table}

\subsection{V406 Virginis}\label{obj:v406vir}

   V406 Vir was originally selected as a CV (SDSS J123813.73$-$033933.0)
during the course of the SDSS \citep{szk03SDSSCV2}.
\citet{szk03SDSSCV2} suggested a high inclination and
an orbital period of 76~min.  \citet{zha06j1238} performed
time-resolved photometric and spectroscopic observations
and obtained an orbital period of 0.05592(35)~d.
\citet{zha06j1238} classified the object a WZ Sge-like one
but with cyclic brightening up to 0.4 mag with periods of
the order of 8--12~hr in quiescence.
\citet{avi10j1238} performed infrared $JHK$ photometry
and optical spectroscopy and indicated that the system
has a L4-type brown dwarf.  The Doppler mapping of the system
showed the permanent presence of a spiral arm pattern
in the accretion disk and \citet{avi10j1238} suggested
that they can be a result of the 2:1 resonance.
\citet{avi10j1238} classified this object to be
a period bouncer.

\begin{figure}
  \begin{center}
    \FigureFile(85mm,110mm){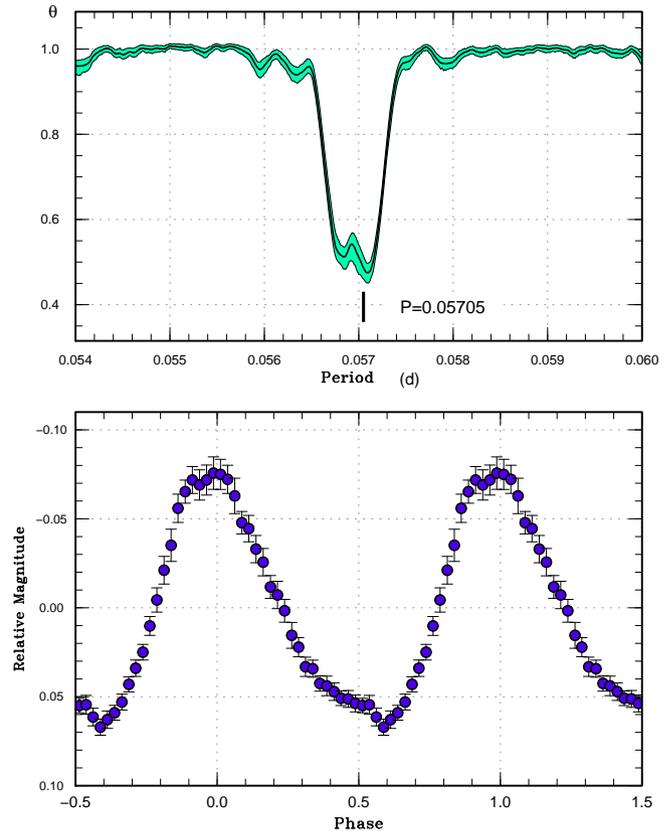}
  \end{center}
  \caption{Superhumps in V406 Vir (2017).
     (Upper): PDM analysis.
     (Lower): Phase-averaged profile.}
  \label{fig:v406virshpdm}
\end{figure}

   Despite that the object has long been suspected to be
a WZ Sge-type dwarf nova, no outburst was recorded until
2017.  The 2017 outburst was detected by the ASAS-SN team
at $V$=11.86 on 2017 July 31 (cf. vsnet-alert 21308).
Despite poor seasonal location in the sky, the outburst
was observed and superhumps were detected
(vsnet-alert 21325, 21328, 21330; e-figure \ref{fig:v406virshpdm}).
These superhumps were detected already on 2017 August 4
(they were already stage B superhumps), indicating
that the waiting time for the appearance of ordinary
superhumps was short.
The times of superhump maxima are listed in
e-table \ref{tab:v406viroc2017}.  All superhumps were
stage B ones and $P_{\rm dot}$ was relatively large
$+8.1(1.5) \times 10^{-5}$

   The short waiting time for ordinary superhumps,
large amplitude of superhumps (e-figure \ref{fig:v406virshpdm})
and the relatively large $P_{\rm dot}$ for stage B
superhumps indicate that this object is not an extreme
WZ Sge-type dwarf nova as expected from the conclusion
by \citet{avi10j1238}.  By using the relation between
$q$ and $P_{\rm dot}$ for WZ Sge-type dwarf novae
(equation 6 in \cite{kat15wzsge}), the $q$ value is
expected to be 0.095(6) (the error corresponds to
the measurement error only).  This value is not particularly
small for WZ Sge-type dwarf novae (cf. figure 17
in \cite{kat15wzsge}), particularly considering
the high-mass white dwarf claimed by \citet{avi10j1238}.
The ASAS-SN observations in 2017 only consisted of
three nights and there was 7~d gap before the outburst
detection.  The waiting time for the appearance of ordinary
superhumps may have been longer if the true maximum
was missed.  In any case, the conclusion by \citet{avi10j1238}
would be worth revisiting using higher quality observations.

\begin{table}
\caption{Superhump maxima of V406 Vir (2017)}\label{tab:v406viroc2017}
\begin{center}
\begin{tabular}{rp{55pt}p{40pt}r@{.}lr}
\hline
\multicolumn{1}{c}{$E$} & \multicolumn{1}{c}{max\commenta} & \multicolumn{1}{c}{error} & \multicolumn{2}{c}{$O-C$\commentb} & \multicolumn{1}{c}{$N$\commentc} \\
\hline
0 & 57970.8436 & 0.0024 & 0&0026 & 18 \\
1 & 57970.8993 & 0.0005 & 0&0013 & 34 \\
7 & 57971.2403 & 0.0002 & 0&0005 & 131 \\
12 & 57971.5255 & 0.0006 & 0&0009 & 13 \\
18 & 57971.8666 & 0.0003 & 0&0003 & 129 \\
24 & 57972.2081 & 0.0002 & 0&0000 & 92 \\
25 & 57972.2652 & 0.0005 & 0&0002 & 87 \\
29 & 57972.4925 & 0.0004 & $-$0&0004 & 31 \\
36 & 57972.8893 & 0.0003 & $-$0&0024 & 55 \\
42 & 57973.2308 & 0.0002 & $-$0&0026 & 131 \\
47 & 57973.5153 & 0.0007 & $-$0&0029 & 22 \\
53 & 57973.8572 & 0.0005 & $-$0&0027 & 70 \\
60 & 57974.2562 & 0.0005 & $-$0&0024 & 86 \\
64 & 57974.4852 & 0.0005 & $-$0&0013 & 38 \\
70 & 57974.8316 & 0.0035 & 0&0033 & 14 \\
71 & 57974.8862 & 0.0004 & 0&0009 & 85 \\
77 & 57975.2273 & 0.0003 & 0&0003 & 131 \\
82 & 57975.5133 & 0.0019 & 0&0016 & 14 \\
88 & 57975.8562 & 0.0006 & 0&0027 & 68 \\
\hline
  \multicolumn{6}{l}{\commenta BJD$-$2400000.} \\
  \multicolumn{6}{l}{\commentb Against max $= 2457970.8411 + 0.056960 E$.} \\
  \multicolumn{6}{l}{\commentc Number of points used to determine the maximum.} \\
\end{tabular}
\end{center}
\end{table}

\subsection{NSV 35}\label{obj:nsv35}

   NSV 35 was discovered as a variable (HV 8001 = AN 97.1933)
with a photographic range of 14.5--16.5
during a proper-motion survey \citep{luy38propermotion2}.
The object is located in the region of the Small Magellanic Cloud
and is also given a variable star name of SMC V2.
\citet{aug10CTCVCV2} found a cataclysmic variable
(CTCV J0006$-$6900) at this location and gave a possible
identification with NSV 35.  \citet{aug10CTCVCV2} obtained
an orbital period of 0.0790(12)~d by a radial-velocity study
with rather limited baselines.  \citet{aug10CTCVCV2}
also noted the presence of outbursts in the ASAS-3 data
\citep{ASAS3}.

   A bright outburst was detected on 2017 October 15
by S. Hovell at a visual magnitude of 13.3 (vsnet-alert 21533).
G. Myers detected superhumps, confirming the SU UMa-type
nature of this object.  This superoutburst reached
$V$=12.6 on 2017 October 18.  The times of superhump maxima
are listed in e-table \ref{tab:nsv35oc2017}.
Although there were observations after BJD 2458054 (rapid fading
from the superoutburst), individual superhump maxima could
not be measured.

\begin{figure}
  \begin{center}
    \FigureFile(85mm,110mm){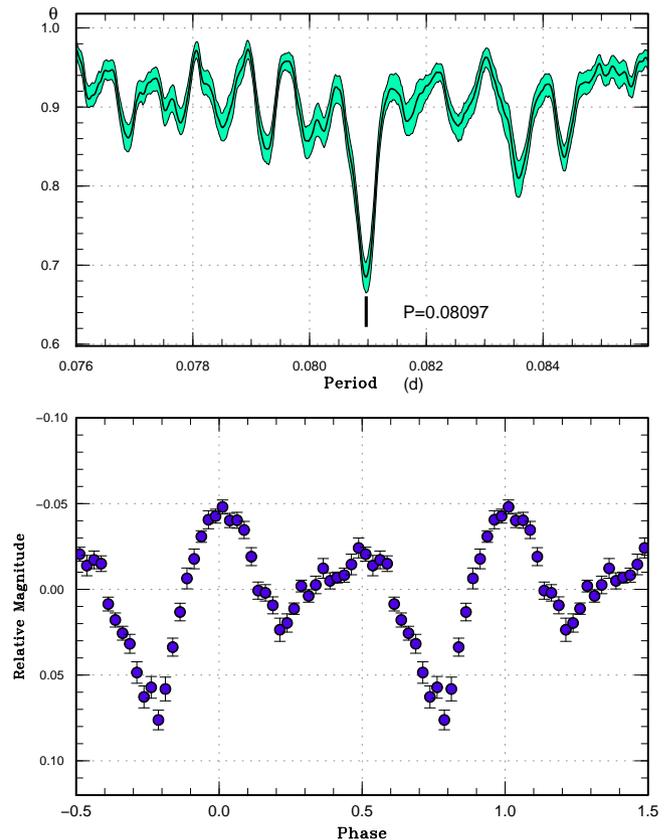}
  \end{center}
  \caption{Superhumps in NSV 35 (2017).
     (Upper): PDM analysis.  The interval BJD 2458044--2458055
     was used.
     (Lower): Phase-averaged profile.}
  \label{fig:nsv35shpdm}
\end{figure}

   The mean profile of the superhumps is given in
e-figure \ref{fig:nsv35shpdm}.  There is a prominent
secondary maximum.  The profile is similar to the one
in the late phase of V344 Lyr in Kepler data
(cf. \cite{woo11v344lyr}).  \citet{woo11v344lyr} interpreted
the secondary maximum as the accretion stream bright spot
sweeping around the rim of the non-axisymmetric disk.
This signal corresponds to the traditional late superhumps,
and it is associated with a high mass-transfer rate
from the secondary.  Both V344 Lyr and NSV 35 have
frequent normal outbursts, and the origin of the secondary
maximum in these systems is likely the same.

\begin{figure*}
  \begin{center}
    \FigureFile(160mm,100mm){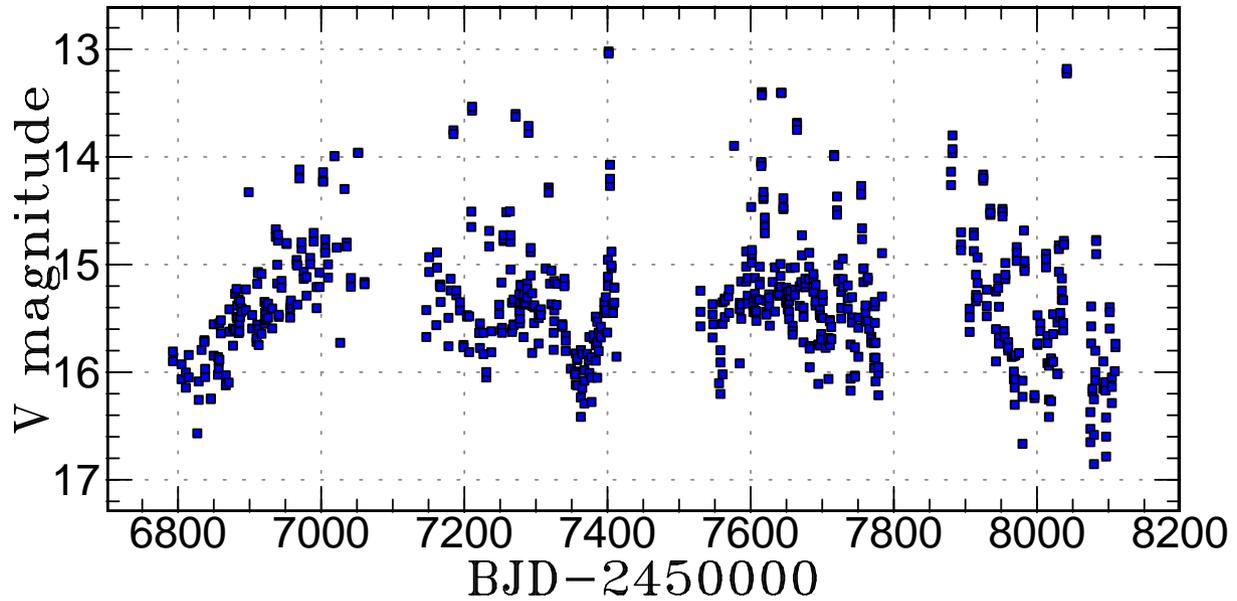}
  \end{center}
  \caption{Long-term light curve of NSV 35 from ASAS-SN
     observations from 2014 through 2017.  The only superoutburst
     was around BJD 2458042, which was not followed by
     ASAS-SN observations.}
  \label{fig:nsv35lc}
\end{figure*}

   This object has been monitored by the ASAS-SN team
since 2014 May (e-figure \ref{fig:nsv35lc}).
There were no definite superoutburst
in the past data.  It may have been either that
this object rarely showed superoutbursts despite frequent
normal outbursts or that past superoutbursts escaped
detection around solar conjunctions.  The mean brightness
(in quiescence) in 2014 gradually rose from 16 mag to
15 mag, suggesting long-term variation of
the mass-transfer rate.  There was also an interval
between 2015 November and 2016 January, when no
outbursts were recorded with gradually varying quiescent
brightness from 16 mag to 15 mag.  This interval was
difficult to interpret as a standstill since
the brightness was similar to that in 2017.
Outbursts in this system may have been somehow suppressed
in certain epochs, and this system deserves a further
detailed study.

\begin{table}
\caption{Superhump maxima of NSV 35 (2017)}\label{tab:nsv35oc2017}
\begin{center}
\begin{tabular}{rp{55pt}p{40pt}r@{.}lr}
\hline
\multicolumn{1}{c}{$E$} & \multicolumn{1}{c}{max\commenta} & \multicolumn{1}{c}{error} & \multicolumn{2}{c}{$O-C$\commentb} & \multicolumn{1}{c}{$N$\commentc} \\
\hline
0 & 58044.9118 & 0.0004 & $-$0&0017 & 94 \\
37 & 58047.9148 & 0.0012 & 0&0030 & 61 \\
58 & 58049.6117 & 0.0022 & $-$0&0018 & 24 \\
59 & 58049.6872 & 0.0057 & $-$0&0073 & 13 \\
61 & 58049.8509 & 0.0126 & $-$0&0058 & 39 \\
62 & 58049.9398 & 0.0007 & 0&0022 & 236 \\
63 & 58050.0220 & 0.0009 & 0&0033 & 230 \\
64 & 58050.1011 & 0.0008 & 0&0013 & 236 \\
65 & 58050.1844 & 0.0010 & 0&0037 & 235 \\
66 & 58050.2644 & 0.0012 & 0&0026 & 133 \\
70 & 58050.5848 & 0.0040 & $-$0&0011 & 17 \\
71 & 58050.6723 & 0.0017 & 0&0054 & 16 \\
72 & 58050.7474 & 0.0044 & $-$0&0005 & 13 \\
83 & 58051.6379 & 0.0028 & $-$0&0014 & 21 \\
84 & 58051.7272 & 0.0050 & 0&0068 & 13 \\
87 & 58051.9587 & 0.0006 & $-$0&0048 & 228 \\
95 & 58052.6119 & 0.0026 & 0&0001 & 24 \\
96 & 58052.6874 & 0.0068 & $-$0&0054 & 13 \\
100 & 58053.0213 & 0.0007 & 0&0043 & 231 \\
109 & 58053.7392 & 0.0049 & $-$0&0070 & 13 \\
112 & 58053.9936 & 0.0033 & 0&0043 & 232 \\
\hline
  \multicolumn{6}{l}{\commenta BJD$-$2400000.} \\
  \multicolumn{6}{l}{\commentb Against max $= 2458044.9136 + 0.081034 E$.} \\
  \multicolumn{6}{l}{\commentc Number of points used to determine the maximum.} \\
\end{tabular}
\end{center}
\end{table}

\subsection{1RXS J161659.5$+$620014}\label{obj:j1616}

   This object (hereafter 1RXS J161659) was initially
identified as an X-ray selected variable
(also known as MASTER OT J161700.81$+$620024.9),
which was first detected in bright state on
2012 September 11 at an unfiltered CCD magnitude
of 14.4 \citep{bal13j1616}.  The SU UMa-type nature
was confirmed during the 2016 outburst \citep{Pdot9}.
For more information, see \citet{Pdot9}.

   The 2017 superoutburst was detected by the ASAS-SN
team at $V$=14.9 on 2017 August 31.  
Subsequent observations detected superhump (vsnet-alert 21429).
The time of superhump maxima are listed in
e-table \ref{tab:j1616oc2017}.
A comparison of the $O-C$ diagram suggests that the 2017
observation recorded mostly stage C.  The 2017
superoutburst apparently was not detected early
enough, since it started fading rapidly on August 6,
only 6~d since the initial outburst detection.

\begin{figure}
  \begin{center}
    \FigureFile(88mm,70mm){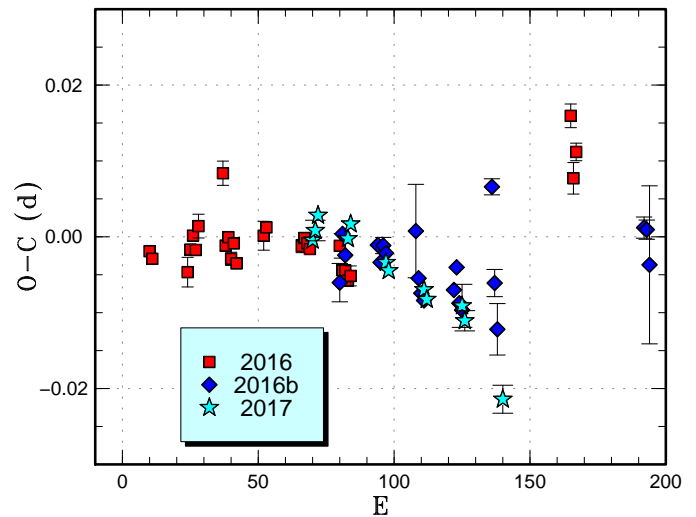}
  \end{center}
  \caption{Comparison of $O-C$ diagrams of 1RXS J161659
  between different superoutbursts.
  A period of 0.07130~d was used to draw this figure.
  Approximate cycle counts ($E$) after the start of the superoutburst
  were used.  The 2017 curve was shifted by 46 cycles to
  best match the others.
  }
  \label{fig:j1616comp2}
\end{figure}

\begin{table}
\caption{Superhump maxima of 1RXS J161659 (2017)}\label{tab:j1616oc2017}
\begin{center}
\begin{tabular}{rp{55pt}p{40pt}r@{.}lr}
\hline
\multicolumn{1}{c}{$E$} & \multicolumn{1}{c}{max\commenta} & \multicolumn{1}{c}{error} & \multicolumn{2}{c}{$O-C$\commentb} & \multicolumn{1}{c}{$N$\commentc} \\
\hline
0 & 57998.4361 & 0.0007 & $-$0&0034 & 75 \\
1 & 57998.5087 & 0.0014 & $-$0&0018 & 52 \\
2 & 57998.5820 & 0.0006 & 0&0005 & 72 \\
13 & 57999.3632 & 0.0008 & 0&0004 & 60 \\
14 & 57999.4365 & 0.0008 & 0&0026 & 74 \\
27 & 58000.3583 & 0.0008 & 0&0010 & 50 \\
28 & 58000.4285 & 0.0006 & 0&0002 & 59 \\
41 & 58001.3529 & 0.0011 & 0&0013 & 55 \\
42 & 58001.4229 & 0.0010 & 0&0003 & 62 \\
55 & 58002.3490 & 0.0028 & 0&0029 & 51 \\
56 & 58002.4183 & 0.0013 & 0&0013 & 75 \\
70 & 58003.4062 & 0.0018 & $-$0&0053 & 58 \\
\hline
  \multicolumn{6}{l}{\commenta BJD$-$2400000.} \\
  \multicolumn{6}{l}{\commentb Against max $= 2457998.4395 + 0.071028 E$.} \\
  \multicolumn{6}{l}{\commentc Number of points used to determine the maximum.} \\
\end{tabular}
\end{center}
\end{table}

\subsection{ASASSN-13ce}\label{obj:asassn13ce}

\begin{figure}
  \begin{center}
    \FigureFile(85mm,110mm){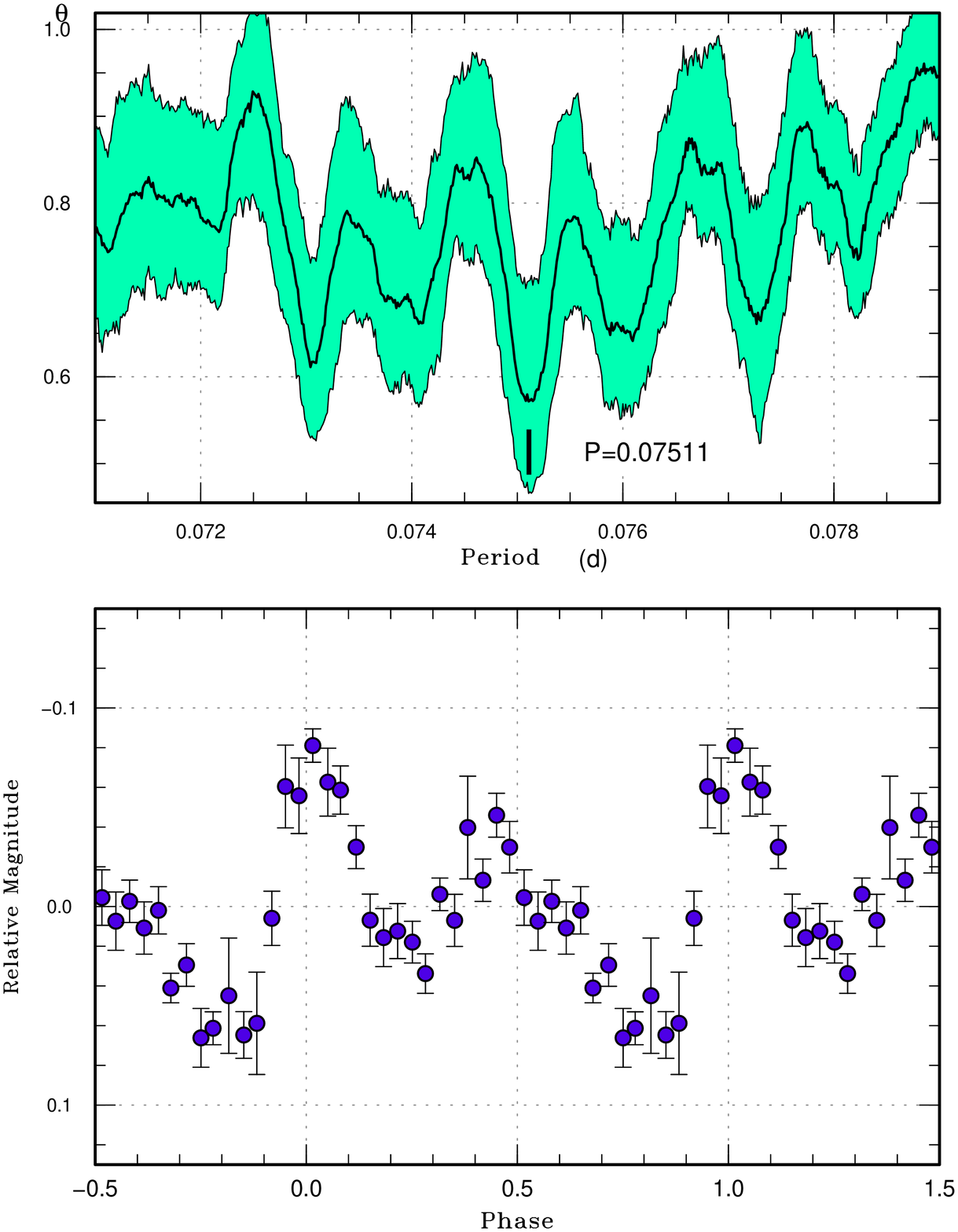}
  \end{center}
  \caption{Superhumps in ASASSN-13ce (2017).
     (Upper): PDM analysis.
     (Lower): Phase-averaged profile.}
  \label{fig:asassn13ceshpdm}
\end{figure}

   This object was detected as a transient
at $V$=16.27 on 2013 August 19 by the ASAS-SN team.
The 2017 outburst was detected by the ASAS-SN team
at $V$=15.55 on 2017 September 21 and announced after
the observation of $V$=15.68 on 2017 September 24.
The object had a blue SDSS counterpart and was
suggested to be an SU UMa-type dwarf nova (vsnet-alert 21466).
Subsequent observations indeed detected superhumps
(vsnet-alert 21471; e-figure \ref{fig:asassn13ceshpdm}).
Although only two superhump maxima were measured,
we could reasonably choose the superhump period
since observations were relatively well spaced.
The maxima were BJD 2458023.4409(16) ($N$=43) and
2458023.6647(11) ($N$=62).  The period determined by
the PDM method was 0.07511(3)~d.

\subsection{ASASSN-13dh}\label{obj:asassn13dh}

   This object was detected as a transient
at $V$=15.61 on 2013 October 2 by the ASAS-SN team.
This outburst was retrospectively detected on
MASTER-Amur images on 2013 September 30 (13.1--13.3
unfiltered CCD magnitudes, vsnet-alert 16504).
This outburst appears to be a normal outburst.

\begin{figure}
  \begin{center}
    \FigureFile(85mm,110mm){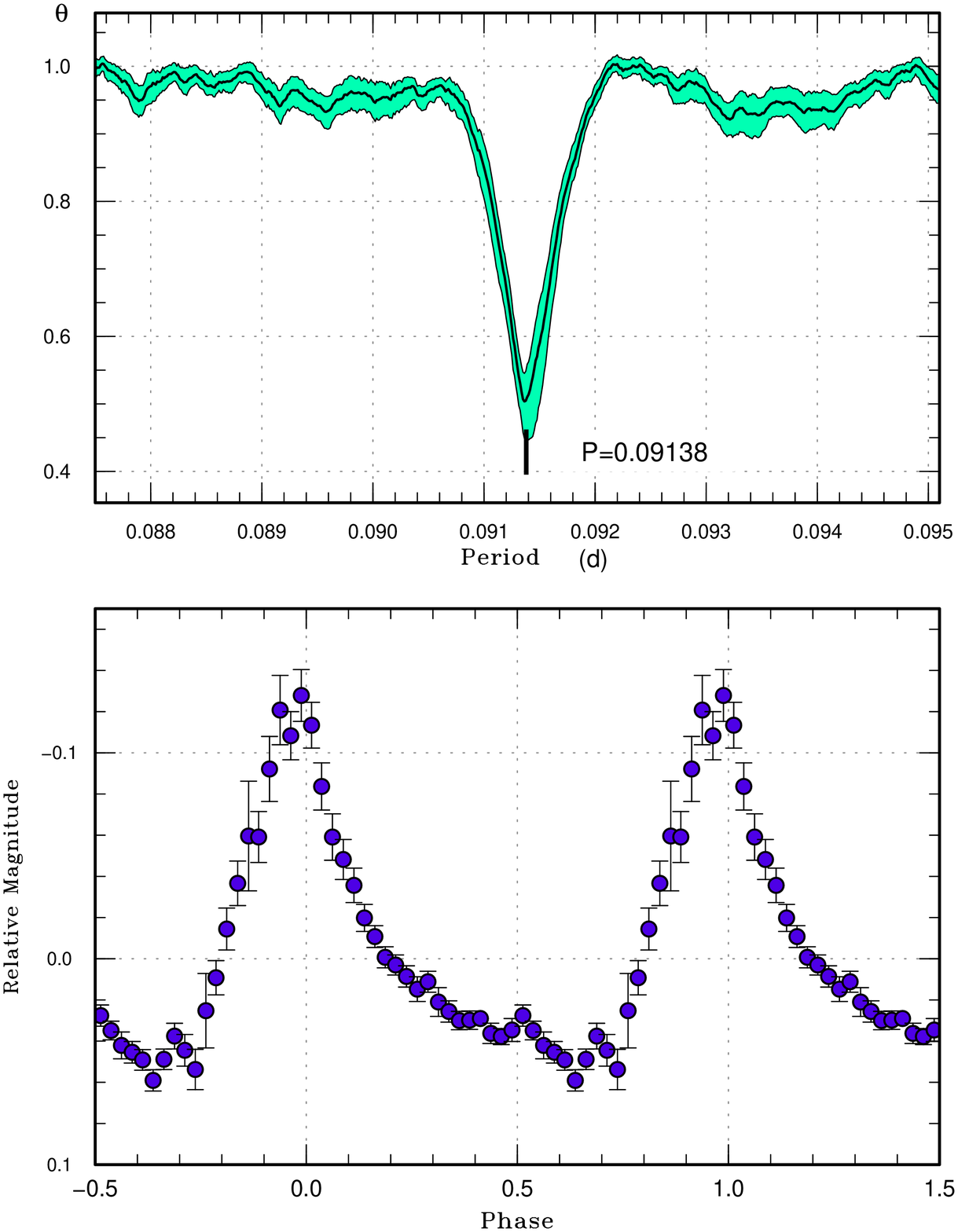}
  \end{center}
  \caption{Superhumps in ASASSN-13dh (2017).
     (Upper): PDM analysis.
     (Lower): Phase-averaged profile.}
  \label{fig:asassn13dhshpdm}
\end{figure}

   The 2017 outburst was detected at an unfiltered
CCD magnitude of 13.62 by E. Muyllaert on
2017 September 8 (cvnet-outburst 7688).
The outburst was immediately confirmed (vsnet-alert 21417)
and time-resolved photometry detected superhumps
(vsnet-alert 21419, 21432; figure \ref{fig:asassn13dhshpdm}).
There was a post-superoutburst rebrightening at $V$=15.87
on 2017 September 25 (ASAS-SN data).

   The times of superhump maxima are listed in
e-table \ref{tab:asassn13dhoc2017}.
Although initial two maxima may have been stage B
superhumps, we could not determine the period.
We consider that the later maxima correspond to
stage C superhumps since they were recorded
before termination of the superoutburst.

   This object have undergone superoutbursts relatively
regularly (e-table \ref{tab:asassn13dhout}).  These
superoutburst can be expressed by a supercycle of 450(8)~d.

\begin{table}
\caption{Superhump maxima of ASASSN-13dh (2017)}\label{tab:asassn13dhoc2017}
\begin{center}
\begin{tabular}{rp{55pt}p{40pt}r@{.}lr}
\hline
\multicolumn{1}{c}{$E$} & \multicolumn{1}{c}{max\commenta} & \multicolumn{1}{c}{error} & \multicolumn{2}{c}{$O-C$\commentb} & \multicolumn{1}{c}{$N$\commentc} \\
\hline
0 & 58005.4910 & 0.0004 & $-$0&0020 & 85 \\
1 & 58005.5821 & 0.0004 & $-$0&0022 & 94 \\
38 & 58008.9676 & 0.0003 & 0&0013 & 91 \\
49 & 58009.9730 & 0.0004 & 0&0013 & 95 \\
59 & 58010.8929 & 0.0034 & 0&0072 & 39 \\
60 & 58010.9771 & 0.0004 & $-$0&0001 & 88 \\
75 & 58012.3493 & 0.0006 & 0&0011 & 81 \\
76 & 58012.4403 & 0.0009 & 0&0007 & 92 \\
92 & 58013.8953 & 0.0073 & $-$0&0068 & 38 \\
93 & 58013.9920 & 0.0006 & $-$0&0015 & 82 \\
100 & 58014.6342 & 0.0019 & 0&0009 & 62 \\
\hline
  \multicolumn{6}{l}{\commenta BJD$-$2400000.} \\
  \multicolumn{6}{l}{\commentb Against max $= 2458005.4929 + 0.091404 E$.} \\
  \multicolumn{6}{l}{\commentc Number of points used to determine the maximum.} \\
\end{tabular}
\end{center}
\end{table}

\begin{table}
\caption{List of superoutbursts (including possible) of ASASSN-13df in the ASAS-SN data}\label{tab:asassn13dhout}
\begin{center}
\begin{tabular}{ccccc}
\hline
Year & Month & Day & max\commenta & $V$ mag \\
\hline
2012 & 10 &  3 & 56204 & 13.4\commentb \\
2013 & 12 & 17 & 56644 & 13.3 \\
2015 &  2 & 11 & 57065 & 14.1 \\
2017 &  9 &  8 & 58005 & 13.3 \\
\hline
  \multicolumn{5}{l}{\commenta JD$-$2400000.} \\
  \multicolumn{5}{l}{\commentb Single detection.} \\
\end{tabular}
\end{center}
\end{table}

\subsection{ASASSN-14ca}\label{obj:asassn14ca}

   This object was detected as a transient at $V$=15.5
on 2014 June 7 by the ASAS-SN team \citep{dav14asassn14caatel6211}.
The object was confirmed to be an SU UMa-type dwarf nova
during the 2015 superoutburst, but the details were
unknown \citep{Pdot8}.  Refer to \citet{Pdot8} for more history.

\begin{figure}
  \begin{center}
    \FigureFile(85mm,110mm){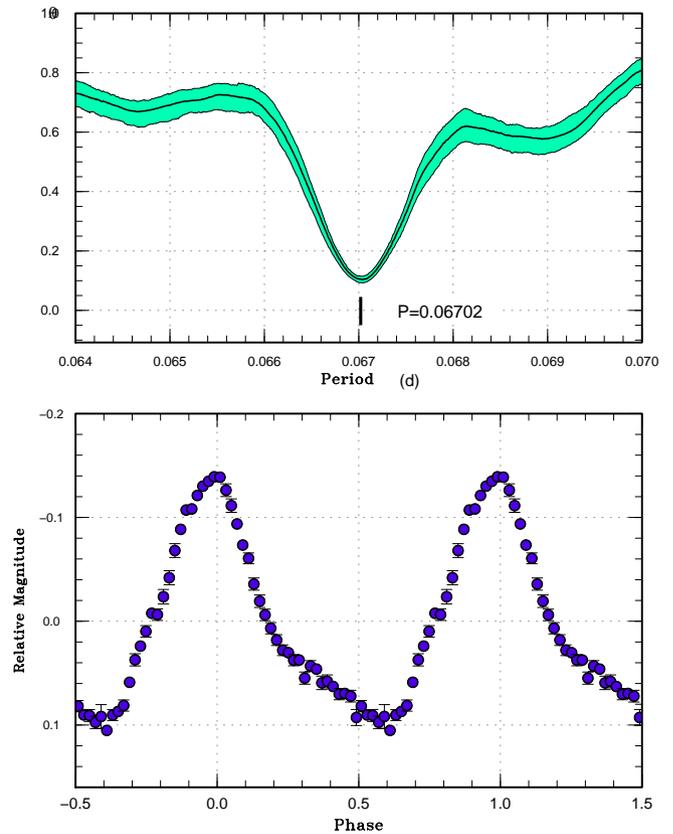}
  \end{center}
  \caption{Superhumps in ASASSN-14ca (2017).
     (Upper): PDM analysis.
     (Lower): Phase-averaged profile.}
  \label{fig:asassn14cashpdm}
\end{figure}

   The 2017 superoutburst was detected by the ASAS-SN
team at $V$=14.87 on 2017 October 17 and was announced
after observation at $V$=14.86 on 2017 October 18.
Subsequent observations detected superhumps
(vsnet-alert 21529, 21538; e-figure \ref{fig:asassn14cashpdm}).
The times of superhump maxima are listed in
e-table \ref{tab:asassn14caoc2017}.

   The object was recorded in superoutburst three times
in the ASAS-SN data (e-table \ref{tab:asassn14caout}).
Two of them showed a separate precursor outburst.
The supercycle is estimated to be 391(20)~d.

\begin{table}
\caption{Superhump maxima of ASASSN-14ca (2017)}\label{tab:asassn14caoc2017}
\begin{center}
\begin{tabular}{rp{55pt}p{40pt}r@{.}lr}
\hline
\multicolumn{1}{c}{$E$} & \multicolumn{1}{c}{max\commenta} & \multicolumn{1}{c}{error} & \multicolumn{2}{c}{$O-C$\commentb} & \multicolumn{1}{c}{$N$\commentc} \\
\hline
0 & 58046.3695 & 0.0003 & $-$0&0009 & 74 \\
1 & 58046.4376 & 0.0004 & 0&0001 & 72 \\
13 & 58047.2422 & 0.0002 & 0&0002 & 173 \\
14 & 58047.3094 & 0.0002 & 0&0004 & 171 \\
15 & 58047.3762 & 0.0002 & 0&0001 & 150 \\
16 & 58047.4435 & 0.0003 & 0&0004 & 124 \\
44 & 58049.3191 & 0.0006 & $-$0&0009 & 48 \\
45 & 58049.3876 & 0.0005 & 0&0005 & 66 \\
\hline
  \multicolumn{6}{l}{\commenta BJD$-$2400000.} \\
  \multicolumn{6}{l}{\commentb Against max $= 2458046.3705 + 0.067036 E$.} \\
  \multicolumn{6}{l}{\commentc Number of points used to determine the maximum.} \\
\end{tabular}
\end{center}
\end{table}

\begin{table}
\caption{List of superoutbursts of ASASSN-14ca in the ASAS-SN data}\label{tab:asassn14caout}
\begin{center}
\begin{tabular}{ccccc}
\hline
Year & Month & Day & max\commenta & $V$ mag \\
\hline
2014 &  8 & 12 & 56882 & 15.4\commentb \\
2015 &  7 & 11 & 57215 & 15.4\commentb \\
2017 & 10 & 16 & 58043 & 14.9 \\
\hline
  \multicolumn{5}{l}{\commenta JD$-$2400000.} \\
  \multicolumn{5}{l}{\commentb Precursor and superoutburst.} \\
\end{tabular}
\end{center}
\end{table}

\subsection{ASASSN-14cr}\label{obj:asassn14cr}

\begin{figure}
  \begin{center}
    \FigureFile(85mm,110mm){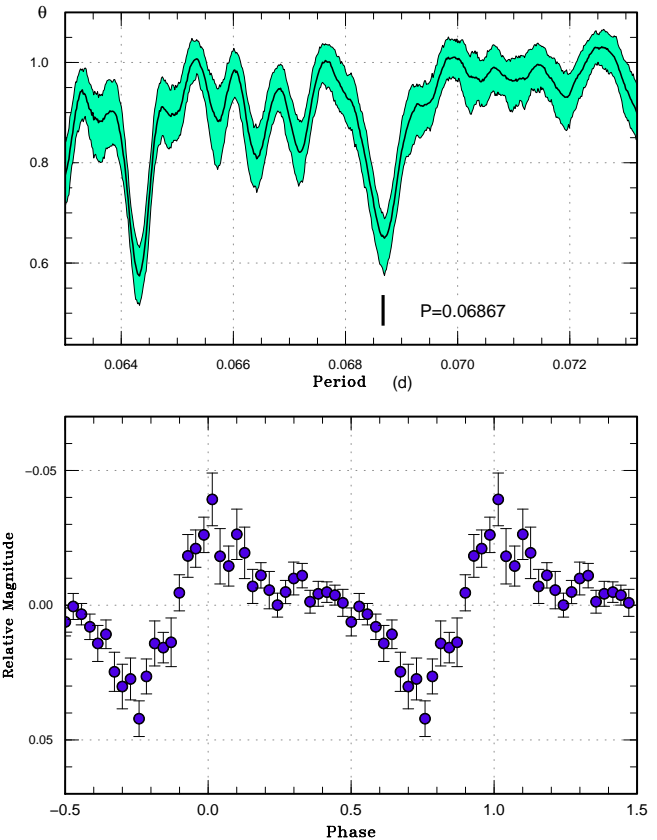}
  \end{center}
  \caption{Superhumps in ASASSN-14cr (2017).
     (Upper): PDM analysis.
     (Lower): Phase-averaged profile.}
  \label{fig:asassn14crshpdm}
\end{figure}

   This object was detected as a transient
at $V$=15.20 on 2014 June 19 by the ASAS-SN team.
Although there were time-resolved observations
during long outbursts on 2016 August 5 (T. Vanmunster),
observations were too short to detect superhumps
(there were also observations on 2015 August 22
by T. Vanmunster during a normal outburst).
The 2017 outburst was detected by the ASAS-SN team
at $V$=15.08 on 2017 July 26 (ASAS-SN data indicated
that the object was already in outburst 8~d before).
Subsequent observations detected superhumps (vsnet-alert 21301).
The times of superhump maxima are listed in
e-table \ref{tab:asassn14croc2017}.  Although
a PDM analysis prefers a period around 0.0646~d
(as in vsnet-alert 21301; e-figure \ref{fig:asassn14crshpdm}),
an alias of 0.0687~d gives much smaller $O-C$ values
and we adopted it.  The superhump stage is likely C
since the observation covered the final part of
the superoutburst.

   The object showed relatively regular superoutbursts
in the past (e-table \ref{tab:asassn14crout}).
The detection 2014 December 13 looks likely a superoutburst
as judged from the brightness.  If this was indeed
a superoutburst, the supercycle is 190(3)~d.

\begin{table}
\caption{Superhump maxima of ASASSN-14cr (2017)}\label{tab:asassn14croc2017}
\begin{center}
\begin{tabular}{rp{55pt}p{40pt}r@{.}lr}
\hline
\multicolumn{1}{c}{$E$} & \multicolumn{1}{c}{max\commenta} & \multicolumn{1}{c}{error} & \multicolumn{2}{c}{$O-C$\commentb} & \multicolumn{1}{c}{$N$\commentc} \\
\hline
0 & 57964.3638 & 0.0010 & $-$0&0003 & 34 \\
1 & 57964.4313 & 0.0013 & $-$0&0015 & 37 \\
2 & 57964.5014 & 0.0035 & $-$0&0001 & 36 \\
3 & 57964.5707 & 0.0007 & 0&0006 & 10 \\
14 & 57965.3267 & 0.0055 & 0&0009 & 23 \\
15 & 57965.3957 & 0.0017 & 0&0012 & 36 \\
16 & 57965.4647 & 0.0038 & 0&0015 & 37 \\
32 & 57966.5577 & 0.0013 & $-$0&0047 & 26 \\
44 & 57967.3920 & 0.0015 & 0&0052 & 37 \\
45 & 57967.4525 & 0.0010 & $-$0&0030 & 37 \\
\hline
  \multicolumn{6}{l}{\commenta BJD$-$2400000.} \\
  \multicolumn{6}{l}{\commentb Against max $= 2457964.3640 + 0.068698 E$.} \\
  \multicolumn{6}{l}{\commentc Number of points used to determine the maximum.} \\
\end{tabular}
\end{center}
\end{table}

\begin{table}
\caption{List of superoutbursts (including possible) of ASASSN-14cr in the ASAS-SN data}\label{tab:asassn14crout}
\begin{center}
\begin{tabular}{ccccc}
\hline
Year & Month & Day & max\commenta & $V$ mag \\
\hline
2014 &  6 & 18 & 56827 & 14.9 \\
2014 & 12 & 13 & 57005 & 15.0\commentb \\
2015 &  6 &  7 & 57181 & 14.9 \\
2016 &  7 & 26 & 57596 & 14.9 \\
2017 &  7 & 17 & 57952 & 15.1 \\
\hline
  \multicolumn{5}{l}{\commenta JD$-$2400000.} \\
  \multicolumn{5}{l}{\commentb Single detection.} \\
\end{tabular}
\end{center}
\end{table}

\subsection{ASASSN-14kb}\label{obj:asassn14kb}

   This object was detected as a transient
at $V$=15.4 on 2014 December 11 by the ASAS-SN team
\citep{pri14asassn14kbatel6688}.
\citet{wyr14asassn14kbatel6690} reported from
OGLE-IV Magellanic System monitoring program observations
that this object (=OGLE-LMC529.30.114) is an eclipsing
SU UMa-type dwarf nova with an orbital period of
0.0681057~d.  Further details of the OGLE-IV
observations were reported in \citet{mro15OGLEDNe},
which provides all photometric observations.\footnote{
CVOM: OGLE Monitoring system of cataclysmic variable stars:
$<$http://ogle.astrouw.edu.pl/ogle4/cvom/cvom.html$>$,
under OGLE-MC-DN-0016.
}
Although the OGLE-IV observations recorded the outburst
pattern characteristic to an SU UMa-type dwarf nova,
no time-resolved photometry was made by the OGLE-IV team.

\begin{figure}
  \begin{center}
    \FigureFile(85mm,110mm){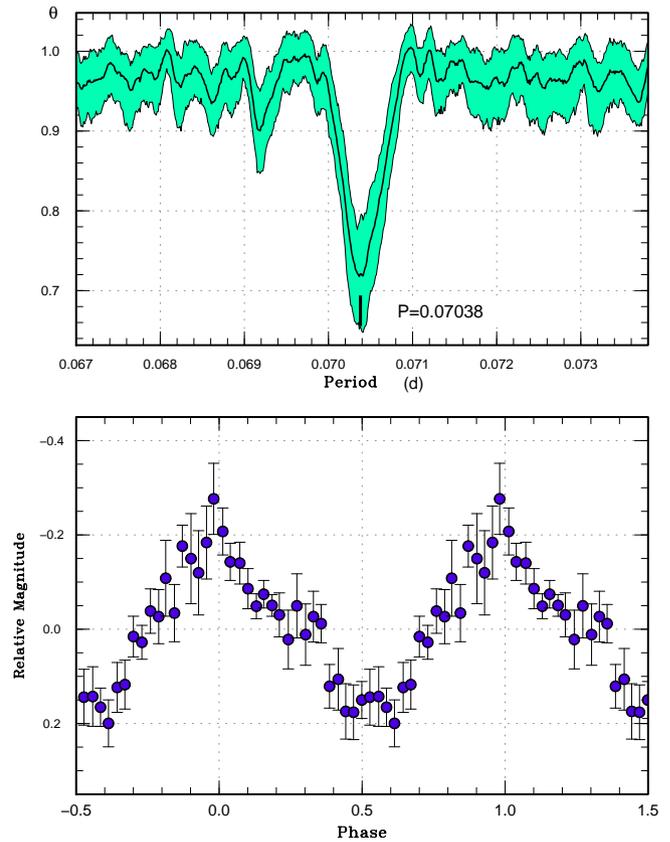}
  \end{center}
  \caption{Superhumps in ASASSN-14kb outside the eclipses (2017).
     (Upper): PDM analysis.
     (Lower): Phase-averaged profile.}
  \label{fig:asassn14kbshpdm}
\end{figure}

   The 2017 superoutburst was detected by the ASAS-SN
team at $V$=15.72 on 2017 April 14.  Subsequent observations
detected superhumps (vsnet-alert 20935;
e-figure \ref{fig:asassn14kbshpdm}).

   By using the OGLE-IV CVOM data (outside outbursts)
and our 2017 observations, we refined the eclipse ephemeris
using the MCMC analysis \citep{Pdot4} as follows:
\begin{equation}
{\rm Min(BJD)} = 2457865.24389(2) + 0.0681057201(10) E .
\label{equ:asassn14kbecl}
\end{equation}
The epoch corresponds to the center of all the observations.

   The times of superhump maxima are listed in
e-table \ref{tab:asassn14kboc2017}.  Since stages were
unclear due to the limited observations (low sampling rate
and low signal-to-noise ratio when the object faded),
we provided a globally averaged period.

   The object shows superoutburst relatively regularly
(e-table \ref{tab:asassn14kbout}).  These superoutburst
can be expressed by a mean supercycle of 150.4(9)~d
with maximum residuals of 42~d.  There appear to have been
variations of supercycles: 162(1)~d for the interval
JD 2455276--2456255 and 144.4(5)~d for the interval
JD 2456255--2457699.  The supercycle lengthened again
until now.

\begin{table}
\caption{Superhump maxima of ASASSN-14kb (2017)}\label{tab:asassn14kboc2017}
\begin{center}
\begin{tabular}{rp{50pt}p{30pt}r@{.}lcr}
\hline
\multicolumn{1}{c}{$E$} & \multicolumn{1}{c}{max\commenta} & \multicolumn{1}{c}{error} & \multicolumn{2}{c}{$O-C$\commentb} & \multicolumn{1}{c}{phase\commentc} & \multicolumn{1}{c}{$N$\commentd} \\
\hline
0 & 57860.5214 & 0.0005 & $-$0&0003 & 0.66 & 20 \\
1 & 57860.5915 & 0.0007 & $-$0&0006 & 0.69 & 14 \\
15 & 57861.5805 & 0.0021 & 0&0025 & 0.21 & 14 \\
43 & 57863.5502 & 0.0017 & 0&0005 & 0.13 & 29 \\
57 & 57864.5366 & 0.0008 & 0&0009 & 0.61 & 29 \\
58 & 57864.6051 & 0.0009 & $-$0&0010 & 0.62 & 14 \\
71 & 57865.5174 & 0.0012 & $-$0&0041 & 0.02 & 22 \\
72 & 57865.5873 & 0.0012 & $-$0&0046 & 0.04 & 13 \\
85 & 57866.5112 & 0.0032 & 0&0038 & 0.61 & 24 \\
86 & 57866.5806 & 0.0031 & 0&0028 & 0.63 & 15 \\
\hline
  \multicolumn{7}{l}{\commenta BJD$-$2400000.} \\
  \multicolumn{7}{l}{\commentb Against max $= 2457860.5217 + 0.070420 E$.} \\
  \multicolumn{7}{l}{\commentc Orbital phase.} \\
  \multicolumn{7}{l}{\commentd Number of points used to determine the maximum.} \\
\end{tabular}
\end{center}
\end{table}

\begin{table*}
\caption{List of past outbursts of ASASSN-14kb}\label{tab:asassn14kbout}
\begin{center}
\begin{tabular}{cccccc}
\hline
Year & Month & Day & max\commenta & mag & Source \\
\hline
2010 &  3 & 22 & 55276 & 15.7I & OGLE-IV \\
2010 &  9 & 15 & 55455 & 15.3I & OGLE-IV \\
2011 &  2 & 14 & 55607 & 15.4I & OGLE-IV \\
2012 &  1 &  3 & 55930 & 15.3I & OGLE-IV \\
2012 & 11 & 23 & 56255 & 15.9I & OGLE-IV \\
2014 & 11 & 10 & 56972 & 15.2V & ASAS-SN \\
2015 &  4 & 13 & 57126 & 15.8V & ASAS-SN, OGLE-IV \\
2015 &  9 &  1 & 57267 & 15.1V & ASAS-SN, OGLE-IV \\
2016 &  1 & 18 & 57406 & 15.7I & OGLE-IV \\
2016 & 11 &  6 & 57699 & 15.5V & ASAS-SN, OGLE-IV \\
2017 &  4 & 14 & 57858 & 15.6V & ASAS-SN \\
2017 &  9 & 24 & 58021 & 15.5V & ASAS-SN \\
\hline
  \multicolumn{5}{l}{\commenta JD$-$2400000.} \\
\end{tabular}
\end{center}
\end{table*}

\subsection{ASASSN-14lk}\label{obj:asassn14lk}

   This object was detected as a transient at $V$=13.48
on 2014 December 1 by ASAS-SN team (vsnet-alert 18032).
It may be identical to NSV 12802 = HV 9672
(see \cite{Pdot7}).  The 2014 observations detected
superhumps \citep{Pdot7}.

\begin{figure}
  \begin{center}
    \FigureFile(85mm,70mm){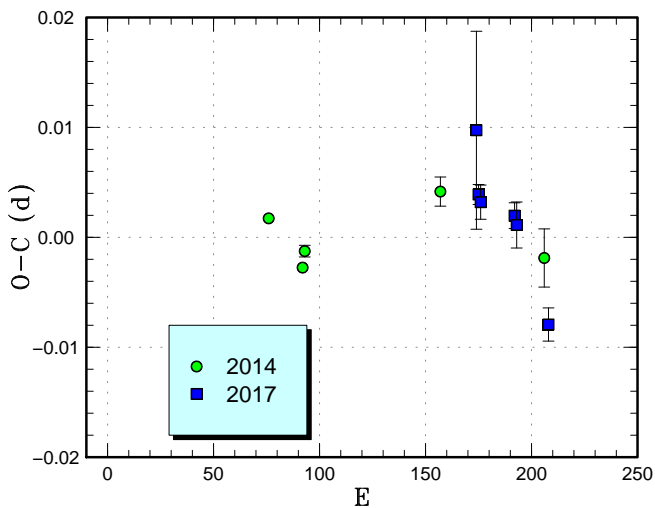}
  \end{center}
  \caption{Comparison of $O-C$ diagrams of ASASSN-14lk between different
  superoutbursts.  A period of 0.06143~d was used to draw this figure.
  Approximate cycle counts ($E$) after the start of the superoutburst
  were used.}
  \label{fig:asassn14lkcomp}
\end{figure}

   The 2017 superoutburst was detected by the ASAS-SN
team at $V$=14.16 on 2017 October 30 and was visually
observed at 14.2 mag on 2017 November 3 by R. Stubbings
(vsnet-alert 21566).  Subsequent observations starting
on 2017 November 9 detected superhumps (vsnet-alert 21575).
The times of superhump maxima are listed in
e-table \ref{tab:asassn14lkoc2017}.
The short superhump period and low amplitudes
(initially 0.11 mag and later decreased to 0.09 mag)
suggest that these superhumps were stage C ones.
The object started fading rapidly on 2017 November 11,
supporting the stage C nature of the superhumps.
The $O-C$ values also support this interpretation
(e-figure \ref{fig:asassn14lkcomp}).

   A list of past superoutbursts recorded in the ASAS-SN
data are listed in e-table \ref{tab:asassn14lkout}.
The supercycle is around 540~d or its $N$-th.

\begin{table}
\caption{Superhump maxima of ASASSN-14lk (2017)}\label{tab:asassn14lkoc2017}
\begin{center}
\begin{tabular}{rp{55pt}p{40pt}r@{.}lr}
\hline
\multicolumn{1}{c}{$E$} & \multicolumn{1}{c}{max\commenta} & \multicolumn{1}{c}{error} & \multicolumn{2}{c}{$O-C$\commentb} & \multicolumn{1}{c}{$N$\commentc} \\
\hline
0 & 58067.2468 & 0.0090 & 0&0031 & 86 \\
1 & 58067.3024 & 0.0009 & $-$0&0024 & 142 \\
2 & 58067.3631 & 0.0016 & $-$0&0027 & 108 \\
18 & 58068.3447 & 0.0012 & 0&0021 & 141 \\
19 & 58068.4053 & 0.0021 & 0&0016 & 73 \\
34 & 58069.3177 & 0.0015 & $-$0&0018 & 94 \\
\hline
  \multicolumn{6}{l}{\commenta BJD$-$2400000.} \\
  \multicolumn{6}{l}{\commentb Against max $= 2458067.2437 + 0.061054 E$.} \\
  \multicolumn{6}{l}{\commentc Number of points used to determine the maximum.} \\
\end{tabular}
\end{center}
\end{table}

\begin{table}
\caption{List of superoutbursts of ASASSN-14lk in the ASAS-SN data}\label{tab:asassn14lkout}
\begin{center}
\begin{tabular}{ccccc}
\hline
Year & Month & Day & max\commenta & $V$ mag \\
\hline
2014 & 11 & 30 & 56992 & 13.6 \\
2016 &  5 & 14 & 57523 & 14.2 \\
2017 & 10 & 30 & 58056 & 14.2 \\
\hline
  \multicolumn{5}{l}{\commenta JD$-$2400000.} \\
\end{tabular}
\end{center}
\end{table}

\subsection{ASASSN-15fu}\label{obj:asassn15fu}

   This object was detected as a transient at $V$=15.6
on 2015 March 27 by the ASAS-SN team.
Observations of superhumps during the 2015 superoutburst
were reported in \citet{Pdot8}.

\begin{figure}
  \begin{center}
    \FigureFile(85mm,70mm){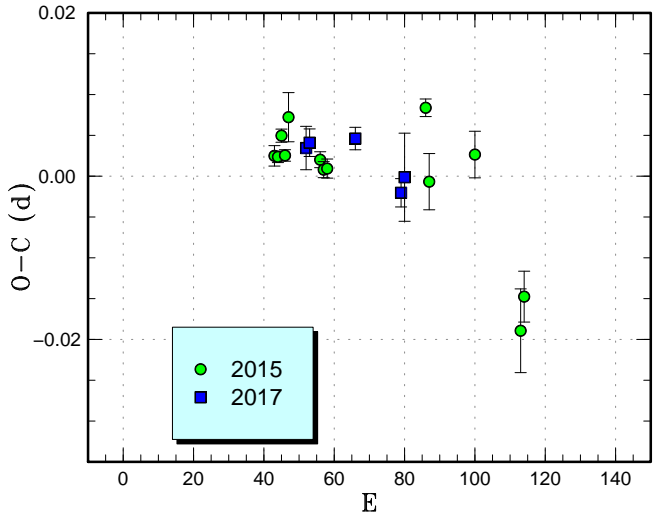}
  \end{center}
  \caption{Comparison of $O-C$ diagrams of ASASSN-15fu between different
  superoutbursts.  A period of 0.07477~d was used to draw this figure.
  Approximate cycle counts ($E$) after the start of the superoutburst
  were used.}
  \label{fig:asassn15fucomp}
\end{figure}

   The 2017 superoutburst was detected by the ASAS-SN
team at $V$=15.4 on 2017 May 29.
The times of superhump maxima are listed in
e-table \ref{tab:asassn15fuoc2017}.  Since the resultant
period was between those of stages B and C in 2015
\citep{Pdot8}, the 2017 observations were likely performed
near stage B-C transition.  A comparison of $O-C$ diagrams
is also consistent with this interpretation
(e-figure \ref{fig:asassn15fucomp}).

   There were three known superoutbursts: 2015 March 26
(JD 2457108), 2015 December 28 (JD 2457385) and
the present one.  The supercycle is around 260--270~d.

\begin{table}
\caption{Superhump maxima of ASASSN-15fu (2017)}\label{tab:asassn15fuoc2017}
\begin{center}
\begin{tabular}{rp{55pt}p{40pt}r@{.}lr}
\hline
\multicolumn{1}{c}{$E$} & \multicolumn{1}{c}{max\commenta} & \multicolumn{1}{c}{error} & \multicolumn{2}{c}{$O-C$\commentb} & \multicolumn{1}{c}{$N$\commentc} \\
\hline
0 & 57906.4982 & 0.0027 & $-$0&0010 & 19 \\
1 & 57906.5736 & 0.0017 & $-$0&0002 & 11 \\
14 & 57907.5461 & 0.0014 & 0&0026 & 13 \\
27 & 57908.5115 & 0.0017 & $-$0&0017 & 23 \\
28 & 57908.5882 & 0.0054 & 0&0004 & 8 \\
\hline
  \multicolumn{6}{l}{\commenta BJD$-$2400000.} \\
  \multicolumn{6}{l}{\commentb Against max $= 2457906.4992 + 0.074592 E$.} \\
  \multicolumn{6}{l}{\commentc Number of points used to determine the maximum.} \\
\end{tabular}
\end{center}
\end{table}

\subsection{ASASSN-15fv}\label{obj:asassn15fv}

\begin{figure}
  \begin{center}
    \FigureFile(85mm,110mm){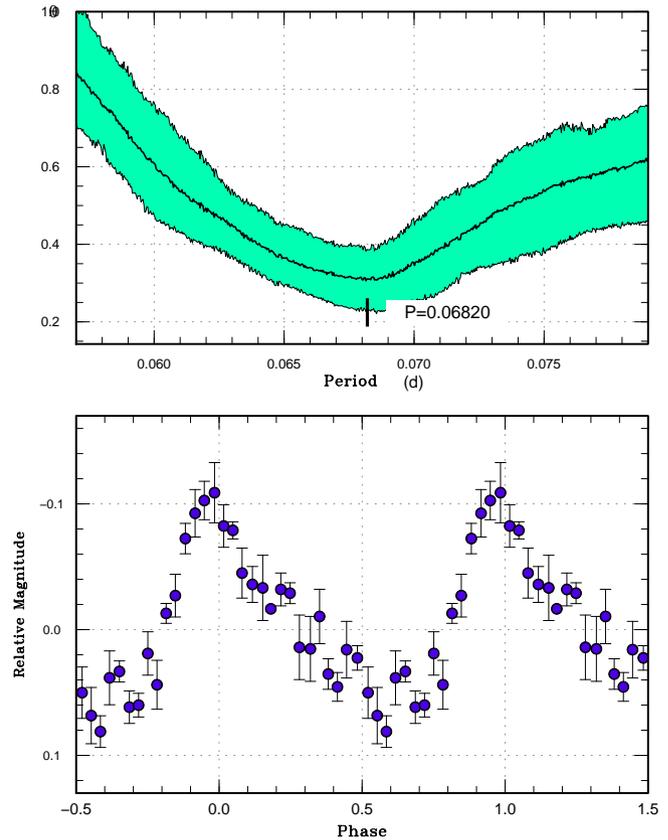}
  \end{center}
  \caption{Ordinary superhumps in ASASSN-15fv (2017).
     (Upper): PDM analysis.
     (Lower): Phase-averaged profile.}
  \label{fig:asassn15fvshpdm}
\end{figure}

   This object was detected as a transient at $V$=15.7
on 2015 March 27 by the ASAS-SN team.
The 2017 outburst was detected by the ASAS-SN team
at $V$=15.76 on 2017 May 19.
Single-night observations detected superhumps
(vsnet-alert 21073; e-figure \ref{fig:asassn15fvshpdm}).
The times of superhump maxima were BJD 57898.4260(6) ($N$=50)
and 2457898.4967(13) ($N$=49).

   There was also a superoutburst in the ASAS-SN data
on 2016 September 26.  The interval between the 2016 and
2017 superoutbursts suggests a supercycle of 235~d.

\subsection{ASASSN-15qu}\label{obj:asassn15qu}

\begin{figure}
  \begin{center}
    \FigureFile(85mm,110mm){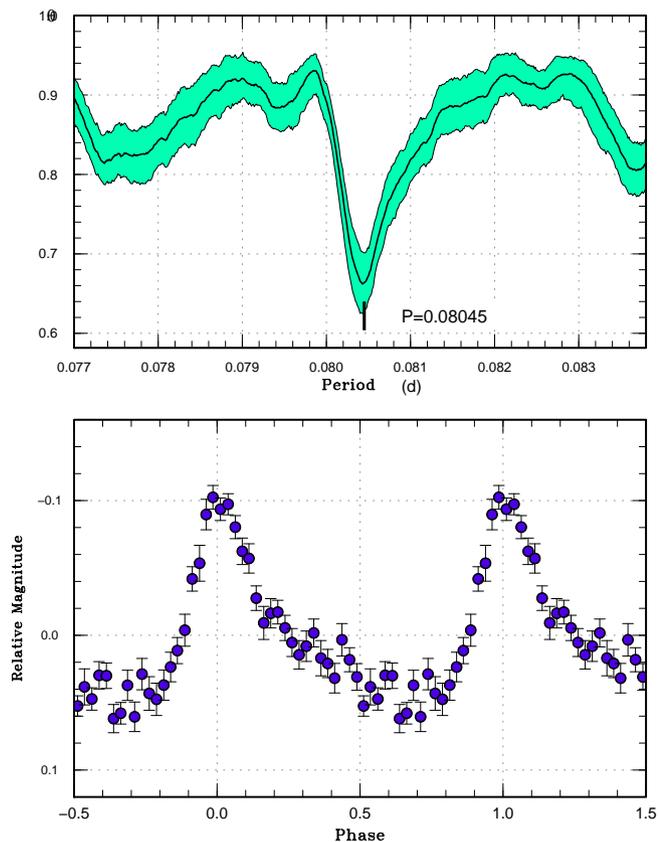}
  \end{center}
  \caption{Ordinary superhumps in ASASSN-15qu (2017).
     (Upper): PDM analysis.
     (Lower): Phase-averaged profile.}
  \label{fig:asassn15qushpdm}
\end{figure}

   This object was detected as a transient at $V$=15.9
on 2015 October 8 by the ASAS-SN team.
The 2017 outburst was detected by the ASAS-SN team
at $V$=14.37 on 2017 August 30.
Subsequent observations detected well-developed superhumps
(vsnet-alert 21390, 21398; e-figure \ref{fig:asassn15qushpdm}).
The times of superhump maxima are listed in
e-table \ref{tab:asassn15quoc2017}.
The period apparently increase between $E$=0 and $E$=28.
We, however, did not adopt $P_{\rm dot}$ from this segment
since the baseline was too short and instead
gave a global $P_{\rm dot}$.

   Although this field was monitored by the ASAS-SN team
since 2014 May, no other outbursts were recorded.
The 2015 outburst was much fainter and it must have been
a normal outburst.  ASAS-3 probably recorded a superoutburst
in 2005 November ($V$=15.14 on 2005 November 3 and
$V$=14.66 on 2015 November 6).

\begin{table}
\caption{Superhump maxima of ASASSN-15qu (2017)}\label{tab:asassn15quoc2017}
\begin{center}
\begin{tabular}{rp{55pt}p{40pt}r@{.}lr}
\hline
\multicolumn{1}{c}{$E$} & \multicolumn{1}{c}{max\commenta} & \multicolumn{1}{c}{error} & \multicolumn{2}{c}{$O-C$\commentb} & \multicolumn{1}{c}{$N$\commentc} \\
\hline
0 & 57998.4052 & 0.0005 & $-$0&0025 & 185 \\
1 & 57998.4880 & 0.0006 & $-$0&0001 & 184 \\
2 & 57998.5677 & 0.0005 & $-$0&0009 & 185 \\
15 & 57999.6124 & 0.0016 & $-$0&0021 & 19 \\
16 & 57999.6925 & 0.0016 & $-$0&0024 & 19 \\
26 & 58000.5036 & 0.0019 & 0&0043 & 163 \\
27 & 58000.5817 & 0.0023 & 0&0019 & 188 \\
28 & 58000.6636 & 0.0024 & 0&0033 & 51 \\
40 & 58001.6261 & 0.0021 & 0&0004 & 23 \\
65 & 58003.6410 & 0.0023 & 0&0041 & 24 \\
77 & 58004.6038 & 0.0016 & 0&0015 & 19 \\
78 & 58004.6752 & 0.0023 & $-$0&0075 & 20 \\
\hline
  \multicolumn{6}{l}{\commenta BJD$-$2400000.} \\
  \multicolumn{6}{l}{\commentb Against max $= 2457998.4077 + 0.080449 E$.} \\
  \multicolumn{6}{l}{\commentc Number of points used to determine the maximum.} \\
\end{tabular}
\end{center}
\end{table}

\subsection{ASASSN-17ei}\label{obj:asassn17ei}

\begin{figure}
  \begin{center}
    \FigureFile(85mm,110mm){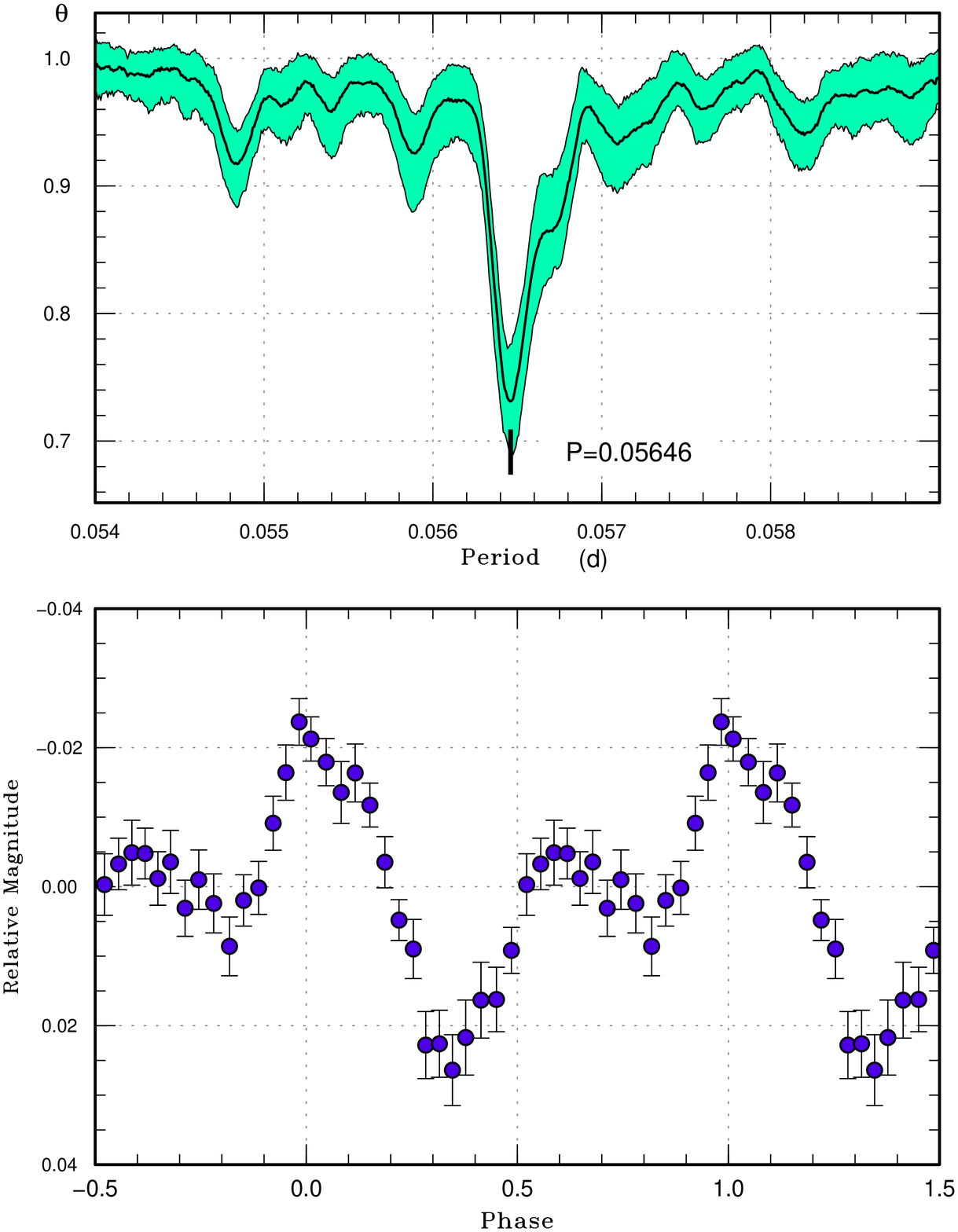}
  \end{center}
  \caption{Early superhumps in ASASSN-17ei (2017).
     (Upper): PDM analysis.
     (Lower): Phase-averaged profile.}
  \label{fig:asassn17eieshpdm}
\end{figure}

\begin{figure}
  \begin{center}
    \FigureFile(85mm,110mm){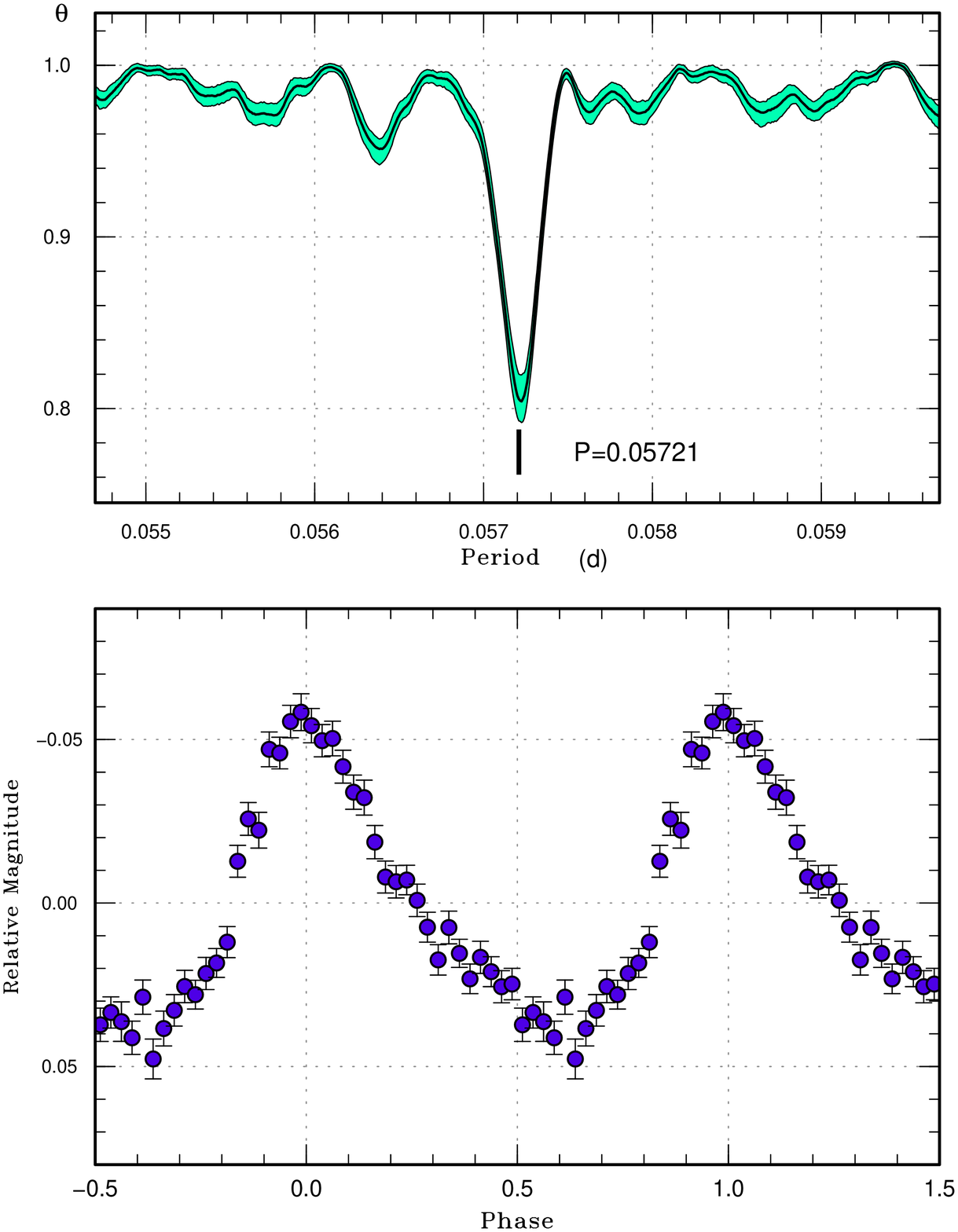}
  \end{center}
  \caption{Ordinary superhumps in ASASSN-17ei (2017).
     (Upper): PDM analysis.
     (Lower): Phase-averaged profile.}
  \label{fig:asassn17eishpdm}
\end{figure}

   This object was detected as a transient
at $V$=13.0 on 2017 April 1 by the ASAS-SN team
(cf. vsnet-alert 20849).
Subsequent observations detected low-amplitude modulations
(vsnet-alert 20857), which were later identified to be
early superhumps (vsnet-alert 20867;
e-figure \ref{fig:asassn17eieshpdm}).
The object started to show ordinary superhumps
on 2017 April 9--10 (vsnet-alert 20894, 20900;
e-figure \ref{fig:asassn17eishpdm}).

\begin{figure*}
  \begin{center}
    \FigureFile(160mm,200mm){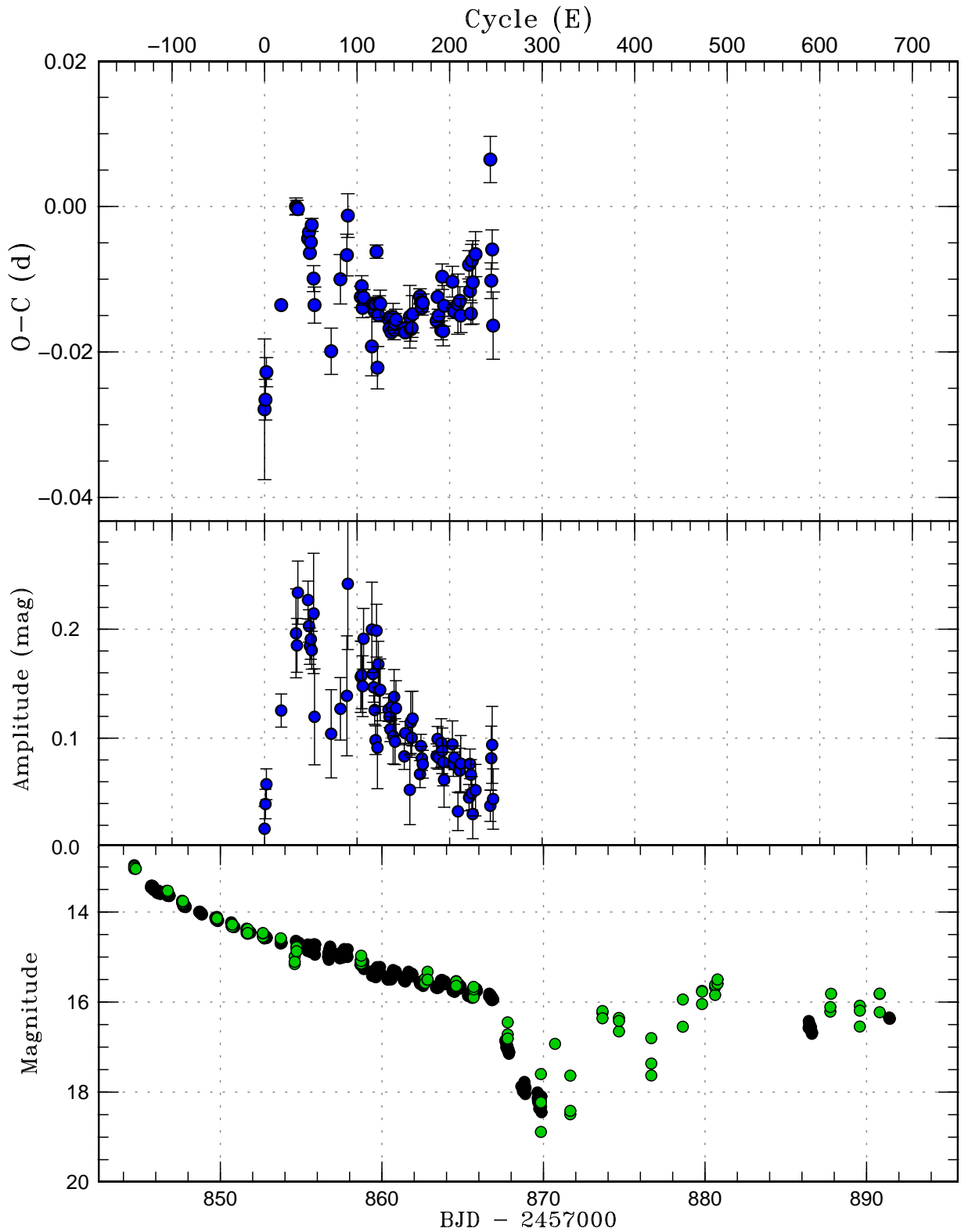}
  \end{center}
  \caption{$O-C$ diagram of superhumps in ASASSN-17ei (2017).
     (Upper:) $O-C$ diagram.
     We used a period of 0.05728~d for calculating the $O-C$ residuals.
     (Middle:) Amplitudes of superhumps.
     (Lower:) Light curve.  The black filled circles are our CCD data
     binned to 0.019~d.  The green filled circles are ASAS-SN
     $V$ measurements.  A constant flux corresponding to
     $V$=16.64 has been subtracted to best match time-series
     CCD observations.
  }
  \label{fig:asassn17eihumpall}
\end{figure*}

   The times of superhump maxima are listed in
e-table \ref{tab:asassn17eioc2017}.  The data between
$E$=53 and $E$=90 were not very good and the times
of superhump maxima had relatively large uncertainties.
It is, however, apparent that the superhump stage
was already B at $E$=34--36, when the amplitudes
of the superhumps may have not yet reached the maximum.
An $O-C$ analysis has shown that times of low-amplitude
maxima on April 9 were on the smooth extension of
the supposed stage A before $E$=34 and we consider 
they were already stage A superhump rather than
early superhumps (cf. e-figure \ref{fig:asassn17eihumpall}).
This identification appears to be
supported by the duration of the phase of
stage A superhumps (2~d), which is a normal value for
a WZ Sge-type dwarf nova.

   The period of early superhumps with the PDM method
was 0.056460(9)~d.  The value for $\epsilon^*$ of
stage A superhumps was 0.0275(9), which corresponds to
$q$=0.074(3).  This small $q$ is consistent with
the small $P_{\rm dot}$ for stage B superhumps
(cf. \cite{kat15wzsge}).

   The entire light curve (e-figure \ref{fig:asassn17eihumpall})
indicates that there was a plateau-type rebrightening.
Note that original ASAS-SN measurements gave
brighter values than our CCD measurements when the object
was faint.  This was due to the contamination since
the aperture size is large in ASAS-SN measurements.
We corrected this difference by subtracting a constant
contribution of $V$=16.64 from neighboring stars
(this value was determined to make the best match to
our CCD measurements).
There was some indication that the object faded slightly
after the initial rise to the rebrightening phase,
particularly at around BJD 2457873.5.
This may be evidence of damping oscillation
as seen in some WZ Sge-type dwarf novae
(ASASSN-15po: \cite{nam17asassn15po}, ASASSN-17el:
in this paper).

   The ASAS-SN started to observe this field on 2014
April 29 and no past outburst was detected.
ASAS-3 recorded no outburst between 2001 and 2009.

\begin{table*}
\caption{Superhump maxima of ASASSN-17ei (2017)}\label{tab:asassn17eioc2017}
\begin{center}
\begin{tabular}{rp{55pt}p{40pt}r@{.}lrrp{55pt}p{40pt}r@{.}lr}
\hline
\multicolumn{1}{c}{$E$} & \multicolumn{1}{c}{max\commenta} & \multicolumn{1}{c}{error} & \multicolumn{2}{c}{$O-C$\commentb} & \multicolumn{1}{c}{$N$\commentc} & \multicolumn{1}{c}{$E$} & \multicolumn{1}{c}{max\commenta} & \multicolumn{1}{c}{error} & \multicolumn{2}{c}{$O-C$\commentb} & \multicolumn{1}{c}{$N$\commentc} \\
\hline
0 & 57852.7066 & 0.0097 & $-$0&0153 & 20 & 140 & 57860.7367 & 0.0013 & $-$0&0045 & 11 \\
1 & 57852.7652 & 0.0028 & $-$0&0139 & 22 & 141 & 57860.7949 & 0.0014 & $-$0&0036 & 21 \\
2 & 57852.8263 & 0.0020 & $-$0&0101 & 22 & 142 & 57860.8527 & 0.0014 & $-$0&0031 & 24 \\
18 & 57853.7520 & 0.0008 & $-$0&0009 & 22 & 151 & 57861.3669 & 0.0010 & $-$0&0043 & 128 \\
34 & 57854.6820 & 0.0012 & 0&0126 & 18 & 152 & 57861.4237 & 0.0007 & $-$0&0048 & 132 \\
35 & 57854.7392 & 0.0010 & 0&0125 & 22 & 157 & 57861.7123 & 0.0043 & $-$0&0027 & 11 \\
36 & 57854.7962 & 0.0009 & 0&0122 & 29 & 158 & 57861.7678 & 0.0016 & $-$0&0044 & 14 \\
47 & 57855.4222 & 0.0006 & 0&0082 & 131 & 159 & 57861.8253 & 0.0013 & $-$0&0042 & 23 \\
48 & 57855.4804 & 0.0005 & 0&0090 & 132 & 160 & 57861.8845 & 0.0026 & $-$0&0023 & 14 \\
49 & 57855.5348 & 0.0006 & 0&0062 & 132 & 168 & 57862.3452 & 0.0010 & 0&0001 & 111 \\
50 & 57855.5936 & 0.0007 & 0&0077 & 133 & 169 & 57862.4016 & 0.0009 & $-$0&0007 & 129 \\
51 & 57855.6532 & 0.0009 & 0&0101 & 113 & 170 & 57862.4581 & 0.0010 & $-$0&0015 & 132 \\
53 & 57855.7604 & 0.0018 & 0&0027 & 21 & 171 & 57862.5161 & 0.0012 & $-$0&0008 & 132 \\
54 & 57855.8140 & 0.0025 & $-$0&0010 & 36 & 186 & 57863.3728 & 0.0009 & $-$0&0033 & 130 \\
72 & 57856.8387 & 0.0032 & $-$0&0073 & 36 & 187 & 57863.4334 & 0.0008 & 0&0000 & 132 \\
82 & 57857.4214 & 0.0034 & 0&0026 & 48 & 188 & 57863.4881 & 0.0009 & $-$0&0026 & 116 \\
89 & 57857.8257 & 0.0029 & 0&0059 & 36 & 191 & 57863.6579 & 0.0013 & $-$0&0046 & 19 \\
90 & 57857.8884 & 0.0030 & 0&0113 & 22 & 192 & 57863.7226 & 0.0018 & 0&0028 & 10 \\
104 & 57858.6792 & 0.0015 & 0&0001 & 17 & 193 & 57863.7724 & 0.0020 & $-$0&0047 & 14 \\
105 & 57858.7379 & 0.0014 & 0&0016 & 11 & 194 & 57863.8331 & 0.0028 & $-$0&0012 & 17 \\
106 & 57858.7922 & 0.0013 & $-$0&0014 & 15 & 203 & 57864.3520 & 0.0021 & 0&0021 & 86 \\
107 & 57858.8509 & 0.0011 & 0&0000 & 18 & 204 & 57864.4056 & 0.0011 & $-$0&0016 & 132 \\
116 & 57859.3597 & 0.0040 & $-$0&0067 & 64 & 205 & 57864.4624 & 0.0010 & $-$0&0020 & 132 \\
117 & 57859.4228 & 0.0005 & $-$0&0009 & 132 & 209 & 57864.6926 & 0.0041 & $-$0&0010 & 9 \\
118 & 57859.4798 & 0.0007 & $-$0&0012 & 131 & 211 & 57864.8076 & 0.0019 & $-$0&0005 & 15 \\
119 & 57859.5364 & 0.0008 & $-$0&0019 & 132 & 212 & 57864.8628 & 0.0023 & $-$0&0026 & 16 \\
120 & 57859.5945 & 0.0011 & $-$0&0010 & 112 & 221 & 57865.3853 & 0.0019 & 0&0044 & 131 \\
121 & 57859.6591 & 0.0009 & 0&0063 & 49 & 222 & 57865.4390 & 0.0013 & 0&0008 & 132 \\
122 & 57859.7005 & 0.0029 & $-$0&0096 & 11 & 223 & 57865.4932 & 0.0015 & $-$0&0023 & 132 \\
123 & 57859.7650 & 0.0009 & $-$0&0024 & 11 & 224 & 57865.5577 & 0.0022 & 0&0050 & 129 \\
124 & 57859.8239 & 0.0010 & $-$0&0007 & 17 & 225 & 57865.6120 & 0.0057 & 0&0020 & 82 \\
125 & 57859.8811 & 0.0019 & $-$0&0009 & 12 & 228 & 57865.7878 & 0.0031 & 0&0059 & 15 \\
134 & 57860.3945 & 0.0006 & $-$0&0030 & 132 & 244 & 57866.7173 & 0.0032 & 0&0189 & 15 \\
135 & 57860.4505 & 0.0006 & $-$0&0043 & 132 & 245 & 57866.7579 & 0.0025 & 0&0022 & 17 \\
136 & 57860.5094 & 0.0008 & $-$0&0027 & 132 & 246 & 57866.8194 & 0.0027 & 0&0065 & 21 \\
137 & 57860.5645 & 0.0007 & $-$0&0048 & 131 & 247 & 57866.8663 & 0.0046 & $-$0&0040 & 23 \\
139 & 57860.6812 & 0.0019 & $-$0&0027 & 14 & \multicolumn{1}{c}{--} & \multicolumn{1}{c}{--} & \multicolumn{1}{c}{--} & \multicolumn{2}{c}{--} & \multicolumn{1}{c}{--}\\
\hline
  \multicolumn{12}{l}{\commenta BJD$-$2400000.} \\
  \multicolumn{12}{l}{\commentb Against max $= 2457852.7218 + 0.057281 E$.} \\
  \multicolumn{12}{l}{\commentc Number of points used to determine the maximum.} \\
\end{tabular}
\end{center}
\end{table*}

\subsection{ASASSN-17el}\label{obj:asassn17el}

\begin{figure}
  \begin{center}
    \FigureFile(85mm,110mm){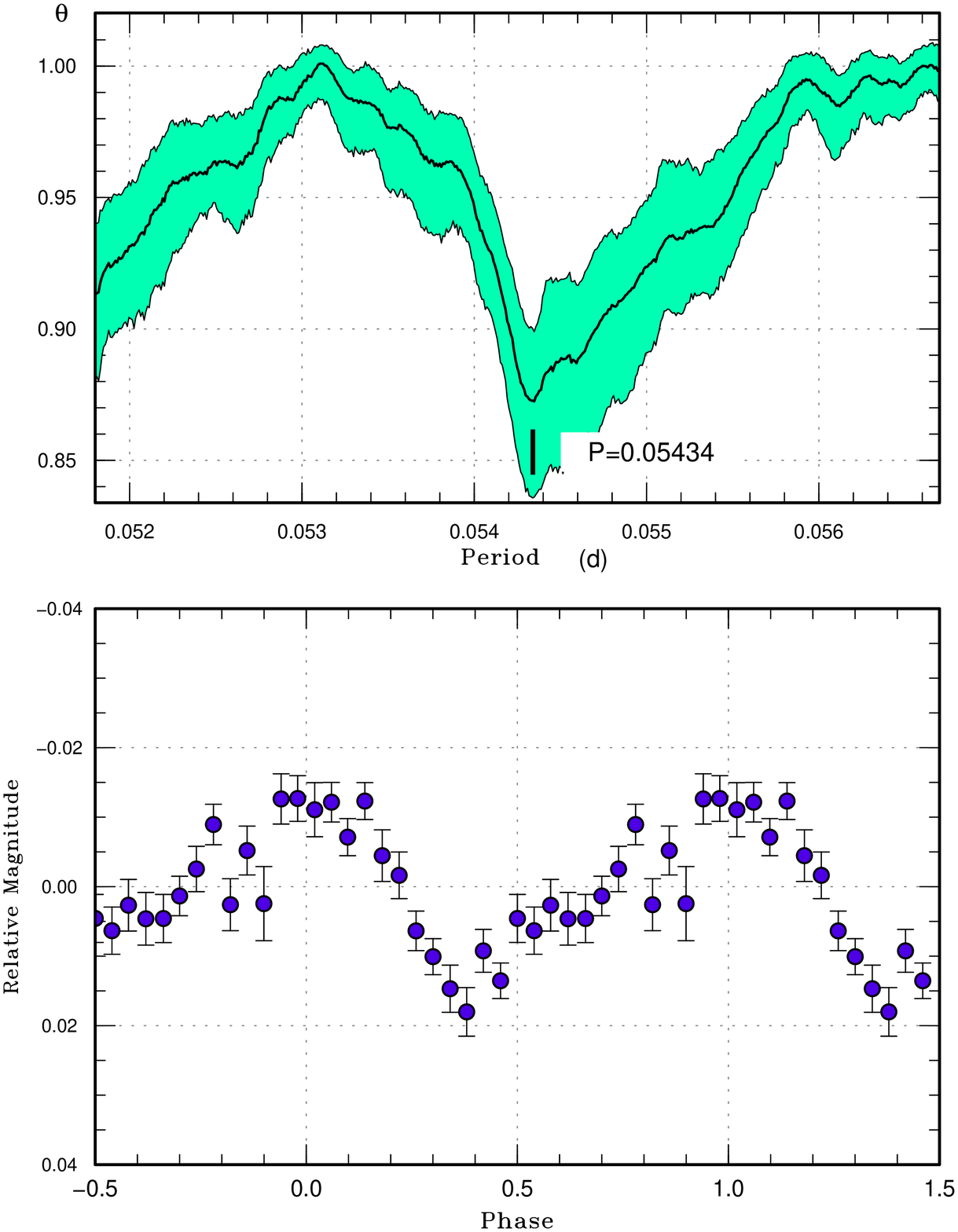}
  \end{center}
  \caption{Early superhumps in ASASSN-17el (2017).
     (Upper): PDM analysis.
     (Lower): Phase-averaged profile.}
  \label{fig:asassn17eleshpdm}
\end{figure}

\begin{figure}
  \begin{center}
    \FigureFile(85mm,110mm){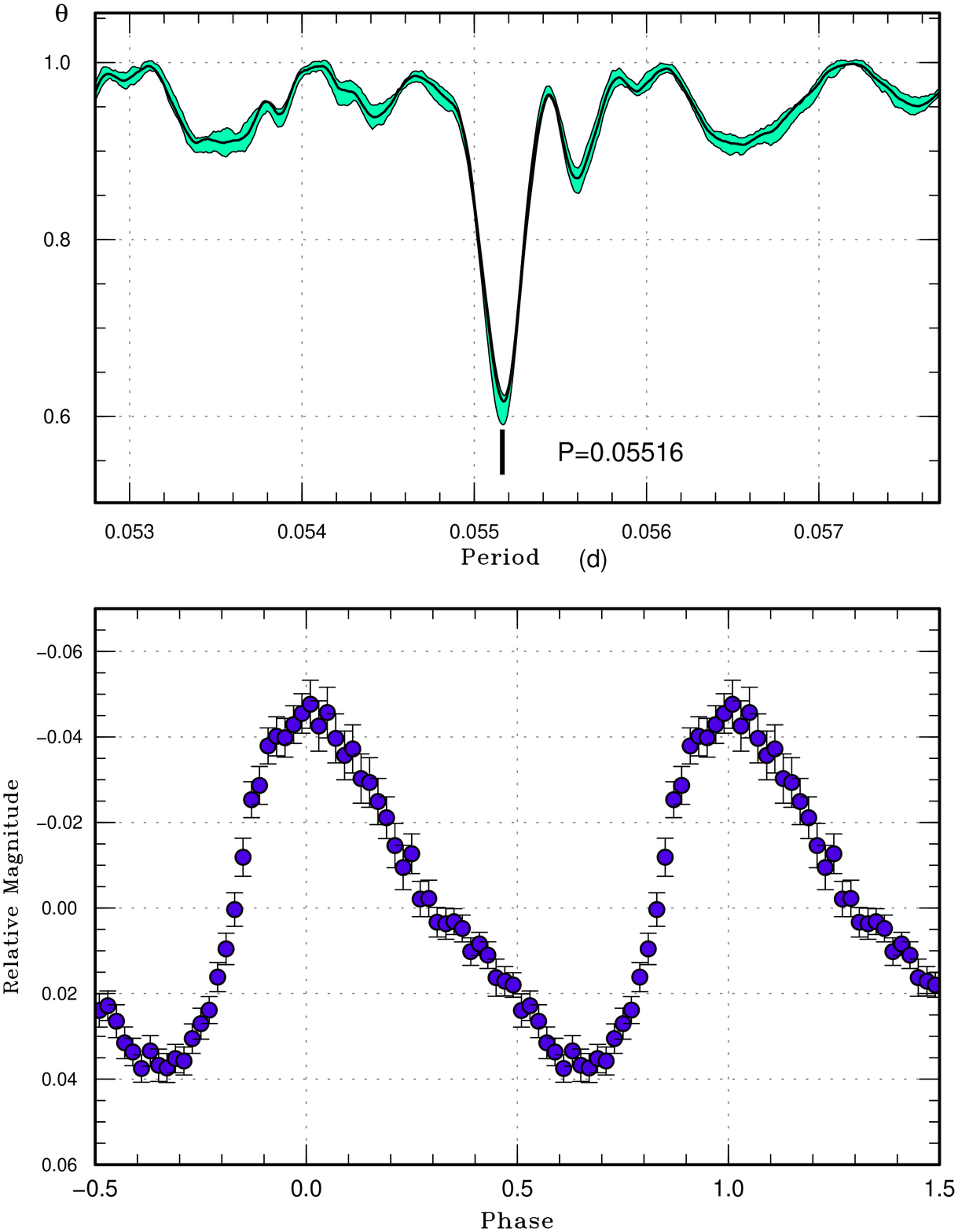}
  \end{center}
  \caption{Ordinary superhumps in ASASSN-17el (2017).
     (Upper): PDM analysis.
     (Lower): Phase-averaged profile.}
  \label{fig:asassn17elshpdm}
\end{figure}

\begin{figure*}
  \begin{center}
    \FigureFile(160mm,200mm){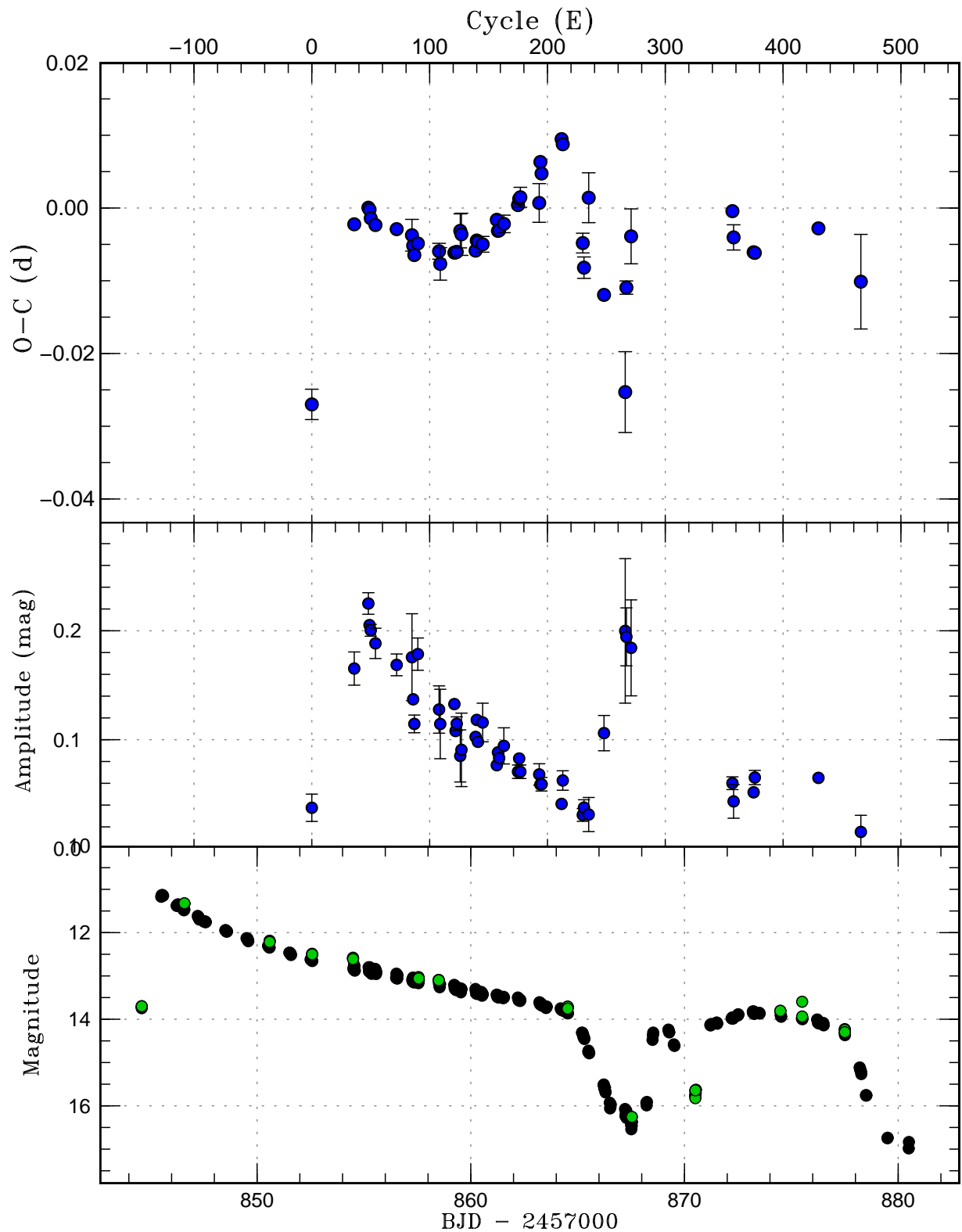}
  \end{center}
  \caption{$O-C$ diagram of superhumps in ASASSN-17el (2017).
     (Upper:) $O-C$ diagram.
     We used a period of 0.055132~d for calculating the $O-C$ residuals.
     (Middle:) Amplitudes of superhumps.
     (Lower:) Light curve.  The black filled circles are our CCD data
     binned to 0.018~d.  The green filled circles are ASAS-SN
     $V$ measurements.  There were also our CCD observations
     at the same locations of a dip in ASAS-SN measurements at
     BJD 2457870.5.
  }
  \label{fig:asassn17elhumpall}
\end{figure*}

   This object was detected as a transient
at $V$=13.7 on 2017 April 1 by the ASAS-SN team
(cf. vsnet-alert 20849).  The object further brightened
to an unfiltered CCD magnitude of 11.1 on April 2.
The object immediately showed likely early superhumps
(vsnet-alert 20866; e-figure \ref{fig:asassn17eleshpdm}).
These modulations were confirmed to be early superhumps
by the detection of ordinary superhumps
(vsnet-alert 20901, 20936; e-figure \ref{fig:asassn17elshpdm}).
The object was thereby confirmed to be a WZ Sge-type dwarf nova.
The times of superhump maxima are listed in
e-table \ref{tab:asassn17eloc2017}, which also includes
post-superoutburst superhumps.
There was typical stage B with a positive $P_{\rm dot}$
between $E$=48 and $E$=213
(e-figure \ref{fig:asassn17elhumpall}).  It looks like that
the epoch $E$=36 was obtained during the final phase
of stage A.  Considering the typical duration of stage A
of WZ Sge-type dwarf novae, we identified a hump maximum
on BJD 2457852 as appearance of stage A superhumps.
As the superoutburst plateau terminated, there was
a apparent phase jump after $E$=213.  We listed a period
derived from $E$=230--271 as stage C superhumps.
There were also superhumps during the rebrightening phase
(vsnet-alert 20961, 20966, 20974)
and the mean period was 0.05509(3)~d ($E$=357--466).

   The period of early superhumps was determined
to be 0.05434(3)~d by the PDM method.  The period of
stage A superhumps gave $\epsilon^*$=0.0265(10),
which corresponds to $q$=0.071(3).

   The light curve was composed of the main superoutburst,
the main dip (BJD 2457865--2457867), a short rebrightening
(BJD 2457868--2457869), a smaller dip (BJD 2457870.5) and
a plateau-type rebrightening (BJD 2457871--2457878)
(e-figure \ref{fig:asassn17elhumpall}).  This type of
phenomenon (damping oscillation when entering the plateau-type
rebrightening) was seen in the WZ Sge-type dwarf nova
ASASSN-15po \citep{nam17asassn15po}.

   The object was recorded in outburst in 2006 by ASAS-3
(D. Denisenko, vsnet-alert 20853).  It was also a superoutburst
and reached at least $V$=11.6 (the peak may have been
missed).  This outburst was not accompanied by a rebrightening
as far as the ASAS-3 data could tell.
There were no other outbursts in the 9-year coverage
in the ASAS-3 data and ASAS-SN observations since
2014 April.  The outburst cycle length of 11~yr is
typical for a WZ Sge-type dwarf nova \citep{kat15wzsge}.

\begin{table}
\caption{Superhump maxima of ASASSN-17el (2017)}\label{tab:asassn17eloc2017}
\begin{center}
\begin{tabular}{rp{55pt}p{40pt}r@{.}lr}
\hline
\multicolumn{1}{c}{$E$} & \multicolumn{1}{c}{max\commenta} & \multicolumn{1}{c}{error} & \multicolumn{2}{c}{$O-C$\commentb} & \multicolumn{1}{c}{$N$\commentc} \\
\hline
0 & 57852.5344 & 0.0021 & $-$0&0231 & 29 \\
36 & 57854.5439 & 0.0006 & 0&0017 & 18 \\
48 & 57855.2077 & 0.0008 & 0&0039 & 67 \\
49 & 57855.2626 & 0.0001 & 0&0037 & 127 \\
50 & 57855.3166 & 0.0002 & 0&0025 & 126 \\
54 & 57855.5362 & 0.0005 & 0&0016 & 17 \\
72 & 57856.5280 & 0.0004 & 0&0010 & 15 \\
85 & 57857.2438 & 0.0022 & 0&0002 & 39 \\
86 & 57857.2975 & 0.0002 & $-$0&0013 & 127 \\
87 & 57857.3514 & 0.0005 & $-$0&0026 & 114 \\
90 & 57857.5184 & 0.0005 & $-$0&0010 & 17 \\
108 & 57858.5097 & 0.0011 & $-$0&0020 & 14 \\
109 & 57858.5631 & 0.0022 & $-$0&0038 & 12 \\
121 & 57859.2262 & 0.0003 & $-$0&0022 & 78 \\
122 & 57859.2814 & 0.0003 & $-$0&0022 & 126 \\
123 & 57859.3366 & 0.0004 & $-$0&0021 & 98 \\
126 & 57859.5049 & 0.0023 & 0&0008 & 12 \\
127 & 57859.5595 & 0.0029 & 0&0003 & 12 \\
139 & 57860.2189 & 0.0004 & $-$0&0019 & 75 \\
140 & 57860.2754 & 0.0002 & $-$0&0005 & 127 \\
141 & 57860.3304 & 0.0005 & $-$0&0007 & 78 \\
145 & 57860.5505 & 0.0011 & $-$0&0011 & 17 \\
157 & 57861.2155 & 0.0004 & 0&0023 & 97 \\
158 & 57861.2691 & 0.0003 & 0&0008 & 127 \\
159 & 57861.3243 & 0.0004 & 0&0009 & 101 \\
163 & 57861.5457 & 0.0012 & 0&0017 & 18 \\
175 & 57862.2099 & 0.0005 & 0&0043 & 96 \\
176 & 57862.2658 & 0.0003 & 0&0052 & 126 \\
177 & 57862.3212 & 0.0014 & 0&0054 & 53 \\
193 & 57863.2025 & 0.0027 & 0&0046 & 61 \\
194 & 57863.2633 & 0.0005 & 0&0103 & 127 \\
195 & 57863.3168 & 0.0007 & 0&0087 & 114 \\
212 & 57864.2588 & 0.0006 & 0&0134 & 127 \\
213 & 57864.3132 & 0.0006 & 0&0127 & 89 \\
230 & 57865.2369 & 0.0014 & $-$0&0009 & 127 \\
231 & 57865.2887 & 0.0015 & $-$0&0042 & 117 \\
235 & 57865.5188 & 0.0034 & 0&0054 & 26 \\
248 & 57866.2222 & 0.0009 & $-$0&0080 & 115 \\
266 & 57867.2012 & 0.0056 & $-$0&0213 & 61 \\
267 & 57867.2707 & 0.0009 & $-$0&0070 & 127 \\
271 & 57867.4982 & 0.0038 & 0&0001 & 15 \\
357 & 57872.2431 & 0.0007 & 0&0036 & 126 \\
358 & 57872.2946 & 0.0017 & $-$0&0001 & 79 \\
375 & 57873.2298 & 0.0005 & $-$0&0021 & 127 \\
376 & 57873.2848 & 0.0007 & $-$0&0022 & 108 \\
430 & 57876.2653 & 0.0006 & 0&0012 & 124 \\
466 & 57878.2427 & 0.0065 & $-$0&0061 & 127 \\
\hline
  \multicolumn{6}{l}{\commenta BJD$-$2400000.} \\
  \multicolumn{6}{l}{\commentb Against max $= 2457852.5575 + 0.055132 E$.} \\
  \multicolumn{6}{l}{\commentc Number of points used to determine the maximum.} \\
\end{tabular}
\end{center}
\end{table}

\subsection{ASASSN-17eq}\label{obj:asassn17eq}

\begin{figure}
  \begin{center}
    \FigureFile(85mm,110mm){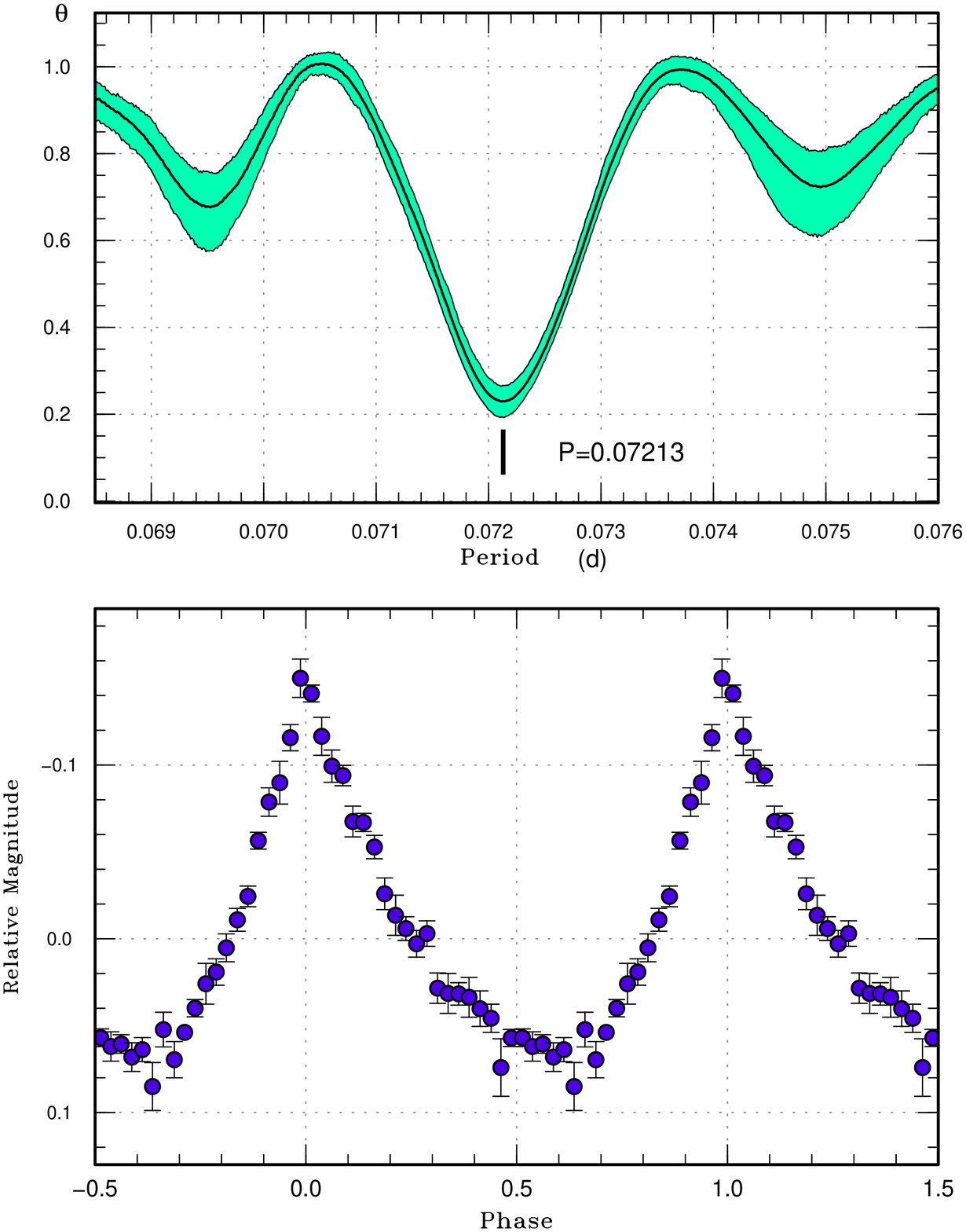}
  \end{center}
  \caption{Superhumps in ASASSN-17eq (2017).
     (Upper): PDM analysis.
     (Lower): Phase-averaged profile.}
  \label{fig:asassn17eqshpdm}
\end{figure}

   This object was detected as a transient
at $V$=13.7 on 2017 April 11 by the ASAS-SN team.
Subsequent observations detected superhumps
(vsnet-alert 20917, 20920; e-figure \ref{fig:asassn17eqshpdm}).
The times of superhump maxima are listed in
e-table \ref{tab:asassn17eqoc2017}.
Well-developed superhumps suggest that they were
most likely stage B ones.

   There was a normal outburst in the ASAS-SN data
at $V$=14.3 on 2016 January 20, which faded to
$V$=15.2 on the next night.  There was no indication
of a past superoutburst in the ASAS-SN data.

\begin{table}
\caption{Superhump maxima of ASASSN-17eq (2017)}\label{tab:asassn17eqoc2017}
\begin{center}
\begin{tabular}{rp{55pt}p{40pt}r@{.}lr}
\hline
\multicolumn{1}{c}{$E$} & \multicolumn{1}{c}{max\commenta} & \multicolumn{1}{c}{error} & \multicolumn{2}{c}{$O-C$\commentb} & \multicolumn{1}{c}{$N$\commentc} \\
\hline
0 & 57857.5031 & 0.0002 & 0&0013 & 111 \\
10 & 57858.2251 & 0.0012 & 0&0013 & 52 \\
11 & 57858.2930 & 0.0029 & $-$0&0030 & 37 \\
12 & 57858.3679 & 0.0005 & $-$0&0002 & 29 \\
25 & 57859.3053 & 0.0024 & $-$0&0015 & 26 \\
26 & 57859.3783 & 0.0002 & $-$0&0006 & 95 \\
27 & 57859.4512 & 0.0003 & 0&0001 & 112 \\
28 & 57859.5259 & 0.0007 & 0&0026 & 44 \\
\hline
  \multicolumn{6}{l}{\commenta BJD$-$2400000.} \\
  \multicolumn{6}{l}{\commentb Against max $= 2457857.5018 + 0.072197 E$.} \\
  \multicolumn{6}{l}{\commentc Number of points used to determine the maximum.} \\
\end{tabular}
\end{center}
\end{table}

\subsection{ASASSN-17es}\label{obj:asassn17es}

\begin{figure}
  \begin{center}
    \FigureFile(85mm,110mm){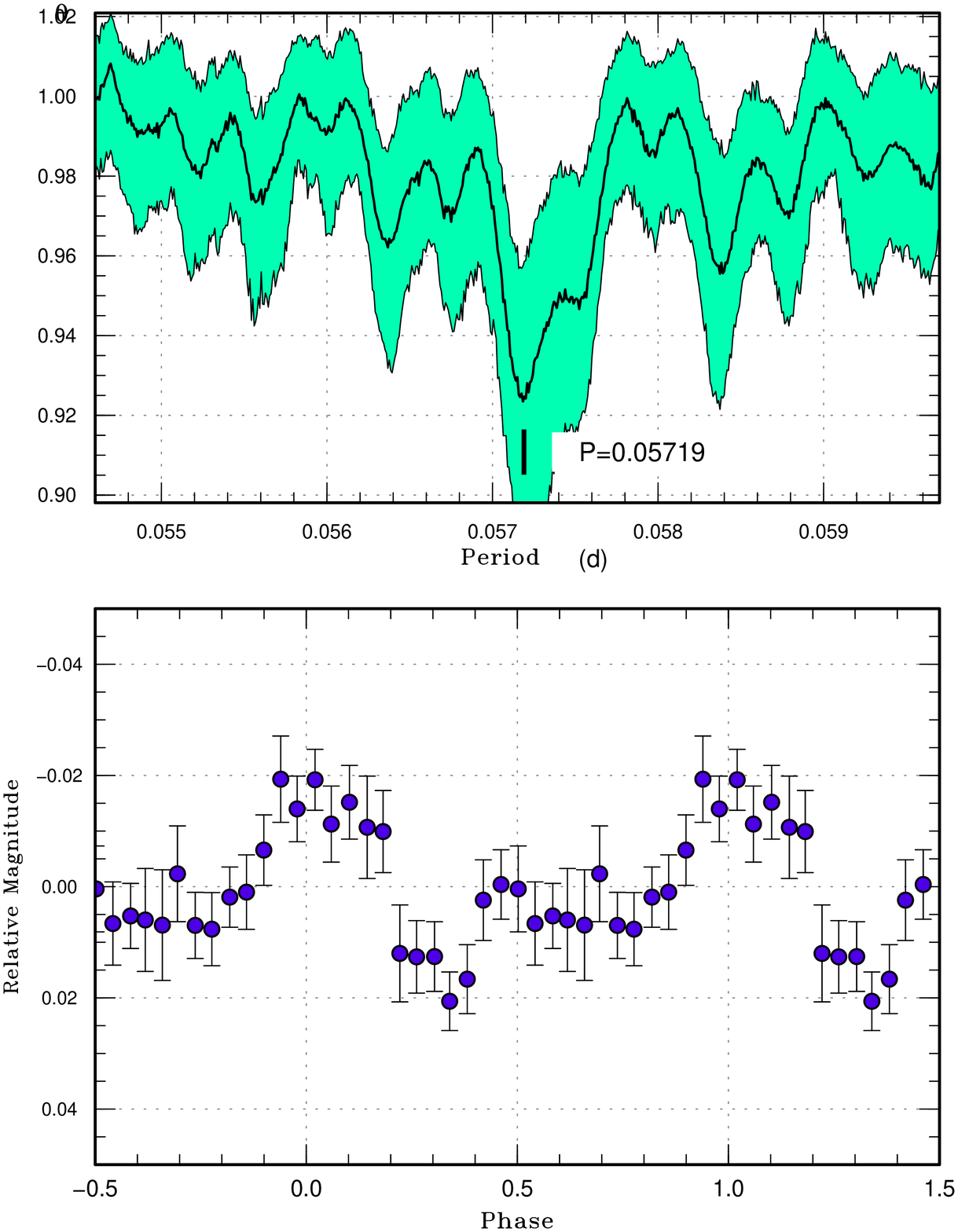}
  \end{center}
  \caption{Early superhumps in ASASSN-17es (2017).
     (Upper): PDM analysis.
     (Lower): Phase-averaged profile.}
  \label{fig:asassn17eseshpdm}
\end{figure}

\begin{figure}
  \begin{center}
    \FigureFile(85mm,110mm){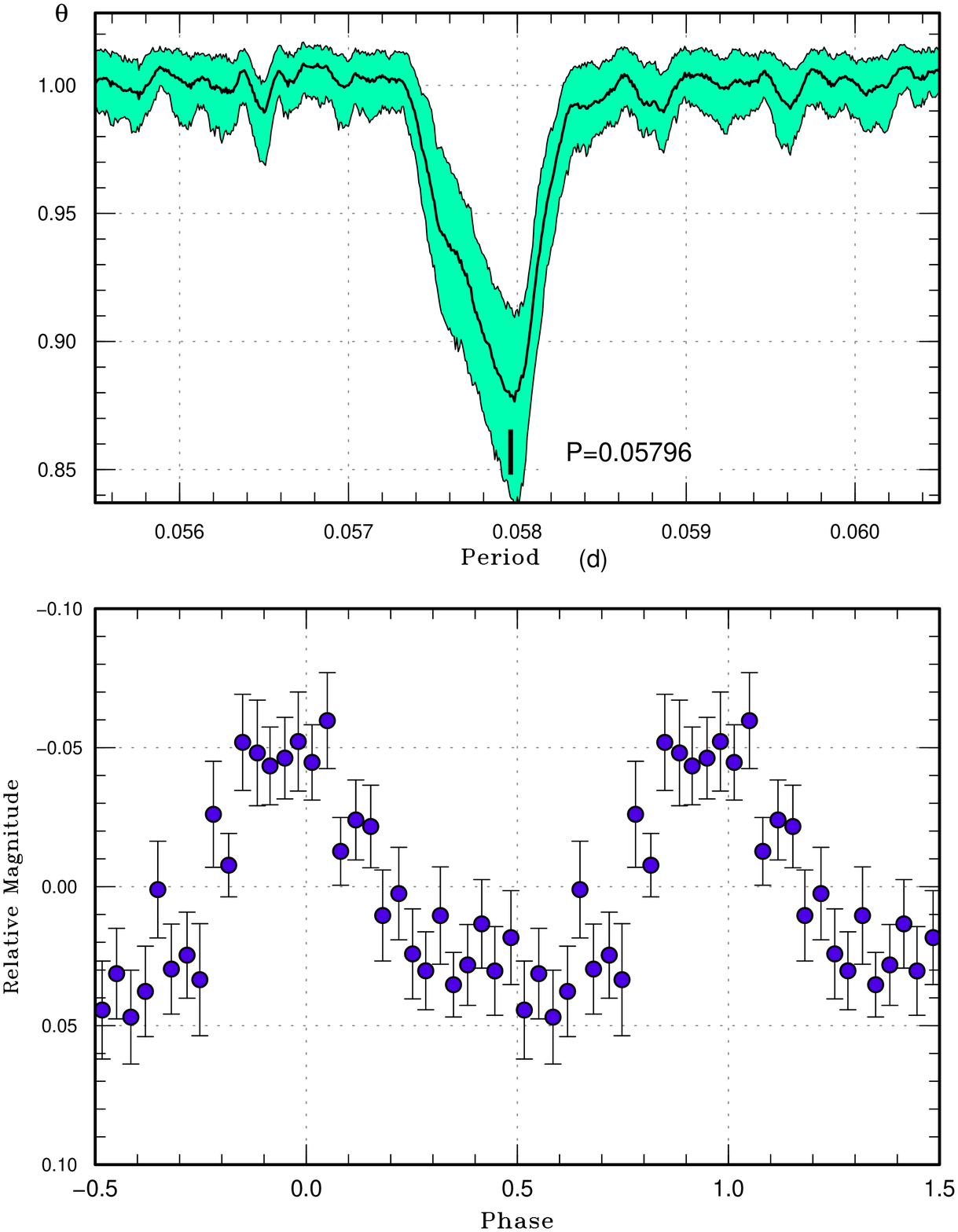}
  \end{center}
  \caption{Ordinary superhumps in ASASSN-17es (2017).
     (Upper): PDM analysis.
     (Lower): Phase-averaged profile.}
  \label{fig:asassn17esshpdm}
\end{figure}

   This object was detected as a transient
at $V$=14.8 on 2017 April 11 by the ASAS-SN team.
Subsequent observations detected early superhumps
(vsnet-alert 20929, 20945; e-figure \ref{fig:asassn17eseshpdm}),
qualifying this object to be a WZ Sge-type dwarf nova.
The object then showed ordinary superhumps
(vsnet-alert 20972; e-figure \ref{fig:asassn17esshpdm}).
The times of superhump maxima are listed in
e-table \ref{tab:asassn17esoc2017}.
There was a rather smooth transition from stage A to B
between $E$=0 and $E$=33.  Assuming that the interval
between $E$=0 and $E$=18 reflected stage A,
we determined the period to be 0.05924(16)~d.
Using the best period of early superhumps
[0.05719(3)~d, PDM method], $\epsilon^*$ for stage A
superhump is 0.037(4), which corresponds to $q$=0.095(9).

   According to the ASAS-SN data, there was no past
outburst and the 2017 outburst was detected up to
May 20 (39~d after the initial rise) at $V$=15.7.
Since the object was close to the detection limit
in late May to June, the actual termination of
the superoutburst was somewhat uncertain.
Although the long duration of the outbursting stage
suggested the combination of the superoutburst
and a plateau-type rebrightening, there was no
strong suggestion of a dip in the ASAS-SN data.
Time-resolved photometry terminated on April 30 and
these observations did not constrain the duration
of the superoutburst.

\begin{table}
\caption{Superhump maxima of ASASSN-17es (2017)}\label{tab:asassn17esoc2017}
\begin{center}
\begin{tabular}{rp{55pt}p{40pt}r@{.}lr}
\hline
\multicolumn{1}{c}{$E$} & \multicolumn{1}{c}{max\commenta} & \multicolumn{1}{c}{error} & \multicolumn{2}{c}{$O-C$\commentb} & \multicolumn{1}{c}{$N$\commentc} \\
\hline
0 & 57864.7608 & 0.0027 & $-$0&0213 & 12 \\
16 & 57865.7118 & 0.0027 & 0&0021 & 9 \\
17 & 57865.7675 & 0.0035 & $-$0&0001 & 13 \\
18 & 57865.8257 & 0.0017 & 0&0002 & 16 \\
33 & 57866.7008 & 0.0023 & 0&0058 & 15 \\
34 & 57866.7592 & 0.0021 & 0&0062 & 18 \\
35 & 57866.8137 & 0.0020 & 0&0028 & 21 \\
36 & 57866.8730 & 0.0021 & 0&0041 & 17 \\
46 & 57867.4518 & 0.0010 & 0&0032 & 19 \\
50 & 57867.6838 & 0.0022 & 0&0034 & 14 \\
51 & 57867.7464 & 0.0033 & 0&0080 & 16 \\
52 & 57867.7944 & 0.0017 & $-$0&0020 & 21 \\
53 & 57867.8555 & 0.0014 & 0&0012 & 24 \\
69 & 57868.7815 & 0.0032 & $-$0&0003 & 15 \\
70 & 57868.8390 & 0.0023 & $-$0&0008 & 16 \\
85 & 57869.7111 & 0.0025 & 0&0019 & 12 \\
86 & 57869.7671 & 0.0036 & $-$0&0001 & 13 \\
87 & 57869.8212 & 0.0025 & $-$0&0039 & 15 \\
88 & 57869.8853 & 0.0181 & 0&0022 & 5 \\
102 & 57870.6927 & 0.0029 & $-$0&0019 & 10 \\
103 & 57870.7468 & 0.0026 & $-$0&0057 & 10 \\
104 & 57870.8058 & 0.0054 & $-$0&0047 & 14 \\
105 & 57870.8683 & 0.0038 & $-$0&0003 & 10 \\
\hline
  \multicolumn{6}{l}{\commenta BJD$-$2400000.} \\
  \multicolumn{6}{l}{\commentb Against max $= 2457864.7822 + 0.057965 E$.} \\
  \multicolumn{6}{l}{\commentc Number of points used to determine the maximum.} \\
\end{tabular}
\end{center}
\end{table}

\subsection{ASASSN-17et}\label{obj:asassn17et}

\begin{figure}
  \begin{center}
    \FigureFile(85mm,110mm){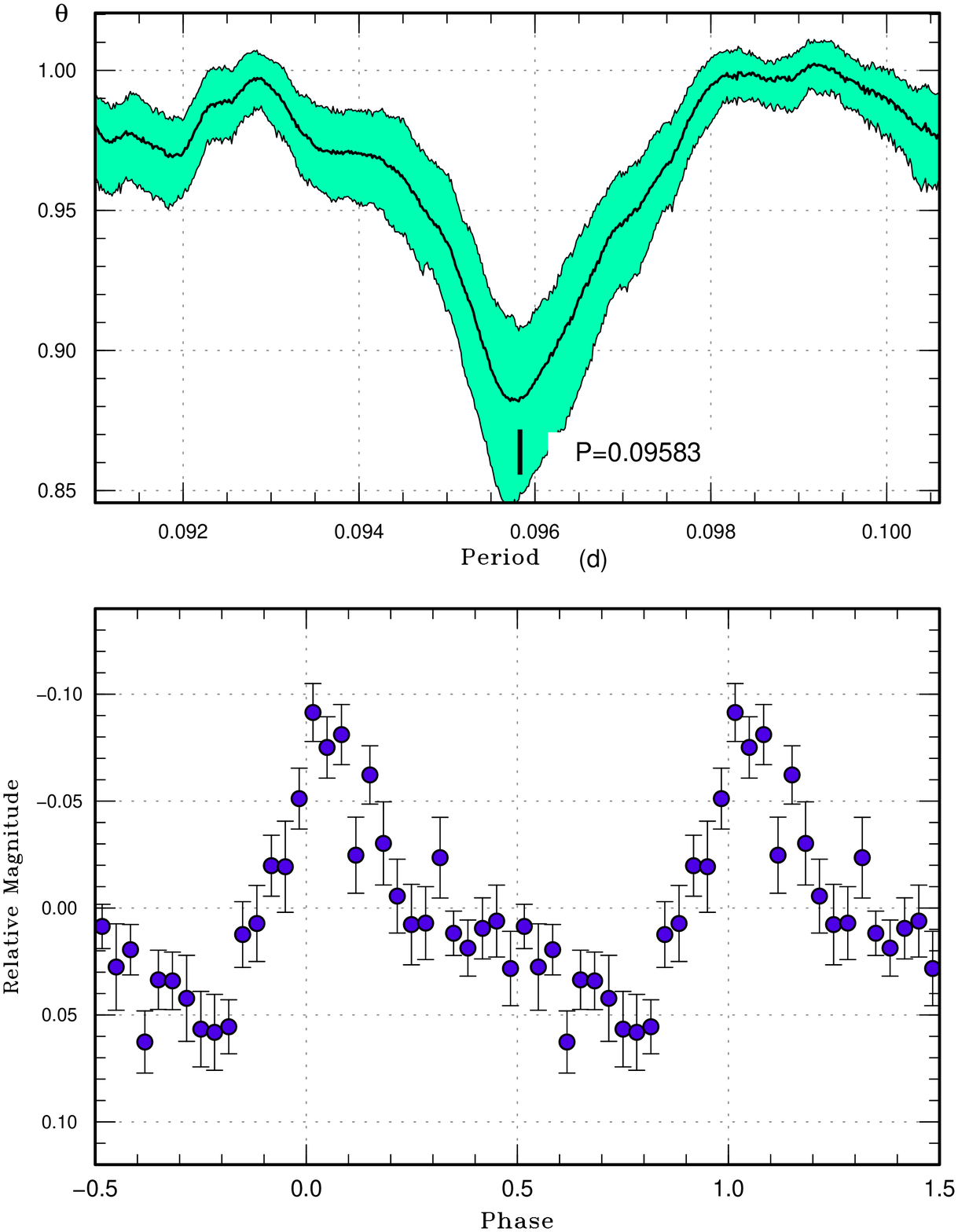}
  \end{center}
  \caption{Superhumps in ASASSN-17et (2017).
     (Upper): PDM analysis.
     (Lower): Phase-averaged profile.}
  \label{fig:asassn17etshpdm}
\end{figure}

   This object was detected as a transient
at $V$=14.9 on 2017 April 10 by the ASAS-SN team.
The outburst was announced after observation
of $V$=15.1 on 2017 April 13.
There is an X-ray counterpart 2XMM J175924.9$-$231503.
Superhumps were immediately detected (vsnet-alert 20919;
e-figure \ref{fig:asassn17etshpdm}).
Their period qualified this object to be an SU UMa-type
dwarf nova in the period gap.
The times of superhump maxima are listed in
table \ref{tab:asassn17etoc2017}.
Since we observed the final part of the superoutburst,
we tentatively classified the observed superhumps
to be stage C ones.  Since period variations of
superhumps in long-period systems may not be similar
to those in short-period systems (cf. \cite{Pdot3};
\cite{Pdot6}; \cite{kat16v1006cyg}), interpretation
of superhumps in such a system may not be straightforward.

   There was no past outburst detection in the ASAS-SN
data starting on 2015 February 24.

\begin{table}
\caption{Superhump maxima of ASASSN-17et (2017)}\label{tab:asassn17etoc2017}
\begin{center}
\begin{tabular}{rp{55pt}p{40pt}r@{.}lr}
\hline
\multicolumn{1}{c}{$E$} & \multicolumn{1}{c}{max\commenta} & \multicolumn{1}{c}{error} & \multicolumn{2}{c}{$O-C$\commentb} & \multicolumn{1}{c}{$N$\commentc} \\
\hline
0 & 57858.7842 & 0.0015 & $-$0&0001 & 23 \\
1 & 57858.8814 & 0.0027 & 0&0015 & 15 \\
40 & 57862.6077 & 0.0015 & $-$0&0021 & 221 \\
50 & 57863.5650 & 0.0019 & $-$0&0011 & 220 \\
51 & 57863.6580 & 0.0029 & $-$0&0037 & 171 \\
52 & 57863.7583 & 0.0020 & 0&0009 & 21 \\
53 & 57863.8503 & 0.0032 & $-$0&0027 & 28 \\
60 & 57864.5318 & 0.0029 & 0&0094 & 203 \\
61 & 57864.6216 & 0.0042 & 0&0035 & 115 \\
62 & 57864.7144 & 0.0032 & 0&0007 & 14 \\
63 & 57864.8031 & 0.0037 & $-$0&0063 & 25 \\
\hline
  \multicolumn{6}{l}{\commenta BJD$-$2400000.} \\
  \multicolumn{6}{l}{\commentb Against max $= 2457858.7843 + 0.095636 E$.} \\
  \multicolumn{6}{l}{\commentc Number of points used to determine the maximum.} \\
\end{tabular}
\end{center}
\end{table}

\subsection{ASASSN-17ew}\label{obj:asassn17ew}

\begin{figure}
  \begin{center}
    \FigureFile(85mm,110mm){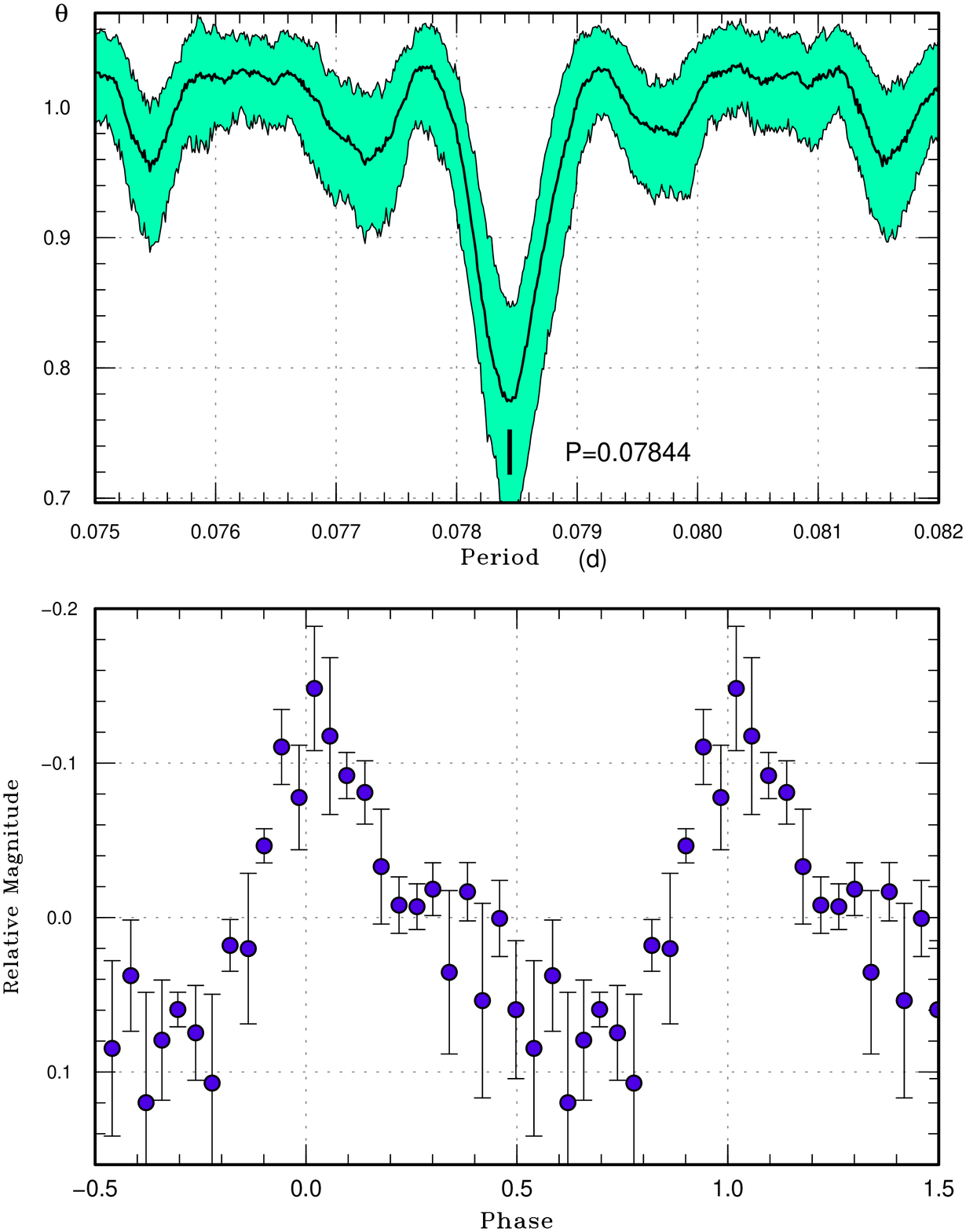}
  \end{center}
  \caption{Superhumps in ASASSN-17ew (2017).
     (Upper): PDM analysis.
     (Lower): Phase-averaged profile.}
  \label{fig:asassn17ewshpdm}
\end{figure}

   This object was detected as a transient
at $V$=14.9 on 2017 April 13 by the ASAS-SN team.
Superhumps were immediately detected (vsnet-alert 20925;
e-figure \ref{fig:asassn17ewshpdm}).
The times of superhump maxima are listed in
table \ref{tab:asassn17ewoc2017}.
Since we observed the final part of the superoutburst,
we tentatively classified the observed superhumps
to be stage C ones.

   There were two past outbursts (most likely normal ones)
in the ASAS-SN data starting on 2016 February 19:
2016 June 30 ($V$=15.5) and 2017 January 7 ($V$=15.2).

\begin{table}
\caption{Superhump maxima of ASASSN-17ew (2017)}\label{tab:asassn17ewoc2017}
\begin{center}
\begin{tabular}{rp{55pt}p{40pt}r@{.}lr}
\hline
\multicolumn{1}{c}{$E$} & \multicolumn{1}{c}{max\commenta} & \multicolumn{1}{c}{error} & \multicolumn{2}{c}{$O-C$\commentb} & \multicolumn{1}{c}{$N$\commentc} \\
\hline
0 & 57858.7842 & 0.0015 & $-$0&0001 & 23 \\
1 & 57858.8814 & 0.0027 & 0&0015 & 15 \\
40 & 57862.6077 & 0.0015 & $-$0&0021 & 221 \\
50 & 57863.5650 & 0.0019 & $-$0&0011 & 220 \\
51 & 57863.6580 & 0.0029 & $-$0&0037 & 171 \\
52 & 57863.7583 & 0.0020 & 0&0009 & 21 \\
53 & 57863.8503 & 0.0032 & $-$0&0027 & 28 \\
60 & 57864.5318 & 0.0029 & 0&0094 & 203 \\
61 & 57864.6216 & 0.0042 & 0&0035 & 115 \\
62 & 57864.7144 & 0.0032 & 0&0007 & 14 \\
63 & 57864.8031 & 0.0037 & $-$0&0063 & 25 \\
\hline
  \multicolumn{6}{l}{\commenta BJD$-$2400000.} \\
  \multicolumn{6}{l}{\commentb Against max $= 2457859.6395 + 0.078497 E$.} \\
  \multicolumn{6}{l}{\commentc Number of points used to determine the maximum.} \\
\end{tabular}
\end{center}
\end{table}

\subsection{ASASSN-17ex}\label{obj:asassn17ex}

\begin{figure}
  \begin{center}
    \FigureFile(85mm,110mm){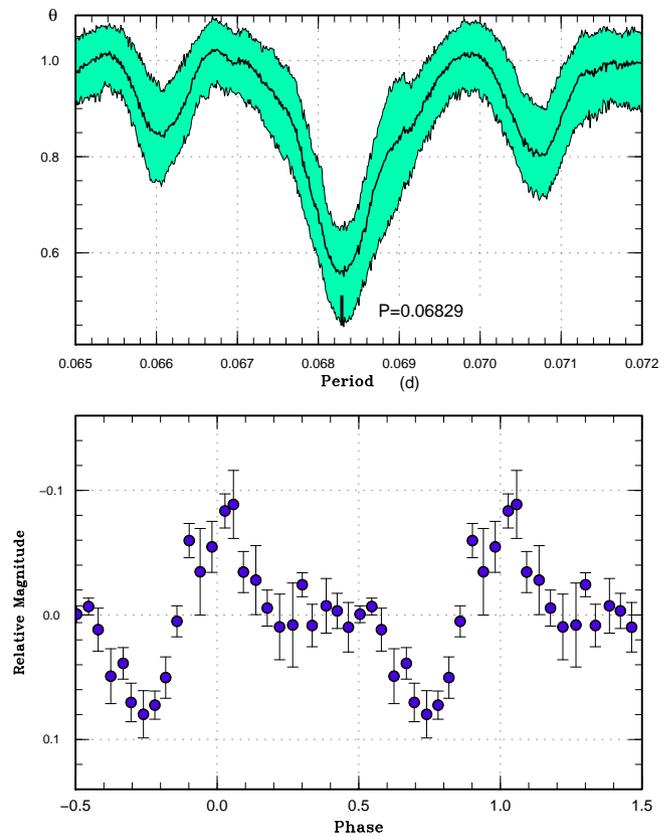}
  \end{center}
  \caption{Superhumps in ASASSN-17ex (2017).
     (Upper): PDM analysis.  The analysis was restricted before
     BJD 2457862 since later observations were too noisy.
     (Lower): Phase-averaged profile.}
  \label{fig:asassn17exshpdm}
\end{figure}

   This object was detected as a transient
at $V$=15.8 on 2017 April 10 by the ASAS-SN team.
The outburst was announced after the observation
at $V$=15.6 on 2017 April 13.
Subsequent observations detected superhumps
(vsnet-alert 20926, 20948; e-figure \ref{fig:asassn17exshpdm}).
The times of superhump maxima are listed in
table \ref{tab:asassn17exoc2017}.
Since we observed the final part of the superoutburst,
we tentatively classified the observed superhumps
to be stage C ones.

   There were no definite past outburst in the ASAS-SN data
starting on 2015 September 18.  There may have been
a borderline detection at $V$=16.3 on 2017 January 28.

\begin{table}
\caption{Superhump maxima of ASASSN-17ex (2017)}\label{tab:asassn17exoc2017}
\begin{center}
\begin{tabular}{rp{55pt}p{40pt}r@{.}lr}
\hline
\multicolumn{1}{c}{$E$} & \multicolumn{1}{c}{max\commenta} & \multicolumn{1}{c}{error} & \multicolumn{2}{c}{$O-C$\commentb} & \multicolumn{1}{c}{$N$\commentc} \\
\hline
0 & 57859.6447 & 0.0020 & $-$0&0003 & 20 \\
1 & 57859.7121 & 0.0016 & $-$0&0012 & 12 \\
15 & 57860.6704 & 0.0014 & 0&0008 & 17 \\
16 & 57860.7399 & 0.0026 & 0&0020 & 13 \\
29 & 57861.6252 & 0.0036 & $-$0&0007 & 14 \\
30 & 57861.6985 & 0.0041 & 0&0043 & 15 \\
31 & 57861.7576 & 0.0035 & $-$0&0049 & 11 \\
\hline
  \multicolumn{6}{l}{\commenta BJD$-$2400000.} \\
  \multicolumn{6}{l}{\commentb Against max $= 2457859.6450 + 0.068306 E$.} \\
  \multicolumn{6}{l}{\commentc Number of points used to determine the maximum.} \\
\end{tabular}
\end{center}
\end{table}

\subsection{ASASSN-17fh}\label{obj:asassn17fh}

\begin{figure}
  \begin{center}
    \FigureFile(85mm,110mm){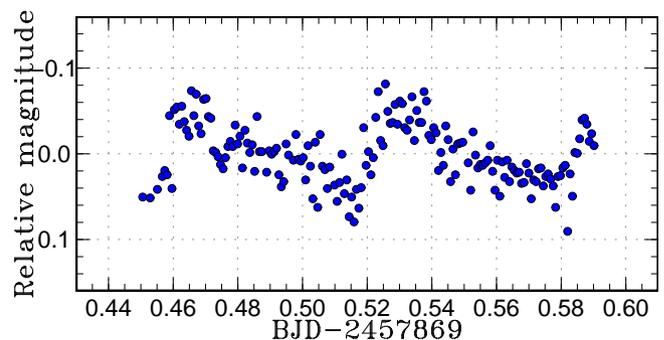}
  \end{center}
  \caption{Superhumps in ASASSN-17fh (2017).
  }
  \label{fig:asassn17fhshlc}
\end{figure}

   This object was detected as a transient
at $V$=16.0 on 2017 April 20 by the ASAS-SN team.
Only two superhumps were recorded in single-night
observations: BJD 2457869.4704(12) ($N$=55) and
2457869.5343(8) ($N$=66) (vsnet-alert 20963;
e-figure \ref{fig:asassn17fhshlc}).  The superhump
stage was unknown.

   The ASAS-SN data since 2013 June 4 did not detect
other superoutbursts.  There was possible single positive
detection at $V$=16.2 on 2015 April 1, but it may have
been a noise.

\subsection{ASASSN-17fi}\label{obj:asassn17fi}

\begin{figure}
  \begin{center}
    \FigureFile(85mm,110mm){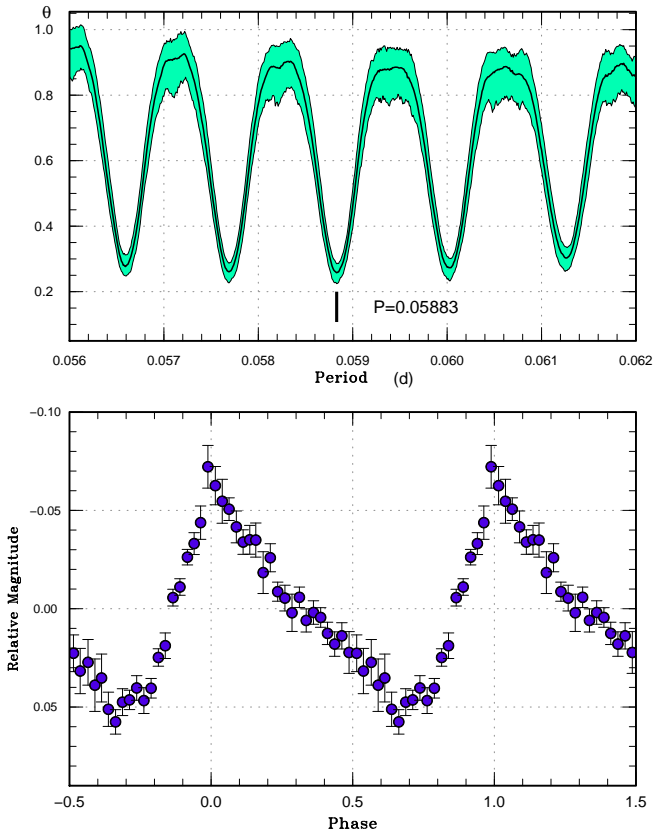}
  \end{center}
  \caption{Superhumps in ASASSN-17fi (2017).
     (Upper): PDM analysis.  The alias selection was based
     on a long single-night observation.
     (Lower): Phase-averaged profile.}
  \label{fig:asassn17fishpdm}
\end{figure}

   This object was detected as a transient
at $V$=14.5 on 2017 April 20 by the ASAS-SN team.
Subsequent observations detected superhumps
(vsnet-alert 20952; e-figure \ref{fig:asassn17fishpdm}).
According to the ASAS-SN data, the outburst lasted
at least up to 2017 April 27.

   The ASAS-SN data covered this region since 2013 October 28.
There is nearby $\sim$14.4 mag (Gaia G magnitude) star
which affected ASAS-SN photometry.  Although the detection
limit was shallow due to this companion star, there was
no indication of an outburst as bright as in the 2017 one.

\begin{table}
\caption{Superhump maxima of ASASSN-17fi (2017)}\label{tab:asassn17fioc2017}
\begin{center}
\begin{tabular}{rp{55pt}p{40pt}r@{.}lr}
\hline
\multicolumn{1}{c}{$E$} & \multicolumn{1}{c}{max\commenta} & \multicolumn{1}{c}{error} & \multicolumn{2}{c}{$O-C$\commentb} & \multicolumn{1}{c}{$N$\commentc} \\
\hline
0 & 57867.4925 & 0.0005 & $-$0&0000 & 62 \\
1 & 57867.5513 & 0.0004 & $-$0&0001 & 58 \\
2 & 57867.6102 & 0.0006 & 0&0001 & 55 \\
51 & 57870.4937 & 0.0004 & 0&0007 & 62 \\
52 & 57870.5511 & 0.0009 & $-$0&0007 & 32 \\
\hline
  \multicolumn{6}{l}{\commenta BJD$-$2400000.} \\
  \multicolumn{6}{l}{\commentb Against max $= 2457867.4925 + 0.058833 E$.} \\
  \multicolumn{6}{l}{\commentc Number of points used to determine the maximum.} \\
\end{tabular}
\end{center}
\end{table}

\subsection{ASASSN-17fj}\label{obj:asassn17fj}

\begin{figure}
  \begin{center}
    \FigureFile(85mm,110mm){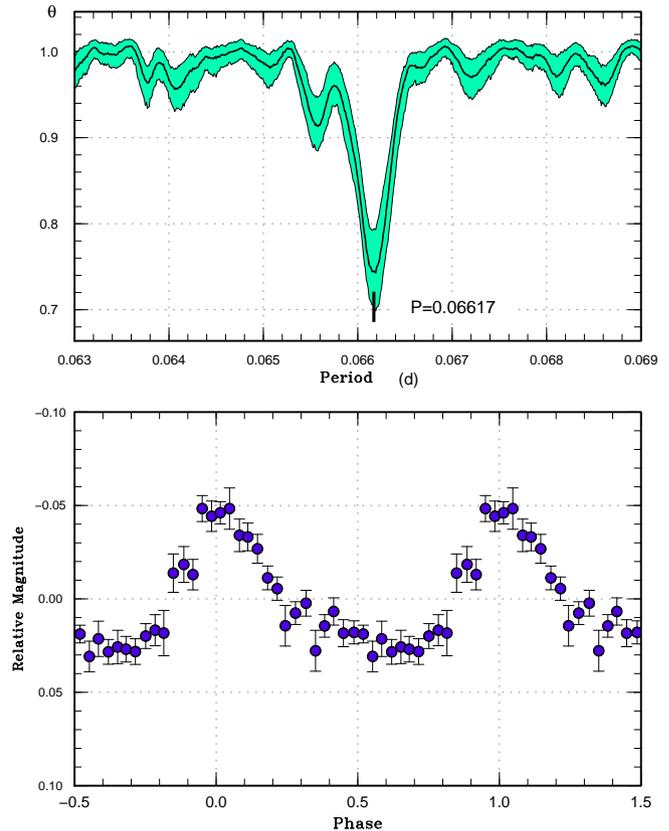}
  \end{center}
  \caption{Superhumps in ASASSN-17fj (2017).
     (Upper): PDM analysis.  The analysis was restricted before
     BJD 2457878 since later observations were too noisy.
     (Lower): Phase-averaged profile.}
  \label{fig:asassn17fjshpdm}
\end{figure}

   This object was detected as a transient
at $V$=15.7 on 2017 April 21 by the ASAS-SN team.
The outburst was announced after the observation
at $V$=15.7 on April 22.
Subsequent observations detected superhumps
(vsnet-alert 20965; e-figure \ref{fig:asassn17fjshpdm}).
The times of superhump maxima are listed in
e-table \ref{tab:asassn17fjoc2017}.
Stages B and C can be identified.

   This field has been observed by the ASAS-SN team since
2016 March 10.  There was an apparent superoutburst
on 2016 March 10 at $V$=15.7 and probably lasted at least
up to March 15.  The supercycle looks likely to be
an order of a year.

\begin{table}
\caption{Superhump maxima of ASASSN-17fj (2017)}\label{tab:asassn17fjoc2017}
\begin{center}
\begin{tabular}{rp{55pt}p{40pt}r@{.}lr}
\hline
\multicolumn{1}{c}{$E$} & \multicolumn{1}{c}{max\commenta} & \multicolumn{1}{c}{error} & \multicolumn{2}{c}{$O-C$\commentb} & \multicolumn{1}{c}{$N$\commentc} \\
\hline
0 & 57868.7539 & 0.0009 & $-$0&0016 & 15 \\
1 & 57868.8200 & 0.0010 & $-$0&0017 & 17 \\
2 & 57868.8903 & 0.0026 & 0&0025 & 9 \\
14 & 57869.6791 & 0.0012 & $-$0&0030 & 11 \\
15 & 57869.7466 & 0.0018 & $-$0&0017 & 13 \\
16 & 57869.8118 & 0.0014 & $-$0&0026 & 17 \\
17 & 57869.8797 & 0.0026 & $-$0&0009 & 10 \\
30 & 57870.7369 & 0.0012 & $-$0&0041 & 10 \\
31 & 57870.8060 & 0.0013 & $-$0&0011 & 16 \\
32 & 57870.8718 & 0.0046 & $-$0&0016 & 12 \\
46 & 57871.8011 & 0.0023 & 0&0011 & 17 \\
47 & 57871.8628 & 0.0036 & $-$0&0033 & 16 \\
59 & 57872.6630 & 0.0037 & 0&0027 & 12 \\
61 & 57872.7966 & 0.0017 & 0&0039 & 23 \\
62 & 57872.8598 & 0.0032 & 0&0010 & 25 \\
75 & 57873.7256 & 0.0017 & 0&0063 & 17 \\
76 & 57873.7914 & 0.0016 & 0&0059 & 23 \\
77 & 57873.8550 & 0.0028 & 0&0033 & 26 \\
90 & 57874.7175 & 0.0014 & 0&0055 & 18 \\
91 & 57874.7842 & 0.0017 & 0&0060 & 23 \\
92 & 57874.8479 & 0.0016 & 0&0035 & 25 \\
105 & 57875.7017 & 0.0037 & $-$0&0030 & 18 \\
106 & 57875.7775 & 0.0026 & 0&0065 & 24 \\
107 & 57875.8349 & 0.0030 & $-$0&0022 & 25 \\
120 & 57876.6957 & 0.0015 & $-$0&0018 & 15 \\
121 & 57876.7561 & 0.0022 & $-$0&0076 & 19 \\
122 & 57876.8256 & 0.0021 & $-$0&0043 & 25 \\
135 & 57877.6826 & 0.0026 & $-$0&0077 & 15 \\
\hline
  \multicolumn{6}{l}{\commenta BJD$-$2400000.} \\
  \multicolumn{6}{l}{\commentb Against max $= 2457868.7555 + 0.066184 E$.} \\
  \multicolumn{6}{l}{\commentc Number of points used to determine the maximum.} \\
\end{tabular}
\end{center}
\end{table}

\subsection{ASASSN-17fl}\label{obj:asassn17fl}

\begin{figure}
  \begin{center}
    \FigureFile(85mm,110mm){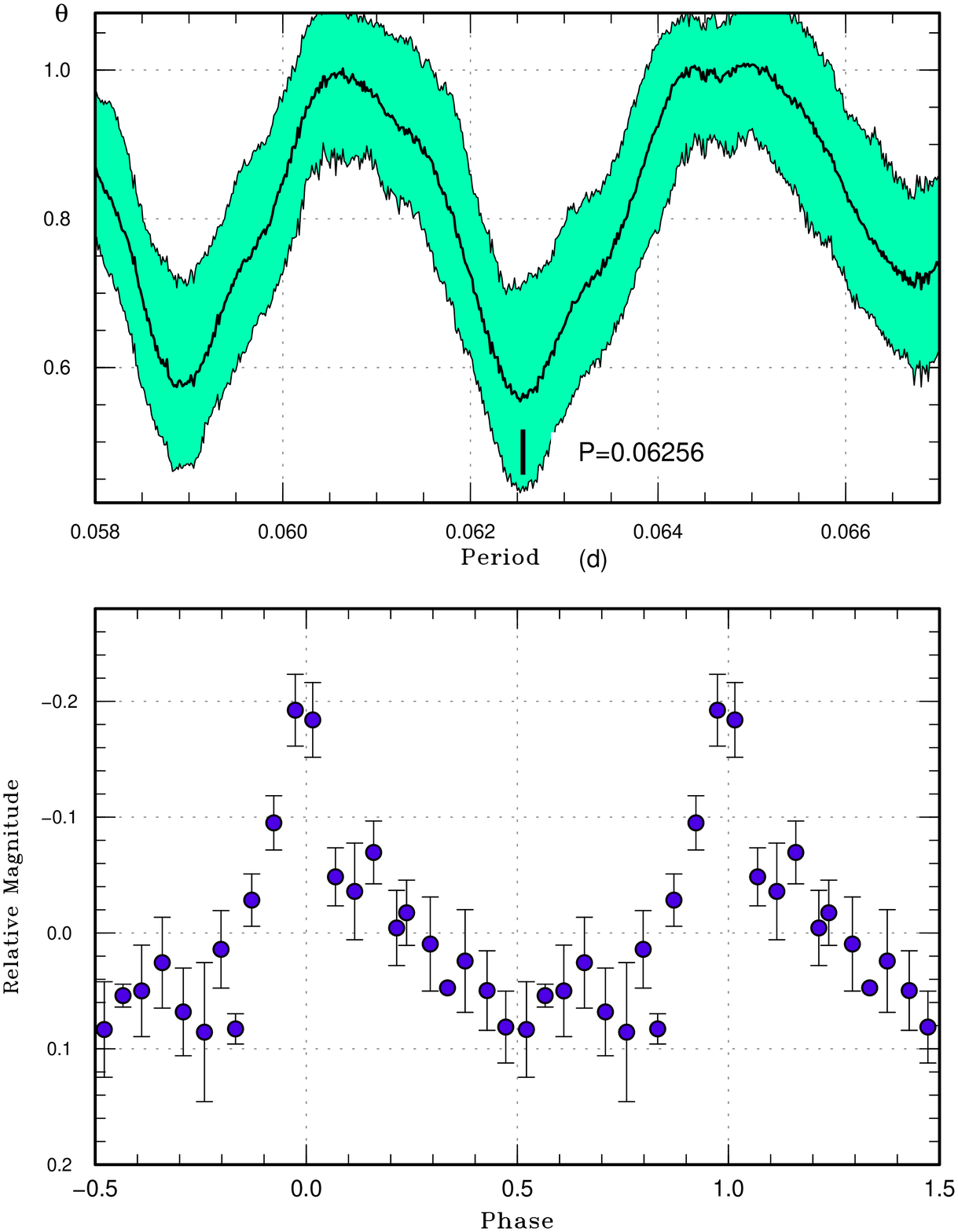}
  \end{center}
  \caption{Superhumps in ASASSN-17fl (2017).
     (Upper): PDM analysis.  The alias selection was
     also supported by $O-C$ analysis.
     (Lower): Phase-averaged profile.}
  \label{fig:asassn17flshpdm}
\end{figure}

   This object was detected as a transient
at $V$=16.3 on 2017 April 21 by the ASAS-SN team.
The outburst was announced after several confirmatory
observations up to 2017 April 26 ($V$=17.3).
Although the object was very faint, superhumps
were detected (vsnet-alert 20973;
e-figure \ref{fig:asassn17flshpdm}).
The times of superhump maxima are listed in
e-table \ref{tab:asassn17floc2017}.

   Although ASAS-SN observations of this field started
on 2016 March 9, no past outburst was detected.
Since the object is very faint, past outbursts may
have escaped detection even if they existed.

\begin{table}
\caption{Superhump maxima of ASASSN-17fl (2017)}\label{tab:asassn17floc2017}
\begin{center}
\begin{tabular}{rp{55pt}p{40pt}r@{.}lr}
\hline
\multicolumn{1}{c}{$E$} & \multicolumn{1}{c}{max\commenta} & \multicolumn{1}{c}{error} & \multicolumn{2}{c}{$O-C$\commentb} & \multicolumn{1}{c}{$N$\commentc} \\
\hline
0 & 57870.7141 & 0.0079 & 0&0014 & 10 \\
1 & 57870.7761 & 0.0018 & 0&0008 & 13 \\
2 & 57870.8353 & 0.0019 & $-$0&0027 & 15 \\
16 & 57871.7179 & 0.0043 & 0&0031 & 11 \\
17 & 57871.7756 & 0.0017 & $-$0&0018 & 14 \\
18 & 57871.8392 & 0.0022 & $-$0&0009 & 17 \\
\hline
  \multicolumn{6}{l}{\commenta BJD$-$2400000.} \\
  \multicolumn{6}{l}{\commentb Against max $= 2457870.7127 + 0.062632 E$.} \\
  \multicolumn{6}{l}{\commentc Number of points used to determine the maximum.} \\
\end{tabular}
\end{center}
\end{table}

\subsection{ASASSN-17fn}\label{obj:asassn17fn}

\begin{figure}
  \begin{center}
    \FigureFile(85mm,110mm){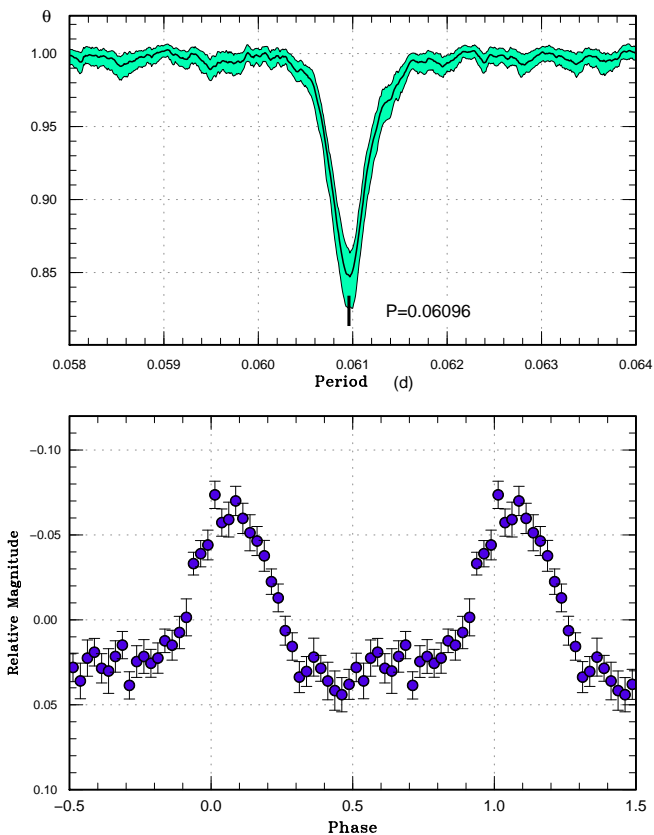}
  \end{center}
  \caption{Early superhumps in ASASSN-17fn (2017).
     (Upper): PDM analysis.
     (Lower): Phase-averaged profile.}
  \label{fig:asassn17fneshpdm}
\end{figure}

\begin{figure}
  \begin{center}
    \FigureFile(85mm,110mm){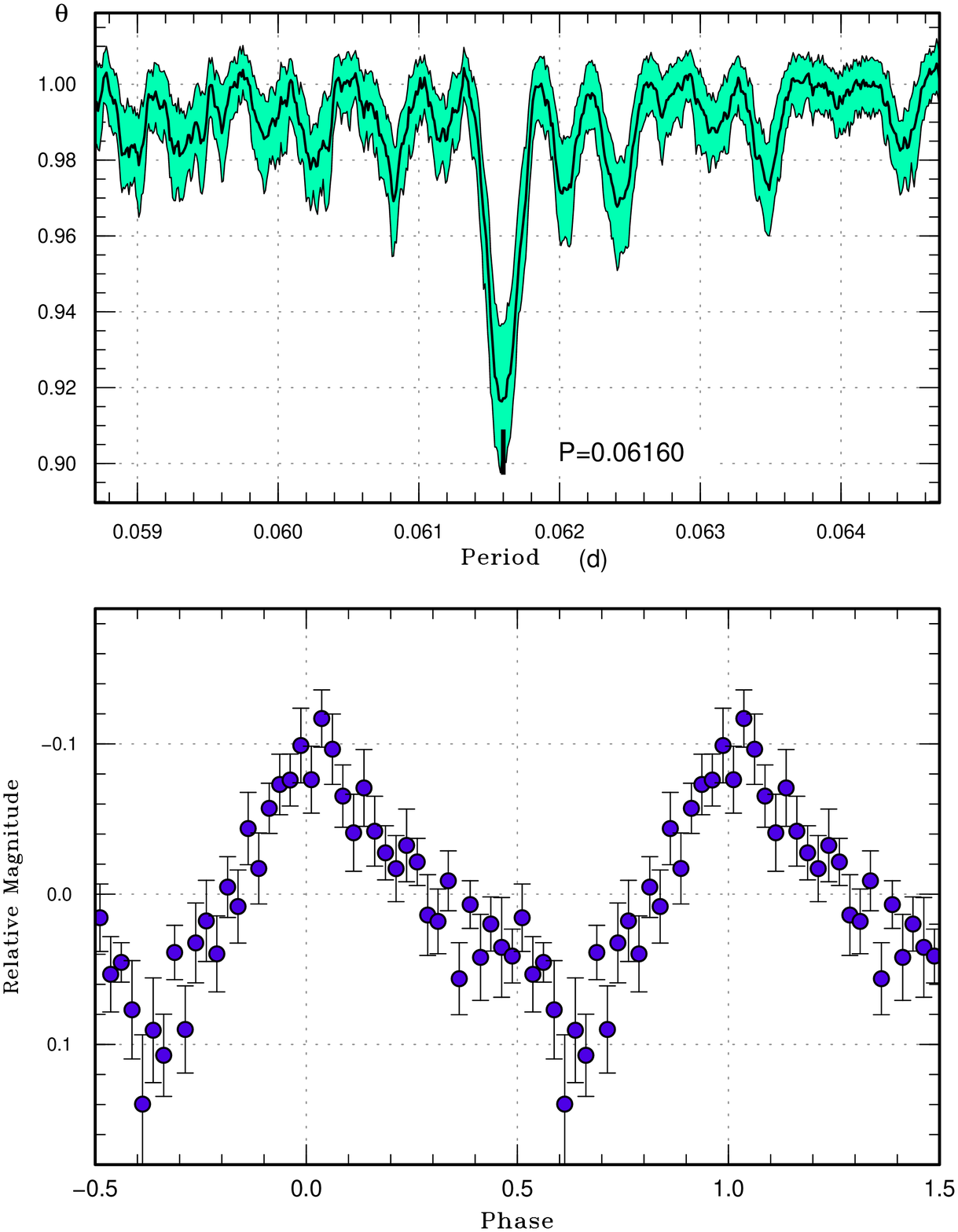}
  \end{center}
  \caption{Ordinary superhumps in ASASSN-17fn (2017).
     (Upper): PDM analysis.
     (Lower): Phase-averaged profile.}
  \label{fig:asassn17fnshpdm}
\end{figure}

   This object was detected as a transient
at $V$=17.4 on 2017 April 22 by the ASAS-SN team.
The outburst was announced after the observation
at $V$=14.8 on 2017 April 26.
There is also an X-ray counterpart 2RXP J103528.7$+$541910
(=2RXP J103527.8$+$541852).
Subsequent observations detected likely early superhumps
(vsnet-alert 20971, 20975; e-figure \ref{fig:asassn17fneshpdm}).
Ordinary superhumps appeared on May 7 and developed
further (vsnet-alert 20997; e-figure \ref{fig:asassn17fnshpdm}).
Although initially detected superhumps were singly-peaked,
the development of ordinary superhumps confirmed
the identification as early superhumps.  The object is
confirmed to be a WZ Sge-type dwarf nova.
The times of superhump maxima are listed in
e-table \ref{tab:asassn17fnoc2017}.

   Although stage A superhumps were only marginally recorded,
the fractional excess $\epsilon^*$ was estimated to be
0.0352(3), which corresponds to $q$=0.097(1).
The object appears to have a rather high $q$ despite
the long orbital period.
The relatively large $q$ appears to be consistent with
the large superhump amplitudes (0.20 mag,
e-figure \ref{fig:asassn17fnshpdm}).
WZ Sge-type dwarf novae with such features tend to show
multiple rebrightenings \citep{nak13j2112j2037}.
In this object, its faintness prevented us from detecting
such rebrightenings.

   There were no definite past outburst in the ASAS-SN data
starting on 2013 October 9.

\begin{table}
\caption{Superhump maxima of ASASSN-17fn (2017)}\label{tab:asassn17fnoc2017}
\begin{center}
\begin{tabular}{rp{55pt}p{40pt}r@{.}lr}
\hline
\multicolumn{1}{c}{$E$} & \multicolumn{1}{c}{max\commenta} & \multicolumn{1}{c}{error} & \multicolumn{2}{c}{$O-C$\commentb} & \multicolumn{1}{c}{$N$\commentc} \\
\hline
0 & 57881.0463 & 0.0028 & $-$0&0353 & 85 \\
21 & 57882.3772 & 0.0018 & 0&0010 & 18 \\
22 & 57882.4362 & 0.0009 & $-$0&0016 & 51 \\
23 & 57882.4982 & 0.0010 & $-$0&0013 & 63 \\
24 & 57882.5620 & 0.0006 & 0&0008 & 63 \\
37 & 57883.3655 & 0.0013 & 0&0029 & 49 \\
38 & 57883.4289 & 0.0008 & 0&0047 & 72 \\
39 & 57883.4915 & 0.0005 & 0&0056 & 76 \\
40 & 57883.5531 & 0.0010 & 0&0056 & 61 \\
41 & 57883.6130 & 0.0100 & 0&0039 & 15 \\
53 & 57884.3553 & 0.0013 & 0&0063 & 27 \\
54 & 57884.4152 & 0.0006 & 0&0047 & 92 \\
55 & 57884.4751 & 0.0008 & 0&0028 & 61 \\
56 & 57884.5389 & 0.0008 & 0&0051 & 55 \\
65 & 57885.0902 & 0.0027 & 0&0015 & 77 \\
103 & 57887.4350 & 0.0008 & 0&0037 & 56 \\
104 & 57887.4966 & 0.0013 & 0&0037 & 40 \\
118 & 57888.3592 & 0.0011 & 0&0032 & 31 \\
119 & 57888.4223 & 0.0012 & 0&0046 & 30 \\
120 & 57888.4781 & 0.0009 & $-$0&0013 & 71 \\
121 & 57888.5398 & 0.0012 & $-$0&0012 & 44 \\
122 & 57888.6043 & 0.0055 & 0&0017 & 17 \\
135 & 57889.4043 & 0.0015 & 0&0003 & 31 \\
136 & 57889.4781 & 0.0036 & 0&0124 & 11 \\
150 & 57890.3236 & 0.0034 & $-$0&0052 & 19 \\
151 & 57890.3861 & 0.0013 & $-$0&0043 & 30 \\
152 & 57890.4492 & 0.0009 & $-$0&0028 & 73 \\
153 & 57890.5096 & 0.0032 & $-$0&0041 & 33 \\
154 & 57890.5690 & 0.0065 & $-$0&0063 & 13 \\
167 & 57891.3711 & 0.0031 & $-$0&0057 & 32 \\
168 & 57891.4329 & 0.0015 & $-$0&0055 & 31 \\
169 & 57891.5001 & 0.0022 & 0&0000 & 31 \\
\hline
  \multicolumn{6}{l}{\commenta BJD$-$2400000.} \\
  \multicolumn{6}{l}{\commentb Against max $= 2457881.0816 + 0.061648 E$.} \\
  \multicolumn{6}{l}{\commentc Number of points used to determine the maximum.} \\
\end{tabular}
\end{center}
\end{table}

\subsection{ASASSN-17fo}\label{obj:asassn17fo}

\begin{figure}
  \begin{center}
    \FigureFile(85mm,110mm){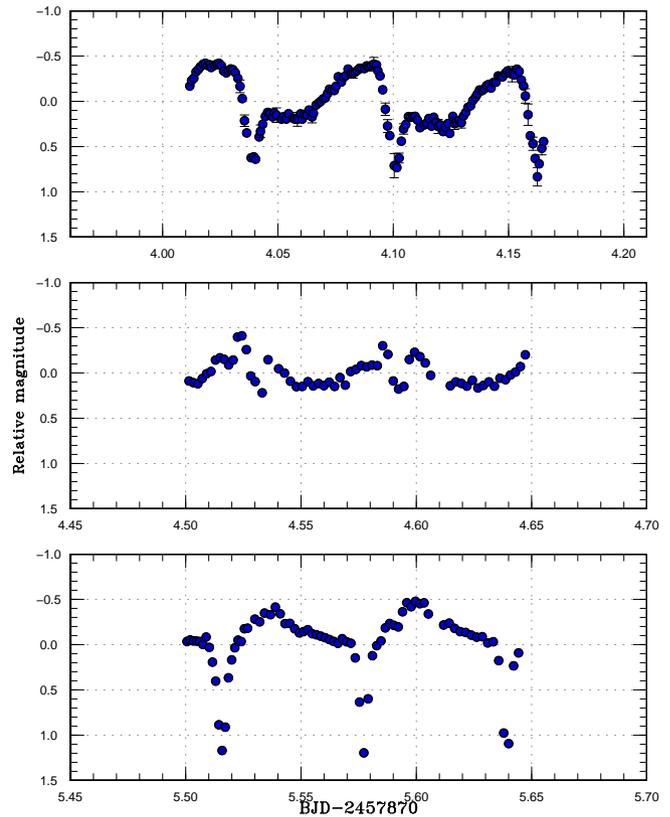}
  \end{center}
  \caption{Eclipses and superhumps in ASASSN-17fo
  in the earliest phase (2017).
  The data were binned to 0.001~d.
  Strong beat phenomenon was present.
  }
  \label{fig:asassn17foshlc}
\end{figure}

\begin{figure}
  \begin{center}
    \FigureFile(85mm,110mm){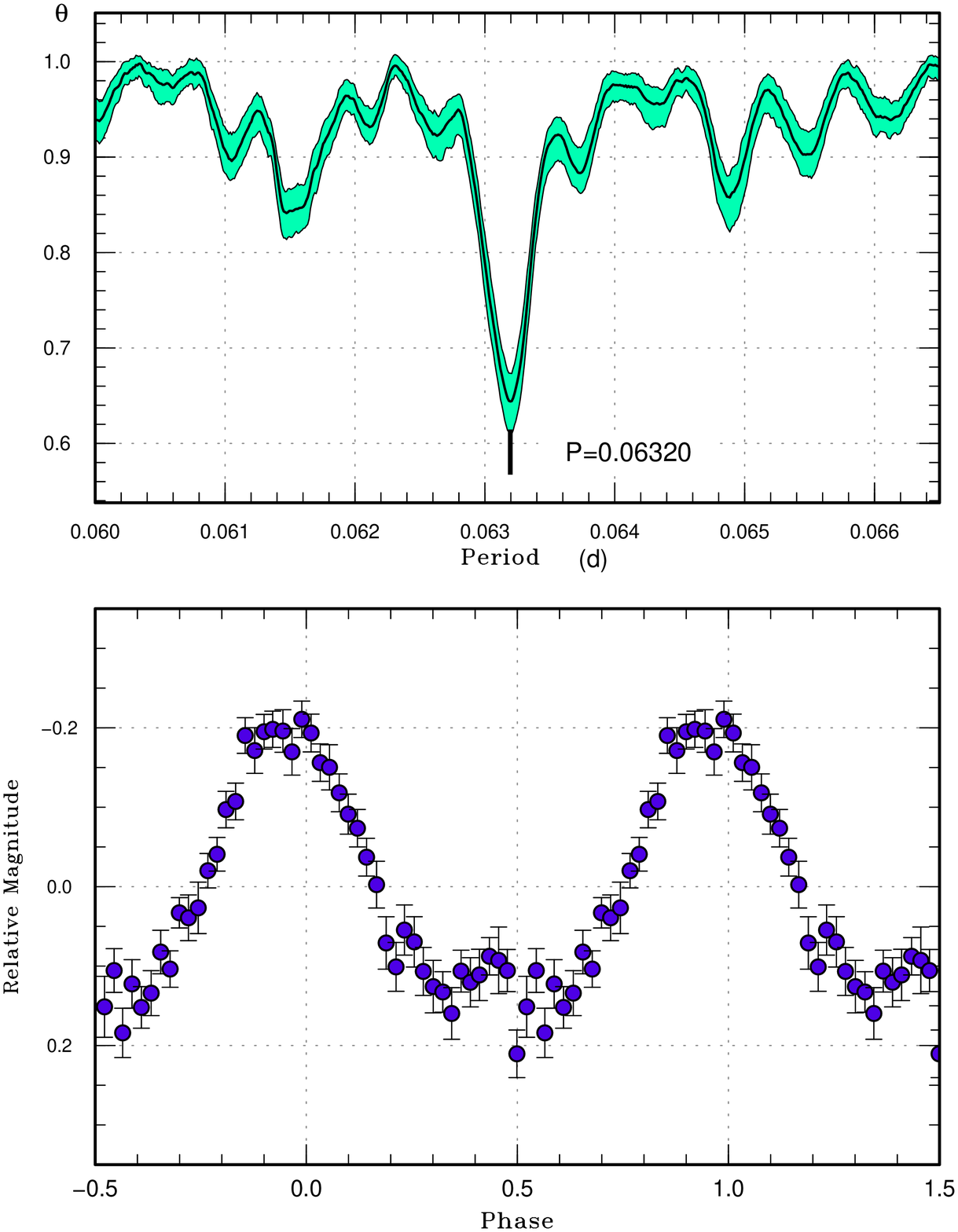}
  \end{center}
  \caption{Superhumps in ASASSN-17fo outside the eclipses (2017).
     (Upper): PDM analysis.
     (Lower): Phase-averaged profile.}
  \label{fig:asassn17foshpdm}
\end{figure}

   This object was detected as a transient
at $V$=16.0 on 2017 April 27 by the ASAS-SN team.
Two previous outbursts were detected in the CRTS data.
Subsequent observations detected deep eclipses and
superhumps (vsnet-alert 20977;
e-figures \ref{fig:asassn17foshlc}, \ref{fig:asassn17foshpdm}).

   The times of superhump maxima are listed in
e-table \ref{tab:asassn17fooc2017}.  Except for $E$=0
all superhumps were probably observed during stage B.

\begin{figure}
  \begin{center}
    \FigureFile(85mm,70mm){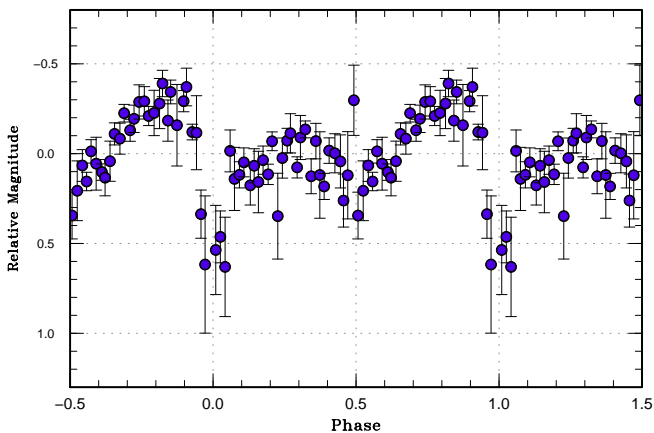}
  \end{center}
  \caption{Mean orbital light curve of ASASSN-17fo.
      The CRTS data in quiescence were used.
      The ephemeris of equation (\ref{equ:asassn17foecl}) is used.}
  \label{fig:asassn17foporb}
\end{figure}

   We obtained the following eclipse ephemeris
using the MCMC analysis \citep{Pdot4}
on the quiescent observations in the CRTS data:
\begin{equation}
{\rm Min(BJD)} = 2457877.05448(13) + 0.061548044(3) E .
\label{equ:asassn17foecl}
\end{equation}
The epoch corresponds to the center of our observations
during the outburst.  This period was uniquely determined
both by the PDM method and the MCMC method, and is consistent
with the period [0.061536~d] determined from
observations during the current outburst.
The orbital light curve (e-figure \ref{fig:asassn17foporb})
indicates the presence of double-wave orbital variations
and an orbital hump.  The depth of eclipsed was probably
underestimated since the object becomes too faint
for CRTS at eclipse centers.  The quiescent orbital
light curve suggests a low-mass transfer rate, which is
compatible with the relatively rare outbursts in
the past data.  Since the object appears to have a very
high orbital inclination as judged from high amplitudes
of superhumps and low outburst amplitudes, detailed analysis
of the quiescent light curve is desired.

\begin{table}
\caption{Superhump maxima of ASASSN-17fo (2017)}\label{tab:asassn17fooc2017}
\begin{center}
\begin{tabular}{rp{50pt}p{30pt}r@{.}lcr}
\hline
\multicolumn{1}{c}{$E$} & \multicolumn{1}{c}{max\commenta} & \multicolumn{1}{c}{error} & \multicolumn{2}{c}{$O-C$\commentb} & \multicolumn{1}{c}{phase\commentc} & \multicolumn{1}{c}{$N$\commentd} \\
\hline
0 & 57873.5174 & 0.0006 & 0&0018 & 0.52 & 16 \\
8 & 57874.0230 & 0.0003 & 0&0015 & 0.74 & 61 \\
9 & 57874.0859 & 0.0004 & 0&0012 & 0.76 & 84 \\
10 & 57874.1502 & 0.0003 & 0&0023 & 0.80 & 79 \\
16 & 57874.5253 & 0.0013 & $-$0&0020 & 0.90 & 17 \\
17 & 57874.5910 & 0.0014 & 0&0005 & 0.97 & 14 \\
32 & 57875.5386 & 0.0005 & $-$0&0004 & 0.37 & 21 \\
33 & 57875.6016 & 0.0005 & $-$0&0006 & 0.39 & 19 \\
39 & 57875.9846 & 0.0009 & 0&0030 & 0.61 & 46 \\
40 & 57876.0414 & 0.0007 & $-$0&0034 & 0.54 & 83 \\
41 & 57876.1054 & 0.0009 & $-$0&0027 & 0.58 & 99 \\
45 & 57876.3583 & 0.0006 & $-$0&0027 & 0.69 & 46 \\
46 & 57876.4215 & 0.0007 & $-$0&0027 & 0.71 & 27 \\
48 & 57876.5494 & 0.0008 & $-$0&0013 & 0.79 & 19 \\
49 & 57876.6113 & 0.0012 & $-$0&0027 & 0.80 & 17 \\
50 & 57876.6764 & 0.0004 & $-$0&0008 & 0.85 & 45 \\
51 & 57876.7473 & 0.0026 & 0&0069 & 0.01 & 38 \\
56 & 57877.0539 & 0.0005 & $-$0&0027 & 0.99 & 75 \\
57 & 57877.1174 & 0.0013 & $-$0&0024 & 0.02 & 72 \\
61 & 57877.3691 & 0.0034 & $-$0&0036 & 0.11 & 18 \\
63 & 57877.5021 & 0.0025 & 0&0029 & 0.27 & 20 \\
64 & 57877.5621 & 0.0022 & $-$0&0003 & 0.25 & 13 \\
72 & 57878.0675 & 0.0027 & $-$0&0008 & 0.46 & 62 \\
73 & 57878.1391 & 0.0028 & 0&0076 & 0.62 & 50 \\
79 & 57878.5120 & 0.0019 & 0&0011 & 0.68 & 19 \\
80 & 57878.5747 & 0.0023 & 0&0006 & 0.70 & 14 \\
\hline
  \multicolumn{7}{l}{\commenta BJD$-$2400000.} \\
  \multicolumn{7}{l}{\commentb Against max $= 2457873.5156 + 0.063232 E$.} \\
  \multicolumn{7}{l}{\commentc Orbital phase.} \\
  \multicolumn{7}{l}{\commentd Number of points used to determine the maximum.} \\
\end{tabular}
\end{center}
\end{table}

\subsection{ASASSN-17fp}\label{obj:asassn17fp}

   This object was detected as a transient
at $V$=16.2 on 2017 April 28 by the ASAS-SN team.
The outburst was announced after the observation
on 2017 April 28 ($V$=15.7).
The object was spectroscopically studied on 2017 April 28
and found to show He I lines but lack hydrogen
\citep{car17asassn17fpatel10334}.  The object was 
then suggested to be a candidate AM CVn-type object.
\citet{mar17asassn17fpatel10354} reported 51.0(1)-min
modulations on 2017 May 6 and suggested that the object
is located at the long period end of the AM CVn period
distribution.  The object faded quickly after
this, but it showed at least one further short outburst
\citep{waa17asassn17fpaan580}.

\begin{figure}
  \begin{center}
    \FigureFile(85mm,110mm){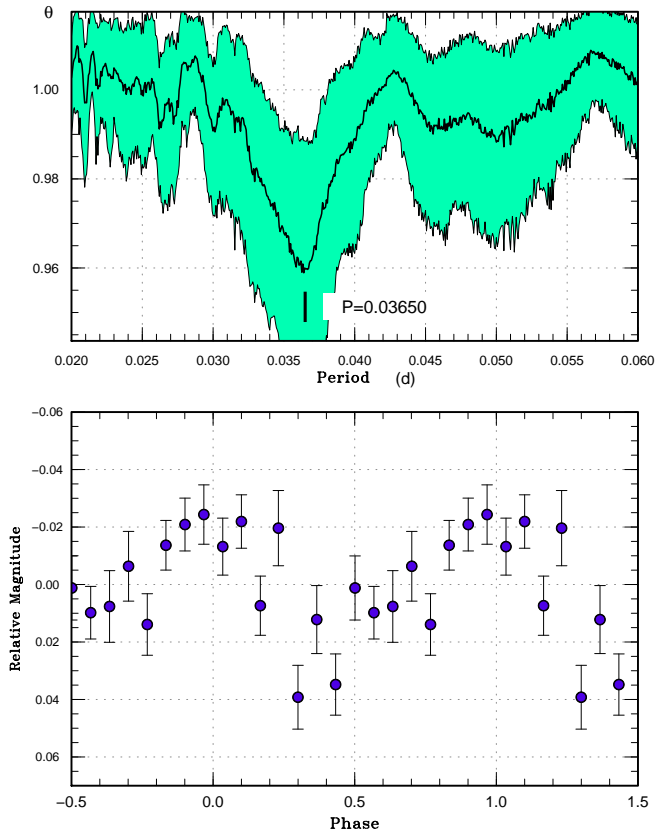}
  \end{center}
  \caption{Likely superhumps in ASASSN-17fp (2017).
     (Upper): PDM analysis.
     (Lower): Phase-averaged profile.}
  \label{fig:asassn17fpshpdm}
\end{figure}

   Our observations during the outburst were performed
on 2017 May 4--5.  A PDM analysis yielded a period
of 0.0365(5)~d [52.6(7)~min], close to the value
obtained by \citet{mar17asassn17fpatel10354}
(e-figure \ref{fig:asassn17fpshpdm}).

   This field has been monitored by the ASAS-SN team
since 2014 April 29 and no past outbursts were
recorded.  An outburst at $V$=16.1 on 2017 May 16 was
the only positive detection by the ASAS-SN team
after the main outburst.  This outburst is thus
considered to be a post-superoutburst rebrightening.

\subsection{ASASSN-17fz}\label{obj:asassn17fz}

\begin{figure}
  \begin{center}
    \FigureFile(85mm,110mm){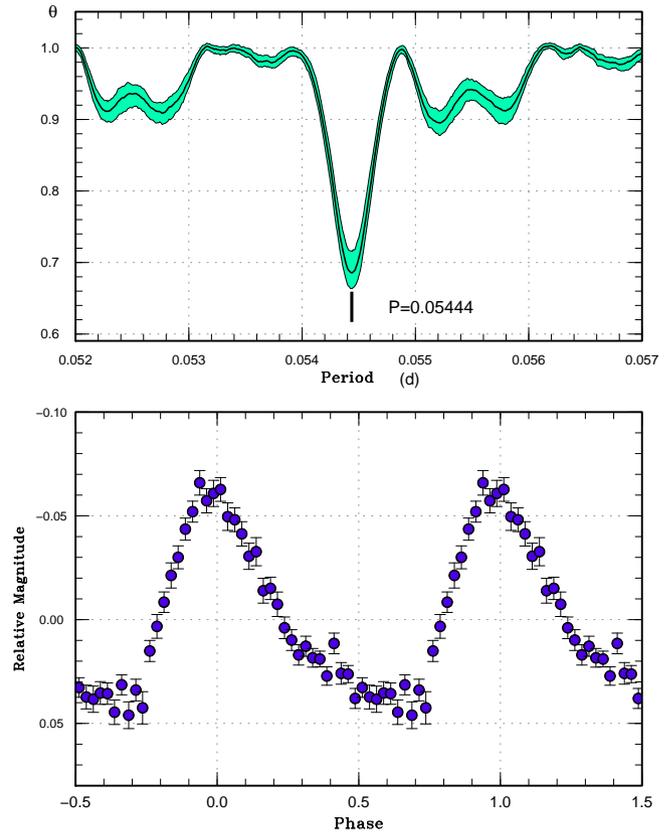}
  \end{center}
  \caption{Ordinary superhumps in ASASSN-17fz (2017).
     (Upper): PDM analysis.
     (Lower): Phase-averaged profile.}
  \label{fig:asassn17fzshpdm}
\end{figure}

   This object was detected as a transient
at $V$=14.2 on 2017 May 5 by the ASAS-SN team.
The large outburst amplitude received attention.
Although superhump-like features were suggested
(vsnet-alert 20992), they were not confirmed by
later observations (vsnet-alert 21003).
The object developed evolving ordinary superhumps
on May 12 (7~d after the initial outburst detection;
vsnet-alert 21017, 21026; e-figure \ref{fig:asassn17fzshpdm}).
The times of superhump maxima are listed in
e-table \ref{tab:asassn17fzoc2017}.  The maxima for
$E \le$41 were clearly stage A superhumps.

   Although a possible period of early superhumps was
reported in vsnet-alert 21026, we could not confirm
it by further analysis.  We consider that the object
is likely a WZ Sge-type dwarf nova based on long
duration of stage A (suggesting a small $q$).
The waiting time (7~d) of the appearance of ordinary
superhump is slight shorter than in most WZ Sge-type
dwarf novae.  This may have been a result of the lack
of observations on three nights preceding the initial
ASAS-SN detection and the true waiting time may have been
longer.

   The object was still in superoutburst plateau
on 2017 May 21.  ASAS-SN observations on subsequent three
nights were not deep enough to see whether the outburst
still continued.  The object was found to have faded
to 19 mag on 2017 May 28.  There was no indication of
a post-superoutburst rebrightening although our CCD
observations were not sufficient and ASAS-SN data
were not deep enough to detect a rebrightening.
ASAS-SN started regularly observing this field
on 2016 February 2 and there were no previous outbursts.

\begin{table}
\caption{Superhump maxima of ASASSN-17fz (2017)}\label{tab:asassn17fzoc2017}
\begin{center}
\begin{tabular}{rp{55pt}p{40pt}r@{.}lr}
\hline
\multicolumn{1}{c}{$E$} & \multicolumn{1}{c}{max\commenta} & \multicolumn{1}{c}{error} & \multicolumn{2}{c}{$O-C$\commentb} & \multicolumn{1}{c}{$N$\commentc} \\
\hline
0 & 57886.2577 & 0.0062 & $-$0&0176 & 95 \\
1 & 57886.3163 & 0.0021 & $-$0&0135 & 125 \\
2 & 57886.3822 & 0.0076 & $-$0&0021 & 106 \\
17 & 57887.2023 & 0.0007 & 0&0007 & 91 \\
18 & 57887.2579 & 0.0005 & 0&0018 & 125 \\
19 & 57887.3101 & 0.0007 & $-$0&0005 & 125 \\
20 & 57887.3665 & 0.0008 & 0&0014 & 125 \\
23 & 57887.5312 & 0.0013 & 0&0027 & 14 \\
36 & 57888.2445 & 0.0003 & 0&0075 & 124 \\
37 & 57888.2989 & 0.0004 & 0&0075 & 124 \\
38 & 57888.3545 & 0.0004 & 0&0086 & 123 \\
41 & 57888.5185 & 0.0009 & 0&0091 & 13 \\
60 & 57889.5480 & 0.0008 & 0&0033 & 12 \\
72 & 57890.2037 & 0.0006 & 0&0051 & 103 \\
73 & 57890.2557 & 0.0006 & 0&0027 & 124 \\
74 & 57890.3085 & 0.0005 & 0&0010 & 124 \\
75 & 57890.3622 & 0.0007 & 0&0002 & 117 \\
78 & 57890.5287 & 0.0018 & 0&0032 & 12 \\
91 & 57891.2335 & 0.0006 & $-$0&0004 & 125 \\
92 & 57891.2885 & 0.0005 & 0&0001 & 125 \\
93 & 57891.3429 & 0.0010 & $-$0&0000 & 96 \\
109 & 57892.2123 & 0.0005 & $-$0&0024 & 118 \\
110 & 57892.2668 & 0.0005 & $-$0&0024 & 125 \\
111 & 57892.3228 & 0.0007 & $-$0&0009 & 125 \\
112 & 57892.3707 & 0.0033 & $-$0&0074 & 63 \\
115 & 57892.5393 & 0.0015 & $-$0&0024 & 11 \\
133 & 57893.5174 & 0.0023 & $-$0&0051 & 12 \\
151 & 57894.4994 & 0.0013 & $-$0&0038 & 17 \\
152 & 57894.5615 & 0.0048 & 0&0037 & 8 \\
\hline
  \multicolumn{6}{l}{\commenta BJD$-$2400000.} \\
  \multicolumn{6}{l}{\commentb Against max $= 2457886.2753 + 0.054490 E$.} \\
  \multicolumn{6}{l}{\commentc Number of points used to determine the maximum.} \\
\end{tabular}
\end{center}
\end{table}

\subsection{ASASSN-17gf}\label{obj:asassn17gf}

\begin{figure}
  \begin{center}
    \FigureFile(85mm,110mm){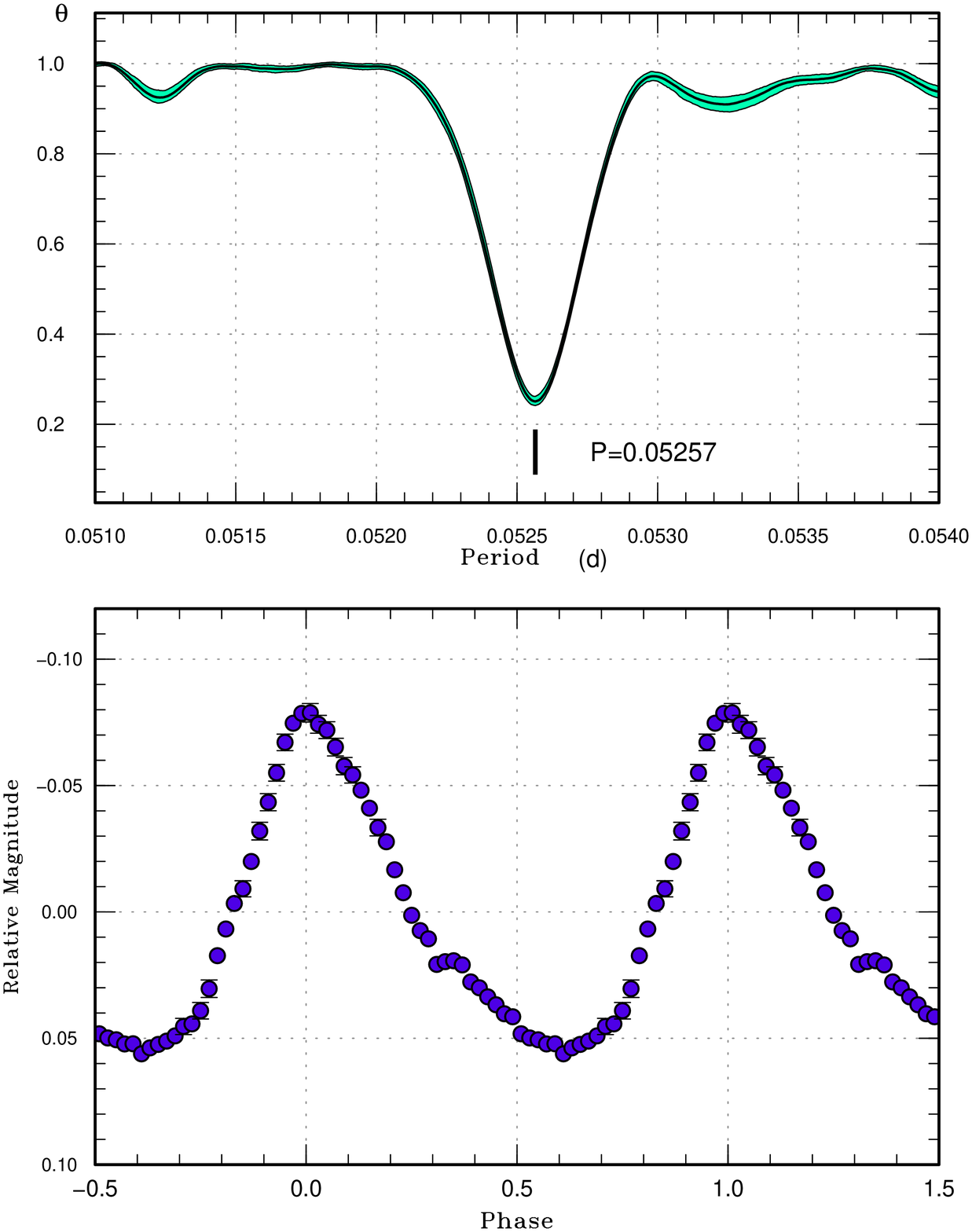}
  \end{center}
  \caption{Superhumps in ASASSN-17gf (2017).
     (Upper): PDM analysis.
     (Lower): Phase-averaged profile.}
  \label{fig:asassn17gfshpdm}
\end{figure}

   This object was detected as a transient
at $V$=13.05 on 2017 May 14 by the ASAS-SN team.
Superhumps soon grew and the object was identified
as an SU UMa-type dwarf nova below the period
minimum (vsnet-alert 21021; e-figure \ref{fig:asassn17gfshpdm}).  
The times of superhump maxima are listed in
e-table \ref{tab:asassn17gfoc2017}.  The presence
of stages A and B is clearly visible.

   Although our CCD observations of the superoutburst
ended on 2017 May 21, the object was confirmed to be bright
at least up to May 26 at a visual magnitude of 14.0
according to the AAVSO observations.
The object was found to be fading rapidly on 2017 May 27.
There was no indication of post-superoutburst rebrightening.
The object was still in quiescence on 2017 May 10.
The duration of the superoutburst was between 13~d and 17~d.
This duration is typical for an SU UMa-type dwarf nova,
and is unlike WZ Sge-type objects below the period minimum,
such as ASASSN-15po \citep{nam17asassn15po} and OV Boo,
which are possibly population II dwarf novae
(e.g. \cite{pat08j1507}; \cite{nam17asassn15po};
\cite{ohn19ovboo}).
ASASSN-17gf is thus most likely an EI Psc-like object.
EI Psc-like systems are CVs below the period minimum
showing hydrogen (likely somewhat reduced in abundance)
in their spectra (cf. \cite{tho02j2329};
\cite{uem02j2329letter}; \cite{lit13sbs1108}) and are
considered to be evolving towards AM CVn-type objects.

   This object has been monitored by the ASAS-SN team since
2016 February 2 and there has been no other outburst
in the ASAS-SN data.  There were, however, rather numerous
outburst detections in the ASAS-3 data
(e-table \ref{tab:asassn17gfout}; since there is a possible
nearby contaminating object, only outbursts with more than
two positive detections are listed).  The listed outbursts
were all likely superoutbursts and the frequency of
superoutbursts appears to be comparable to or even
higher than EI Psc (\Ohtprep).

\begin{table}
\caption{Superhump maxima of ASASSN-17gf (2017)}\label{tab:asassn17gfoc2017}
\begin{center}
\begin{tabular}{rp{55pt}p{40pt}r@{.}lr}
\hline
\multicolumn{1}{c}{$E$} & \multicolumn{1}{c}{max\commenta} & \multicolumn{1}{c}{error} & \multicolumn{2}{c}{$O-C$\commentb} & \multicolumn{1}{c}{$N$\commentc} \\
\hline
0 & 57888.5567 & 0.0013 & $-$0&0124 & 10 \\
12 & 57889.2046 & 0.0011 & 0&0046 & 51 \\
13 & 57889.2498 & 0.0001 & $-$0&0027 & 121 \\
14 & 57889.3025 & 0.0002 & $-$0&0026 & 121 \\
15 & 57889.3569 & 0.0002 & $-$0&0008 & 121 \\
16 & 57889.4083 & 0.0002 & $-$0&0020 & 121 \\
18 & 57889.5174 & 0.0005 & 0&0020 & 11 \\
19 & 57889.5717 & 0.0010 & 0&0037 & 9 \\
20 & 57889.6224 & 0.0003 & 0&0019 & 10 \\
31 & 57890.2035 & 0.0002 & 0&0047 & 92 \\
32 & 57890.2549 & 0.0002 & 0&0035 & 121 \\
33 & 57890.3068 & 0.0001 & 0&0028 & 121 \\
34 & 57890.3588 & 0.0002 & 0&0022 & 121 \\
35 & 57890.4076 & 0.0034 & $-$0&0015 & 32 \\
37 & 57890.5159 & 0.0006 & 0&0016 & 12 \\
38 & 57890.5689 & 0.0009 & 0&0020 & 10 \\
39 & 57890.6219 & 0.0008 & 0&0024 & 10 \\
50 & 57891.1983 & 0.0004 & 0&0005 & 74 \\
51 & 57891.2516 & 0.0002 & 0&0013 & 114 \\
52 & 57891.3030 & 0.0003 & 0&0001 & 106 \\
53 & 57891.3564 & 0.0002 & 0&0009 & 100 \\
54 & 57891.4083 & 0.0003 & 0&0002 & 79 \\
56 & 57891.5126 & 0.0015 & $-$0&0006 & 12 \\
57 & 57891.5649 & 0.0010 & $-$0&0009 & 10 \\
58 & 57891.6162 & 0.0011 & $-$0&0021 & 10 \\
69 & 57892.1970 & 0.0007 & 0&0003 & 62 \\
70 & 57892.2486 & 0.0002 & $-$0&0006 & 121 \\
71 & 57892.3014 & 0.0002 & $-$0&0004 & 121 \\
72 & 57892.3534 & 0.0002 & $-$0&0010 & 121 \\
75 & 57892.5106 & 0.0006 & $-$0&0016 & 12 \\
76 & 57892.5662 & 0.0008 & 0&0015 & 10 \\
77 & 57892.6176 & 0.0006 & 0&0004 & 9 \\
89 & 57893.2461 & 0.0002 & $-$0&0021 & 121 \\
90 & 57893.2995 & 0.0002 & $-$0&0012 & 121 \\
91 & 57893.3518 & 0.0003 & $-$0&0015 & 121 \\
92 & 57893.4036 & 0.0003 & $-$0&0022 & 121 \\
94 & 57893.5108 & 0.0015 & $-$0&0002 & 11 \\
95 & 57893.5597 & 0.0020 & $-$0&0039 & 10 \\
96 & 57893.6183 & 0.0009 & 0&0021 & 9 \\
108 & 57894.2454 & 0.0006 & $-$0&0017 & 68 \\
109 & 57894.2985 & 0.0005 & $-$0&0011 & 84 \\
113 & 57894.5092 & 0.0006 & $-$0&0007 & 18 \\
127 & 57895.2468 & 0.0003 & 0&0009 & 118 \\
128 & 57895.3005 & 0.0005 & 0&0020 & 106 \\
129 & 57895.3534 & 0.0007 & 0&0023 & 56 \\
\hline
  \multicolumn{6}{l}{\commenta BJD$-$2400000.} \\
  \multicolumn{6}{l}{\commentb Against max $= 2457888.5691 + 0.052574 E$.} \\
  \multicolumn{6}{l}{\commentc Number of points used to determine the maximum.} \\
\end{tabular}
\end{center}
\end{table}

\begin{table}
\caption{List of past outbursts of ASASSN-17gf}\label{tab:asassn17gfout}
\begin{center}
\begin{tabular}{ccccc}
\hline
Year & Month & Day & max\commenta & $V$-mag \\
\hline
2002 & 10 & 11 & 52559 & 13.00 \\
2003 &  7 & 10 & 52830 & 13.27 \\
2004 &  1 &  3 & 53008 & 13.61 \\
2004 &  5 & 19 & 53145 & 13.56 \\
2005 &  2 & 11 & 53412 & 13.04 \\
2005 &  1 & 21 & 53656 & 12.76 \\
2008 &  2 &  5 & 54502 & 13.27 \\
\hline
  \multicolumn{5}{l}{\commenta JD$-$2400000.} \\
\end{tabular}
\end{center}
\end{table}

\subsection{ASASSN-17gh}\label{obj:asassn17gh}

\begin{figure}
  \begin{center}
    \FigureFile(85mm,110mm){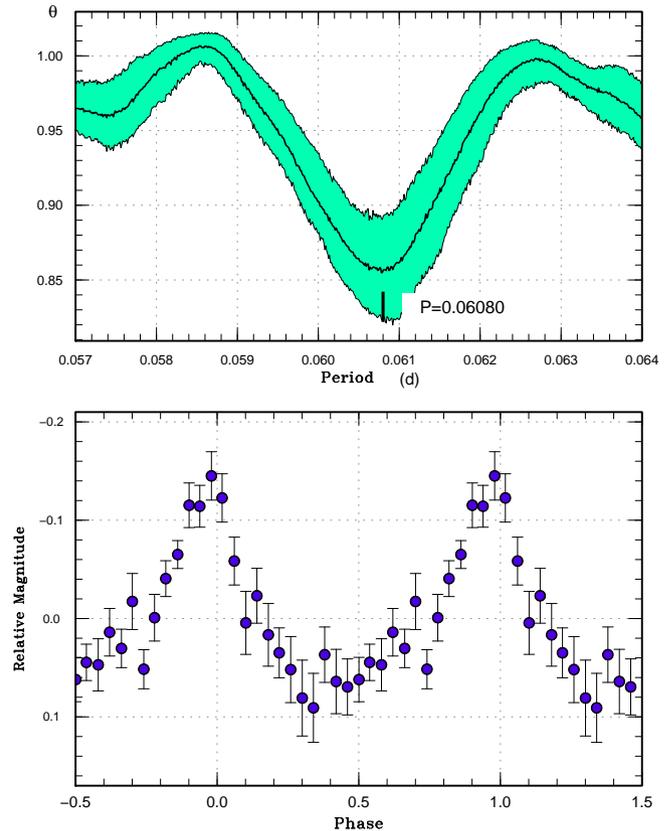}
  \end{center}
  \caption{Superhumps in ASASSN-17gh (2017).
     (Upper): PDM analysis.
     (Lower): Phase-averaged profile.}
  \label{fig:asassn17ghshpdm}
\end{figure}

   This object was detected as a transient
at $V$=16.2 on 2017 May 16 by the ASAS-SN team.
The outburst was announced after the observation
at $V$=15.9 on 2017 May 16.  Subsequent observations
detected superhumps (vsnet-alert 21048;
e-figure \ref{fig:asassn17ghshpdm}).
The times of superhump maxima are listed in
e-table \ref{tab:asassn17ghoc2017}.  This table
does not include observations on the second night
with less quality to determine the times of maxima.
The night, however, was included in the PDM analysis
giving a period of 0.0608(1)~d, which is considered
to be more reliable than the one from single-night
$O-C$ analysis.

\begin{table}
\caption{Superhump maxima of ASASSN-17gh (2017)}\label{tab:asassn17ghoc2017}
\begin{center}
\begin{tabular}{rp{55pt}p{40pt}r@{.}lr}
\hline
\multicolumn{1}{c}{$E$} & \multicolumn{1}{c}{max\commenta} & \multicolumn{1}{c}{error} & \multicolumn{2}{c}{$O-C$\commentb} & \multicolumn{1}{c}{$N$\commentc} \\
\hline
0 & 57894.0326 & 0.0009 & 0&0019 & 91 \\
1 & 57894.0949 & 0.0013 & 0&0028 & 97 \\
2 & 57894.1480 & 0.0037 & $-$0&0054 & 52 \\
6 & 57894.3972 & 0.0009 & $-$0&0018 & 31 \\
7 & 57894.4623 & 0.0004 & 0&0018 & 63 \\
8 & 57894.5218 & 0.0005 & $-$0&0000 & 62 \\
9 & 57894.5839 & 0.0009 & 0&0007 & 31 \\
\hline
  \multicolumn{6}{l}{\commenta BJD$-$2400000.} \\
  \multicolumn{6}{l}{\commentb Against max $= 2457894.0307 + 0.061394 E$.} \\
  \multicolumn{6}{l}{\commentc Number of points used to determine the maximum.} \\
\end{tabular}
\end{center}
\end{table}

\subsection{ASASSN-17gv}\label{obj:asassn17gv}

\begin{figure}
  \begin{center}
    \FigureFile(85mm,110mm){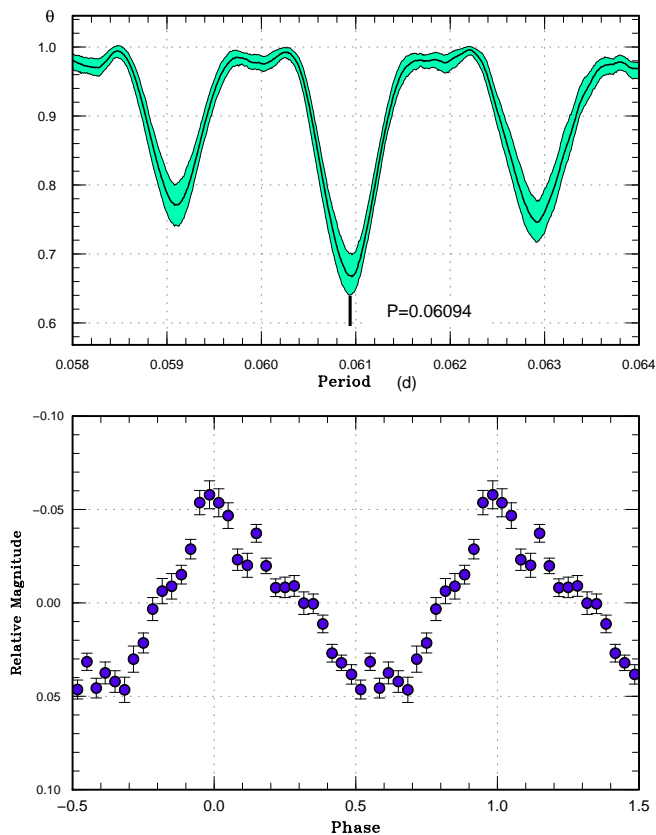}
  \end{center}
  \caption{Superhumps in ASASSN-17gv (2017).
     (Upper): PDM analysis.
     (Lower): Phase-averaged profile.}
  \label{fig:asassn17gvshpdm}
\end{figure}

   This object was detected as a transient
at $V$=14.2 on 2017 May 28 by the ASAS-SN team.
Subsequent observations detected superhumps
(vsnet-alert 21065, 21080; e-figure \ref{fig:asassn17gvshpdm}).
The times of superhump maxima are listed in
e-table \ref{tab:asassn17gvoc2017}.
Although superhumps apparently had not yet fully
developed on the first night and a period of
stage A superhumps was reported in vsnet-alert 21080,
this value was not confirmed by this $O-C$ analysis.
The difference between periods of stage A and B
superhumps was too large (more than 4\%) and we consider
that this period was a spurious detection.
Although a positive $P_{\rm dot}$ is expected for
this superhump period, this $O-C$ analysis did not
give such a tendency.  The data were probably of
limited quality and the data may have also been
contaminated by stage C superhumps.

   The outburst lasted at least up to June 4,
when the ASAS-SN data apparently showed rapid fading.
The object faded to $\sim$18.5 mag on June 9
as recorded by our CCD observations.
The region of this object has been covered
by the ASAS-SN team since 2016 March 8 and no other
outburst was recorded.  Although ASAS-3 did not detect
any outburst, it may have been due to crowding
in this region.

\begin{table}
\caption{Superhump maxima of ASASSN-17gv (2017)}\label{tab:asassn17gvoc2017}
\begin{center}
\begin{tabular}{rp{55pt}p{40pt}r@{.}lr}
\hline
\multicolumn{1}{c}{$E$} & \multicolumn{1}{c}{max\commenta} & \multicolumn{1}{c}{error} & \multicolumn{2}{c}{$O-C$\commentb} & \multicolumn{1}{c}{$N$\commentc} \\
\hline
0 & 57903.2021 & 0.0056 & $-$0&0117 & 45 \\
2 & 57903.3428 & 0.0021 & 0&0072 & 139 \\
3 & 57903.3959 & 0.0005 & $-$0&0006 & 139 \\
6 & 57903.5845 & 0.0008 & 0&0053 & 16 \\
25 & 57904.7355 & 0.0022 & $-$0&0007 & 16 \\
26 & 57904.7965 & 0.0014 & $-$0&0007 & 17 \\
33 & 57905.2243 & 0.0006 & 0&0009 & 109 \\
34 & 57905.2812 & 0.0003 & $-$0&0031 & 129 \\
35 & 57905.3522 & 0.0010 & 0&0070 & 38 \\
41 & 57905.7167 & 0.0036 & 0&0061 & 12 \\
42 & 57905.7702 & 0.0020 & $-$0&0013 & 13 \\
55 & 57906.5626 & 0.0023 & $-$0&0005 & 10 \\
57 & 57906.6801 & 0.0023 & $-$0&0048 & 13 \\
58 & 57906.7395 & 0.0030 & $-$0&0063 & 12 \\
59 & 57906.8069 & 0.0009 & 0&0002 & 13 \\
66 & 57907.2332 & 0.0014 & 0&0002 & 141 \\
67 & 57907.2969 & 0.0009 & 0&0030 & 140 \\
68 & 57907.3536 & 0.0009 & $-$0&0012 & 102 \\
73 & 57907.6543 & 0.0034 & $-$0&0050 & 11 \\
74 & 57907.7217 & 0.0031 & 0&0015 & 12 \\
75 & 57907.7802 & 0.0042 & $-$0&0009 & 13 \\
88 & 57908.5780 & 0.0023 & 0&0053 & 9 \\
\hline
  \multicolumn{6}{l}{\commenta BJD$-$2400000.} \\
  \multicolumn{6}{l}{\commentb Against max $= 2457903.2138 + 0.060897 E$.} \\
  \multicolumn{6}{l}{\commentc Number of points used to determine the maximum.} \\
\end{tabular}
\end{center}
\end{table}

\subsection{ASASSN-17hm}\label{obj:asassn17hm}

\begin{figure}
  \begin{center}
    \FigureFile(85mm,110mm){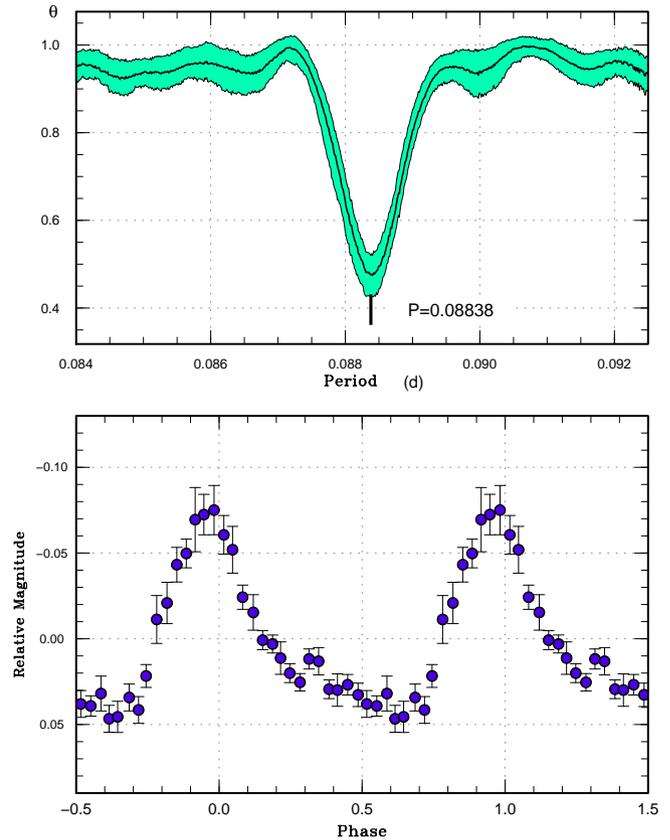}
  \end{center}
  \caption{Superhumps in ASASSN-17hm (2017).
     (Upper): PDM analysis.
     (Lower): Phase-averaged profile.}
  \label{fig:asassn17hmshpdm}
\end{figure}

   This object was detected as a transient
at $V$=14.2 on 2017 June 10 by the ASAS-SN team.
The outburst was announced after observations
on 2017 June 11 ($V$=14.3) and 2017 June 12 ($V$=14.6).
The object was not yet in outburst on June 9.
Subsequent observations detected superhumps
(vsnet-alert 21117; e-figure \ref{fig:asassn17hmshpdm}).
The times of superhump maxima are listed in
e-table \ref{tab:asassn17hmoc2017}.
The epochs for $E \ge$68 refer to post-superoutburst
superhumps.  Although there was an apparent change
in the period around the termination of the superoutburst,
no phase jump was recorded. 

   This field has been monitored by the ASAS-SN team
since 2016 March 9 and there was another outburst
at $V$=14.1 on 2016 July 16.  There were gaps before
and after this observation and the outburst type
is unknown.

\begin{table}
\caption{Superhump maxima of ASASSN-17hm (2017)}\label{tab:asassn17hmoc2017}
\begin{center}
\begin{tabular}{rp{55pt}p{40pt}r@{.}lr}
\hline
\multicolumn{1}{c}{$E$} & \multicolumn{1}{c}{max\commenta} & \multicolumn{1}{c}{error} & \multicolumn{2}{c}{$O-C$\commentb} & \multicolumn{1}{c}{$N$\commentc} \\
\hline
0 & 57918.5083 & 0.0011 & $-$0&0021 & 22 \\
2 & 57918.6894 & 0.0008 & 0&0022 & 18 \\
3 & 57918.7739 & 0.0014 & $-$0&0018 & 16 \\
11 & 57919.4738 & 0.0056 & $-$0&0091 & 10 \\
12 & 57919.5709 & 0.0017 & $-$0&0005 & 10 \\
13 & 57919.6589 & 0.0027 & $-$0&0009 & 18 \\
14 & 57919.7482 & 0.0009 & 0&0000 & 19 \\
23 & 57920.5496 & 0.0021 & 0&0058 & 14 \\
24 & 57920.6349 & 0.0030 & 0&0027 & 20 \\
25 & 57920.7225 & 0.0012 & 0&0019 & 21 \\
34 & 57921.5183 & 0.0018 & 0&0020 & 21 \\
36 & 57921.6985 & 0.0012 & 0&0054 & 20 \\
37 & 57921.7850 & 0.0027 & 0&0034 & 12 \\
45 & 57922.4888 & 0.0040 & 0&0000 & 12 \\
48 & 57922.7553 & 0.0022 & 0&0012 & 16 \\
58 & 57923.6347 & 0.0015 & $-$0&0035 & 17 \\
59 & 57923.7243 & 0.0020 & $-$0&0022 & 16 \\
68 & 57924.5175 & 0.0016 & $-$0&0047 & 20 \\
69 & 57924.6084 & 0.0016 & $-$0&0023 & 13 \\
70 & 57924.6962 & 0.0007 & $-$0&0028 & 16 \\
81 & 57925.6702 & 0.0044 & $-$0&0014 & 16 \\
82 & 57925.7666 & 0.0051 & 0&0066 & 12 \\
\hline
  \multicolumn{6}{l}{\commenta BJD$-$2400000.} \\
  \multicolumn{6}{l}{\commentb Against max $= 2457918.5104 + 0.088409 E$.} \\
  \multicolumn{6}{l}{\commentc Number of points used to determine the maximum.} \\
\end{tabular}
\end{center}
\end{table}

\subsection{ASASSN-17hw}\label{obj:asassn17hw}

\begin{figure}
  \begin{center}
    \FigureFile(85mm,110mm){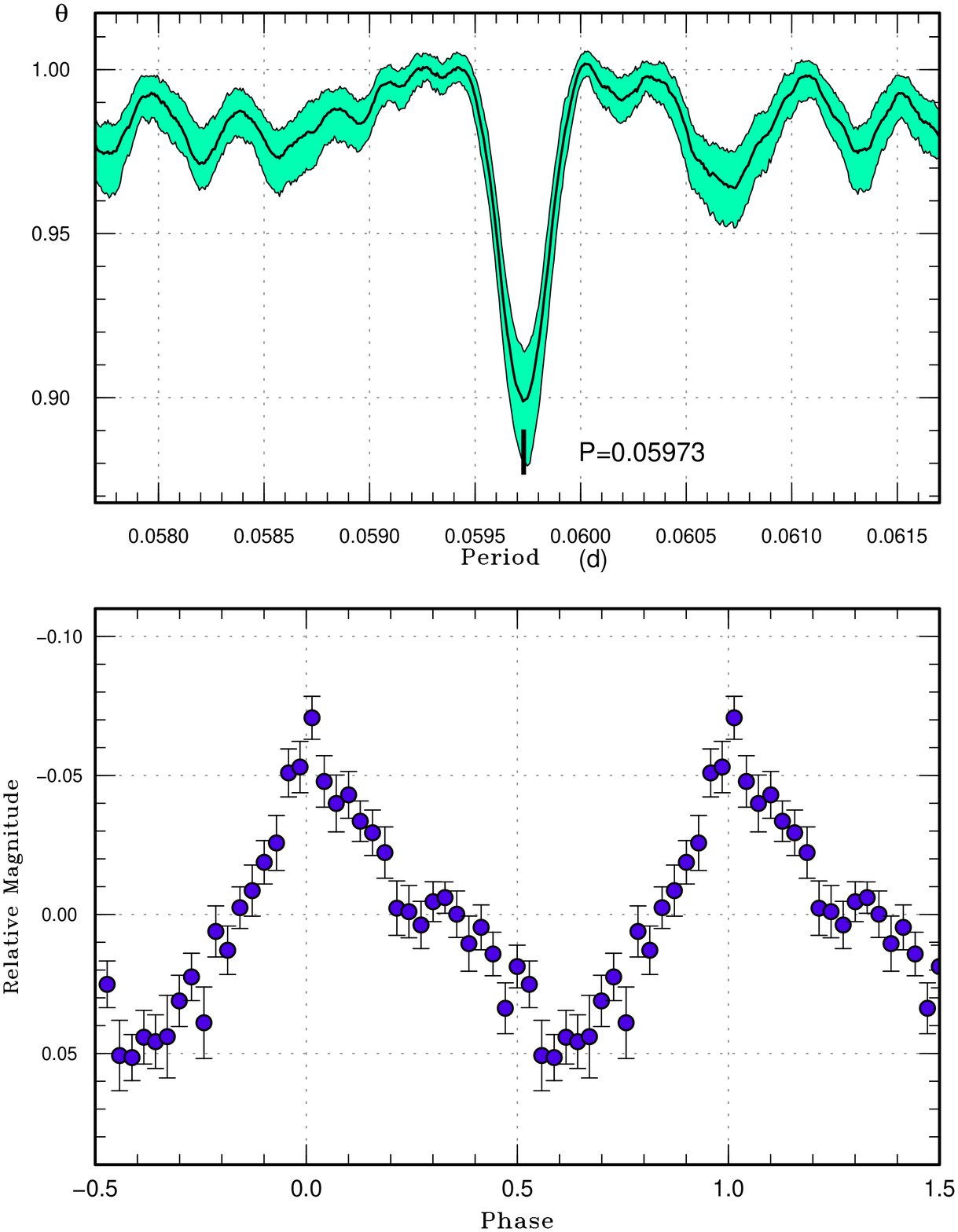}
  \end{center}
  \caption{Ordinary superhumps in ASASSN-17hw (2017).
     (Upper): PDM analysis.
     (Lower): Phase-averaged profile.}
  \label{fig:asassn17hwshpdm}
\end{figure}

   This object was detected as a transient
at $V$=13.2 on 2017 June 20 by the ASAS-SN team.
The object started to show superhumps on 2017 July 2
(vsnet-alert 21194, 21208, 21209, 21219;
e-figure \ref{fig:asassn17hwshpdm}).
The times of superhump maxima are listed in
e-table \ref{tab:asassn17hwoc2017}.  There are
unambiguous stages A and B.  It is remarkable
that $P_{\rm dot}$ is almost zero during stage B
(cf. vsnet-alert 21251).

\begin{figure}
  \begin{center}
    \FigureFile(85mm,110mm){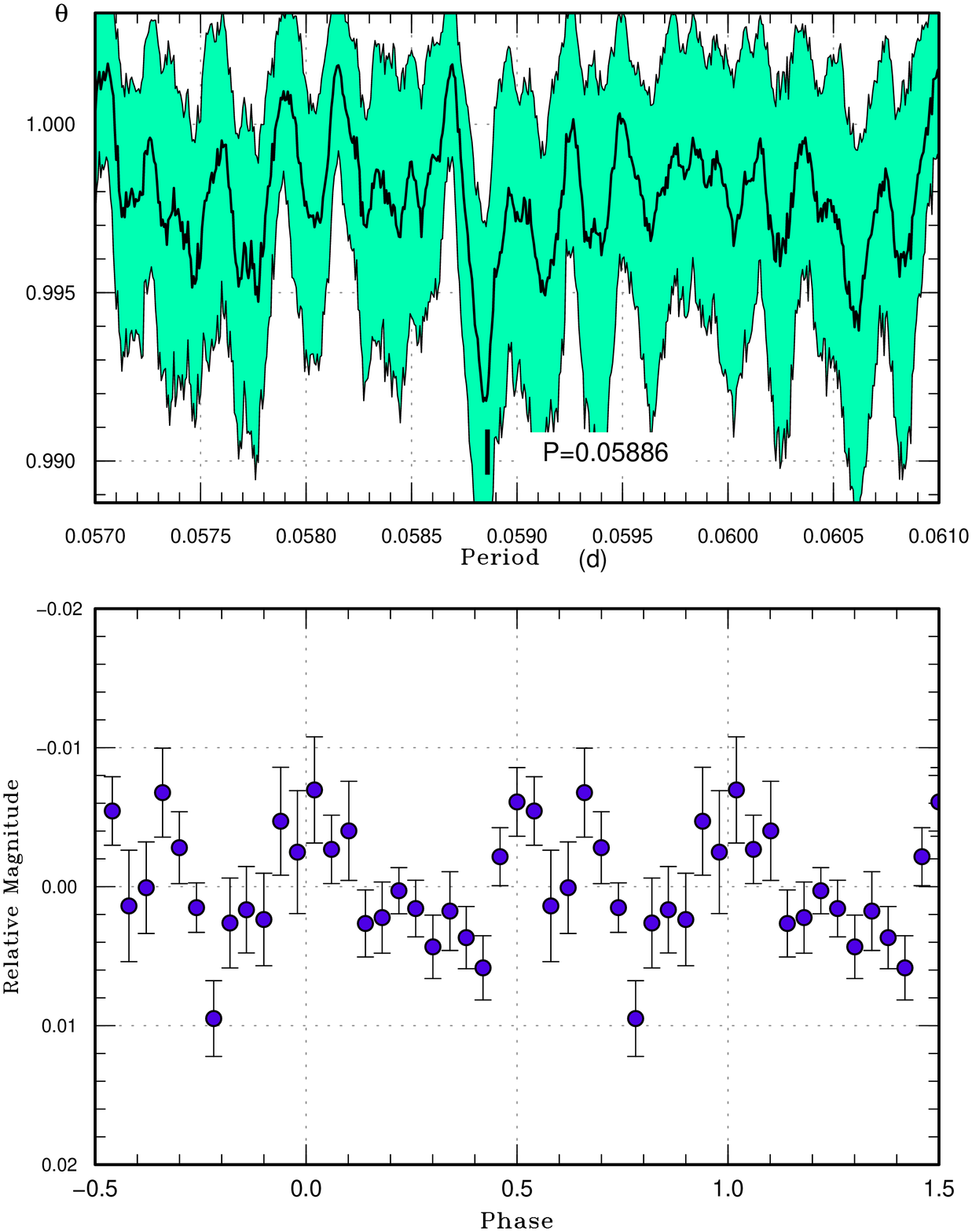}
  \end{center}
  \caption{Early superhumps in ASASSN-17hw (2017).
     (Upper): PDM analysis.
     (Lower): Phase-averaged profile.}
  \label{fig:asassn17hweshpdm}
\end{figure}

   An analysis of the early part of the outburst
yielded small-amplitude early superhumps
(e-figure \ref{fig:asassn17hweshpdm}), confirming
the WZ Sge-type nature of this object.
The period of early superhumps determined by
the PDM method was 0.05886(2)~d.  Combined with
the period of stage A superhumps [0.060617(5)~d],
the value of $\epsilon^*$ for stage A is
0.0290(4), which corresponds to $q$=0.078(1).

   Although the object was well monitored after
the termination of the outburst, no rebrightening
was recorded.

\begin{table}
\caption{Superhump maxima of ASASSN-17hw (2017)}\label{tab:asassn17hwoc2017}
\begin{center}
\begin{tabular}{rp{55pt}p{40pt}r@{.}lr}
\hline
\multicolumn{1}{c}{$E$} & \multicolumn{1}{c}{max\commenta} & \multicolumn{1}{c}{error} & \multicolumn{2}{c}{$O-C$\commentb} & \multicolumn{1}{c}{$N$\commentc} \\
\hline
0 & 57936.7251 & 0.0012 & $-$0&0170 & 13 \\
1 & 57936.7829 & 0.0007 & $-$0&0189 & 10 \\
11 & 57937.3901 & 0.0014 & $-$0&0093 & 43 \\
16 & 57937.6948 & 0.0004 & $-$0&0033 & 11 \\
17 & 57937.7565 & 0.0007 & $-$0&0014 & 13 \\
27 & 57938.3594 & 0.0006 & 0&0039 & 159 \\
28 & 57938.4200 & 0.0006 & 0&0048 & 175 \\
29 & 57938.4830 & 0.0011 & 0&0080 & 81 \\
33 & 57938.7196 & 0.0006 & 0&0056 & 13 \\
34 & 57938.7779 & 0.0010 & 0&0041 & 10 \\
44 & 57939.3763 & 0.0020 & 0&0050 & 23 \\
45 & 57939.4353 & 0.0008 & 0&0042 & 36 \\
46 & 57939.4966 & 0.0026 & 0&0057 & 14 \\
60 & 57940.3362 & 0.0031 & 0&0088 & 16 \\
61 & 57940.3925 & 0.0016 & 0&0052 & 18 \\
62 & 57940.4410 & 0.0061 & $-$0&0059 & 22 \\
63 & 57940.4977 & 0.0022 & $-$0&0090 & 13 \\
66 & 57940.6895 & 0.0011 & 0&0035 & 12 \\
67 & 57940.7509 & 0.0017 & 0&0052 & 12 \\
83 & 57941.7080 & 0.0018 & 0&0061 & 12 \\
84 & 57941.7617 & 0.0040 & 0&0000 & 10 \\
93 & 57942.3023 & 0.0028 & 0&0029 & 137 \\
94 & 57942.3632 & 0.0021 & 0&0040 & 137 \\
95 & 57942.4198 & 0.0021 & 0&0008 & 105 \\
100 & 57942.7209 & 0.0022 & 0&0031 & 13 \\
117 & 57943.7331 & 0.0023 & $-$0&0006 & 12 \\
133 & 57944.6956 & 0.0024 & 0&0059 & 12 \\
134 & 57944.7522 & 0.0015 & 0&0027 & 11 \\
143 & 57945.2867 & 0.0013 & $-$0&0007 & 137 \\
144 & 57945.3468 & 0.0009 & $-$0&0003 & 138 \\
145 & 57945.4042 & 0.0016 & $-$0&0026 & 137 \\
146 & 57945.4630 & 0.0042 & $-$0&0037 & 79 \\
150 & 57945.7034 & 0.0019 & $-$0&0022 & 13 \\
176 & 57947.2585 & 0.0010 & $-$0&0009 & 129 \\
177 & 57947.3197 & 0.0010 & 0&0006 & 136 \\
178 & 57947.3807 & 0.0011 & 0&0018 & 134 \\
179 & 57947.4415 & 0.0010 & 0&0029 & 128 \\
198 & 57948.5587 & 0.0023 & $-$0&0153 & 15 \\
199 & 57948.6312 & 0.0039 & $-$0&0026 & 17 \\
201 & 57948.7551 & 0.0057 & 0&0018 & 15 \\
214 & 57949.5249 & 0.0025 & $-$0&0052 & 16 \\
218 & 57949.7717 & 0.0046 & 0&0025 & 10 \\
\hline
  \multicolumn{6}{l}{\commenta BJD$-$2400000.} \\
  \multicolumn{6}{l}{\commentb Against max $= 2457936.7420 + 0.059758 E$.} \\
  \multicolumn{6}{l}{\commentc Number of points used to determine the maximum.} \\
\end{tabular}
\end{center}
\end{table}

\subsection{ASASSN-17hy}\label{obj:asassn17hy}

\begin{figure}
  \begin{center}
    \FigureFile(85mm,110mm){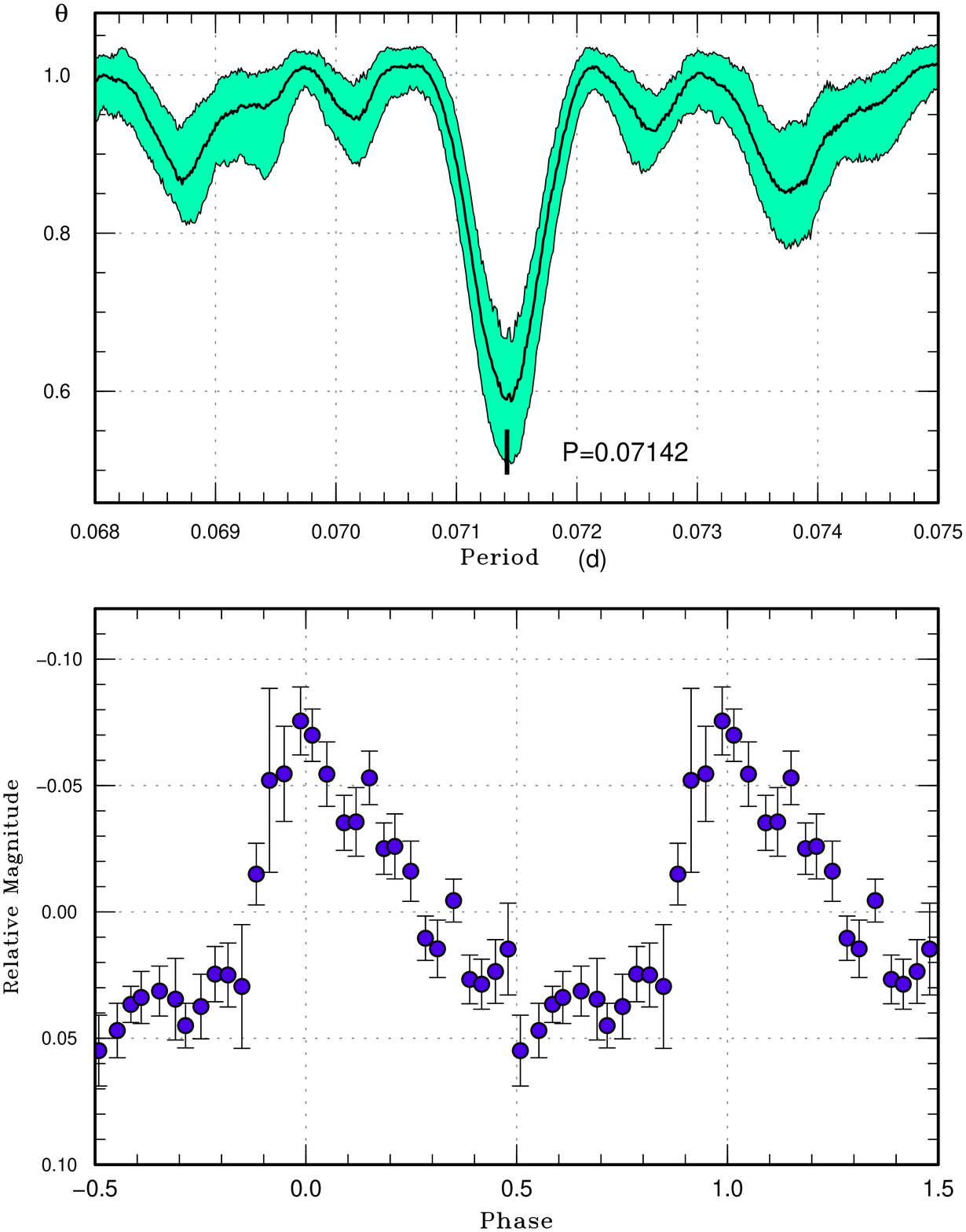}
  \end{center}
  \caption{Superhumps in ASASSN-17hy (2017).
     (Upper): PDM analysis.
     (Lower): Phase-averaged profile.}
  \label{fig:asassn17hyshpdm}
\end{figure}

   This object was detected as a transient
at $V$=16.1 on 2017 June 18 by the ASAS-SN team.
The outburst was announced after observations
on 2017 June 19 ($V$=16.2).
Subsequent observations detected superhumps
(vsnet-alert 21157; e-figure \ref{fig:asassn17hyshpdm}).
The times of superhump maxima are listed in
e-table \ref{tab:asassn17hyoc2017}.
The $O-C$ data suggest a positive $P_{\rm dot}$
although the data were relatively well sampled only
on four nights.

\begin{table}
\caption{Superhump maxima of ASASSN-17hy (2017)}\label{tab:asassn17hyoc2017}
\begin{center}
\begin{tabular}{rp{55pt}p{40pt}r@{.}lr}
\hline
\multicolumn{1}{c}{$E$} & \multicolumn{1}{c}{max\commenta} & \multicolumn{1}{c}{error} & \multicolumn{2}{c}{$O-C$\commentb} & \multicolumn{1}{c}{$N$\commentc} \\
\hline
0 & 57929.4948 & 0.0018 & 0&0019 & 22 \\
1 & 57929.5704 & 0.0012 & 0&0060 & 20 \\
2 & 57929.6395 & 0.0013 & 0&0036 & 21 \\
28 & 57931.4910 & 0.0012 & $-$0&0038 & 17 \\
29 & 57931.5633 & 0.0009 & $-$0&0031 & 15 \\
30 & 57931.6346 & 0.0018 & $-$0&0033 & 15 \\
42 & 57932.4927 & 0.0012 & $-$0&0031 & 16 \\
43 & 57932.5663 & 0.0013 & $-$0&0011 & 14 \\
56 & 57933.4965 & 0.0054 & $-$0&0004 & 16 \\
57 & 57933.5645 & 0.0021 & $-$0&0039 & 15 \\
58 & 57933.6342 & 0.0029 & $-$0&0057 & 14 \\
70 & 57934.4953 & 0.0023 & $-$0&0026 & 16 \\
71 & 57934.5719 & 0.0020 & 0&0025 & 17 \\
72 & 57934.6502 & 0.0054 & 0&0093 & 10 \\
100 & 57936.6466 & 0.0042 & 0&0037 & 18 \\
\hline
  \multicolumn{6}{l}{\commenta BJD$-$2400000.} \\
  \multicolumn{6}{l}{\commentb Against max $= 2457929.4929 + 0.071500 E$.} \\
  \multicolumn{6}{l}{\commentc Number of points used to determine the maximum.} \\
\end{tabular}
\end{center}
\end{table}

\subsection{ASASSN-17id}\label{obj:asassn17id}

\begin{figure}
  \begin{center}
    \FigureFile(85mm,110mm){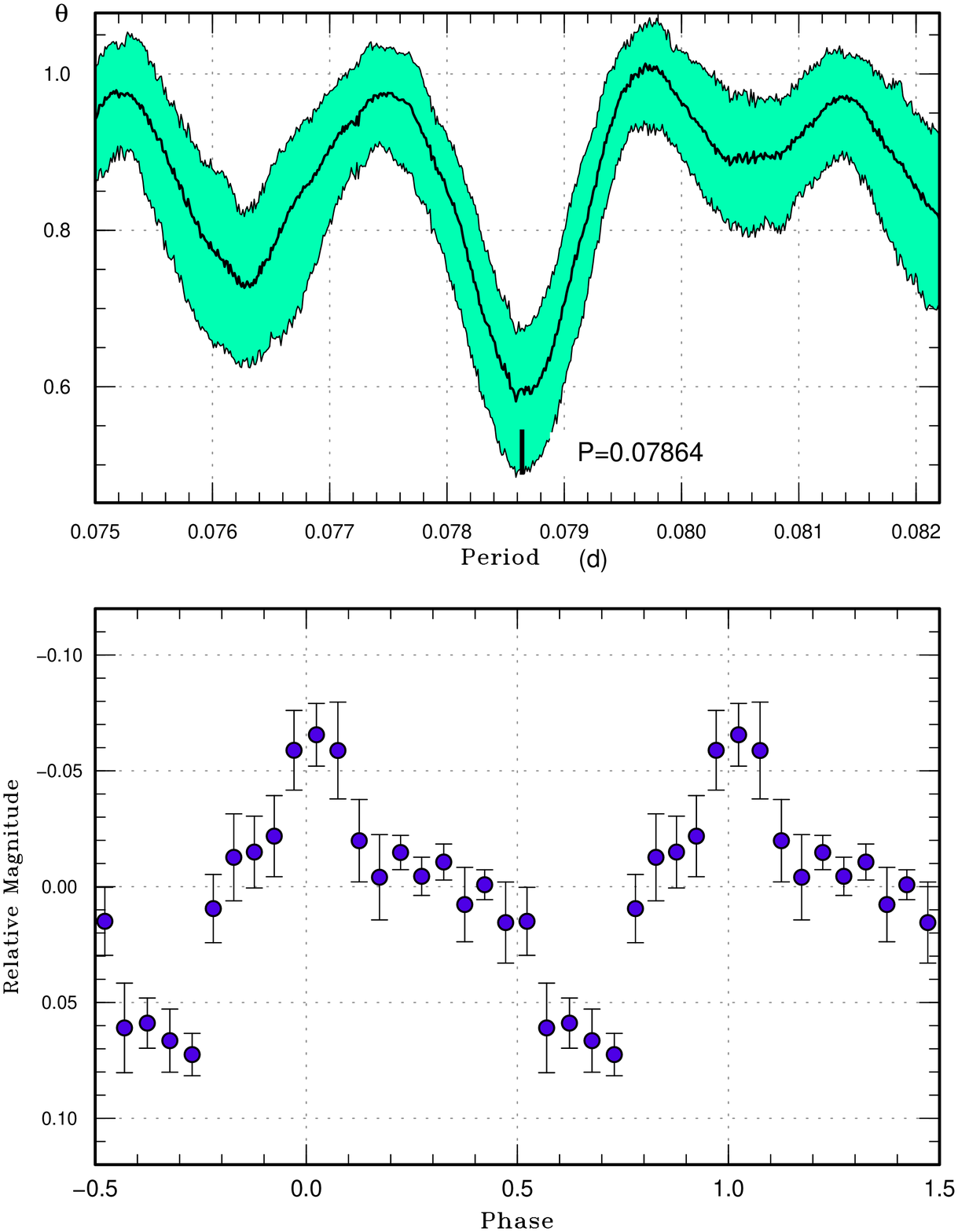}
  \end{center}
  \caption{Superhumps in ASASSN-17id (2017).
     (Upper): PDM analysis.  Data before BJD 2457933
     were used.
     (Lower): Phase-averaged profile.}
  \label{fig:asassn17idshpdm}
\end{figure}

   This object was detected as a transient
at $V$=16.8 on 2017 June 21 by the ASAS-SN team.
The outburst was announced after observations
on 2017 June 23 ($V$=16.8).  Subsequent observations detected
superhumps (vsnet-alert 21158, 21172; e-figure \ref{fig:asassn17idshpdm}).
The times of superhump maxima are listed in
e-table \ref{tab:asassn17idoc2017}.  Although there were
observations after BJD 2457933, the object became too
faint to detect superhumps.  The object faded to 19.6 mag
on 2017 July 2.

\begin{table}
\caption{Superhump maxima of ASASSN-17id (2017)}\label{tab:asassn17idoc2017}
\begin{center}
\begin{tabular}{rp{55pt}p{40pt}r@{.}lr}
\hline
\multicolumn{1}{c}{$E$} & \multicolumn{1}{c}{max\commenta} & \multicolumn{1}{c}{error} & \multicolumn{2}{c}{$O-C$\commentb} & \multicolumn{1}{c}{$N$\commentc} \\
\hline
0 & 57929.4795 & 0.0025 & $-$0&0019 & 22 \\
1 & 57929.5587 & 0.0027 & $-$0&0013 & 23 \\
2 & 57929.6406 & 0.0034 & 0&0020 & 22 \\
26 & 57931.5270 & 0.0023 & 0&0017 & 18 \\
27 & 57931.6065 & 0.0039 & 0&0026 & 18 \\
38 & 57932.4634 & 0.0039 & $-$0&0052 & 12 \\
39 & 57932.5493 & 0.0019 & 0&0021 & 17 \\
\hline
  \multicolumn{6}{l}{\commenta BJD$-$2400000.} \\
  \multicolumn{6}{l}{\commentb Against max $= 2457929.4814 + 0.078613 E$.} \\
  \multicolumn{6}{l}{\commentc Number of points used to determine the maximum.} \\
\end{tabular}
\end{center}
\end{table}

\subsection{ASASSN-17if}\label{obj:asassn17if}

\begin{figure}
  \begin{center}
    \FigureFile(85mm,110mm){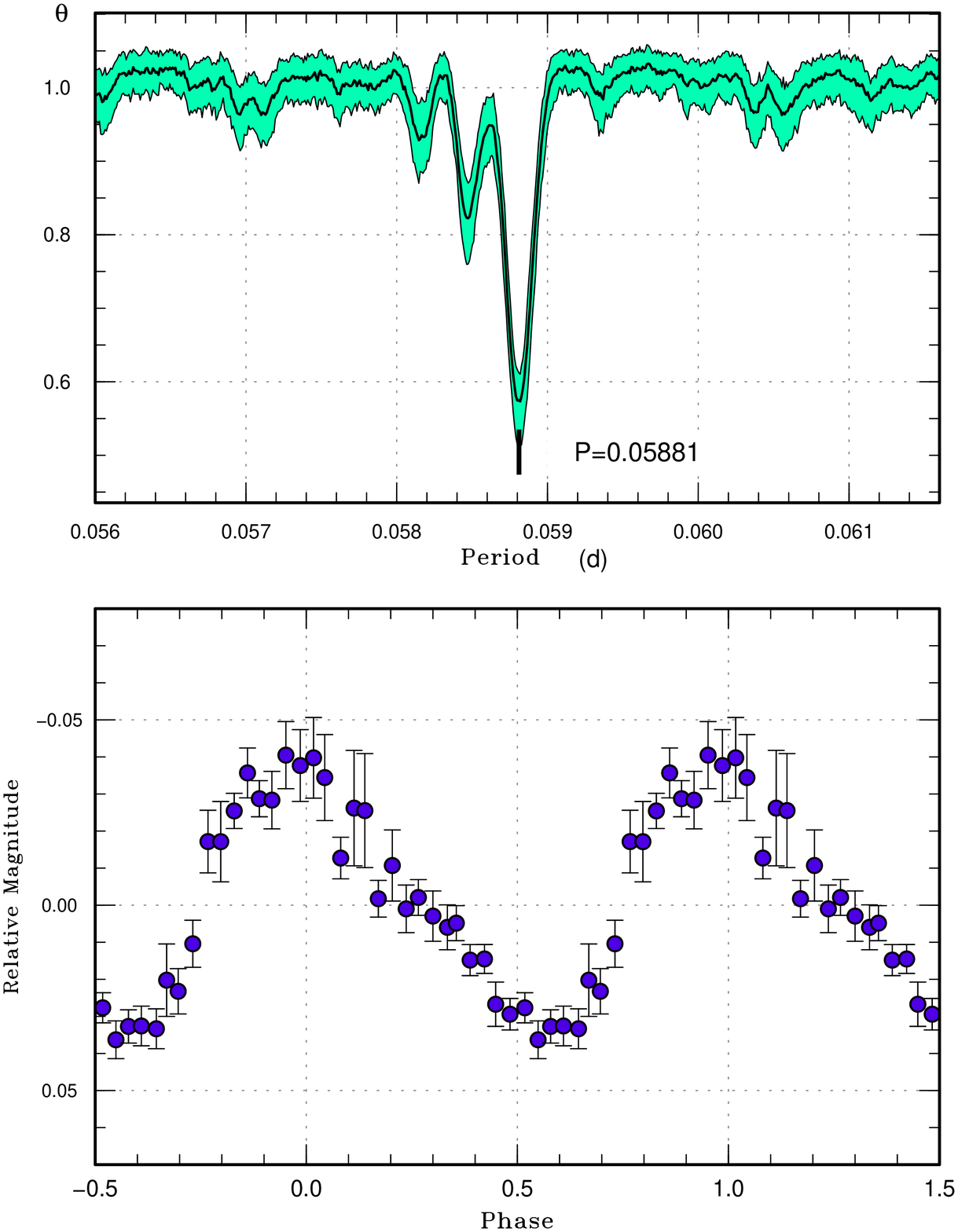}
  \end{center}
  \caption{Superhumps in ASASSN-17if (2017).
     (Upper): PDM analysis.
     (Lower): Phase-averaged profile.}
  \label{fig:asassn17ifshpdm}
\end{figure}

   This object was detected as a transient
at $V$=14.0 on 2017 June 25 by the ASAS-SN team.
The outburst was announced after observations
on 2017 June 26 ($V$=14.1).  Subsequent observations
detected superhumps (vsnet-alert 21183;
e-figure \ref{fig:asassn17ifshpdm}).
The times of superhump maxima are listed in
e-table \ref{tab:asassn17ifoc2017}.
The stages B and C were clearly recorded despite
relatively low sampling rates.  The positive $P_{\rm dot}$
for stage B superhumps was typical for this $P_{\rm SH}$.

   This field has been monitored by the ASAS-SN team
since 2013 June 27 and no past outbursts were recorded.
It may be that outbursts were relatively rare or
past outbursts occurred when the object was invisible
near the Sun.

\begin{table}
\caption{Superhump maxima of ASASSN-17if (2017)}\label{tab:asassn17ifoc2017}
\begin{center}
\begin{tabular}{rp{55pt}p{40pt}r@{.}lr}
\hline
\multicolumn{1}{c}{$E$} & \multicolumn{1}{c}{max\commenta} & \multicolumn{1}{c}{error} & \multicolumn{2}{c}{$O-C$\commentb} & \multicolumn{1}{c}{$N$\commentc} \\
\hline
0 & 57931.8329 & 0.0004 & 0&0074 & 13 \\
17 & 57932.8273 & 0.0004 & 0&0022 & 16 \\
34 & 57933.8234 & 0.0003 & $-$0&0014 & 16 \\
35 & 57933.8810 & 0.0003 & $-$0&0026 & 18 \\
68 & 57935.8179 & 0.0016 & $-$0&0063 & 21 \\
69 & 57935.8776 & 0.0020 & $-$0&0054 & 17 \\
86 & 57936.8767 & 0.0008 & $-$0&0059 & 21 \\
102 & 57937.8201 & 0.0012 & $-$0&0034 & 15 \\
103 & 57937.8789 & 0.0008 & $-$0&0034 & 17 \\
119 & 57938.8203 & 0.0011 & $-$0&0029 & 16 \\
120 & 57938.8808 & 0.0011 & $-$0&0012 & 18 \\
121 & 57938.9408 & 0.0025 & $-$0&0000 & 11 \\
137 & 57939.8838 & 0.0034 & 0&0021 & 11 \\
153 & 57940.8364 & 0.0034 & 0&0138 & 9 \\
154 & 57940.8868 & 0.0018 & 0&0055 & 11 \\
170 & 57941.8285 & 0.0041 & 0&0063 & 9 \\
171 & 57941.8877 & 0.0013 & 0&0067 & 11 \\
187 & 57942.8249 & 0.0027 & 0&0030 & 9 \\
188 & 57942.8854 & 0.0024 & 0&0048 & 11 \\
204 & 57943.8218 & 0.0041 & 0&0003 & 8 \\
205 & 57943.8810 & 0.0024 & 0&0006 & 10 \\
221 & 57944.8093 & 0.0030 & $-$0&0119 & 9 \\
222 & 57944.8782 & 0.0017 & $-$0&0018 & 14 \\
223 & 57944.9320 & 0.0022 & $-$0&0068 & 10 \\
\hline
  \multicolumn{6}{l}{\commenta BJD$-$2400000.} \\
  \multicolumn{6}{l}{\commentb Against max $= 2457931.8255 + 0.058804 E$.} \\
  \multicolumn{6}{l}{\commentc Number of points used to determine the maximum.} \\
\end{tabular}
\end{center}
\end{table}

\subsection{ASASSN-17ig}\label{obj:asassn17ig}

\begin{figure}
  \begin{center}
    \FigureFile(85mm,110mm){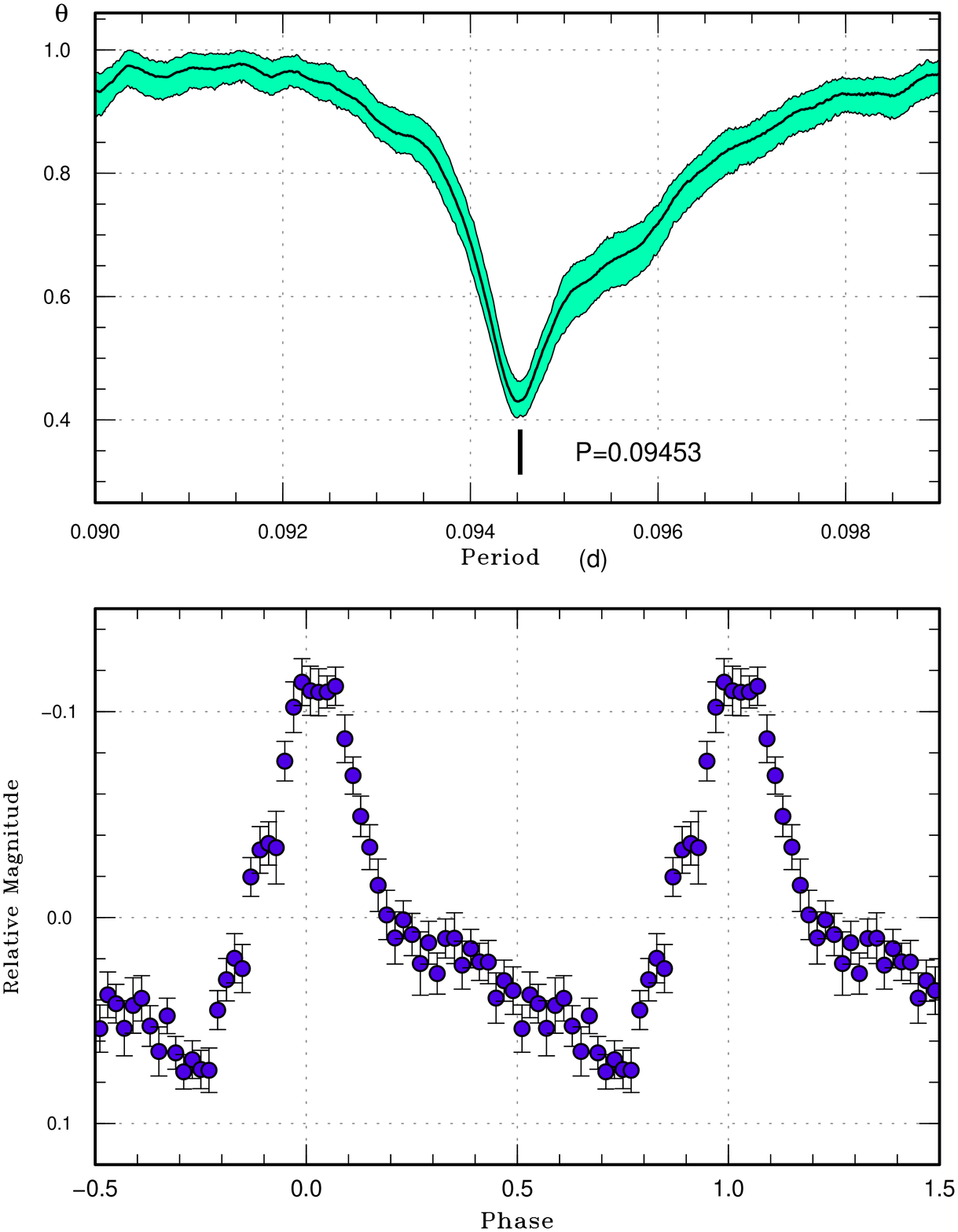}
  \end{center}
  \caption{Superhumps in ASASSN-17ig (2017).
     (Upper): PDM analysis.
     (Lower): Phase-averaged profile.}
  \label{fig:asassn17igshpdm}
\end{figure}

   This object was detected as a transient
at $V$=14.8 on 2017 June 24 by the ASAS-SN team.
Subsequent observations detected long-period superhumps
(vsnet-alert 21168, 21184; e-figure \ref{fig:asassn17igshpdm}).
The times of superhump maxima are listed in
e-table \ref{tab:asassn17igoc2017}.  The superhump
period was initially longer, suggesting that they were
stage B superhumps (less likely stage A ones, since
the amplitudes were already large).

   This field has been monitored by the ASAS-SN team
since 2015 February 1 and no past outburst was detected.
Since the object is an SU UMa-type dwarf nova
in the period gap, further observations would be
interesting.

\begin{table}
\caption{Superhump maxima of ASASSN-17ig (2017)}\label{tab:asassn17igoc2017}
\begin{center}
\begin{tabular}{rp{55pt}p{40pt}r@{.}lr}
\hline
\multicolumn{1}{c}{$E$} & \multicolumn{1}{c}{max\commenta} & \multicolumn{1}{c}{error} & \multicolumn{2}{c}{$O-C$\commentb} & \multicolumn{1}{c}{$N$\commentc} \\
\hline
0 & 57931.6531 & 0.0024 & $-$0&0084 & 12 \\
1 & 57931.7519 & 0.0006 & $-$0&0042 & 30 \\
12 & 57932.7977 & 0.0016 & 0&0021 & 19 \\
14 & 57932.9861 & 0.0008 & 0&0015 & 123 \\
15 & 57933.0785 & 0.0005 & $-$0&0006 & 136 \\
22 & 57933.7444 & 0.0007 & 0&0037 & 30 \\
25 & 57934.0283 & 0.0005 & 0&0041 & 190 \\
32 & 57934.6898 & 0.0007 & 0&0040 & 29 \\
54 & 57936.7688 & 0.0009 & 0&0039 & 17 \\
64 & 57937.7102 & 0.0010 & 0&0002 & 20 \\
75 & 57938.7479 & 0.0013 & $-$0&0016 & 20 \\
85 & 57939.6946 & 0.0022 & $-$0&0001 & 19 \\
96 & 57940.7296 & 0.0043 & $-$0&0046 & 19 \\
\hline
  \multicolumn{6}{l}{\commenta BJD$-$2400000.} \\
  \multicolumn{6}{l}{\commentb Against max $= 2457931.6615 + 0.094507 E$.} \\
  \multicolumn{6}{l}{\commentc Number of points used to determine the maximum.} \\
\end{tabular}
\end{center}
\end{table}

\subsection{ASASSN-17il}\label{obj:asassn17il}

\begin{figure}
  \begin{center}
    \FigureFile(85mm,80mm){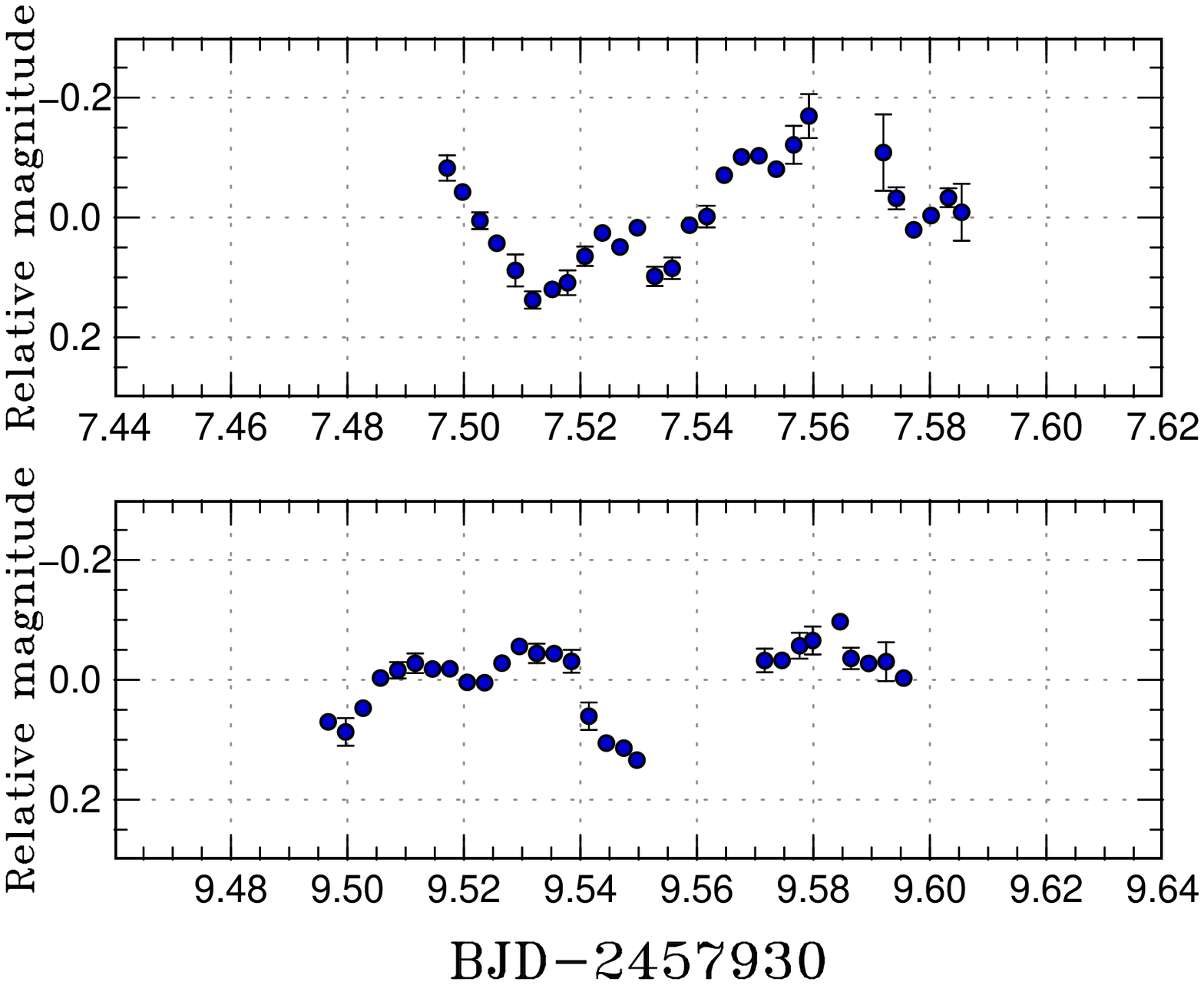}
  \end{center}
  \caption{Possible superhumps in ASASSN-17il (2017).
  The data were binned to 0.003~d.
  }
  \label{fig:asassn17illc}
\end{figure}

   This object was detected as a transient
at $V$=15.3 on 2017 June 30 by the ASAS-SN team.
Although observations detected superhump-like
variations (vsnet-alert 21202; e-figure \ref{fig:asassn17illc}), 
we have not been able to determine a unique period using
the data on two nights.
Observations on July 6 could not detect similar variations.
Although the SDSS colors in quiescence suggest an object
below the period gap (cf. \cite{kat12DNSDSS}), the variations
detected on the first night may have been large-amplitude
random variations. 
Further observations are needed to clarify the nature of
this object.

\subsection{ASASSN-17iv}\label{obj:asassn17iv}

\begin{figure}
  \begin{center}
    \FigureFile(85mm,110mm){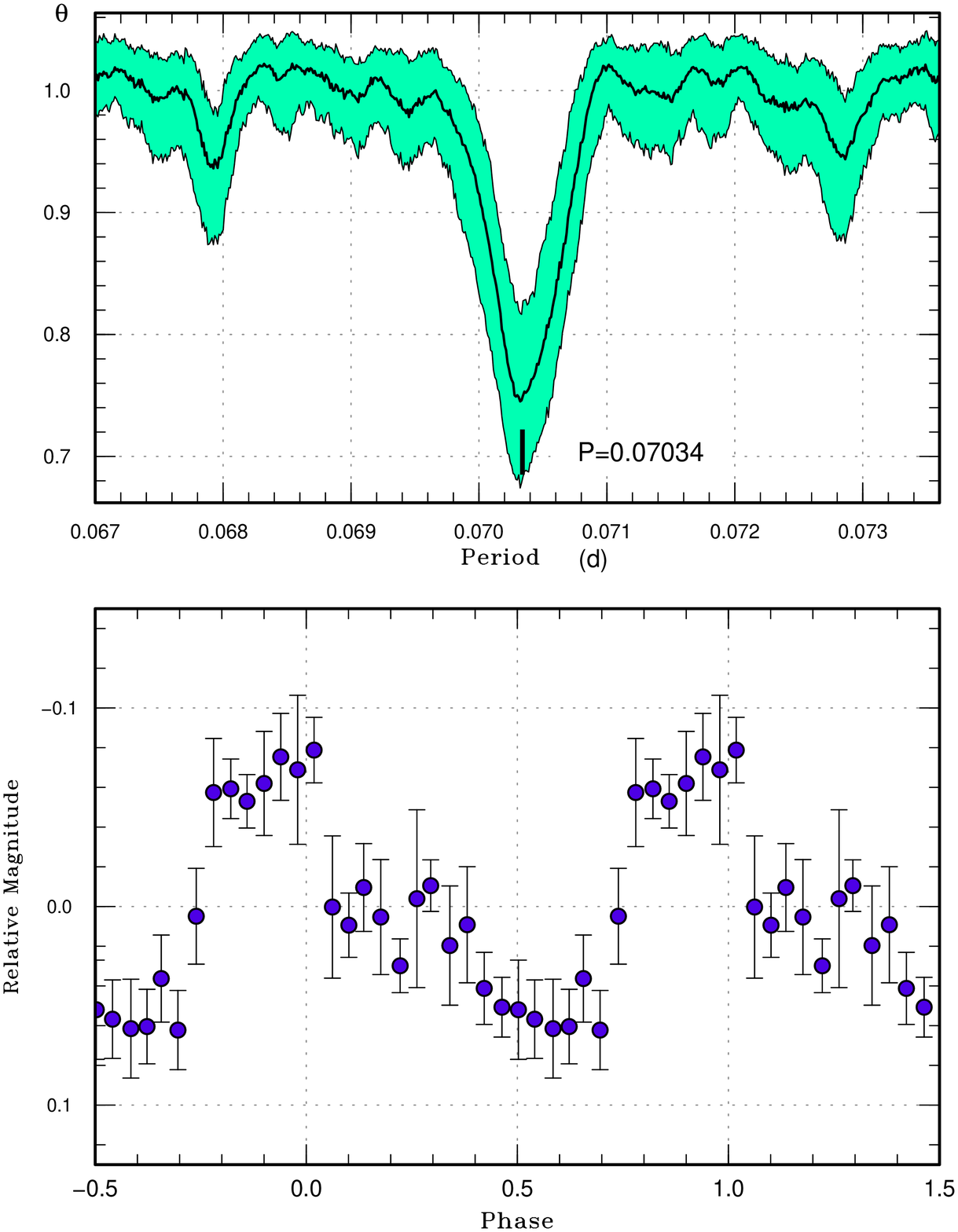}
  \end{center}
  \caption{Superhumps in ASASSN-17iv (2017).
     (Upper): PDM analysis.
     (Lower): Phase-averaged profile.}
  \label{fig:asassn17ivshpdm}
\end{figure}

   This object was detected as a transient
at $V$=16.8 on 2017 June 30 by the ASAS-SN team.
The outburst was announced after observations on
2017 July 2 ($V$=16.4) and 2017 July 3 ($V$=16.4).
Subsequent observations starting on July 5 detected
superhumps (vsnet-alert 21245; e-figure \ref{fig:asassn17ivshpdm}).
The times of superhump maxima are listed in
e-table \ref{tab:asassn17ivoc2017}.
Maxima for $E \le$1 had negative $O-C$ values and
there appears to have been a stage transition
between $E$=1 and $E$=15.  Since the observation
started relatively late after the initial outburst
detection, the majority of observations apparently
recorded stage C superhumps.  We adopted a period
for $E \ge$15 following this interpretation.

\begin{table}
\caption{Superhump maxima of ASASSN-17iv (2017)}\label{tab:asassn17ivoc2017}
\begin{center}
\begin{tabular}{rp{55pt}p{40pt}r@{.}lr}
\hline
\multicolumn{1}{c}{$E$} & \multicolumn{1}{c}{max\commenta} & \multicolumn{1}{c}{error} & \multicolumn{2}{c}{$O-C$\commentb} & \multicolumn{1}{c}{$N$\commentc} \\
\hline
0 & 57939.8022 & 0.0014 & $-$0&0061 & 9 \\
1 & 57939.8725 & 0.0011 & $-$0&0061 & 12 \\
15 & 57940.8666 & 0.0030 & 0&0036 & 12 \\
16 & 57940.9377 & 0.0036 & 0&0044 & 8 \\
29 & 57941.8466 & 0.0045 & $-$0&0008 & 11 \\
30 & 57941.9190 & 0.0031 & 0&0013 & 13 \\
43 & 57942.8348 & 0.0017 & 0&0030 & 11 \\
44 & 57942.9078 & 0.0069 & 0&0057 & 14 \\
57 & 57943.8225 & 0.0015 & 0&0063 & 10 \\
58 & 57943.8829 & 0.0014 & $-$0&0036 & 14 \\
71 & 57944.8045 & 0.0026 & 0&0038 & 13 \\
72 & 57944.8646 & 0.0062 & $-$0&0064 & 16 \\
73 & 57944.9431 & 0.0044 & 0&0018 & 12 \\
85 & 57945.7878 & 0.0038 & 0&0027 & 11 \\
86 & 57945.8479 & 0.0023 & $-$0&0075 & 14 \\
87 & 57945.9234 & 0.0041 & $-$0&0023 & 15 \\
\hline
  \multicolumn{6}{l}{\commenta BJD$-$2400000.} \\
  \multicolumn{6}{l}{\commentb Against max $= 2457939.8083 + 0.070315 E$.} \\
  \multicolumn{6}{l}{\commentc Number of points used to determine the maximum.} \\
\end{tabular}
\end{center}
\end{table}

\subsection{ASASSN-17iw}\label{obj:asassn17iw}

\begin{figure}
  \begin{center}
    \FigureFile(85mm,110mm){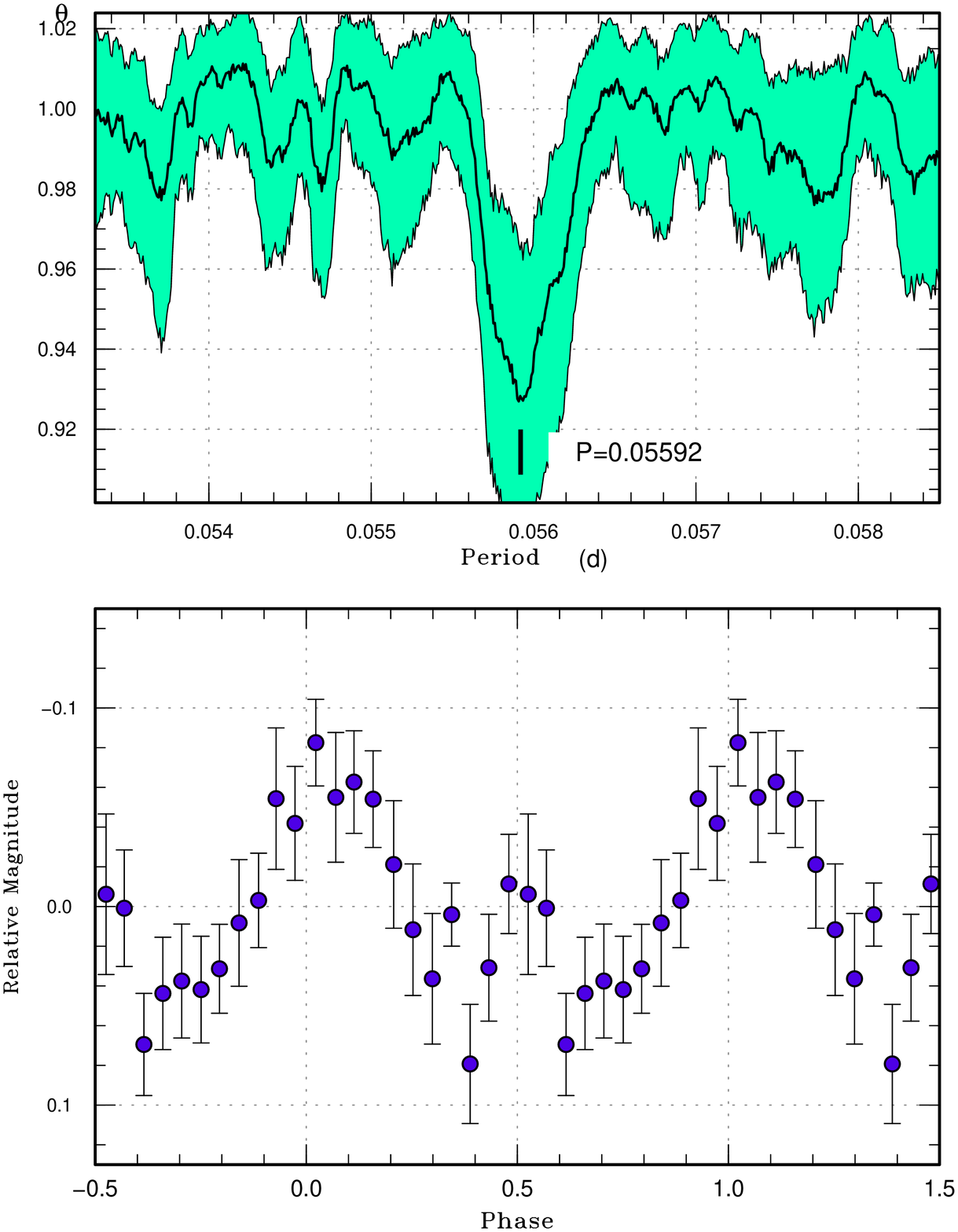}
  \end{center}
  \caption{Superhumps in ASASSN-17iw (2017).
     (Upper): PDM analysis.
     (Lower): Phase-averaged profile.}
  \label{fig:asassn17iwshpdm}
\end{figure}

   This object was detected as a transient
at $V$=16.6 on 2017 June 28 by the ASAS-SN team.
The outburst was announced after observations on
2017 June 30 ($V$=16.5) and 2017 July 1 ($V$=17.0).
Subsequent observations starting on July 4 detected
superhumps (vsnet-alert 21246; e-figure \ref{fig:asassn17iwshpdm}).
The times of superhump maxima are listed in
e-table \ref{tab:asassn17iwoc2017}.
Although the $O-C$ diagram is noisy due to the faintness
of the object, there was apparently stage B-C transition
around $E$=90.

\begin{table}
\caption{Superhump maxima of ASASSN-17iw (2017)}\label{tab:asassn17iwoc2017}
\begin{center}
\begin{tabular}{rp{55pt}p{40pt}r@{.}lr}
\hline
\multicolumn{1}{c}{$E$} & \multicolumn{1}{c}{max\commenta} & \multicolumn{1}{c}{error} & \multicolumn{2}{c}{$O-C$\commentb} & \multicolumn{1}{c}{$N$\commentc} \\
\hline
0 & 57939.5829 & 0.0031 & 0&0077 & 14 \\
1 & 57939.6286 & 0.0022 & $-$0&0025 & 13 \\
18 & 57940.5839 & 0.0020 & 0&0028 & 14 \\
19 & 57940.6338 & 0.0022 & $-$0&0031 & 13 \\
35 & 57941.5252 & 0.0019 & $-$0&0058 & 17 \\
52 & 57942.4720 & 0.0021 & $-$0&0090 & 15 \\
53 & 57942.5387 & 0.0038 & 0&0018 & 17 \\
53 & 57942.5394 & 0.0071 & 0&0026 & 17 \\
55 & 57942.6448 & 0.0024 & $-$0&0038 & 12 \\
70 & 57943.4820 & 0.0026 & $-$0&0049 & 17 \\
71 & 57943.5418 & 0.0027 & $-$0&0009 & 17 \\
72 & 57943.5997 & 0.0026 & 0&0011 & 13 \\
73 & 57943.6646 & 0.0040 & 0&0101 & 6 \\
88 & 57944.4920 & 0.0018 & $-$0&0007 & 17 \\
89 & 57944.5504 & 0.0034 & 0&0018 & 16 \\
90 & 57944.6134 & 0.0024 & 0&0089 & 13 \\
108 & 57945.6045 & 0.0043 & $-$0&0059 & 13 \\
\hline
  \multicolumn{6}{l}{\commenta BJD$-$2400000.} \\
  \multicolumn{6}{l}{\commentb Against max $= 2457939.5752 + 0.055881 E$.} \\
  \multicolumn{6}{l}{\commentc Number of points used to determine the maximum.} \\
\end{tabular}
\end{center}
\end{table}

\subsection{ASASSN-17ix}\label{obj:asassn17ix}

\begin{figure}
  \begin{center}
    \FigureFile(85mm,110mm){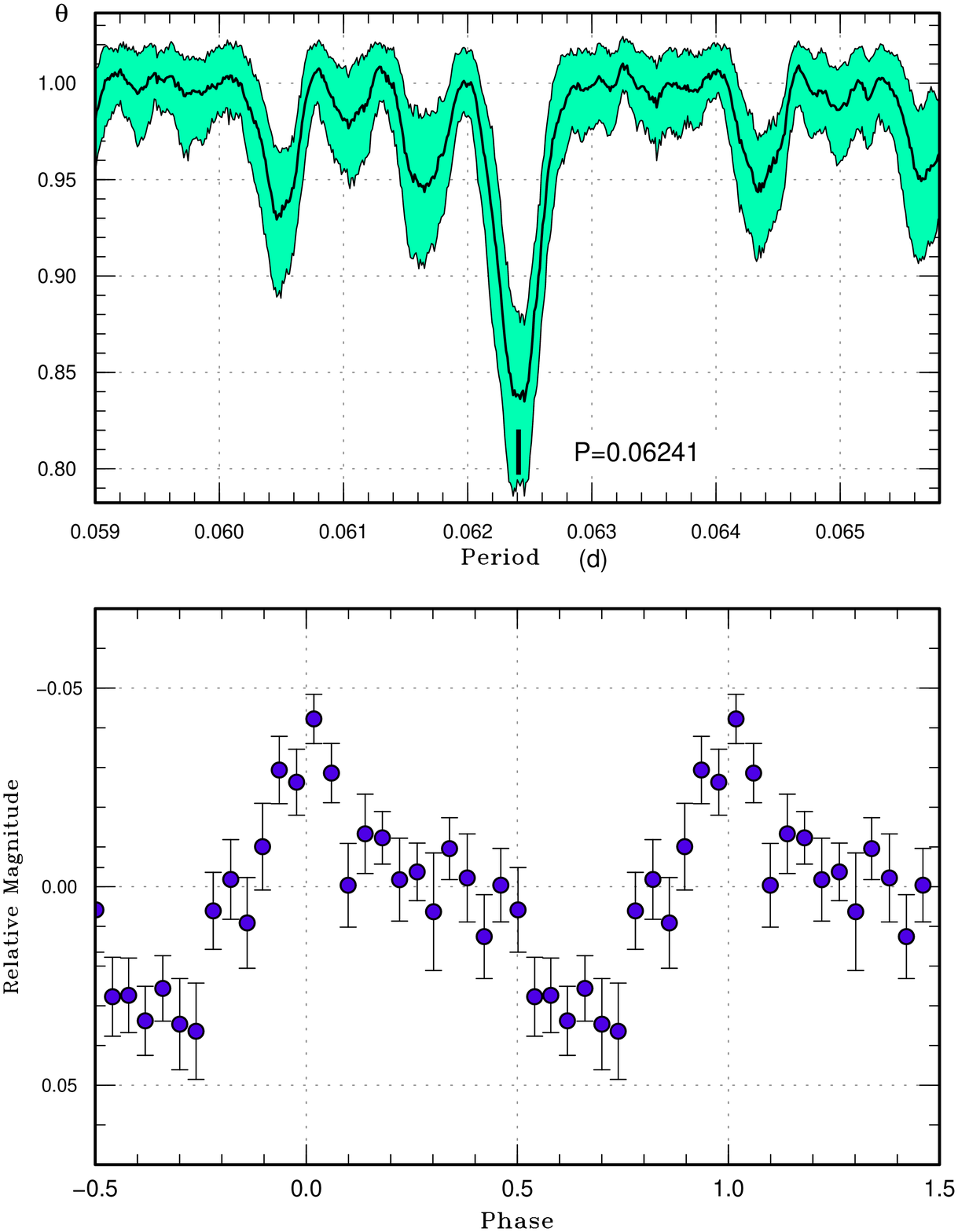}
  \end{center}
  \caption{Superhumps in ASASSN-17ix (2017).
     (Upper): PDM analysis.
     (Lower): Phase-averaged profile.}
  \label{fig:asassn17ixshpdm}
\end{figure}

   This object was detected as a transient
at $V$=15.2 on 2017 June 29 by the ASAS-SN team.
The outburst was announced after observations between
2017 June 30 ($V$=15.5) and 2017 July 2 ($V$=15.6).
Subsequent observations starting on 2017 July 4 detected
superhumps (vsnet-alert 21247; e-figure \ref{fig:asassn17ixshpdm}).
The times of superhump maxima are listed in
e-table \ref{tab:asassn17ixoc2017}.
Although the $O-C$ diagram is noisy due to the faintness
of the object, there was apparently stage B-C transition
around $E$=82.

\begin{table}
\caption{Superhump maxima of ASASSN-17ix (2017)}\label{tab:asassn17ixoc2017}
\begin{center}
\begin{tabular}{rp{55pt}p{40pt}r@{.}lr}
\hline
\multicolumn{1}{c}{$E$} & \multicolumn{1}{c}{max\commenta} & \multicolumn{1}{c}{error} & \multicolumn{2}{c}{$O-C$\commentb} & \multicolumn{1}{c}{$N$\commentc} \\
\hline
0 & 57939.4779 & 0.0013 & 0&0043 & 19 \\
1 & 57939.5362 & 0.0024 & 0&0001 & 19 \\
2 & 57939.6021 & 0.0030 & 0&0036 & 16 \\
3 & 57939.6608 & 0.0020 & $-$0&0001 & 10 \\
16 & 57940.4733 & 0.0016 & 0&0007 & 17 \\
17 & 57940.5332 & 0.0020 & $-$0&0018 & 19 \\
32 & 57941.4713 & 0.0024 & $-$0&0002 & 16 \\
33 & 57941.5276 & 0.0017 & $-$0&0062 & 19 \\
34 & 57941.5930 & 0.0032 & $-$0&0033 & 16 \\
49 & 57942.5306 & 0.0077 & $-$0&0021 & 19 \\
65 & 57943.5294 & 0.0044 & $-$0&0023 & 19 \\
66 & 57943.5919 & 0.0043 & $-$0&0021 & 14 \\
82 & 57944.6047 & 0.0031 & 0&0117 & 14 \\
96 & 57945.4739 & 0.0016 & 0&0069 & 16 \\
97 & 57945.5235 & 0.0022 & $-$0&0059 & 19 \\
98 & 57945.5885 & 0.0021 & $-$0&0034 & 15 \\
\hline
  \multicolumn{6}{l}{\commenta BJD$-$2400000.} \\
  \multicolumn{6}{l}{\commentb Against max $= 2457939.4736 + 0.062431 E$.} \\
  \multicolumn{6}{l}{\commentc Number of points used to determine the maximum.} \\
\end{tabular}
\end{center}
\end{table}

\subsection{ASASSN-17ji}\label{obj:asassn17ji}

\begin{figure}
  \begin{center}
    \FigureFile(85mm,110mm){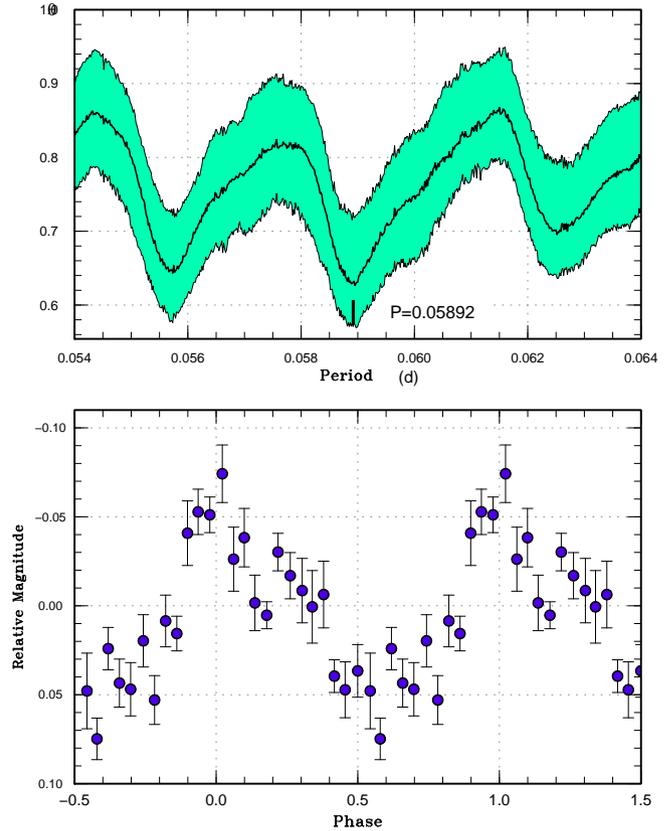}
  \end{center}
  \caption{Superhumps in ASASSN-17ji (2017).
     (Upper): PDM analysis.  The data after BJD 2457959
     were used.
     (Lower): Phase-averaged profile.}
  \label{fig:asassn17jishpdm}
\end{figure}

   This object was detected as a transient
at $V$=15.2 on 2017 July 14 by the ASAS-SN team.
The outburst was announced after the observation
on 2017 July 14 ($V$=15.0).
Observations on 2017 July 17 did not show superhumps.
Superhumps were later found to have developed
at least on 2017 July 24 (vsnet-alert 21275;
e-figure \ref{fig:asassn17jishpdm}).
The times of superhump maxima are listed in
e-table \ref{tab:asassn17jioc2017}.
Since the observations on 2017 July 25 were not of
good quality, the maximum at $E$=18 was relatively
uncertain.  The cycle count, however, appears to be
certain from a continuous run on July 24.
A PDM analysis of the July 24 data yielded 0.0576(8)~d.
An $O-C$ analysis favored a longer period of
0.061(1)~d.  It is likely that the true period
lies between 0.057~d and 0.061~d.
The combined data of July 24 and 25 yielded
0.0589(1)~d.  The one-day alias of 0.0558(1)~d appears
to be ruled out (see e-figure \ref{fig:asassn17jishpdm}).
The superhump stage is unknown.

   This field has been monitored by the ASAS-SN team
since 2013 June 26 and no past outbursts were detected.

\begin{table}
\caption{Superhump maxima of ASASSN-17ji (2017)}\label{tab:asassn17jioc2017}
\begin{center}
\begin{tabular}{rp{55pt}p{40pt}r@{.}lr}
\hline
\multicolumn{1}{c}{$E$} & \multicolumn{1}{c}{max\commenta} & \multicolumn{1}{c}{error} & \multicolumn{2}{c}{$O-C$\commentb} & \multicolumn{1}{c}{$N$\commentc} \\
\hline
0 & 57959.4454 & 0.0019 & $-$0&0016 & 56 \\
1 & 57959.5048 & 0.0010 & $-$0&0010 & 61 \\
2 & 57959.5675 & 0.0039 & 0&0029 & 23 \\
18 & 57960.5051 & 0.0019 & $-$0&0003 & 55 \\
\hline
  \multicolumn{6}{l}{\commenta BJD$-$2400000.} \\
  \multicolumn{6}{l}{\commentb Against max $= 2457959.4470 + 0.058798 E$.} \\
  \multicolumn{6}{l}{\commentc Number of points used to determine the maximum.} \\
\end{tabular}
\end{center}
\end{table}

\subsection{ASASSN-17jr}\label{obj:asassn17jr}

   This object was detected as a transient
at $V$=15.9 on 2017 July 25 by the ASAS-SN team.
Subsequent observations detected superhumps
(vsnet-alert 21280, 21297).
The times of superhump maxima are listed in
e-table \ref{tab:asassn17jroc2017}.
Only stage B with a positive $P_{\rm dot}$ was
observed.

\begin{figure}
  \begin{center}
    \FigureFile(85mm,110mm){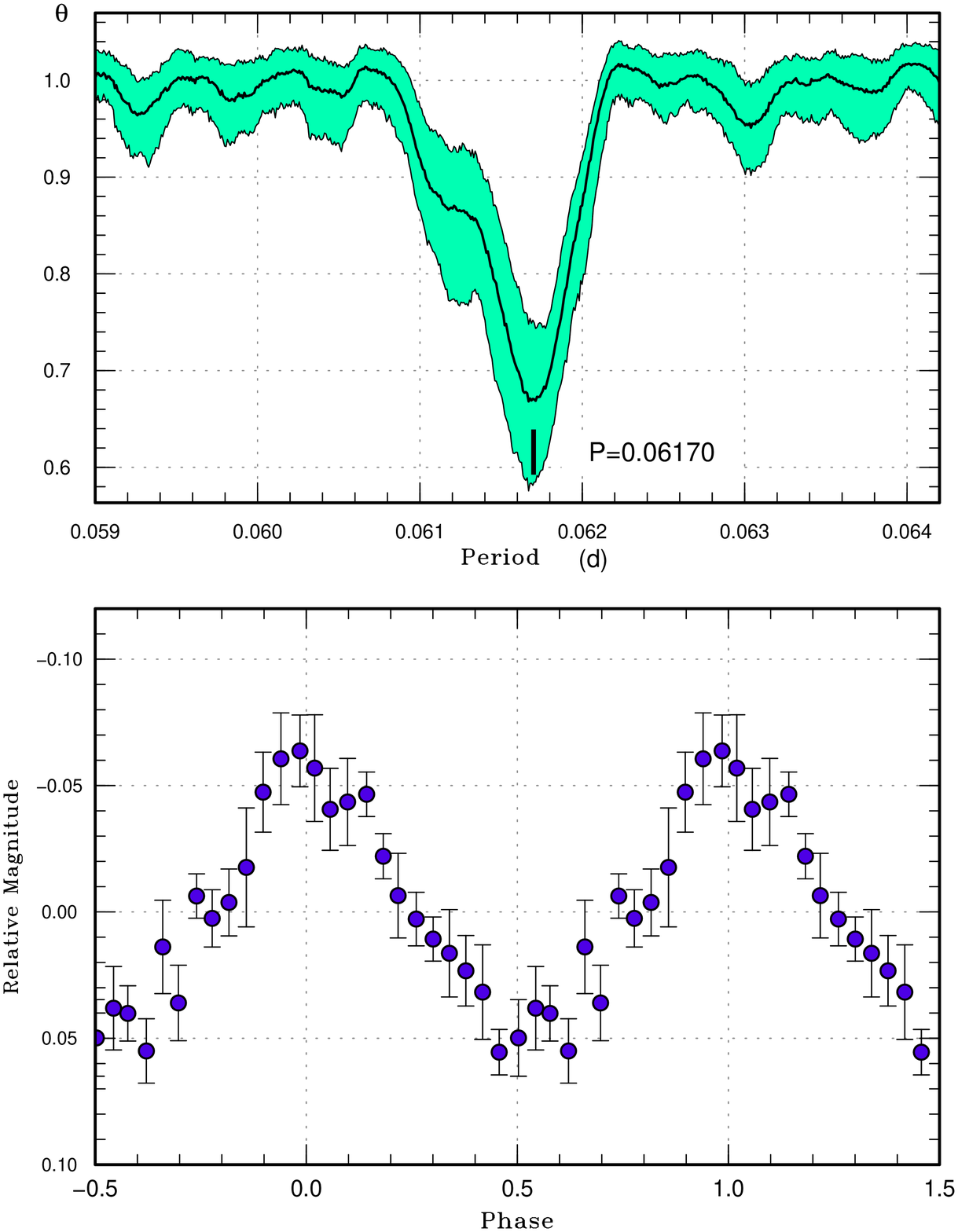}
  \end{center}
  \caption{Superhumps in ASASSN-17jr (2017).
     (Upper): PDM analysis.
     (Lower): Phase-averaged profile.}
  \label{fig:asassn17jrshpdm}
\end{figure}

\begin{table}
\caption{Superhump maxima of ASASSN-17jr (2017)}\label{tab:asassn17jroc2017}
\begin{center}
\begin{tabular}{rp{55pt}p{40pt}r@{.}lr}
\hline
\multicolumn{1}{c}{$E$} & \multicolumn{1}{c}{max\commenta} & \multicolumn{1}{c}{error} & \multicolumn{2}{c}{$O-C$\commentb} & \multicolumn{1}{c}{$N$\commentc} \\
\hline
0 & 57960.5348 & 0.0012 & 0&0034 & 15 \\
1 & 57960.5951 & 0.0009 & 0&0020 & 14 \\
16 & 57961.5198 & 0.0009 & 0&0011 & 16 \\
17 & 57961.5816 & 0.0013 & 0&0012 & 16 \\
32 & 57962.5017 & 0.0017 & $-$0&0043 & 15 \\
33 & 57962.5669 & 0.0015 & $-$0&0008 & 16 \\
48 & 57963.4925 & 0.0018 & $-$0&0008 & 12 \\
49 & 57963.5505 & 0.0024 & $-$0&0045 & 15 \\
51 & 57963.6727 & 0.0034 & $-$0&0057 & 13 \\
65 & 57964.5416 & 0.0028 & $-$0&0007 & 15 \\
67 & 57964.6713 & 0.0085 & 0&0056 & 13 \\
98 & 57966.5821 & 0.0021 & 0&0035 & 13 \\
\hline
  \multicolumn{6}{l}{\commenta BJD$-$2400000.} \\
  \multicolumn{6}{l}{\commentb Against max $= 2457960.5314 + 0.061706 E$.} \\
  \multicolumn{6}{l}{\commentc Number of points used to determine the maximum.} \\
\end{tabular}
\end{center}
\end{table}

\subsection{ASASSN-17kc}\label{obj:asassn17kc}

\begin{figure}
  \begin{center}
    \FigureFile(85mm,110mm){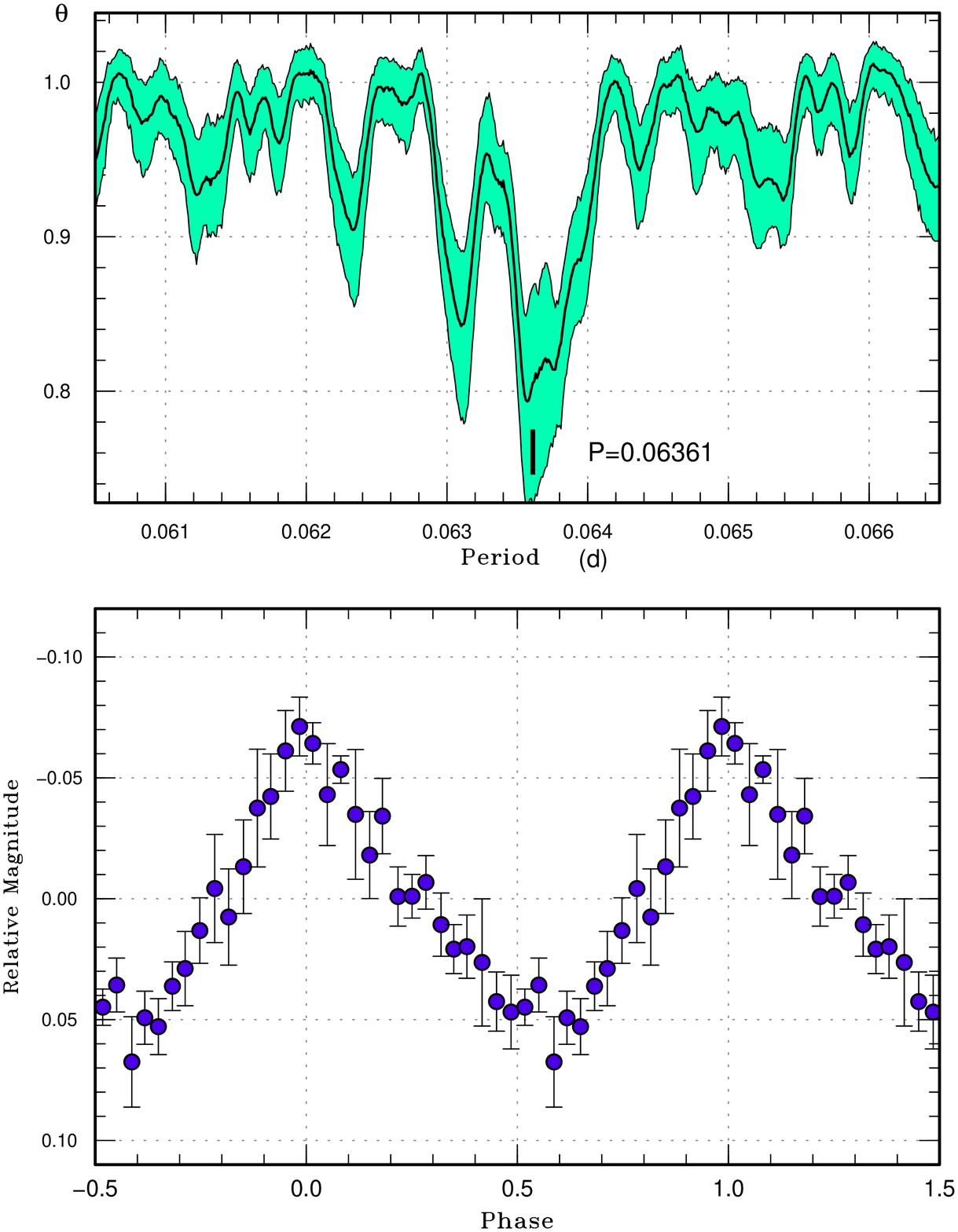}
  \end{center}
  \caption{Superhumps in ASASSN-17kc (2017).
     (Upper): PDM analysis.
     (Lower): Phase-averaged profile.}
  \label{fig:asassn17kcshpdm}
\end{figure}

   This object was detected as a transient
at $V$=13.5 on 2017 July 30 by the ASAS-SN team.
There were no past outbursts in the CRTS data
and ASAS-3 data.
Subsequent observations detected superhumps
(vsnet-alert 21316, 21340; e-figure \ref{fig:asassn17kcshpdm}).
The times of superhump maxima are listed in
e-table \ref{tab:asassn17kcoc2017}.
The maxima for $E \ge$158 refer to post-superoutburst
observations.
Stages B and C can be clearly recognized and
the large positive $P_{\rm dot}$ for stage B
is apparent.

\begin{table}
\caption{Superhump maxima of ASASSN-17kc (2017)}\label{tab:asassn17kcoc2017}
\begin{center}
\begin{tabular}{rp{55pt}p{40pt}r@{.}lr}
\hline
\multicolumn{1}{c}{$E$} & \multicolumn{1}{c}{max\commenta} & \multicolumn{1}{c}{error} & \multicolumn{2}{c}{$O-C$\commentb} & \multicolumn{1}{c}{$N$\commentc} \\
\hline
0 & 57968.6996 & 0.0007 & $-$0&0018 & 12 \\
1 & 57968.7626 & 0.0005 & $-$0&0025 & 15 \\
2 & 57968.8260 & 0.0004 & $-$0&0027 & 24 \\
3 & 57968.8896 & 0.0004 & $-$0&0027 & 19 \\
16 & 57969.7159 & 0.0008 & $-$0&0032 & 14 \\
17 & 57969.7788 & 0.0008 & $-$0&0039 & 17 \\
18 & 57969.8423 & 0.0007 & $-$0&0040 & 22 \\
19 & 57969.9068 & 0.0011 & $-$0&0031 & 13 \\
33 & 57970.7958 & 0.0010 & $-$0&0046 & 18 \\
34 & 57970.8607 & 0.0012 & $-$0&0033 & 21 \\
47 & 57971.6924 & 0.0013 & 0&0015 & 12 \\
48 & 57971.7510 & 0.0013 & $-$0&0035 & 15 \\
49 & 57971.8172 & 0.0006 & $-$0&0008 & 24 \\
50 & 57971.8830 & 0.0014 & 0&0014 & 20 \\
79 & 57973.7367 & 0.0018 & 0&0105 & 14 \\
80 & 57973.8003 & 0.0013 & 0&0104 & 23 \\
81 & 57973.8643 & 0.0013 & 0&0109 & 19 \\
94 & 57974.6885 & 0.0012 & 0&0083 & 13 \\
95 & 57974.7515 & 0.0012 & 0&0076 & 16 \\
96 & 57974.8150 & 0.0009 & 0&0075 & 23 \\
97 & 57974.8779 & 0.0010 & 0&0069 & 20 \\
126 & 57976.7174 & 0.0014 & 0&0018 & 15 \\
127 & 57976.7811 & 0.0020 & 0&0019 & 21 \\
128 & 57976.8466 & 0.0018 & 0&0038 & 20 \\
158 & 57978.7398 & 0.0033 & $-$0&0112 & 15 \\
159 & 57978.8029 & 0.0020 & $-$0&0117 & 24 \\
160 & 57978.8647 & 0.0038 & $-$0&0135 & 20 \\
\hline
  \multicolumn{6}{l}{\commenta BJD$-$2400000.} \\
  \multicolumn{6}{l}{\commentb Against max $= 2457968.7014 + 0.063605 E$.} \\
  \multicolumn{6}{l}{\commentc Number of points used to determine the maximum.} \\
\end{tabular}
\end{center}
\end{table}

\subsection{ASASSN-17kd}\label{obj:asassn17kd}

\begin{figure}
  \begin{center}
    \FigureFile(85mm,110mm){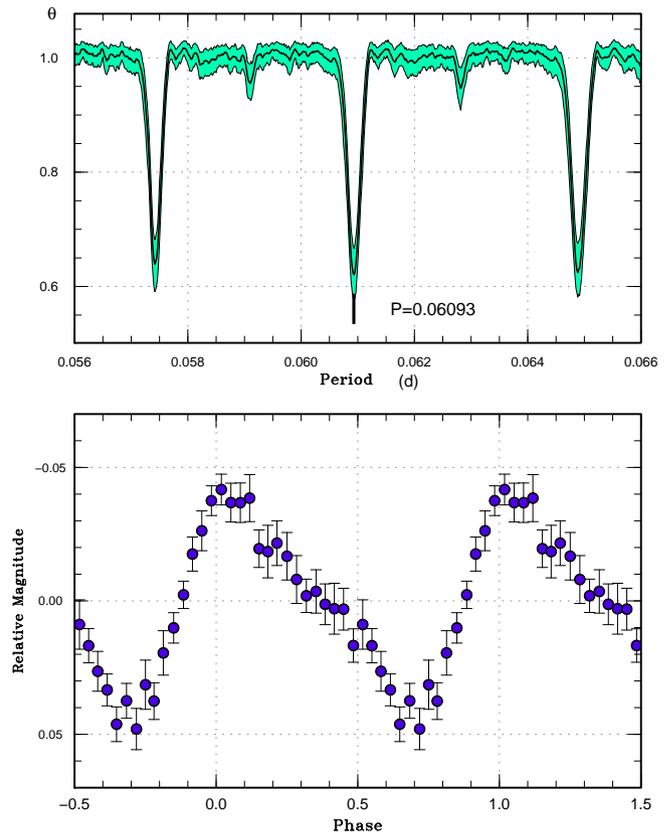}
  \end{center}
  \caption{Ordinary superhumps in ASASSN-17kd (2017).
     (Upper): PDM analysis.  The alias was selected
     by $O-C$ analysis.
     (Lower): Phase-averaged profile.}
  \label{fig:asassn17kdshpdm}
\end{figure}

   This object was detected as a transient
at $V$=12.4 on 2017 July 29 by the ASAS-SN team.
The outburst was announced after the observation
of $V$=12.8 on 2017 July 30.
The object started to show superhumps on
2017 August 10 (vsnet-alert 21339;
e-figure \ref{fig:asassn17kdshpdm}).
The times of superhump maxima are listed in
e-table \ref{tab:asassn17kdoc2017}.
Although stage A was recorded ($E \le$33),
observations in the stage was not long enough
to measure the period accurately.

   We could not detect significant early superhumps
to an upper limit of 0.01 mag.  Although early
superhumps were not observationally confirmed,
we consider that the object is a WZ Sge-type
dwarf nova since the waiting time to appear
ordinary superhumps was long (at least 12~d),
which is comparable to typical WZ Sge-type dwarf novae
(cf. \cite{kat15wzsge}).
At the time of our initial observation on 2017 August 3,
the object already faded to 13.5 mag, indicating
that the object faded rapidly since the peak
brightness.  Such a rapid fading is also characteristic
to a WZ Sge-type superoutburst \citep{kat15wzsge}.

   No past outbursts are known.  Our observations
indicated that the object faded close to 18 mag
on August 28.  According to ASAS-SN observations,
the object showed a rebrightening to $V$=15.6
on 2017 September 3--4.  Although there was
an observational gap between 2017 August 18 and 31
in the ASAS-SN data, the rebrightening was likely
a short one.

\begin{table}
\caption{Superhump maxima of ASASSN-17kd (2017)}\label{tab:asassn17kdoc2017}
\begin{center}
\begin{tabular}{rp{55pt}p{40pt}r@{.}lr}
\hline
\multicolumn{1}{c}{$E$} & \multicolumn{1}{c}{max\commenta} & \multicolumn{1}{c}{error} & \multicolumn{2}{c}{$O-C$\commentb} & \multicolumn{1}{c}{$N$\commentc} \\
\hline
0 & 57974.8537 & 0.0033 & $-$0&0129 & 17 \\
17 & 57975.8959 & 0.0012 & $-$0&0070 & 16 \\
33 & 57976.8877 & 0.0013 & 0&0095 & 16 \\
65 & 57978.8350 & 0.0015 & 0&0063 & 16 \\
66 & 57978.8946 & 0.0007 & 0&0049 & 13 \\
82 & 57979.8690 & 0.0008 & 0&0039 & 25 \\
98 & 57980.8423 & 0.0010 & 0&0020 & 25 \\
99 & 57980.9029 & 0.0022 & 0&0016 & 14 \\
115 & 57981.8770 & 0.0011 & 0&0004 & 25 \\
131 & 57982.8481 & 0.0014 & $-$0&0038 & 25 \\
147 & 57983.8250 & 0.0013 & $-$0&0022 & 23 \\
148 & 57983.8866 & 0.0023 & $-$0&0016 & 21 \\
164 & 57984.8694 & 0.0025 & 0&0060 & 25 \\
180 & 57985.8377 & 0.0038 & $-$0&0010 & 25 \\
181 & 57985.9001 & 0.0030 & 0&0004 & 13 \\
196 & 57986.8129 & 0.0025 & $-$0&0011 & 21 \\
197 & 57986.8729 & 0.0025 & $-$0&0020 & 25 \\
213 & 57987.8557 & 0.0030 & 0&0054 & 20 \\
229 & 57988.8236 & 0.0030 & $-$0&0020 & 20 \\
230 & 57988.8799 & 0.0029 & $-$0&0067 & 18 \\
\hline
  \multicolumn{6}{l}{\commenta BJD$-$2400000.} \\
  \multicolumn{6}{l}{\commentb Against max $= 2457974.8666 + 0.060956 E$.} \\
  \multicolumn{6}{l}{\commentc Number of points used to determine the maximum.} \\
\end{tabular}
\end{center}
\end{table}

\subsection{ASASSN-17kg}\label{obj:asassn17kg}

\begin{figure}
  \begin{center}
    \FigureFile(85mm,110mm){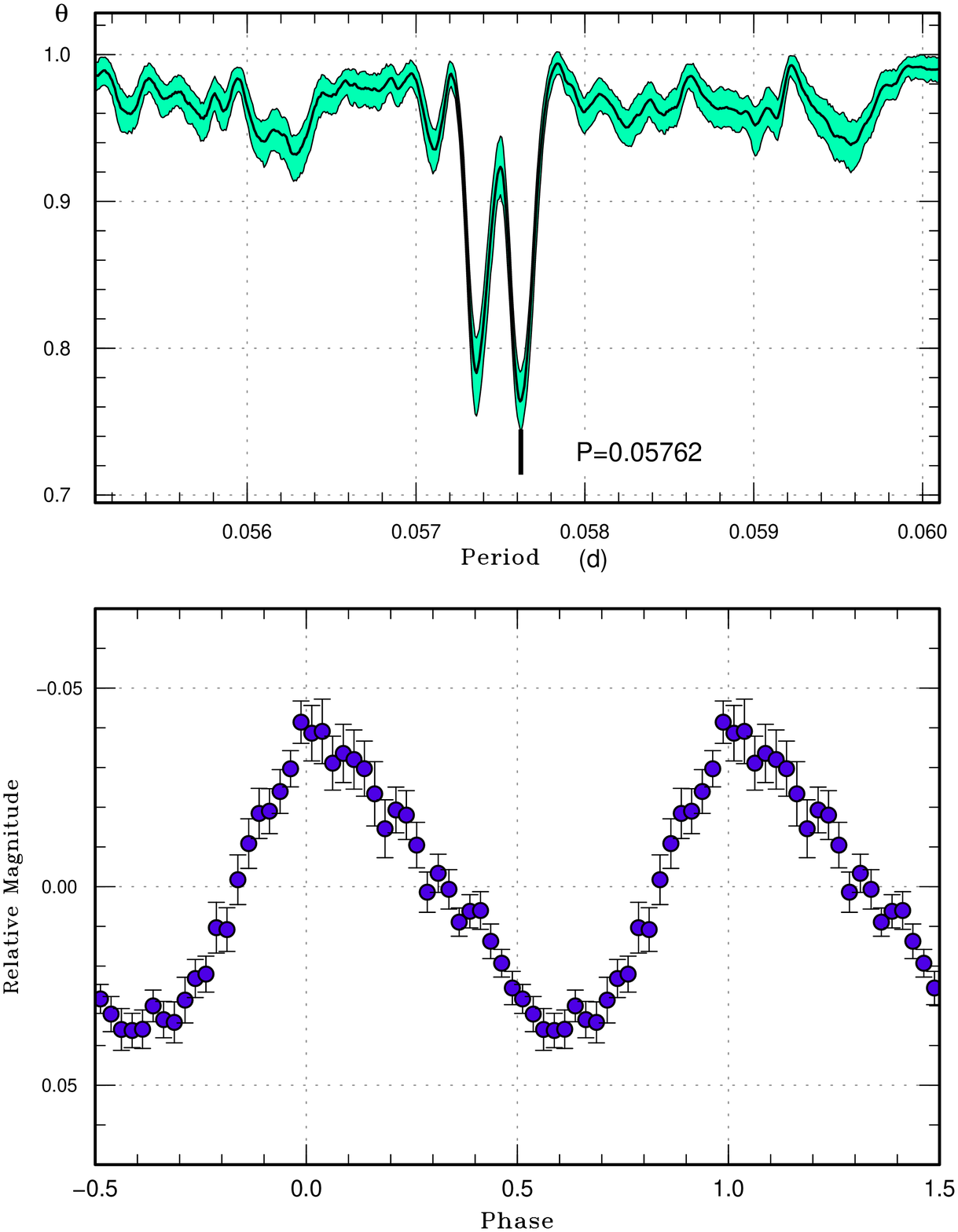}
  \end{center}
  \caption{Ordinary superhumps in ASASSN-17kg (2017).
     The data before the dip were used.
     (Upper): PDM analysis.
     (Lower): Phase-averaged profile.}
  \label{fig:asassn17kgshpdm}
\end{figure}

\begin{figure*}
  \begin{center}
    \FigureFile(160mm,200mm){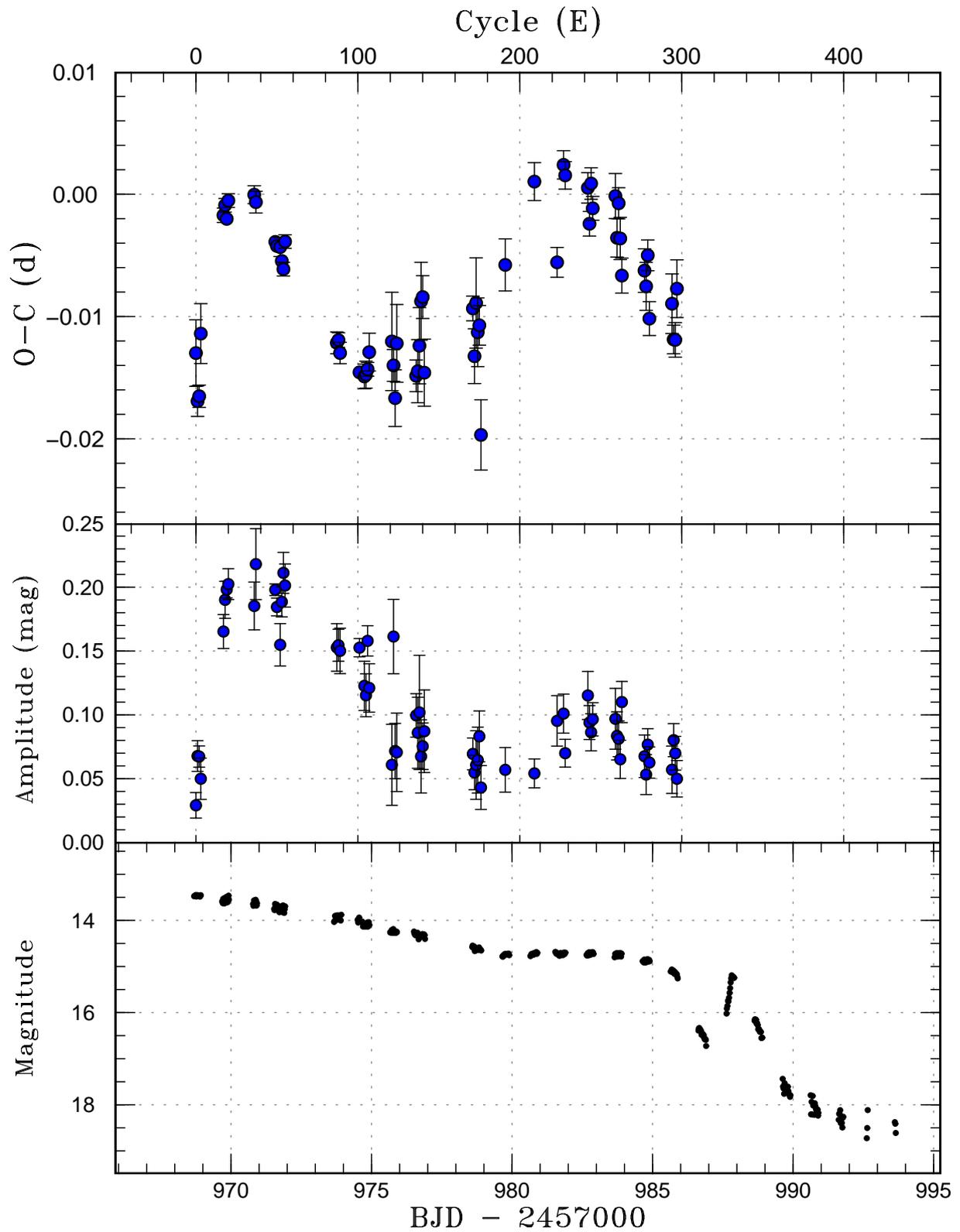}
  \end{center}
  \caption{$O-C$ diagram of superhumps in ASASSN-17kg (2017).
     (Upper:) $O-C$ diagram.
     We used a period of 0.05762~d for calculating the $O-C$ residuals.
     (Middle:) Amplitudes of superhumps.
     (Lower:) Light curve.  The data were binned to 0.019~d.
     Note the presence of a dip just before the termination
     of the superoutburst.
  }
  \label{fig:asassn17kghumpall}
\end{figure*}

   This object was detected as a transient
at $V$=13.0 on 2017 August 1 by the ASAS-SN team.
The object has an X-ray counterpart of 1RXP J003152$+$0841.1.
Subsequent observations detected superhumps
(vsnet-alert 21317, 21326; e-figure \ref{fig:asassn17kgshpdm}).
The times of superhump maxima are listed in
e-table \ref{tab:asassn17kgoc2017}.
There were ``textbook'' stage A-B-C superhumps
(e-figure \ref{fig:asassn17kghumpall}).

   Before the end of the superoutburst, the object
showed a dip of $\sim$1.5 mag on 2017 August 21
(e-figure \ref{fig:asassn17kghumpall}).
This behavior is very similar to the WZ Sge-type
dwarf nova KK Cnc (=OT J080714.2$+$113812, \cite{Pdot}).
Early superhumps, however, were not detected in the case
of ASASSN-17kg.  The outburst light curve and
a relatively small $P_{\rm dot}$ resemble those of
WZ Sge-type dwarf novae.  ASASSN-17kg may be
a WZ Sge-type dwarf nova which failed to develop
early superhumps when the stored mass in the disk
was not sufficient (e.g. AL Com, \cite{kim16alcom}).
There was no further post-superoutburst rebrightening
in the ASAS-SN data.

\begin{table*}
\caption{Superhump maxima of ASASSN-17kg (2017)}\label{tab:asassn17kgoc2017}
\begin{center}
\begin{tabular}{rp{55pt}p{40pt}r@{.}lrrp{55pt}p{40pt}r@{.}lr}
\hline
\multicolumn{1}{c}{$E$} & \multicolumn{1}{c}{max\commenta} & \multicolumn{1}{c}{error} & \multicolumn{2}{c}{$O-C$\commentb} & \multicolumn{1}{c}{$N$\commentc} & \multicolumn{1}{c}{$E$} & \multicolumn{1}{c}{max\commenta} & \multicolumn{1}{c}{error} & \multicolumn{2}{c}{$O-C$\commentb} & \multicolumn{1}{c}{$N$\commentc} \\
\hline
0 & 57968.7404 & 0.0027 & $-$0&0033 & 13 & 139 & 57976.7538 & 0.0032 & $-$0&0007 & 15 \\
1 & 57968.7941 & 0.0012 & $-$0&0072 & 17 & 140 & 57976.8118 & 0.0017 & $-$0&0003 & 20 \\
2 & 57968.8521 & 0.0009 & $-$0&0068 & 20 & 141 & 57976.8632 & 0.0027 & $-$0&0065 & 18 \\
3 & 57968.9148 & 0.0024 & $-$0&0017 & 10 & 171 & 57978.5971 & 0.0010 & $-$0&0016 & 52 \\
17 & 57969.7312 & 0.0006 & 0&0078 & 13 & 172 & 57978.6508 & 0.0022 & $-$0&0055 & 43 \\
18 & 57969.7896 & 0.0006 & 0&0086 & 17 & 173 & 57978.7127 & 0.0037 & $-$0&0012 & 13 \\
19 & 57969.8461 & 0.0003 & 0&0075 & 20 & 174 & 57978.7680 & 0.0028 & $-$0&0036 & 17 \\
20 & 57969.9053 & 0.0006 & 0&0090 & 13 & 175 & 57978.8262 & 0.0016 & $-$0&0030 & 18 \\
36 & 57970.8277 & 0.0007 & 0&0092 & 21 & 176 & 57978.8748 & 0.0029 & $-$0&0120 & 18 \\
37 & 57970.8847 & 0.0009 & 0&0086 & 18 & 191 & 57979.7530 & 0.0021 & 0&0017 & 21 \\
49 & 57971.5729 & 0.0002 & 0&0052 & 140 & 209 & 57980.7970 & 0.0015 & 0&0083 & 35 \\
50 & 57971.6301 & 0.0003 & 0&0049 & 86 & 223 & 57981.5971 & 0.0012 & 0&0015 & 27 \\
52 & 57971.7453 & 0.0008 & 0&0048 & 14 & 227 & 57981.8355 & 0.0012 & 0&0095 & 24 \\
53 & 57971.8018 & 0.0004 & 0&0036 & 20 & 228 & 57981.8923 & 0.0011 & 0&0086 & 20 \\
54 & 57971.8587 & 0.0006 & 0&0030 & 18 & 242 & 57982.6979 & 0.0012 & 0&0074 & 16 \\
55 & 57971.9186 & 0.0006 & 0&0052 & 8 & 243 & 57982.7526 & 0.0010 & 0&0045 & 25 \\
87 & 57973.7542 & 0.0009 & $-$0&0035 & 14 & 244 & 57982.8135 & 0.0013 & 0&0078 & 30 \\
88 & 57973.8120 & 0.0006 & $-$0&0032 & 21 & 245 & 57982.8691 & 0.0010 & 0&0057 & 24 \\
89 & 57973.8686 & 0.0009 & $-$0&0043 & 18 & 259 & 57983.6768 & 0.0018 & 0&0066 & 15 \\
101 & 57974.5584 & 0.0004 & $-$0&0060 & 46 & 260 & 57983.7310 & 0.0016 & 0&0031 & 17 \\
104 & 57974.7310 & 0.0010 & $-$0&0064 & 13 & 261 & 57983.7915 & 0.0013 & 0&0059 & 35 \\
105 & 57974.7887 & 0.0011 & $-$0&0063 & 19 & 262 & 57983.8462 & 0.0017 & 0&0031 & 24 \\
106 & 57974.8468 & 0.0006 & $-$0&0058 & 18 & 263 & 57983.9008 & 0.0014 & 0&0000 & 15 \\
107 & 57974.9058 & 0.0015 & $-$0&0044 & 11 & 277 & 57984.7079 & 0.0018 & 0&0003 & 17 \\
121 & 57975.7134 & 0.0040 & $-$0&0037 & 13 & 278 & 57984.7642 & 0.0020 & $-$0&0011 & 31 \\
122 & 57975.7690 & 0.0013 & $-$0&0057 & 16 & 279 & 57984.8244 & 0.0013 & 0&0015 & 25 \\
123 & 57975.8240 & 0.0023 & $-$0&0084 & 19 & 280 & 57984.8768 & 0.0014 & $-$0&0037 & 23 \\
124 & 57975.8861 & 0.0032 & $-$0&0039 & 18 & 294 & 57985.6847 & 0.0024 & $-$0&0027 & 16 \\
136 & 57976.5749 & 0.0013 & $-$0&0067 & 50 & 295 & 57985.7394 & 0.0012 & $-$0&0056 & 23 \\
137 & 57976.6328 & 0.0026 & $-$0&0064 & 31 & 296 & 57985.7970 & 0.0014 & $-$0&0056 & 31 \\
138 & 57976.6926 & 0.0031 & $-$0&0043 & 12 & 297 & 57985.8588 & 0.0024 & $-$0&0015 & 24 \\
\hline
  \multicolumn{12}{l}{\commenta BJD$-$2400000.} \\
  \multicolumn{12}{l}{\commentb Against max $= 2457968.7437 + 0.057632 E$.} \\
  \multicolumn{12}{l}{\commentc Number of points used to determine the maximum.} \\
\end{tabular}
\end{center}
\end{table*}

\subsection{ASASSN-17kp}\label{obj:asassn17kp}

\begin{figure}
  \begin{center}
    \FigureFile(85mm,110mm){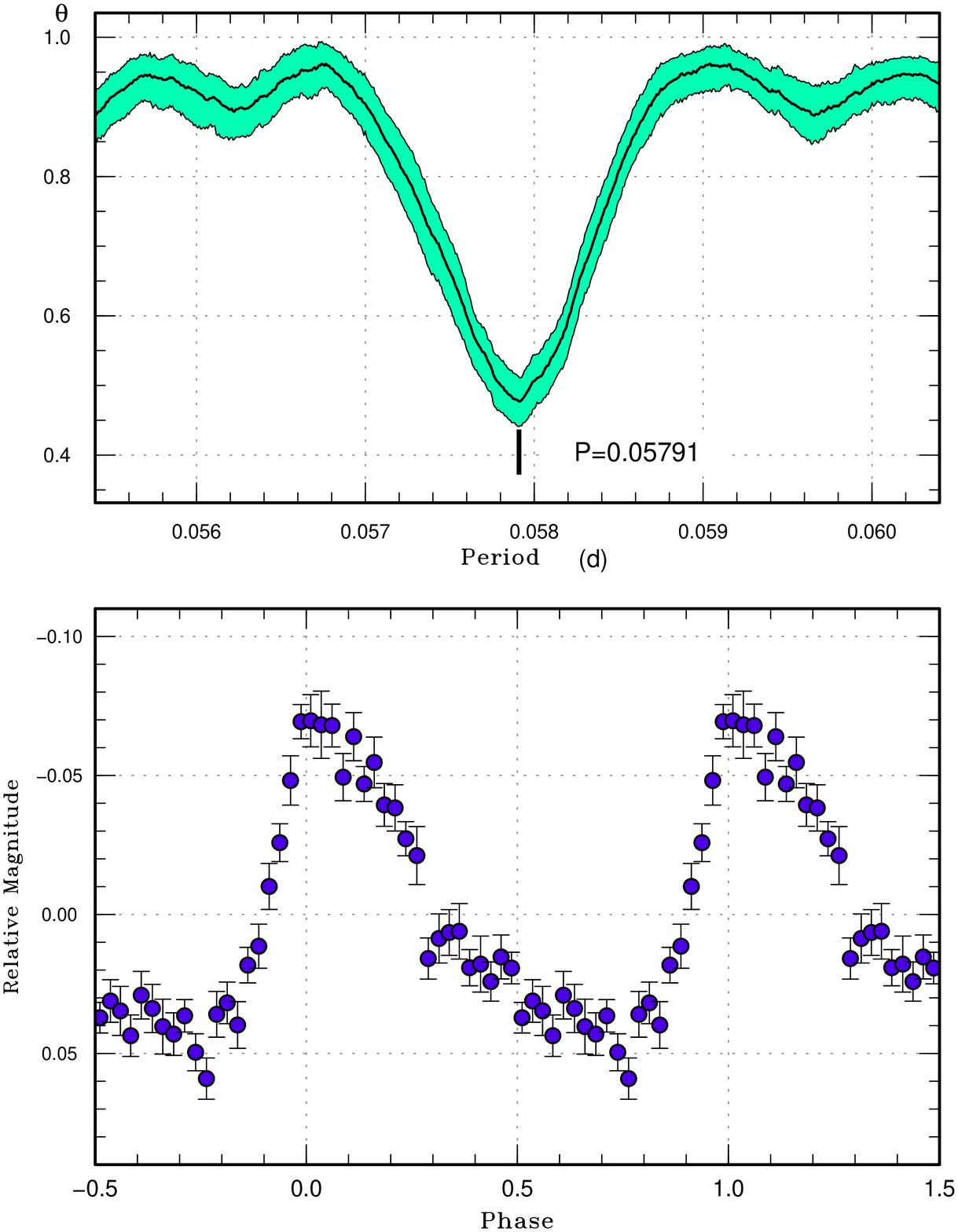}
  \end{center}
  \caption{Ordinary superhumps in ASASSN-17kp (2017).
     The data before the dip were used.
     (Upper): PDM analysis.
     (Lower): Phase-averaged profile.}
  \label{fig:asassn17kpshpdm}
\end{figure}

   This object was detected as a transient
at $V$=14.8 on 2017 August 6 by the ASAS-SN team.
The outburst was announced after observation of
$V$=15.2 on 2017 August 9.  Subsequent observations
detected superhumps (vsnet-alert 21338;
e-figure \ref{fig:asassn17kpshpdm}).
The times of superhump maxima are listed in
e-table \ref{tab:asassn17kpoc2017}.
The fading on 2017 August 9 may suggest that the outburst
on 2017 August 6 was a precursor one.

\begin{table}
\caption{Superhump maxima of ASASSN-17kp (2017)}\label{tab:asassn17kpoc2017}
\begin{center}
\begin{tabular}{rp{55pt}p{40pt}r@{.}lr}
\hline
\multicolumn{1}{c}{$E$} & \multicolumn{1}{c}{max\commenta} & \multicolumn{1}{c}{error} & \multicolumn{2}{c}{$O-C$\commentb} & \multicolumn{1}{c}{$N$\commentc} \\
\hline
0 & 57978.5360 & 0.0007 & 0&0012 & 48 \\
15 & 57979.4042 & 0.0010 & 0&0001 & 63 \\
16 & 57979.4596 & 0.0026 & $-$0&0026 & 24 \\
17 & 57979.5217 & 0.0010 & 0&0016 & 62 \\
32 & 57980.3889 & 0.0005 & $-$0&0005 & 102 \\
33 & 57980.4468 & 0.0007 & $-$0&0006 & 119 \\
34 & 57980.5045 & 0.0006 & $-$0&0009 & 60 \\
35 & 57980.5639 & 0.0017 & 0&0006 & 38 \\
36 & 57980.6211 & 0.0017 & $-$0&0001 & 37 \\
51 & 57981.4919 & 0.0013 & 0&0012 & 41 \\
\hline
  \multicolumn{6}{l}{\commenta BJD$-$2400000.} \\
  \multicolumn{6}{l}{\commentb Against max $= 2457978.5348 + 0.057957 E$.} \\
  \multicolumn{6}{l}{\commentc Number of points used to determine the maximum.} \\
\end{tabular}
\end{center}
\end{table}

\subsection{ASASSN-17la}\label{obj:asassn17la}

\begin{figure}
  \begin{center}
    \FigureFile(85mm,110mm){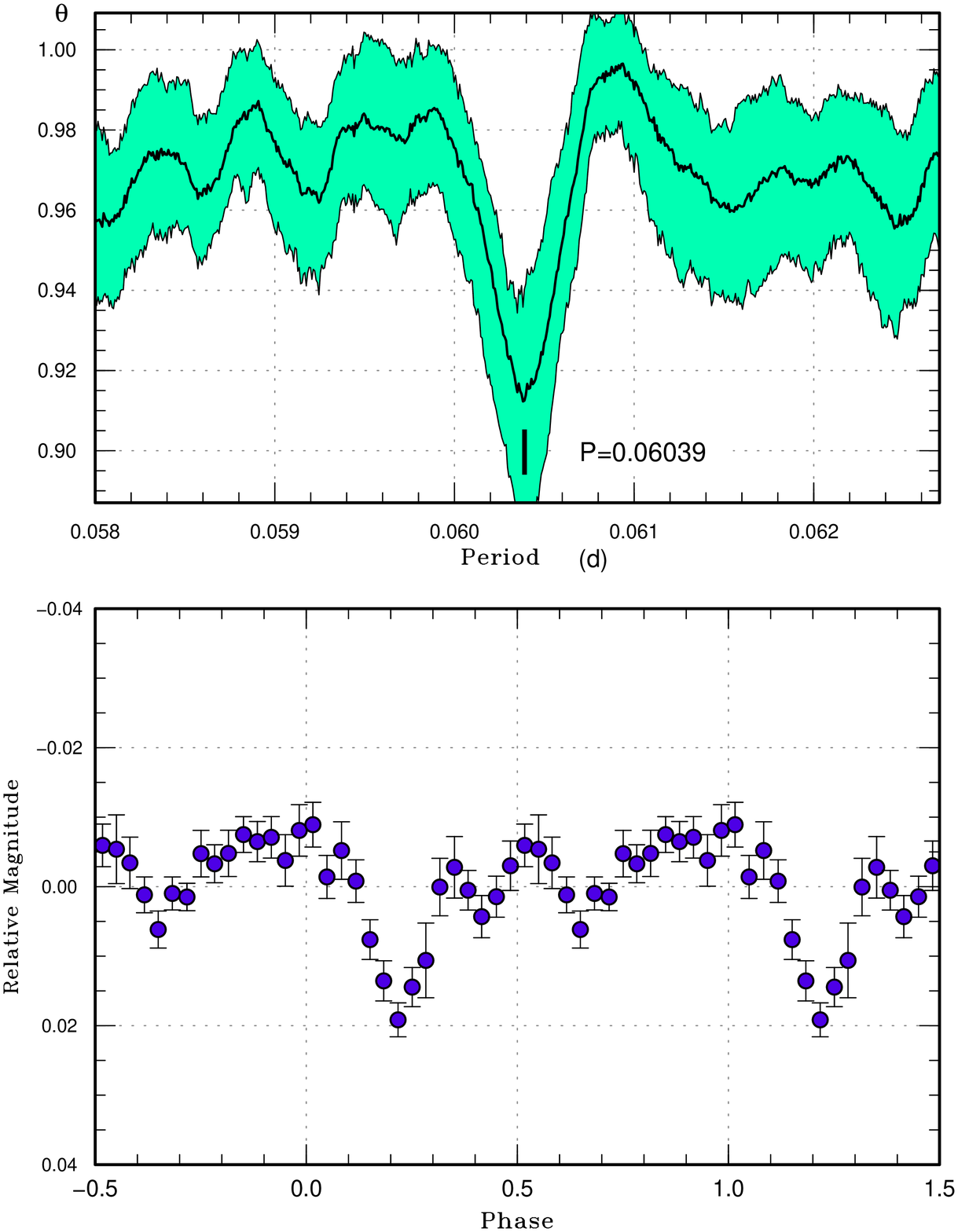}
  \end{center}
  \caption{Early superhumps in ASASSN-17la (2017).
     (Upper): PDM analysis.
     (Lower): Phase-averaged profile.}
  \label{fig:asassn17laeshpdm}
\end{figure}

\begin{figure}
  \begin{center}
    \FigureFile(85mm,110mm){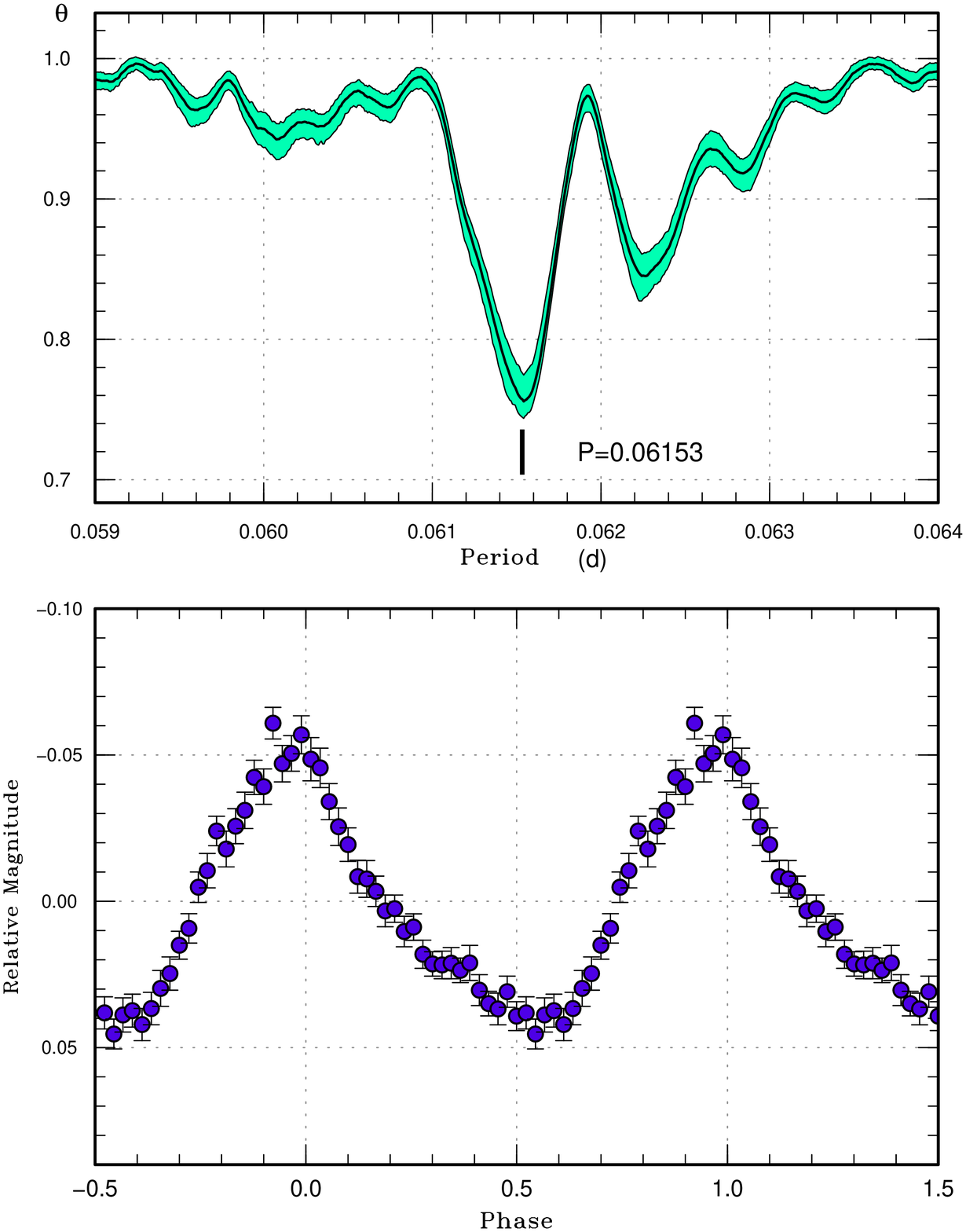}
  \end{center}
  \caption{Ordinary superhumps in ASASSN-17la (2017).
     (Upper): PDM analysis.
     (Lower): Phase-averaged profile.}
  \label{fig:asassn17lashpdm}
\end{figure}

   This object was detected as a transient
at $V$=14.5 on 2017 August 17 by the ASAS-SN team.
Early superhumps were detected, indicating that
this object is a WZ Sge-type dwarf nova
(vsnet-alert 21349; e-figure \ref{fig:asassn17laeshpdm}).
Ordinary superhump developed on 2017 October 23
(vsnet-alert 21362, 6~d after the outburst detection;
e-figure \ref{fig:asassn17lashpdm}).
The times of superhump maxima are listed in
e-table \ref{tab:asassn17laoc2017}.  This object
showed ``textbook'' stages A and B.

   The period of early superhump determined by the PDM
method was 0.06039(3)~d.  The $\epsilon^*$ for stage A
superhumps was 0.031(2), corresponding to $q$=0.084(5).
This value is somewhat smaller than those of ordinary
dwarf novae with the corresponding orbital period
(cf. figure 17 in \cite{kat15wzsge}), but somewhat
higher than those of typical period bouncers.
The resultant parameters were similar to WZ Sge-type
dwarf novae with multiple rebrightenings
[see \citet{nak13j2112j2037}; \citet{kat15wzsge}].
The $P_{\rm dot}$ for stage B superhumps was not
unusually small as in period bouncers.
Due to the faintness of the object, we did not have
observations after the superoutburst.

\begin{table}
\caption{Superhump maxima of ASASSN-17la (2017)}\label{tab:asassn17laoc2017}
\begin{center}
\begin{tabular}{rp{55pt}p{40pt}r@{.}lr}
\hline
\multicolumn{1}{c}{$E$} & \multicolumn{1}{c}{max\commenta} & \multicolumn{1}{c}{error} & \multicolumn{2}{c}{$O-C$\commentb} & \multicolumn{1}{c}{$N$\commentc} \\
\hline
0 & 57988.7336 & 0.0010 & $-$0&0064 & 115 \\
1 & 57988.7917 & 0.0012 & $-$0&0098 & 115 \\
15 & 57989.6640 & 0.0012 & 0&0004 & 76 \\
16 & 57989.7291 & 0.0005 & 0&0039 & 116 \\
17 & 57989.7909 & 0.0005 & 0&0041 & 114 \\
18 & 57989.8546 & 0.0005 & 0&0062 & 92 \\
27 & 57990.4106 & 0.0004 & 0&0081 & 85 \\
28 & 57990.4712 & 0.0003 & 0&0071 & 149 \\
29 & 57990.5328 & 0.0003 & 0&0071 & 121 \\
30 & 57990.5938 & 0.0012 & 0&0065 & 46 \\
31 & 57990.6547 & 0.0023 & 0&0059 & 39 \\
32 & 57990.7172 & 0.0004 & 0&0068 & 83 \\
33 & 57990.7775 & 0.0004 & 0&0055 & 92 \\
34 & 57990.8395 & 0.0003 & 0&0059 & 81 \\
65 & 57992.7391 & 0.0005 & $-$0&0033 & 90 \\
66 & 57992.8011 & 0.0006 & $-$0&0029 & 80 \\
77 & 57993.4765 & 0.0005 & $-$0&0048 & 91 \\
78 & 57993.5392 & 0.0006 & $-$0&0037 & 102 \\
79 & 57993.6030 & 0.0007 & $-$0&0016 & 98 \\
90 & 57994.2747 & 0.0006 & $-$0&0072 & 26 \\
91 & 57994.3367 & 0.0007 & $-$0&0067 & 33 \\
92 & 57994.4005 & 0.0006 & $-$0&0045 & 34 \\
93 & 57994.4616 & 0.0008 & $-$0&0050 & 33 \\
94 & 57994.5231 & 0.0007 & $-$0&0051 & 33 \\
95 & 57994.5876 & 0.0010 & $-$0&0022 & 28 \\
97 & 57994.7084 & 0.0008 & $-$0&0045 & 63 \\
98 & 57994.7713 & 0.0009 & $-$0&0032 & 59 \\
99 & 57994.8322 & 0.0013 & $-$0&0039 & 46 \\
106 & 57995.2674 & 0.0041 & 0&0003 & 18 \\
107 & 57995.3243 & 0.0009 & $-$0&0044 & 33 \\
108 & 57995.3867 & 0.0015 & $-$0&0036 & 32 \\
109 & 57995.4481 & 0.0009 & $-$0&0037 & 33 \\
110 & 57995.5133 & 0.0016 & $-$0&0001 & 33 \\
111 & 57995.5716 & 0.0013 & $-$0&0034 & 29 \\
113 & 57995.6961 & 0.0010 & $-$0&0021 & 63 \\
114 & 57995.7573 & 0.0014 & $-$0&0024 & 64 \\
115 & 57995.8207 & 0.0015 & $-$0&0006 & 60 \\
116 & 57995.8775 & 0.0012 & $-$0&0054 & 64 \\
123 & 57996.3116 & 0.0016 & $-$0&0023 & 33 \\
124 & 57996.3726 & 0.0013 & $-$0&0029 & 34 \\
125 & 57996.4345 & 0.0024 & $-$0&0025 & 33 \\
126 & 57996.4985 & 0.0022 & $-$0&0001 & 33 \\
127 & 57996.5630 & 0.0035 & 0&0028 & 33 \\
141 & 57997.4170 & 0.0019 & $-$0&0053 & 37 \\
172 & 57999.3422 & 0.0013 & 0&0110 & 39 \\
173 & 57999.4026 & 0.0011 & 0&0099 & 43 \\
174 & 57999.4644 & 0.0011 & 0&0101 & 42 \\
175 & 57999.5281 & 0.0017 & 0&0122 & 44 \\
\hline
  \multicolumn{6}{l}{\commenta BJD$-$2400000.} \\
  \multicolumn{6}{l}{\commentb Against max $= 2457988.7400 + 0.061577 E$.} \\
  \multicolumn{6}{l}{\commentc Number of points used to determine the maximum.} \\
\end{tabular}
\end{center}
\end{table}

\subsection{ASASSN-17lr}\label{obj:asassn17lr}

\begin{figure}
  \begin{center}
    \FigureFile(85mm,110mm){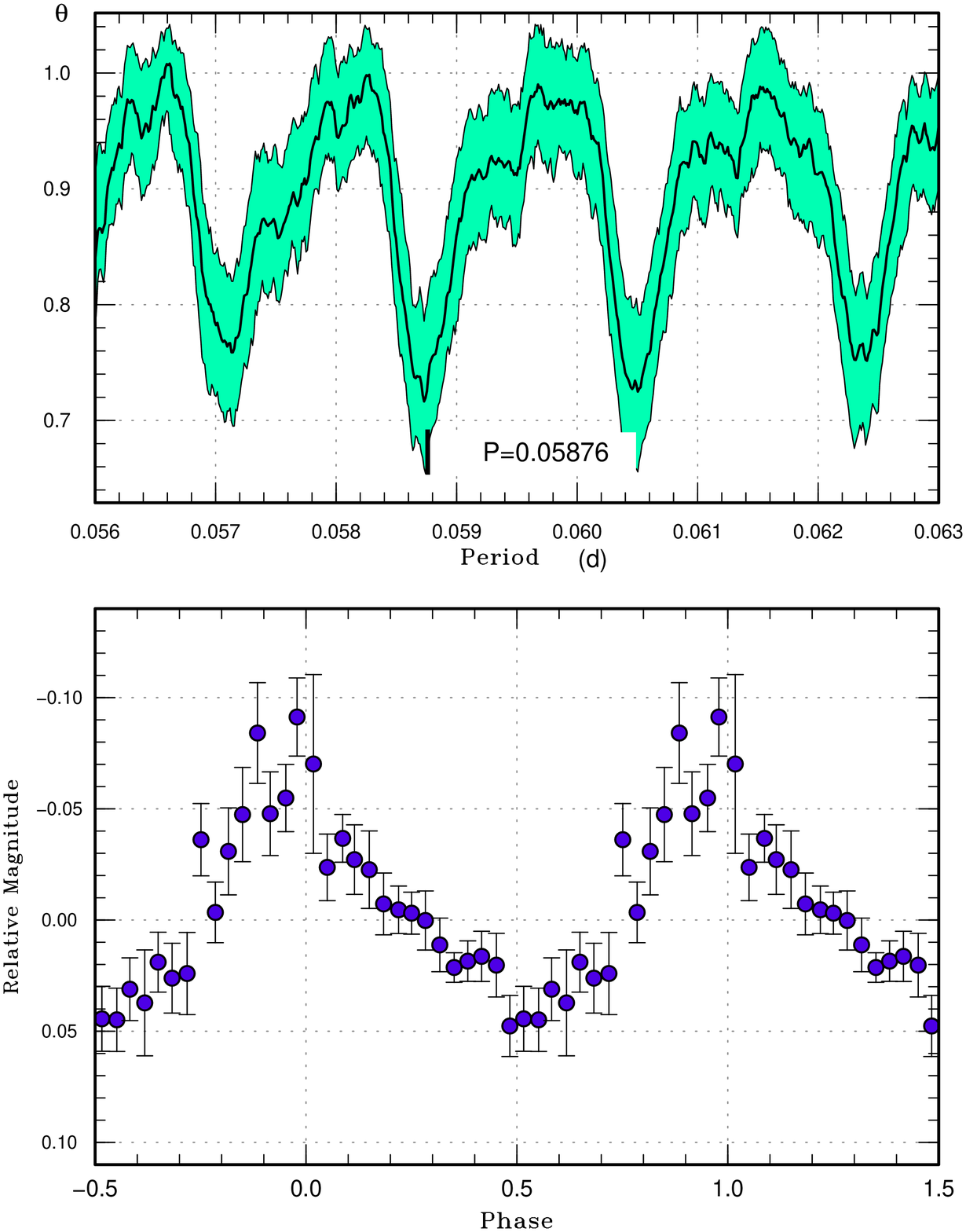}
  \end{center}
  \caption{Superhumps in ASASSN-17lr (2017).
     (Upper): PDM analysis.  The alias selection was based on
     $O-C$ analysis.
     (Lower): Phase-averaged profile.}
  \label{fig:asassn17lrshpdm}
\end{figure}

   This object was detected as a transient
at $V$=14.6 on 2017 September 4 by the ASAS-SN team.
The outburst was announced after the observation
of $V$=14.9 on 2017 Semtember 5.
Observations on 2017 September 9 and 11 did not show
superhump-like modulations.  Superhumps were observed
since 2017 September 15 (vsnet-alert 21443;
e-figure \ref{fig:asassn17lrshpdm}).
The times of superhump maxima are listed in
e-table \ref{tab:asassn17lroc2017}.  Although a one-day
alias 0.06047(3)~d was stronger by the PDM method,
we adopted this alias based on $O-C$ analysis of
the continuous observations on 2017 September 17--18.
Although the object was initially suggested to be
a WZ Sge-type dwarf nova (vsnet-alert 21443), this
was based on the absence of superhumps for a week
after the discovery and not based on detection of
early superhumps.

\begin{table}
\caption{Superhump maxima of ASASSN-17lr (2017)}\label{tab:asassn17lroc2017}
\begin{center}
\begin{tabular}{rp{55pt}p{40pt}r@{.}lr}
\hline
\multicolumn{1}{c}{$E$} & \multicolumn{1}{c}{max\commenta} & \multicolumn{1}{c}{error} & \multicolumn{2}{c}{$O-C$\commentb} & \multicolumn{1}{c}{$N$\commentc} \\
\hline
0 & 58012.3909 & 0.0010 & $-$0&0046 & 37 \\
34 & 58014.3925 & 0.0019 & 0&0034 & 42 \\
36 & 58014.5103 & 0.0015 & 0&0039 & 38 \\
37 & 58014.5647 & 0.0014 & $-$0&0003 & 48 \\
102 & 58018.3738 & 0.0037 & $-$0&0024 & 57 \\
\hline
  \multicolumn{6}{l}{\commenta BJD$-$2400000.} \\
  \multicolumn{6}{l}{\commentb Against max $= 2458012.3955 + 0.058634 E$.} \\
  \multicolumn{6}{l}{\commentc Number of points used to determine the maximum.} \\
\end{tabular}
\end{center}
\end{table}

\subsection{ASASSN-17me}\label{obj:asassn17me}

\begin{figure}
  \begin{center}
    \FigureFile(85mm,110mm){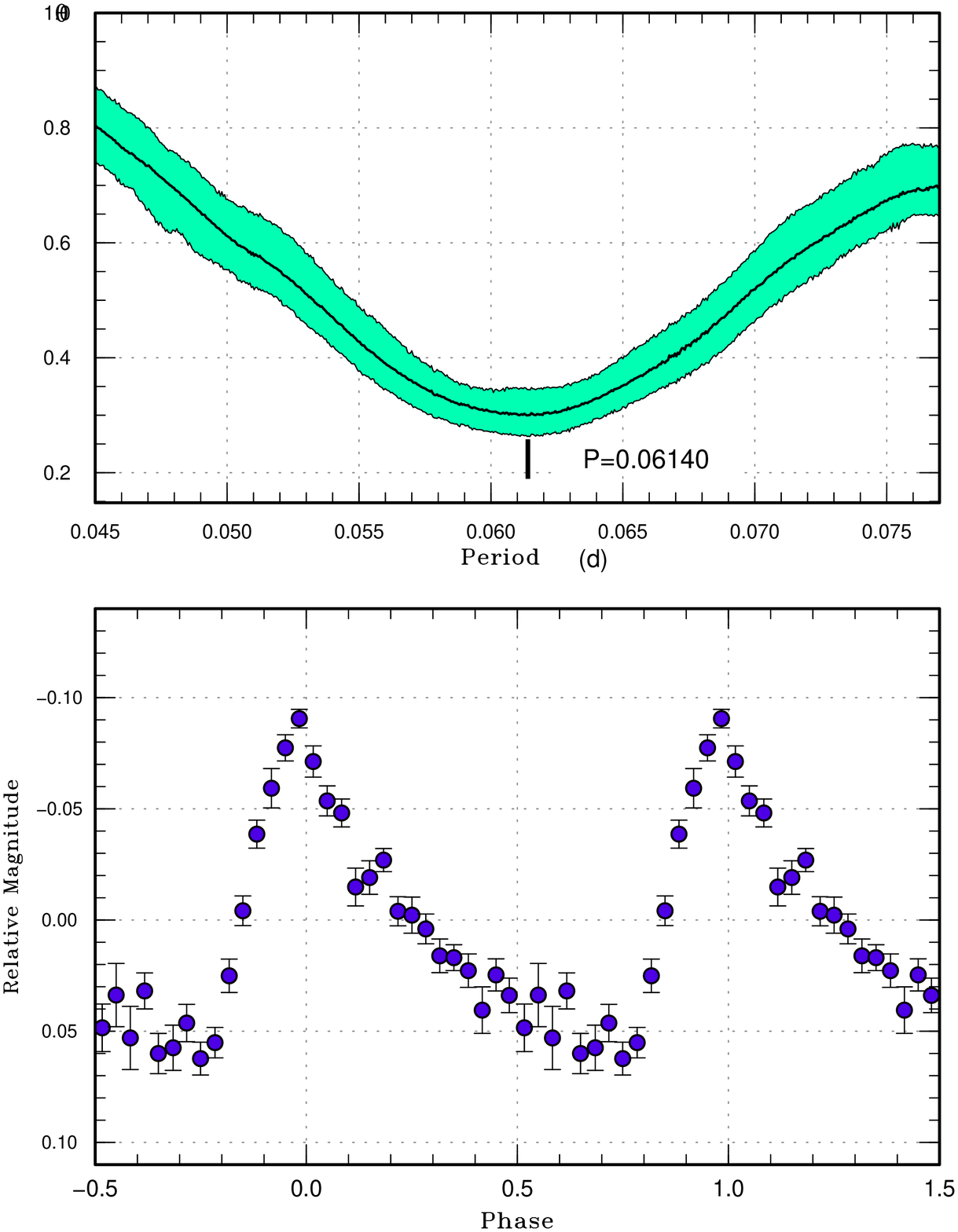}
  \end{center}
  \caption{Superhumps in ASASSN-17me (2017).
     (Upper): PDM analysis.
     (Lower): Phase-averaged profile.}
  \label{fig:asassn17meshpdm}
\end{figure}

   This object was detected as a transient
at $V$=15.2 on 2017 September 13 by the ASAS-SN team.
The outburst was announced after the observation
of $V$=15.4 on 2017 Semtember 15.  There is a likely
quiescent counterpart $g$=21.45 mag object
in Pan-STARRS (vsnet-alert 21440).
Subsequent observations detected superhumps
(vsnet-alert 21447, e-figure \ref{fig:asassn17meshpdm}).
Although there were observations on two additional nights,
the object already started fading rapidly (2017 September
22 and 23) and superhumps were not detected.
The superhump period based on single-nights observations
was 0.0614(4)~d (PDM method).  The times of superhump
maxima were BJD 2458013.6486(5) ($N$=162) and
2458013.7130(4) ($N$=186).

\subsection{ASASSN-17np}\label{obj:asassn17np}

\begin{figure}
  \begin{center}
    \FigureFile(85mm,110mm){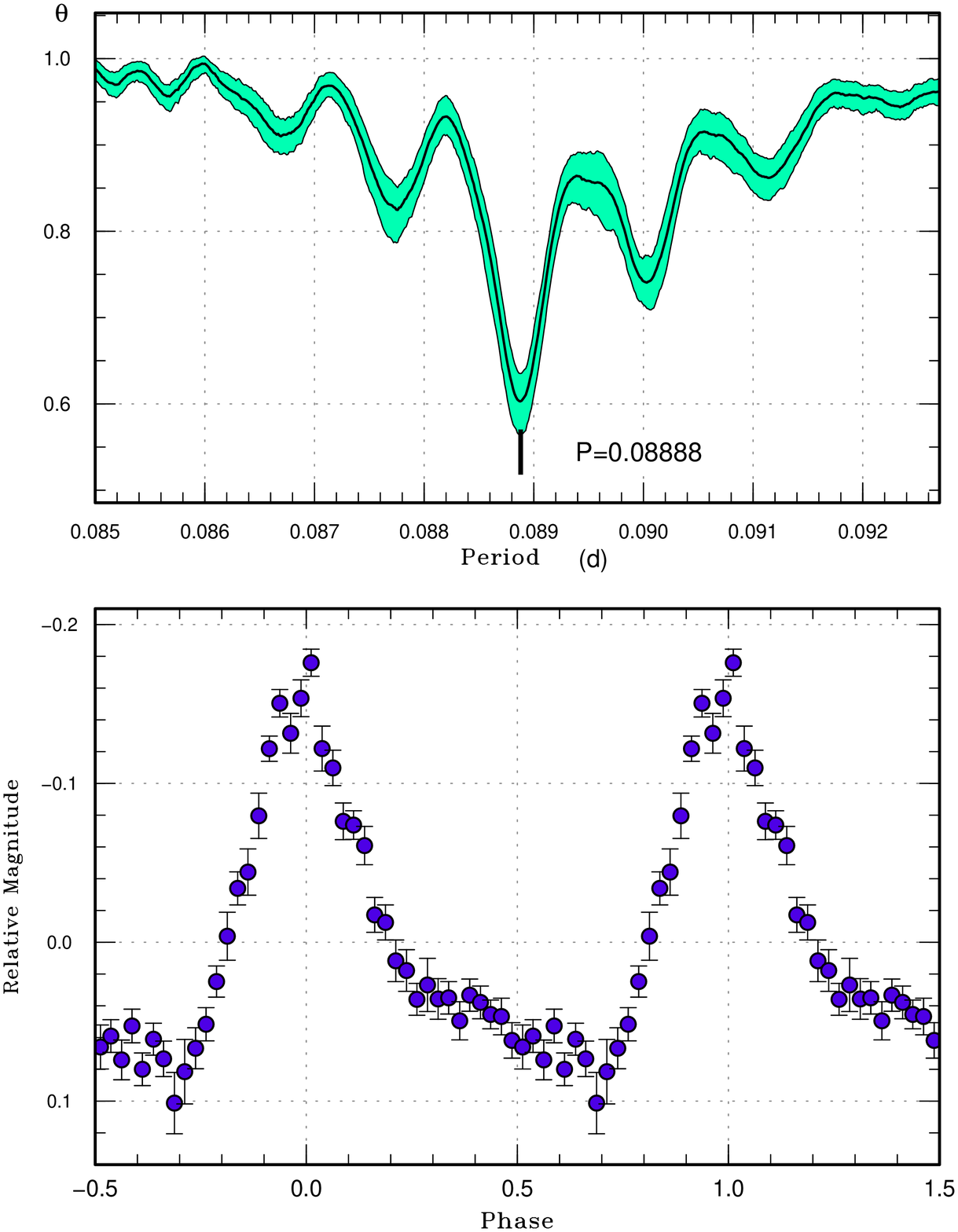}
  \end{center}
  \caption{Superhumps in ASASSN-17np (2017).
     (Upper): PDM analysis.
     (Lower): Phase-averaged profile.}
  \label{fig:asassn17npshpdm}
\end{figure}

   This object was detected as a transient
at $g$=15.0 on 2017 October 18 by the ASAS-SN team.
The outburst was announced after the observation
of $g$=15.2 on 2017 October 20.
Subsequent observations detected superhumps
(vsnet-alert 21534, 21535, 21545;
e-figure \ref{fig:asassn17npshpdm}).
The times of superhump maxima are listed in
e-table \ref{tab:asassn17np2017}.  Although observations
were rather sparse, stages B and C can be recognized.
The object called attention due to its absence
of a known quiescent counterpart and hence a likely
large outburst amplitude (cf. vsnet-alert 21531),
the resultant superhump period suggests a rather
ordinary SU UMa-type dwarf nova.  The outburst amplitude
may not be as large as initially suspected, or the object
may be indeed unusual with a large outburst amplitude
despite the long superhump period.  The large amplitude
of superhumps suggests an ordinary object rather than
a WZ Sge-like object with a long superhump period such
as ASASSN-16eg \citep{wak17asassn16eg}.

\begin{table}
\caption{Superhump maxima of ASASSN-17np (2017)}\label{tab:asassn17np2017}
\begin{center}
\begin{tabular}{rp{55pt}p{40pt}r@{.}lr}
\hline
\multicolumn{1}{c}{$E$} & \multicolumn{1}{c}{max\commenta} & \multicolumn{1}{c}{error} & \multicolumn{2}{c}{$O-C$\commentb} & \multicolumn{1}{c}{$N$\commentc} \\
\hline
0 & 58047.4968 & 0.0005 & $-$0&0055 & 199 \\
1 & 58047.5872 & 0.0004 & $-$0&0039 & 205 \\
10 & 58048.3899 & 0.0003 & $-$0&0009 & 205 \\
11 & 58048.4791 & 0.0003 & $-$0&0006 & 205 \\
12 & 58048.5701 & 0.0004 & 0&0016 & 131 \\
25 & 58049.7293 & 0.0013 & 0&0056 & 11 \\
26 & 58049.8161 & 0.0009 & 0&0036 & 26 \\
37 & 58050.7943 & 0.0016 & 0&0043 & 17 \\
48 & 58051.7706 & 0.0012 & 0&0031 & 16 \\
49 & 58051.8592 & 0.0016 & 0&0029 & 15 \\
59 & 58052.7441 & 0.0014 & $-$0&0008 & 14 \\
60 & 58052.8311 & 0.0019 & $-$0&0027 & 21 \\
70 & 58053.7185 & 0.0026 & $-$0&0039 & 11 \\
71 & 58053.8096 & 0.0016 & $-$0&0016 & 20 \\
79 & 58054.5201 & 0.0010 & $-$0&0020 & 180 \\
82 & 58054.7896 & 0.0059 & 0&0009 & 19 \\
\hline
  \multicolumn{6}{l}{\commenta BJD$-$2400000.} \\
  \multicolumn{6}{l}{\commentb Against max $= 2458047.5022 + 0.088859 E$.} \\
  \multicolumn{6}{l}{\commentc Number of points used to determine the maximum.} \\
\end{tabular}
\end{center}
\end{table}

\subsection{ASASSN-17nr}\label{obj:asassn17nr}

\begin{figure}
  \begin{center}
    \FigureFile(85mm,110mm){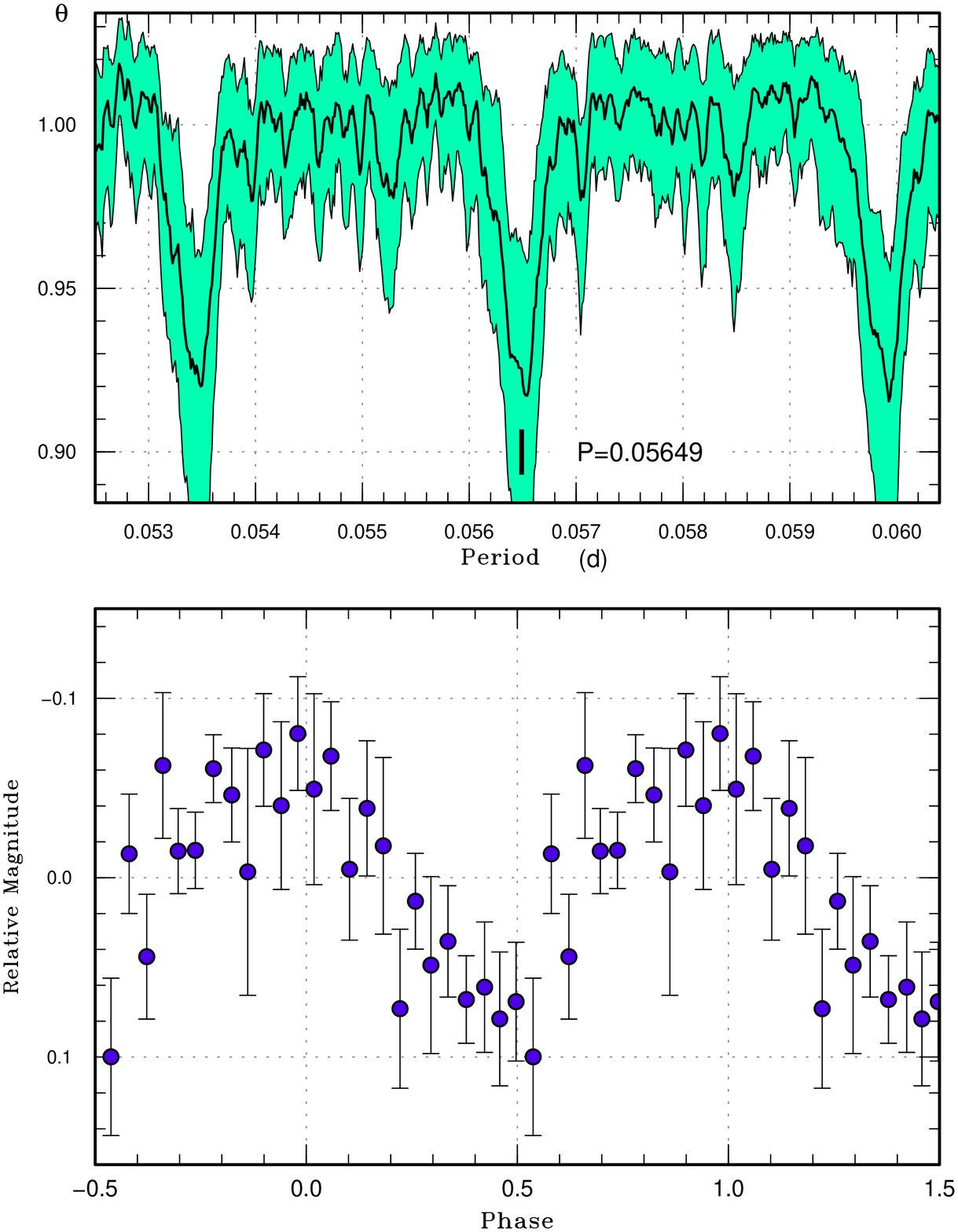}
  \end{center}
  \caption{Superhumps in ASASSN-17nr (2017).
     (Upper): PDM analysis.  The alias selection
     was based on $O-C$ analysis.
     (Lower): Phase-averaged profile.}
  \label{fig:asassn17nrshpdm}
\end{figure}

   This object was detected as a transient
at $V$=14.6 on 2017 October 18 by the ASAS-SN team.
The outburst was announced after the observation
of $V$=15.4 on 2017 October 21.
Subsequent observations detected superhumps
(vsnet-alert 21556; e-figure \ref{fig:asassn17nrshpdm}).
The times of superhump maxima are listed in
e-table \ref{tab:asassn17nroc2017}.
Although $O-C$ data suggested a positive $P_{\rm dot}$,
the value should be treated with caution due to
the limited quality of observations.
There remains possibilities of one-day aliases,
although the $O-C$ analysis favors this selection
(e-figure \ref{fig:asassn17nrshpdm}).

\begin{table}
\caption{Superhump maxima of ASASSN-17nr (2017)}\label{tab:asassn17nroc2017}
\begin{center}
\begin{tabular}{rp{55pt}p{40pt}r@{.}lr}
\hline
\multicolumn{1}{c}{$E$} & \multicolumn{1}{c}{max\commenta} & \multicolumn{1}{c}{error} & \multicolumn{2}{c}{$O-C$\commentb} & \multicolumn{1}{c}{$N$\commentc} \\
\hline
0 & 58052.8080 & 0.0029 & 0&0026 & 13 \\
36 & 58054.8301 & 0.0017 & $-$0&0036 & 13 \\
53 & 58055.7903 & 0.0021 & $-$0&0013 & 11 \\
54 & 58055.8449 & 0.0020 & $-$0&0030 & 13 \\
71 & 58056.8071 & 0.0043 & 0&0014 & 16 \\
72 & 58056.8618 & 0.0015 & $-$0&0003 & 11 \\
89 & 58057.8210 & 0.0011 & 0&0011 & 18 \\
106 & 58058.7824 & 0.0051 & 0&0047 & 15 \\
107 & 58058.8364 & 0.0015 & 0&0023 & 19 \\
142 & 58060.8021 & 0.0045 & $-$0&0040 & 17 \\
\hline
  \multicolumn{6}{l}{\commenta BJD$-$2400000.} \\
  \multicolumn{6}{l}{\commentb Against max $= 2458052.8054 + 0.056343 E$.} \\
  \multicolumn{6}{l}{\commentc Number of points used to determine the maximum.} \\
\end{tabular}
\end{center}
\end{table}

\subsection{ASASSN-17of}\label{obj:asassn17of}

\begin{figure}
  \begin{center}
    \FigureFile(85mm,110mm){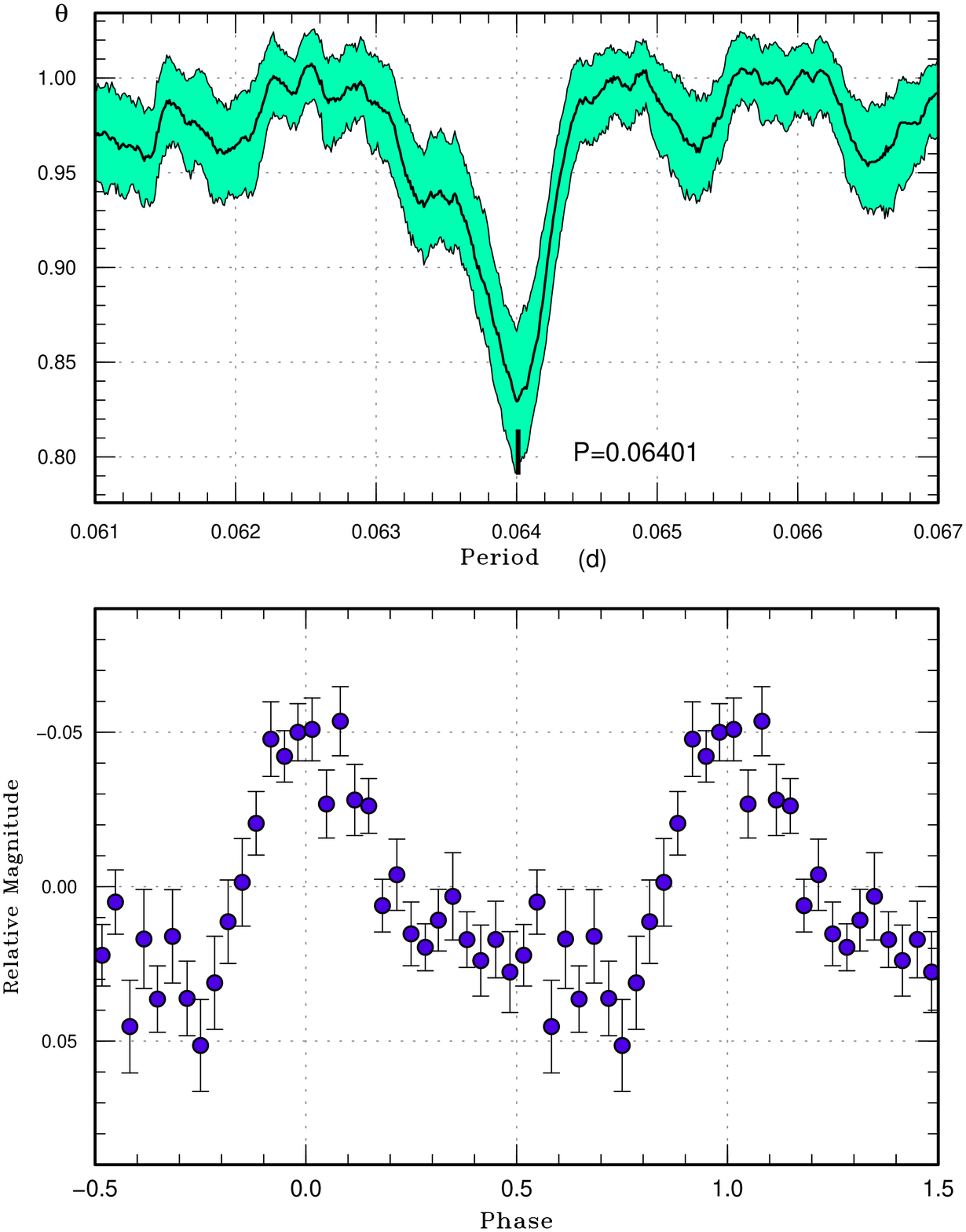}
  \end{center}
  \caption{Superhumps in ASASSN-17of (2017).
     (Upper): PDM analysis.
     (Lower): Phase-averaged profile.}
  \label{fig:asassn17ofshpdm}
\end{figure}

   This object was detected as a transient
at $g$=16.2 on 2017 November 3 by the ASAS-SN team.
The outburst was announced after observations
of $g$=17.0 on 2017 November 6 and $g$=16.8
on 2017 November 7.  Subsequent observations
detected superhumps (vsnet-alert 21567, 21580;
e-figure \ref{fig:asassn17ofshpdm}).
The times of superhump maxima are listed in
e-table \ref{tab:asassn17ofoc2017}.
Stages B and C can be recognized.

\begin{table}
\caption{Superhump maxima of ASASSN-17of (2017)}\label{tab:asassn17ofoc2017}
\begin{center}
\begin{tabular}{rp{55pt}p{40pt}r@{.}lr}
\hline
\multicolumn{1}{c}{$E$} & \multicolumn{1}{c}{max\commenta} & \multicolumn{1}{c}{error} & \multicolumn{2}{c}{$O-C$\commentb} & \multicolumn{1}{c}{$N$\commentc} \\
\hline
0 & 58065.2348 & 0.0059 & $-$0&0121 & 18 \\
1 & 58065.3140 & 0.0022 & 0&0032 & 21 \\
47 & 58068.2582 & 0.0011 & 0&0021 & 55 \\
57 & 58068.8988 & 0.0028 & 0&0024 & 53 \\
58 & 58068.9647 & 0.0017 & 0&0043 & 62 \\
59 & 58069.0280 & 0.0025 & 0&0036 & 61 \\
74 & 58069.9920 & 0.0024 & 0&0071 & 60 \\
79 & 58070.3085 & 0.0023 & 0&0035 & 48 \\
80 & 58070.3719 & 0.0019 & 0&0028 & 48 \\
94 & 58071.2616 & 0.0016 & $-$0&0038 & 46 \\
94 & 58071.2616 & 0.0016 & $-$0&0038 & 45 \\
109 & 58072.2166 & 0.0016 & $-$0&0093 & 24 \\
\hline
  \multicolumn{6}{l}{\commenta BJD$-$2400000.} \\
  \multicolumn{6}{l}{\commentb Against max $= 2458065.2468 + 0.064027 E$.} \\
  \multicolumn{6}{l}{\commentc Number of points used to determine the maximum.} \\
\end{tabular}
\end{center}
\end{table}

\subsection{ASASSN-17oo}\label{obj:asassn17oo}

\begin{figure}
  \begin{center}
    \FigureFile(85mm,110mm){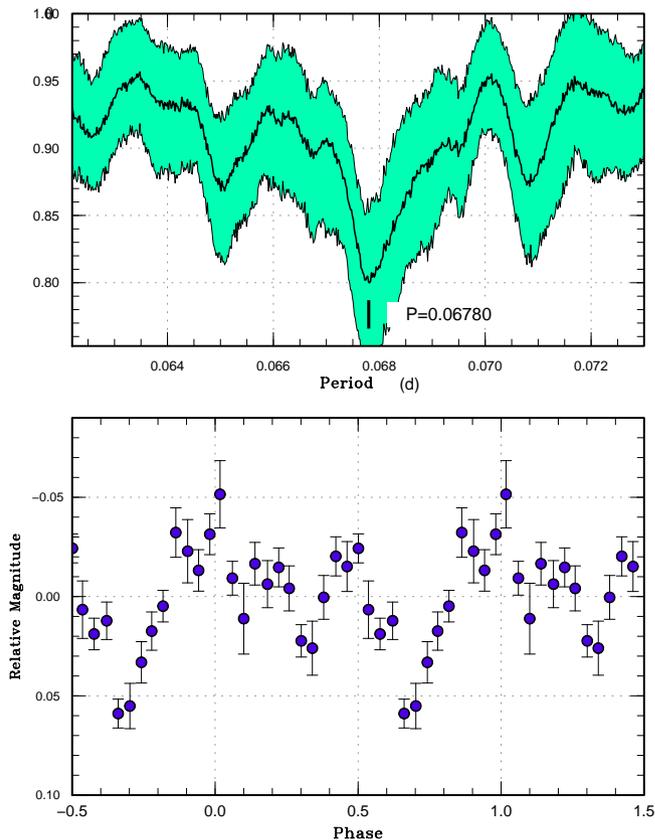}
  \end{center}
  \caption{Superhumps in ASASSN-17oo (2017).
     (Upper): PDM analysis.
     (Lower): Phase-averaged profile.}
  \label{fig:asassn17ooshpdm}
\end{figure}

   This object was detected as a transient
at $g$=15.0 on 2017 November 1 by the ASAS-SN team.
The outburst was announced after an observation
of $g$=15.05 on 2017 November 10.
Although subsequent observations detected superhumps
(vsnet-alert 21591), individual maxima were difficult
to measure.  The presence of superhumps was, however,
secure as shown in the PDM analysis
(e-figure \ref{fig:asassn17ooshpdm}).  It was likely
that we observed superhumps with diminished amplitudes
near the end of stage B or in stage C (due to the delay
in confirmation of the outburst, our observations
started 14~d after the initial outburst detection).
The superhump period was measured to be 0.06781(5)~d
by the PDM method.

\subsection{ASASSN-17ou}\label{obj:asassn17ou}

\begin{figure}
  \begin{center}
    \FigureFile(85mm,110mm){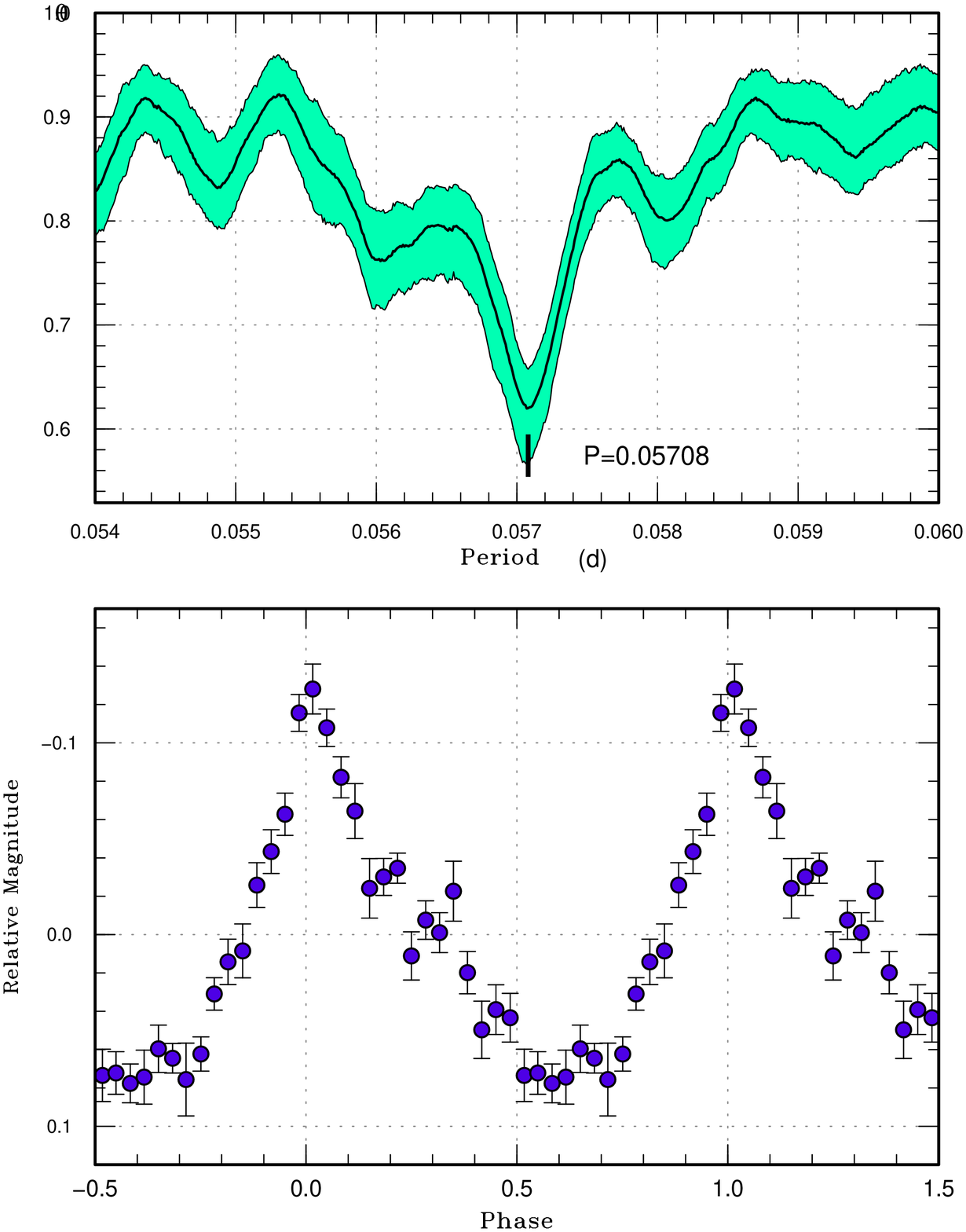}
  \end{center}
  \caption{Superhumps in ASASSN-17ou (2017).
     (Upper): PDM analysis.
     (Lower): Phase-averaged profile.}
  \label{fig:asassn17oushpdm}
\end{figure}

   This object was detected as a transient
at $g$=16.6 on 2017 November 10 by the ASAS-SN team.
The outburst was announced after observations
of $g$=16.9 on 2017 November 11, $g$=17.1 on 2017 November 12
and $g$=17.1 on 2017 November 13.  Subsequent observations
detected superhumps (vsnet-alert 21588, 21589, 21590;
e-figure \ref{fig:asassn17oushpdm}).
The times of superhump maxima are listed in
e-table \ref{tab:asassn17ouoc2017}.

\begin{table}
\caption{Superhump maxima of ASASSN-17ou (2017)}\label{tab:asassn17ouoc2017}
\begin{center}
\begin{tabular}{rp{55pt}p{40pt}r@{.}lr}
\hline
\multicolumn{1}{c}{$E$} & \multicolumn{1}{c}{max\commenta} & \multicolumn{1}{c}{error} & \multicolumn{2}{c}{$O-C$\commentb} & \multicolumn{1}{c}{$N$\commentc} \\
\hline
0 & 58072.2353 & 0.0014 & 0&0037 & 21 \\
1 & 58072.2901 & 0.0002 & 0&0014 & 262 \\
2 & 58072.3471 & 0.0003 & 0&0012 & 236 \\
12 & 58072.9174 & 0.0029 & 0&0003 & 43 \\
13 & 58072.9753 & 0.0017 & 0&0011 & 58 \\
14 & 58073.0226 & 0.0013 & $-$0&0088 & 49 \\
23 & 58073.5451 & 0.0017 & $-$0&0004 & 19 \\
51 & 58075.1502 & 0.0016 & 0&0051 & 20 \\
52 & 58075.1996 & 0.0006 & $-$0&0027 & 40 \\
53 & 58075.2584 & 0.0008 & $-$0&0010 & 55 \\
54 & 58075.3137 & 0.0008 & $-$0&0027 & 51 \\
70 & 58076.2332 & 0.0004 & 0&0027 & 49 \\
\hline
  \multicolumn{6}{l}{\commenta BJD$-$2400000.} \\
  \multicolumn{6}{l}{\commentb Against max $= 2458072.2316 + 0.057128 E$.} \\
  \multicolumn{6}{l}{\commentc Number of points used to determine the maximum.} \\
\end{tabular}
\end{center}
\end{table}

\subsection{ASASSN-17pb}\label{obj:asassn17pb}

\begin{figure}
  \begin{center}
    \FigureFile(85mm,110mm){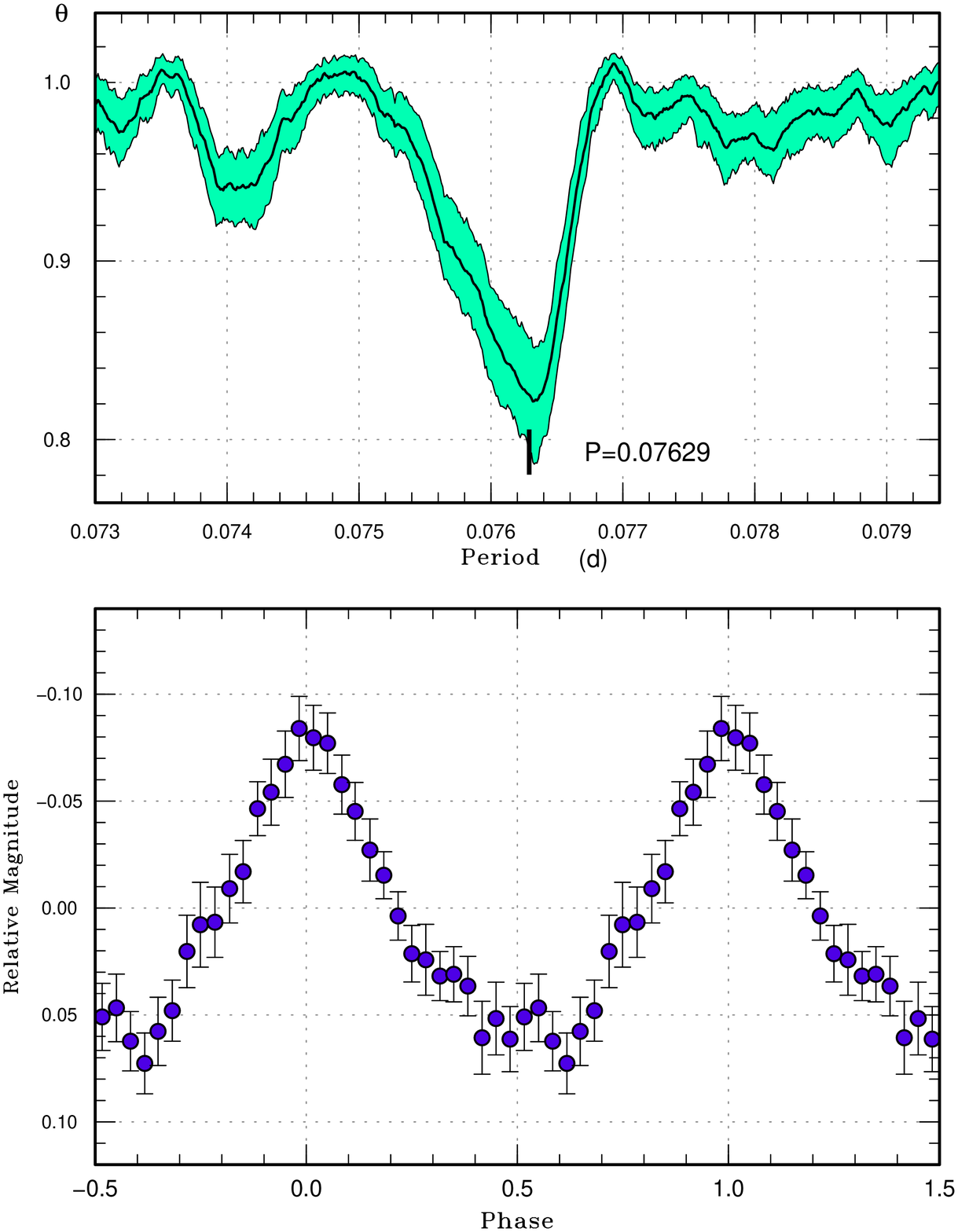}
  \end{center}
  \caption{Superhumps in ASASSN-17pb (2017).
     (Upper): PDM analysis.
     (Lower): Phase-averaged profile.}
  \label{fig:asassn17pbshpdm}
\end{figure}

\begin{figure}
  \begin{center}
    \FigureFile(85mm,110mm){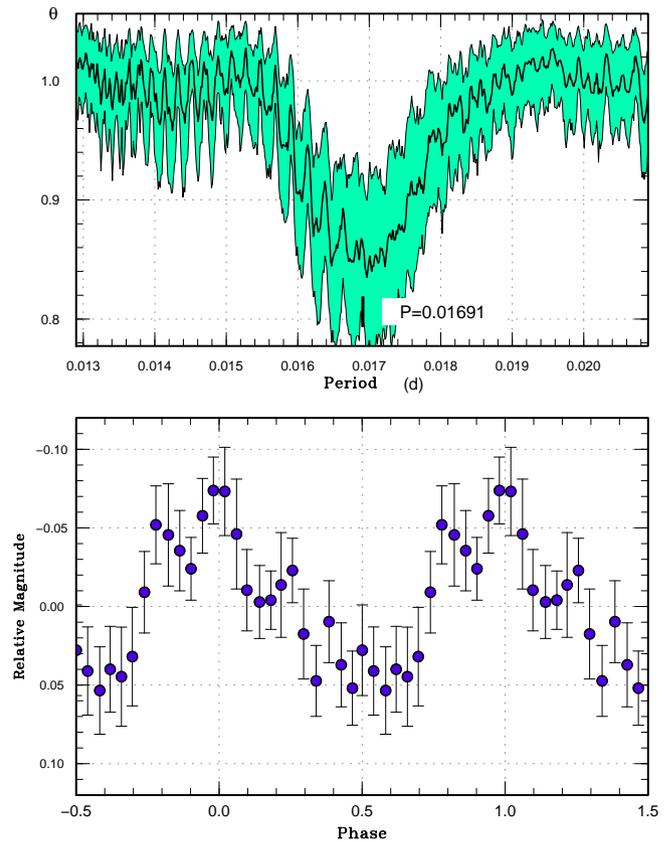}
  \end{center}
  \caption{Short-term modulations on the first two nights
     in ASASSN-17pb (2017).
     (Upper): PDM analysis.
     (Lower): Phase-averaged profile.}
  \label{fig:asassn17earlyshpdm}
\end{figure}

   This object was detected as a transient
at $V$=15.8 on 2017 November 13 by the ASAS-SN team.
The outburst was announced after an observation
of $V$=16.1 on 2017 November 14.
Subsequent observations detected superhumps
(vsnet-alert 21599, 21605, 21642;
e-figure \ref{fig:asassn17pbshpdm}).
The times of superhump maxima are listed in
e-table \ref{tab:asassn17pboc2017}.
The epoch for $E$=0 was a stage A superhump.
The epochs for $E$=23 and 24 may be during the transition
to stage B.  The epoch for $E$=128 probably corresponds to
a stage C superhump.
On the first two nights (2017 November 15 and 16),
the object did not show superhump-like modulations.
There was a suggestion of low-amplitude modulations
with a period of 0.0169(2)~d (e-figure \ref{fig:asassn17earlyshpdm}).
This period is not related to the superhump one and
the origin of this variation is unclear.

\begin{table}
\caption{Superhump maxima of ASASSN-17pb (2017)}\label{tab:asassn17pboc2017}
\begin{center}
\begin{tabular}{rp{55pt}p{40pt}r@{.}lr}
\hline
\multicolumn{1}{c}{$E$} & \multicolumn{1}{c}{max\commenta} & \multicolumn{1}{c}{error} & \multicolumn{2}{c}{$O-C$\commentb} & \multicolumn{1}{c}{$N$\commentc} \\
\hline
0 & 58075.5735 & 0.0009 & $-$0&0129 & 56 \\
23 & 58077.3332 & 0.0005 & $-$0&0032 & 68 \\
24 & 58077.4116 & 0.0007 & $-$0&0009 & 53 \\
47 & 58079.1640 & 0.0034 & 0&0014 & 11 \\
48 & 58079.2417 & 0.0012 & 0&0030 & 19 \\
61 & 58080.2283 & 0.0017 & 0&0005 & 47 \\
62 & 58080.3070 & 0.0010 & 0&0031 & 70 \\
63 & 58080.3847 & 0.0007 & 0&0047 & 102 \\
64 & 58080.4646 & 0.0023 & 0&0086 & 13 \\
74 & 58081.2182 & 0.0022 & 0&0012 & 32 \\
75 & 58081.2894 & 0.0025 & $-$0&0037 & 46 \\
84 & 58081.9851 & 0.0020 & 0&0072 & 15 \\
85 & 58082.0581 & 0.0190 & 0&0042 & 7 \\
88 & 58082.2825 & 0.0021 & 0&0003 & 28 \\
89 & 58082.3616 & 0.0007 & 0&0034 & 22 \\
97 & 58082.9731 & 0.0019 & 0&0062 & 12 \\
100 & 58083.1950 & 0.0038 & $-$0&0003 & 33 \\
101 & 58083.2748 & 0.0016 & 0&0035 & 58 \\
102 & 58083.3417 & 0.0017 & $-$0&0057 & 28 \\
128 & 58085.3051 & 0.0021 & $-$0&0206 & 20 \\
\hline
  \multicolumn{6}{l}{\commenta BJD$-$2400000.} \\
  \multicolumn{6}{l}{\commentb Against max $= 2458075.5865 + 0.076088 E$.} \\
  \multicolumn{6}{l}{\commentc Number of points used to determine the maximum.} \\
\end{tabular}
\end{center}
\end{table}

\subsection{CRTS J044027.1$+$023301}\label{obj:j0440}

   This object (=CSS090219:044027$+$023301, hereafter
CRTS J044027) was detected as a transient by CRTS
on 2009 February 19.  The object has an X-ray
counterpart of 1RXS J044027.0$+$023300.

\begin{figure}
  \begin{center}
    \FigureFile(85mm,110mm){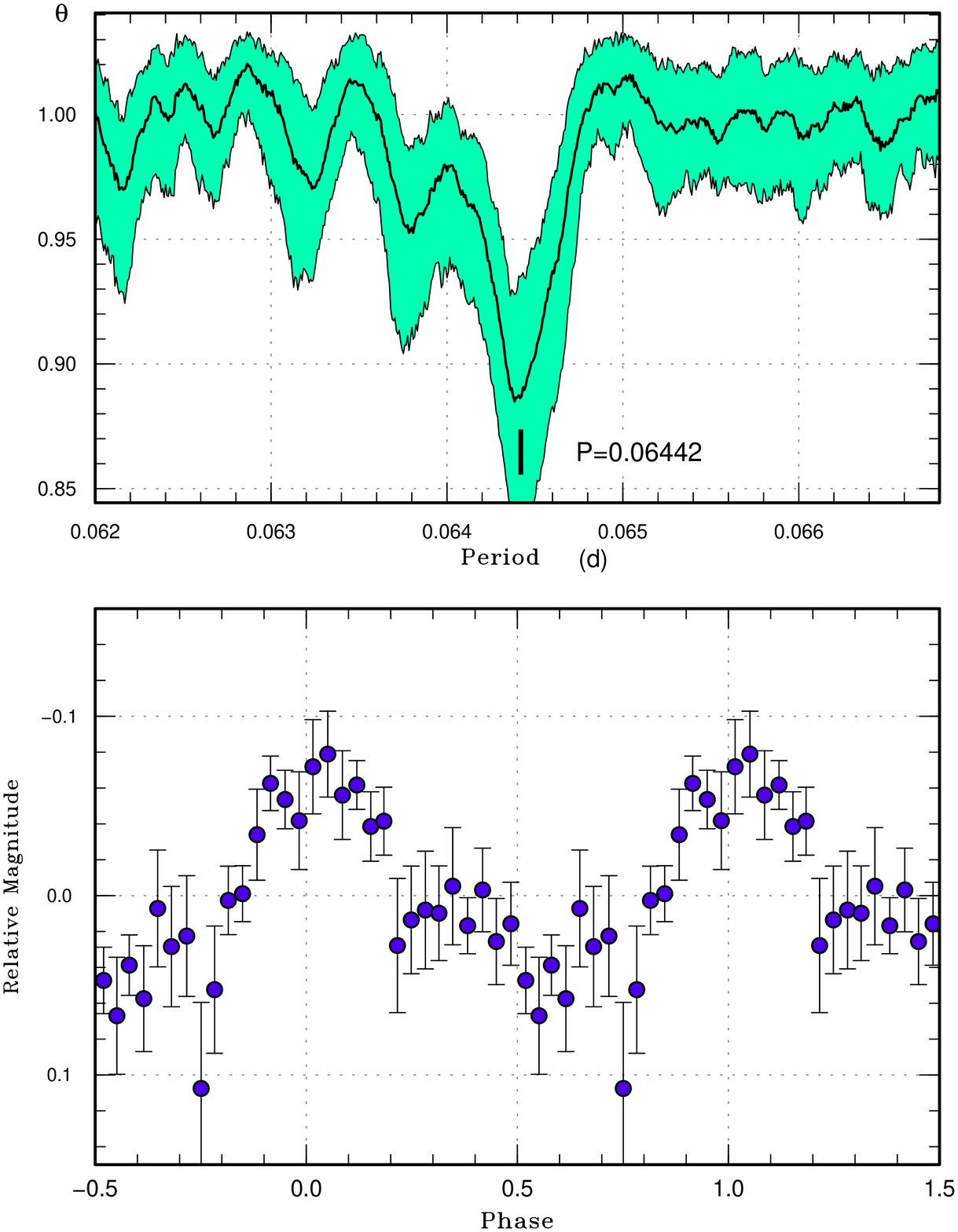}
  \end{center}
  \caption{Superhumps in CRTS J044027 (2017).
     (Upper): PDM analysis.
     (Lower): Phase-averaged profile.}
  \label{fig:j0440shpdm}
\end{figure}

   The 2017 outburst was detected by the ASAS-SN team
at $V$=14.56 on 2017 August 30.
This outburst was originally suspected to be a normal
one due to its large fading rate (vsnet-alert 21388).
Subsequent observations, however, detected superhumps
(vsnet-alert 21397, 21411; figure \ref{fig:j0440shpdm}).
The detection on 2017 August 30 was probably of
a precursor outburst.
The times of superhump maxima are listed in
e-table \ref{tab:j0440oc2017}.  Although $E$=1 appears
to be a stage A superhump, we could not determine the period.
Later observations probably recorded stage B superhumps
since the object faded soon after these observations.

   There was another superoutburst in the ASAS-SN data
peaking on 2016 February 27 at $V$=14.56.
Another detection on 2012 September 22 at $V$=15.37
probably corresponds to a normal outburst.
CRTS detected additional two faint outbursts in addition
to the 2009 one (all of them were apparently normal
outbursts).

\begin{table}
\caption{Superhump maxima of CRTS J044027 (2017)}\label{tab:j0440oc2017}
\begin{center}
\begin{tabular}{rp{55pt}p{40pt}r@{.}lr}
\hline
\multicolumn{1}{c}{$E$} & \multicolumn{1}{c}{max\commenta} & \multicolumn{1}{c}{error} & \multicolumn{2}{c}{$O-C$\commentb} & \multicolumn{1}{c}{$N$\commentc} \\
\hline
0 & 57999.6124 & 0.0005 & $-$0&0031 & 47 \\
49 & 58002.7769 & 0.0012 & 0&0035 & 23 \\
50 & 58002.8377 & 0.0016 & $-$0&0002 & 22 \\
64 & 58003.7440 & 0.0024 & 0&0039 & 22 \\
65 & 58003.8044 & 0.0014 & $-$0&0002 & 22 \\
66 & 58003.8702 & 0.0020 & 0&0011 & 22 \\
80 & 58004.7722 & 0.0024 & 0&0008 & 34 \\
81 & 58004.8343 & 0.0027 & $-$0&0015 & 31 \\
96 & 58005.8013 & 0.0022 & $-$0&0012 & 23 \\
97 & 58005.8641 & 0.0021 & $-$0&0029 & 31 \\
\hline
  \multicolumn{6}{l}{\commenta BJD$-$2400000.} \\
  \multicolumn{6}{l}{\commentb Against max $= 2457999.6154 + 0.064449 E$.} \\
  \multicolumn{6}{l}{\commentc Number of points used to determine the maximum.} \\
\end{tabular}
\end{center}
\end{table}

\subsection{CRTS J080941.3$+$171528}\label{obj:j0809}

   This object (=CSS120120:080941$+$171528, hereafter
CRTS J080941) was detected as a transient by CRTS
on 2012 January 20.

\begin{figure}
  \begin{center}
    \FigureFile(85mm,110mm){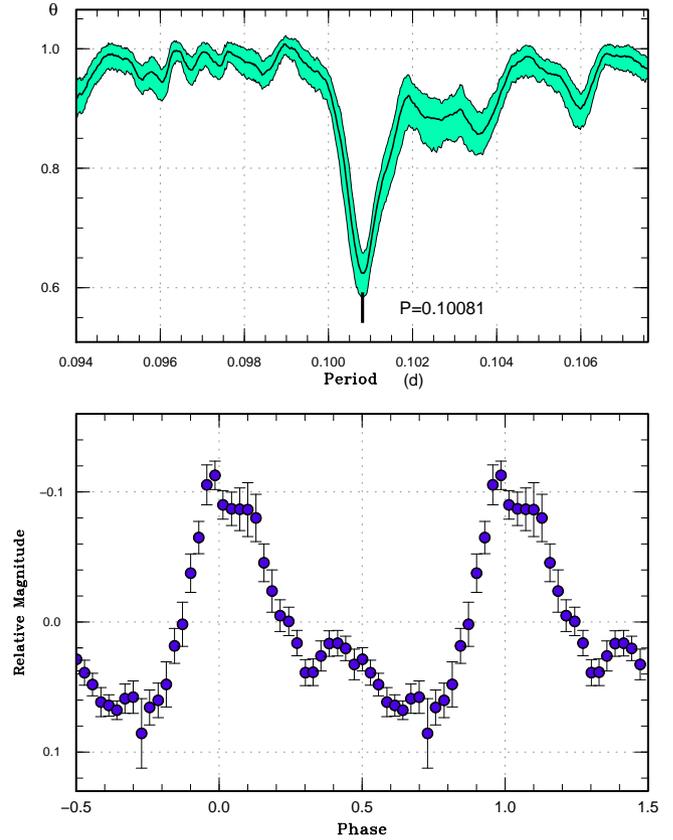}
  \end{center}
  \caption{Superhumps in CRTS J080941 (2017).
     (Upper): PDM analysis.
     (Lower): Phase-averaged profile.}
  \label{fig:j0809shpdm}
\end{figure}

   The 2017 outburst was detected by the ASAS-SN team
at $V$=15.3 on 2017 April 8 and announced after observation
of $V$=15.4 on 2017 April 9.  The large outburst amplitude
and past behavior suggested a superoutburst (vsnet-alert 20888).
Subsequent observations detected long-period superhumps
(vsnet-alert 20893, 20906, 20912; e-figure \ref{fig:j0809shpdm}).
The times of superhump maxima are listed in
e-table \ref{tab:j0809oc2017}.
The period for $E \le$12 was substantially longer.
The difference of the periods before and after $E$=12
was 1.2\%, which is too large to be considered as
stage B-C transition.  We rather consider that this
reflects stage A-B transition, as have been recently recorded
in many SU UMa-type dwarf novae with long superhump periods
(V1006 Cyg and MN Dra: \cite{kat16v1006cyg}; 
CRTS J214738.4$+$244554 and OT J064833.4$+$065624:
\cite{Pdot7}, KK Tel, possibly V452 Cas and ASASSN-15cl: \cite{Pdot8},
MASTER OT J021315.37$+$533822.7: \cite{Pdot9},
OT J002656.6$+$284933 = CSS101212:002657$+$284933:
\cite{kat17j0026}).

   CRTS J080941 is not only an SU UMa-type dwarf nova
in the middle of the period gap but also shows
superhump evolution common to many long-period
SU UMa-type dwarf novae.  Since (supposed) stage A
superhumps were detected, determination of the orbital
period is highly desired to determine the mass ratio
in such a system in the middle of the period gap.

   CRTS recorded two past outbursts: 2007 May 11 (15.8 mag)
and 2012 January 20 (16.3 mag).  No past outbursts were
detected in the ASAS-SN data starting from 2012 January.

\begin{table}
\caption{Superhump maxima of CRTS J080941 (2017)}\label{tab:j0809oc2017}
\begin{center}
\begin{tabular}{rp{55pt}p{40pt}r@{.}lr}
\hline
\multicolumn{1}{c}{$E$} & \multicolumn{1}{c}{max\commenta} & \multicolumn{1}{c}{error} & \multicolumn{2}{c}{$O-C$\commentb} & \multicolumn{1}{c}{$N$\commentc} \\
\hline
0 & 57853.3307 & 0.0008 & $-$0&0056 & 84 \\
1 & 57853.4269 & 0.0004 & $-$0&0101 & 104 \\
10 & 57854.3463 & 0.0005 & 0&0030 & 97 \\
11 & 57854.4452 & 0.0009 & 0&0011 & 63 \\
12 & 57854.5495 & 0.0009 & 0&0047 & 35 \\
20 & 57855.3607 & 0.0048 & 0&0103 & 18 \\
22 & 57855.5545 & 0.0011 & 0&0027 & 33 \\
32 & 57856.5585 & 0.0016 & $-$0&0004 & 27 \\
42 & 57857.5622 & 0.0014 & $-$0&0037 & 35 \\
49 & 57858.2770 & 0.0051 & 0&0060 & 54 \\
52 & 57858.5710 & 0.0016 & $-$0&0021 & 36 \\
62 & 57859.5741 & 0.0040 & $-$0&0060 & 34 \\
\hline
  \multicolumn{6}{l}{\commenta BJD$-$2400000.} \\
  \multicolumn{6}{l}{\commentb Against max $= 2457853.3363 + 0.100707 E$.} \\
  \multicolumn{6}{l}{\commentc Number of points used to determine the maximum.} \\
\end{tabular}
\end{center}
\end{table}

\subsection{CRTS J120052.9$-$152620}\label{obj:j1200}

   This object (=CSS110205:120053$-$152620, hereafter
CRTS J120052) was discovered by the CRTS on 2011 February 5. 
The 2011 and 2016 superoutbursts were reported in
\citet{Pdot3} and \citet{Pdot8}, respectively.
The 2017 superoutburst was detected by the ASAS-SN team
at $V$=13.8 on 2017 April 24.  Single-night observations
were performed and obtained two superhump maxima:
BJD 2457867.0363(3) ($N$=159) and 2457867.1236(4) ($N$=167).

   This object showed many outbursts.  Among them, we listed
likely/possible superoutbursts in e-table \ref{tab:j1200out}
(in the case of ASAS-3 and ASAS-SN, the identifications
were based on durations; in the case of CRTS, we selected
possible ones based on brightness since they were
single-night detections).
The maxima after 2006 appear to be well expressed by
a supercycle of 200.7(9)~d.  The period was stable
at 192(2)~d between 2006 and 2009 and apparently
lengthened to 203(1)~d after that.  The cycle count
between 2001 and 2006 was unclear and we couldn't
determine the supercycle uniquely.  This object appears
to be a relatively ordinary SU UMa-type dwarf nova
with frequent outbursts.

\begin{table*}
\caption{List of possible superoutbursts of CRTS J120052}\label{tab:j1200out}
\begin{center}
\begin{tabular}{cccccc}
\hline
Year & Month & Day & max\commenta & mag\commentb & Source \\
\hline
2001 &  2 & 14 & 51954 & 13.4V & ASAS-3 \\
2006 &  5 & 21 & 53876 & 14.0V & ASAS-3 \\
2007 &  6 &  2 & 54254 & 14.1V & ASAS-3 \\
2007 & 12 & 22 & 54457 & 13.9V & ASAS-3 \\
2008 &  6 &  7 & 54625 & 14.0C & CRTS, ASAS-3 \\
2009 &  1 &  8 & 54840 & 13.7V & ASAS-3 \\
2009 &  7 & 18 & 55031 & 13.9C & CRTS \\
2010 &  1 & 18 & 55215 & 13.8C & CRTS \\
2015 &  2 &  1 & 57055 & 13.6V & ASAS-SN \\
2016 &  3 & 14 & 57462 & 13.8V & ASAS-SN \\
2017 &  4 & 24 & 57867 & 13.7V & ASAS-SN \\
\hline
  \multicolumn{5}{l}{\commenta JD$-$2400000.} \\
  \multicolumn{5}{l}{\commentb C means unfiltered CCD.} \\
\end{tabular}
\end{center}
\end{table*}

\subsection{CRTS J122221.6$-$311524}\label{obj:j1222}

   Although we have already reported the 2013 superoutburst
and claimed the object to be a best candidate for
a period bouncer \citep{kat13j1222}, we treat this object
again since \citet{neu17j1222} reported a spectroscopic
orbital period of 109.80(7)~min [0.07625(5)~d].
Since we have a period of stage A superhumps
[0.07721(1)~d], we can now directly estimate the mass ratio.
The value of $\epsilon^*$ for stage A superhumps is
0.0124(6), which corresponds to $q$=0.032(2).  This value
supersedes our previous constraint \citep{kat13j1222}
and the $q$ value based on stage B superhumps with a large
uncertainty \citep{neu17j1222}.  The value is sufficiently
low to give strong credence to the period-bouncer status.

\subsection{CRTS J162806.2$+$065316}\label{obj:j1628}

   The object was detected as a transient (=CSS110611:162806$+$065316;
hereafter CRTS J162806) by CRTS on 2011 June 11.
The 2011 superoutburst was studied in \citet{Pdot4}.
The 2017 superoutburst was detected by the ASAS-SN team
at $V$=14.64 on 2017 March 11.
Only one superhump was recorded: BJD 2457825.5805(9) ($N$=76).

   Although this field has been monitored by the ASAS-SN
team since 2012 April 1, no other secure outburst was
recorded.  The 2013 October outburst fell in the gap
of the ASAS-SN observations.  The three outbursts detected
by CRTS since 2007 were likely superoutbursts
(e-table \ref{tab:j1628out}).  The intervals of known
(likely) superoutbursts were 1522~d, 844~d and 1256~d.
Considering the sparse past observations of CRTS,
the frequency of outbursts in this system probably
is not very low.

\begin{table}
\caption{List of past outbursts of CRTS J162806}\label{tab:j1628out}
\begin{center}
\begin{tabular}{cccccc}
\hline
Year & Month & Day & max\commenta & mag\commentb & Source \\
\hline
2005 &  9 & 14 & 53628 & 15.3C & CRTS \\
2006 &  6 & 20 & 53907 & 16.0C & CRTS \\
2007 &  4 & 10 & 54201 & 14.1C & CRTS \\
2011 &  6 & 10 & 55723 & 14.2C & CRTS \\
2013 & 10 &  1 & 56567 & 14.1C & CRTS \\
2017 &  3 & 11 & 57823 & 14.6V & ASAS-SN \\
\hline
  \multicolumn{5}{l}{\commenta JD$-$2400000.} \\
  \multicolumn{5}{l}{\commentb C means unfiltered CCD.} \\
\end{tabular}
\end{center}
\end{table}

\subsection{CRTS J214934.1$-$121908}\label{obj:j2149}

   The object was detected as a transient (=CSS120922:214934$-$121908;
hereafter CRTS J214934) by CRTS on 2012 September 22.
The SU UMa-type nature was confirmed by single-night observations
in \citet{Pdot5}.

\begin{figure}
  \begin{center}
    \FigureFile(85mm,110mm){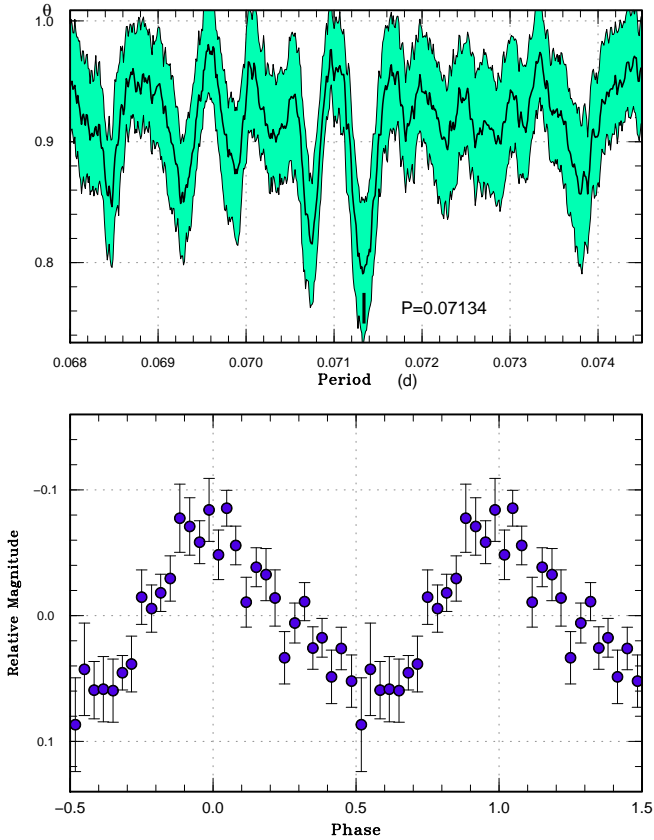}
  \end{center}
  \caption{Superhumps in CRTS J214934 (2017).
     (Upper): PDM analysis.
     (Lower): Phase-averaged profile.}
  \label{fig:j2149shpdm}
\end{figure}

   The 2017 superoutburst was detected by the ASAS-SN team
at $V$=15.69 on 2017 October 8.  Subsequent observations
detected superhumps (vsnet-alert 21511; e-figure \ref{fig:j2149shpdm}).
The times of superhump maxima are listed in
e-table \ref{tab:j2149oc2017}.  Although observations were
insufficient, the data suggest stage B-C transition.
The period of stage B superhumps given in the table
should refer to a lower limit since stage B likely
ended before $E$=64.

   We listed superoutbursts in the ASAS-SN data in
e-table \ref{tab:j2149out}.  The interval between the 2012
superoutburst and the 2013 one was 284~d.  The interval
between the 2013 and 2014 superoutbursts was 332~d.
Assuming that there were four supercycles between
2014 and 2017, the supercycle was estimated to be 308(3)~d.

\begin{table}
\caption{Superhump maxima of CRTS J214934 (2017)}\label{tab:j2149oc2017}
\begin{center}
\begin{tabular}{rp{55pt}p{40pt}r@{.}lr}
\hline
\multicolumn{1}{c}{$E$} & \multicolumn{1}{c}{max\commenta} & \multicolumn{1}{c}{error} & \multicolumn{2}{c}{$O-C$\commentb} & \multicolumn{1}{c}{$N$\commentc} \\
\hline
0 & 58036.9845 & 0.0019 & $-$0&0020 & 63 \\
1 & 58037.0564 & 0.0015 & $-$0&0015 & 69 \\
64 & 58041.5593 & 0.0035 & 0&0040 & 16 \\
65 & 58041.6313 & 0.0030 & 0&0045 & 27 \\
92 & 58043.5531 & 0.0033 & $-$0&0011 & 16 \\
93 & 58043.6280 & 0.0019 & 0&0024 & 28 \\
106 & 58044.5494 & 0.0033 & $-$0&0043 & 15 \\
107 & 58044.6231 & 0.0095 & $-$0&0019 & 20 \\
\hline
  \multicolumn{6}{l}{\commenta BJD$-$2400000.} \\
  \multicolumn{6}{l}{\commentb Against max $= 2458036.9865 + 0.071388 E$.} \\
  \multicolumn{6}{l}{\commentc Number of points used to determine the maximum.} \\
\end{tabular}
\end{center}
\end{table}

\begin{table}
\caption{List of superoutbursts of CRTS J214934 in the ASAS-SN data}\label{tab:j2149out}
\begin{center}
\begin{tabular}{ccccc}
\hline
Year & Month & Day & max\commenta & $V$ mag \\
\hline
2013 &  7 &  3 & 56477 & 15.7 \\
2014 &  5 & 31 & 56809 & 15.4 \\
2017 & 10 &  7 & 58034 & 15.6 \\
\hline
  \multicolumn{5}{l}{\commenta JD$-$2400000.} \\
\end{tabular}
\end{center}
\end{table}

\subsection{CRTS J223235.4$+$304105}\label{obj:j2232}

\begin{figure}
  \begin{center}
    \FigureFile(85mm,110mm){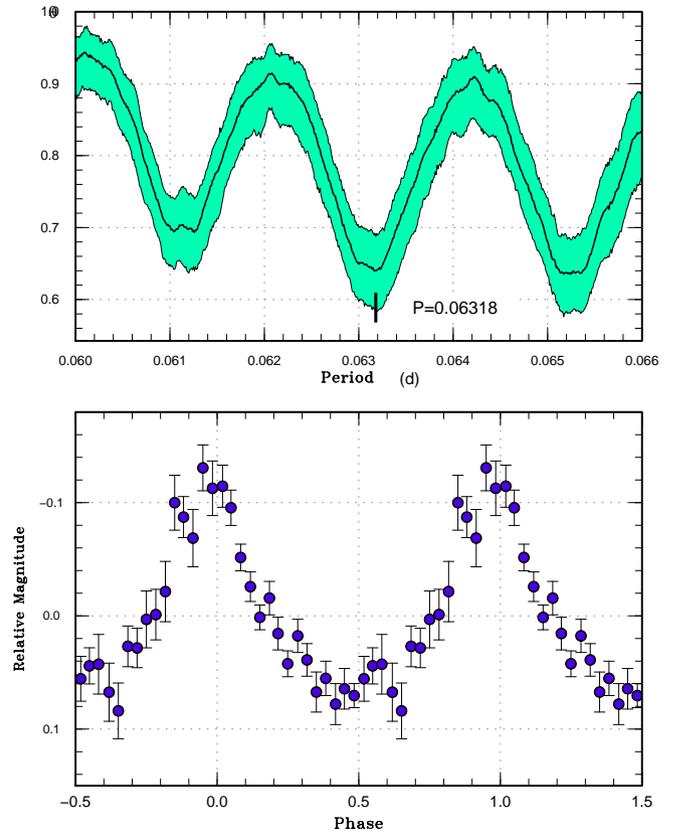}
  \end{center}
  \caption{Superhumps in CRTS J223235 (2017).
     (Upper): PDM analysis.  The alias was selected by
     $O-C$ analysis.
     (Lower): Phase-averaged profile.}
  \label{fig:j2232shpdm}
\end{figure}

   The object was detected as a transient (=CSS081107:223235$+$304105;
hereafter CRTS J223235) by CRTS on 2008 November 7
\citep{dra14CRTSCVs}.  The 2017 outburst was detected
by the ASAS-SN team (the name ASASSN-17nw was assigned)
at $V$=17.7 on 2017 October 18 and was announced after
the object brightened to $V$=15.9 on 2017 October 19.
Subsequent observations detected superhumps
(vsnet-alert 21551, 21553; e-figure \ref{fig:j2232shpdm}).
The times of superhump maxima are listed in
e-table \ref{tab:j2232oc2017}.

\begin{table}
\caption{Superhump maxima of CRTS J223235 (2017)}\label{tab:j2232oc2017}
\begin{center}
\begin{tabular}{rp{55pt}p{40pt}r@{.}lr}
\hline
\multicolumn{1}{c}{$E$} & \multicolumn{1}{c}{max\commenta} & \multicolumn{1}{c}{error} & \multicolumn{2}{c}{$O-C$\commentb} & \multicolumn{1}{c}{$N$\commentc} \\
\hline
0 & 58054.5328 & 0.0017 & 0&0004 & 47 \\
28 & 58056.2899 & 0.0011 & $-$0&0063 & 44 \\
29 & 58056.3626 & 0.0017 & 0&0034 & 72 \\
30 & 58056.4225 & 0.0017 & 0&0003 & 73 \\
31 & 58056.4874 & 0.0011 & 0&0022 & 107 \\
32 & 58056.5483 & 0.0015 & 0&0001 & 60 \\
\hline
  \multicolumn{6}{l}{\commenta BJD$-$2400000.} \\
  \multicolumn{6}{l}{\commentb Against max $= 2458054.5324 + 0.062994 E$.} \\
  \multicolumn{6}{l}{\commentc Number of points used to determine the maximum.} \\
\end{tabular}
\end{center}
\end{table}

\subsection{CTCV J1940$-$4724}\label{obj:j1940}

   This object (hereafter CTCV J1940) was detected
as a CV by Cal\'an-Tololo Survey \citep{aug10CTCVCV2}.
\citet{aug10CTCVCV2} obtained an orbital period of
0.0809(30)~d using a set of 18 spectroscopic observations
taken over a baseline of 0.11~d.  Since the period suggested
an SU UMa-type dwarf nova, a systematic search for
outbursts was conducted (cf. vsnet-alert 21188).
Although the outburst detected on 2017 July 1 turned out
to be a normal outburst, a precursor outburst (2017 August
19--20) and an apparent superoutburst starting on
2017 August 27 were visually detected by
R. Stubbings (vsnet-alert 21370).
Subsequent observations detected superhumps (vsnet-alert 21378).
The resultant data, however, were not ideally spaced
and there remained an ambiguity in choosing the alias.
Although the PDM analysis favored a period of 0.07124(8)~d,
an $O-C$ analysis gave a systematic trend and preferred
0.07667(7)~d.  We adopted the latter, which is also marginally
consistent with the approximate orbital period
by \citet{aug10CTCVCV2} in estimating cycle counts
in e-table \ref{tab:j1940oc2017}.

   We listed superoutbursts in the ASAS-SN data in
e-table \ref{tab:j1940out}.  The data suggest a supercycle
of 168(6)~d.  This supercycle is also consistent with
the ASAS-3 data, which detected several long outbursts
though they were not as well sampled as in the ASAS-SN data.

\begin{table}
\caption{List of superoutbursts of CTCV J1940 in the ASAS-SN data}\label{tab:j1940out}
\begin{center}
\begin{tabular}{ccccc}
\hline
Year & Month & Day & max\commenta & $V$ mag \\
\hline
2015 & 11 &  3 & 57330 & 13.0 \\
2016 &  5 & 28 & 57537 & 13.9 \\
2016 & 10 & 25 & 57687 & 13.3 \\
2017 &  4 &  7 & 57851 & 13.5 \\
2017 &  8 & 29 & 57995 & 13.4\commentb \\
\hline
  \multicolumn{5}{l}{\commenta JD$-$2400000.} \\
  \multicolumn{5}{l}{\commentb Visually detected 2~d earlier.} \\
\end{tabular}
\end{center}
\end{table}

\begin{table}
\caption{Superhump maxima of CTCV J1940 (2017)}\label{tab:j1940oc2017}
\begin{center}
\begin{tabular}{rp{55pt}p{40pt}r@{.}lr}
\hline
\multicolumn{1}{c}{$E$} & \multicolumn{1}{c}{max\commenta} & \multicolumn{1}{c}{error} & \multicolumn{2}{c}{$O-C$\commentb} & \multicolumn{1}{c}{$N$\commentc} \\
\hline
0 & 57994.6257 & 0.0016 & $-$0&0029 & 19 \\
13 & 57995.6269 & 0.0017 & 0&0016 & 18 \\
26 & 57996.6239 & 0.0016 & 0&0019 & 24 \\
39 & 57997.6202 & 0.0020 & 0&0016 & 25 \\
52 & 57998.6136 & 0.0019 & $-$0&0017 & 27 \\
65 & 57999.6099 & 0.0012 & $-$0&0021 & 18 \\
66 & 57999.6912 & 0.0018 & 0&0025 & 20 \\
78 & 58000.6071 & 0.0026 & $-$0&0016 & 25 \\
79 & 58000.6860 & 0.0029 & 0&0006 & 19 \\
\hline
  \multicolumn{6}{l}{\commenta BJD$-$2400000.} \\
  \multicolumn{6}{l}{\commentb Against max $= 2457994.6286 + 0.076668 E$.} \\
  \multicolumn{6}{l}{\commentc Number of points used to determine the maximum.} \\
\end{tabular}
\end{center}
\end{table}

\subsection{DDE 51}\label{obj:dde51}

\begin{figure}
  \begin{center}
    \FigureFile(85mm,110mm){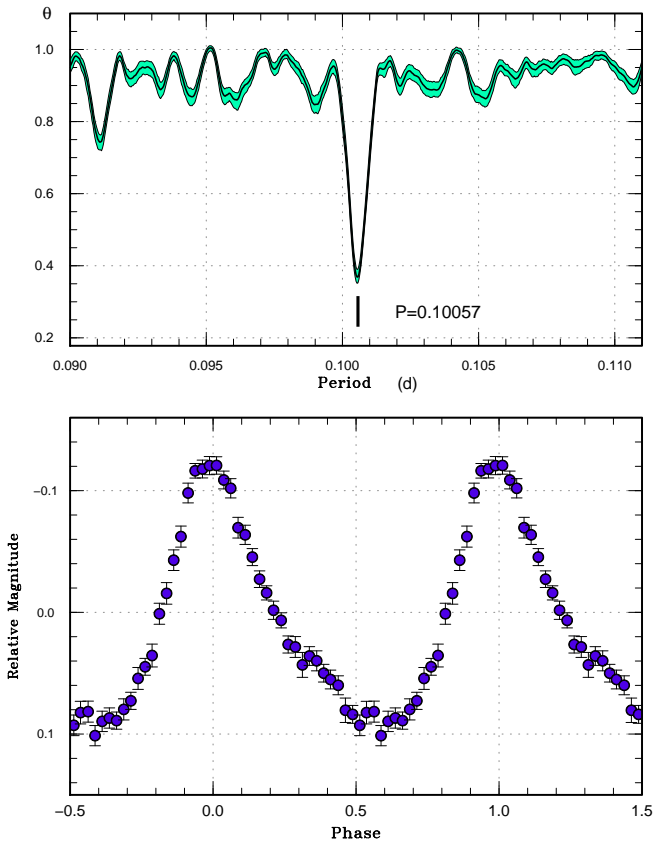}
  \end{center}
  \caption{Superhumps in DDE 51 (2017).
     (Upper): PDM analysis.
     (Lower): Phase-averaged profile.}
  \label{fig:dde51shpdm}
\end{figure}

   DDE 51 was discovered as a dwarf nova by D. Denisenko.\footnote{
$<$http://scan.sai.msu.ru/$^\sim$denis/VarDDE.html$>$.}
Denisenko monitored this object and found an outburst
at an unfiltered CCD magnitude of 16.52 on 2017 September 28
(vsnet-alert 21475).  The object further brightened
and reached an unfiltered CCD magnitude of 15.58 on
2017 September 30.  The initial superhump detection was
made by Oleg Milantiev (vsnet-alert 21489).
Further observations clarified that this object is
an SU UMa-type in the period gap by detecting long-period
superhumps (vsnet-alert 21490, 21491, 21493;
e-figure \ref{fig:dde51shpdm}).
The times of superhump maxima are listed in
e-table \ref{tab:dde51oc2017}.
It is apparent that there was a stage transition between
$E$=12 and $E$=30.  Considering the similarity with
the SU UMa-type dwarf novae in the period gap V1006 Cyg and
MN Dra \citep{kat16v1006cyg}, we consider that superhumps
for $E \le$12 were stage A superhumps despite large
superhump amplitudes (see also a discussion in
subsection \ref{obj:j0809}).
Determination of the orbital period
will be a crucial test for this interpretation.

   Among outbursts recorded by the ASAS-SN team
(e-table \ref{tab:dde51out}), only the 2017 September--October
one appears to be a superoutburst.  Observations are not
yet sufficient to determine the supercycle.

\begin{table}
\caption{Superhump maxima of DDE 51 (2017)}\label{tab:dde51oc2017}
\begin{center}
\begin{tabular}{rp{55pt}p{40pt}r@{.}lr}
\hline
\multicolumn{1}{c}{$E$} & \multicolumn{1}{c}{max\commenta} & \multicolumn{1}{c}{error} & \multicolumn{2}{c}{$O-C$\commentb} & \multicolumn{1}{c}{$N$\commentc} \\
\hline
0 & 58028.4266 & 0.0005 & $-$0&0099 & 112 \\
1 & 58028.5285 & 0.0021 & $-$0&0085 & 50 \\
9 & 58029.3409 & 0.0008 & $-$0&0002 & 81 \\
10 & 58029.4476 & 0.0014 & 0&0059 & 46 \\
11 & 58029.5420 & 0.0005 & $-$0&0001 & 100 \\
12 & 58029.6449 & 0.0009 & 0&0022 & 45 \\
30 & 58031.4572 & 0.0007 & 0&0052 & 87 \\
49 & 58033.3685 & 0.0013 & 0&0066 & 106 \\
50 & 58033.4669 & 0.0010 & 0&0044 & 93 \\
57 & 58034.1696 & 0.0005 & 0&0035 & 180 \\
70 & 58035.4750 & 0.0004 & 0&0022 & 92 \\
71 & 58035.5733 & 0.0005 & $-$0&0001 & 96 \\
76 & 58036.0759 & 0.0005 & $-$0&0001 & 148 \\
77 & 58036.1743 & 0.0007 & $-$0&0022 & 164 \\
108 & 58039.2839 & 0.0005 & $-$0&0087 & 85 \\
\hline
  \multicolumn{6}{l}{\commenta BJD$-$2400000.} \\
  \multicolumn{6}{l}{\commentb Against max $= 2458028.4365 + 0.100520 E$.} \\
  \multicolumn{6}{l}{\commentc Number of points used to determine the maximum.} \\
\end{tabular}
\end{center}
\end{table}

\begin{table}
\caption{List of outbursts of DDE 51 in the ASAS-SN data}\label{tab:dde51out}
\begin{center}
\begin{tabular}{ccccc}
\hline
Year & Month & Day & max\commenta & $V$ mag \\
\hline
2016 &  5 & 12 & 57521 & 16.5 \\
2016 & 10 & 26 & 57688 & 16.0 \\
2017 &  6 & 16 & 57921 & 15.8 \\
2017 &  9 & 29 & 58026 & 15.2 \\
\hline
  \multicolumn{5}{l}{\commenta JD$-$2400000.} \\
\end{tabular}
\end{center}
\end{table}

\subsection{MASTER OT J132501.00$+$431846.1}\label{obj:j1325}

   Although this object has not been identified as an
SU UMa-type dwarf nova, we include it due to its
special characteristics.  This object (hereafter MASTER J132501)
was detected as a transient at an unfiltered CCD magnitude
of 15.4 on 2017 June 5 \citep{bal17j1325atel10470}.
The object was initially proposed to be a large-amplitude
WZ Sge-type dwarf nova.  There were, however, no early
superhumps (vsnet-alert 21107).  Based on this negative
detection and further brightening, \citet{den17j1325atel10480}
suggested this object to be a likely supernova in NGC 5145.
Although we had observations on three nights between
2017 June 7 and 11, no periodic signal was detected.
The object, however, faded quickly (vsnet-alert 21130),
which appears to be inconsistent with the supernova
classification.  Although the object was again suspected
to be a dwarf nova, the behavior is unusual (fading
at a rate of $\sim$0.5 mag d$^{-1}$, which suggests
an SS Cyg-type dwarf nova, while the amplitude exceeds
8 mag).  No spectroscopic observation was reported.
This object probably would require future deep observations
to clarify the nature.

\subsection{MASTER OT J174305.70$+$231107.8}\label{obj:j1743}

   This object (hereafter MASTER J174305) is a transient detected
at an unfiltered CCD magnitude of 15.6 on 2012 April 5
\citep{bal12j1743atel4022}.  Although 2012 observations
established the SU UMa-type classification, only two
superhumps were recorded \citep{Pdot4}.

\begin{figure}
  \begin{center}
    \FigureFile(85mm,110mm){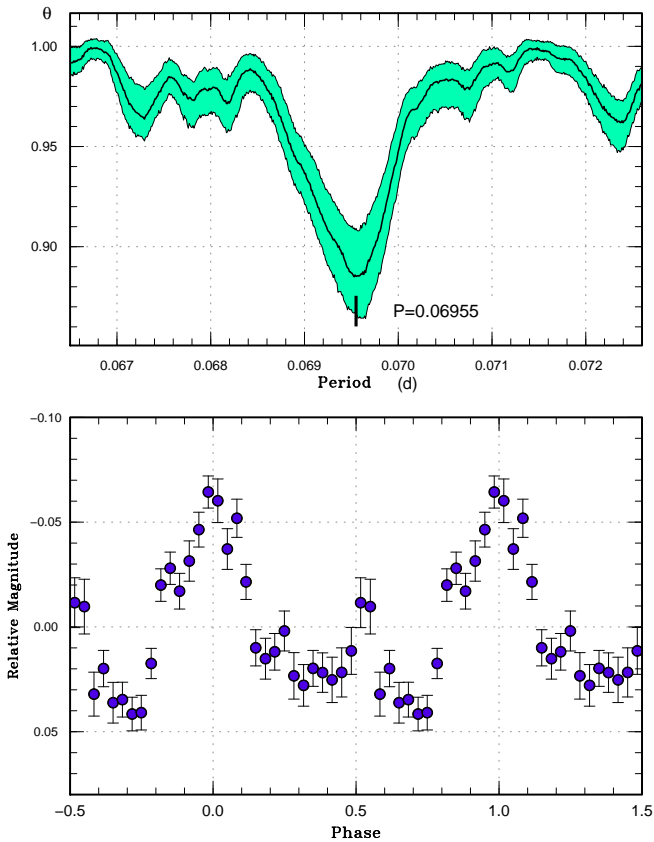}
  \end{center}
  \caption{Superhumps in MASTER J174305 (2017).
     (Upper): PDM analysis.
     (Lower): Phase-averaged profile.}
  \label{fig:j1743shpdm}
\end{figure}

   The 2017 superoutburst was recorded by the ASAS-SN team
(cf. vsnet-alert 21024).  Superhumps were detected
(vsnet-alert 21034, 21043; e-figure \ref{fig:j1743shpdm}).
The times of superhump maxima are listed in
e-table \ref{tab:j1743oc2017}.  Although stages B and C
were apparently recorded, the determined periods were not
precise due to the limited quality of the data.
The mean superhump period with the PDM method
[0.06955(3)~d], however, is sufficiently reliable.

   The field of this object has been monitored since
2012 March 17 by the ASAS-SN team and the coverage
has been particularly good since 2012 September.
The object has been regularly caught in outbursts
and they are listed in e-table \ref{tab:j1743out}.
All these outbursts were superoutburst as judged from
outburst durations.  In additions to them, there were
single-night outburst detections at an unfiltered
CCD magnitude of 17.3 on 2015 October 6 (E. Muyllaert)
and $V$=16.84 (multiple detections on a single night,
ASAS-SN).  These outbursts were likely normal outbursts.
The recorded superoutbursts were well expressed
by a supercycle of 208(1)~d.  These data suggest
that MASTER J174305 is a relatively ordinary SU UMa-type
dwarf nova with rather frequent superoutbursts.

\begin{table}
\caption{Superhump maxima of MASTER J174305 (2017)}\label{tab:j1743oc2017}
\begin{center}
\begin{tabular}{rp{55pt}p{40pt}r@{.}lr}
\hline
\multicolumn{1}{c}{$E$} & \multicolumn{1}{c}{max\commenta} & \multicolumn{1}{c}{error} & \multicolumn{2}{c}{$O-C$\commentb} & \multicolumn{1}{c}{$N$\commentc} \\
\hline
0 & 57891.5471 & 0.0014 & $-$0&0025 & 58 \\
7 & 57892.0348 & 0.0006 & $-$0&0022 & 129 \\
12 & 57892.3855 & 0.0010 & 0&0002 & 75 \\
13 & 57892.4563 & 0.0009 & 0&0014 & 75 \\
14 & 57892.5261 & 0.0007 & 0&0015 & 32 \\
27 & 57893.4328 & 0.0009 & 0&0028 & 113 \\
28 & 57893.5019 & 0.0007 & 0&0023 & 112 \\
29 & 57893.5715 & 0.0010 & 0&0022 & 71 \\
38 & 57894.1926 & 0.0011 & $-$0&0035 & 122 \\
39 & 57894.2665 & 0.0026 & 0&0007 & 94 \\
41 & 57894.4039 & 0.0028 & $-$0&0011 & 75 \\
42 & 57894.4744 & 0.0018 & $-$0&0003 & 77 \\
43 & 57894.5439 & 0.0015 & $-$0&0004 & 76 \\
44 & 57894.6130 & 0.0056 & $-$0&0010 & 42 \\
\hline
  \multicolumn{6}{l}{\commenta BJD$-$2400000.} \\
  \multicolumn{6}{l}{\commentb Against max $= 2457891.5495 + 0.069647 E$.} \\
  \multicolumn{6}{l}{\commentc Number of points used to determine the maximum.} \\
\end{tabular}
\end{center}
\end{table}

\begin{table}
\caption{Superoutbursts of MASTER J174305 in the ASAS-SN data}\label{tab:j1743out}
\begin{center}
\begin{tabular}{ccccc}
\hline
Year & Month & Day & max\commenta & $V$-mag \\
\hline
2012 &  4 &  2 & 56020 & 16.03 \\
2012 & 10 & 30 & 56231 & 15.66 \\
2013 &  5 & 27 & 56440 & 15.91 \\
2014 &  8 & 14 & 56884 & 16.16 \\
2015 &  2 & 15 & 57069 & 15.98 \\
2015 &  9 &  5 & 57271 & 16.14 \\
2016 &  4 &  9 & 57488 & 16.09 \\
2017 &  5 & 14 & 57888 & 16.08 \\
\hline
  \multicolumn{5}{l}{\commenta JD$-$2400000.} \\
\end{tabular}
\end{center}
\end{table}

\subsection{MASTER OT J192757.03$+$404042.8}\label{obj:j1927}

\begin{figure}
  \begin{center}
    \FigureFile(85mm,110mm){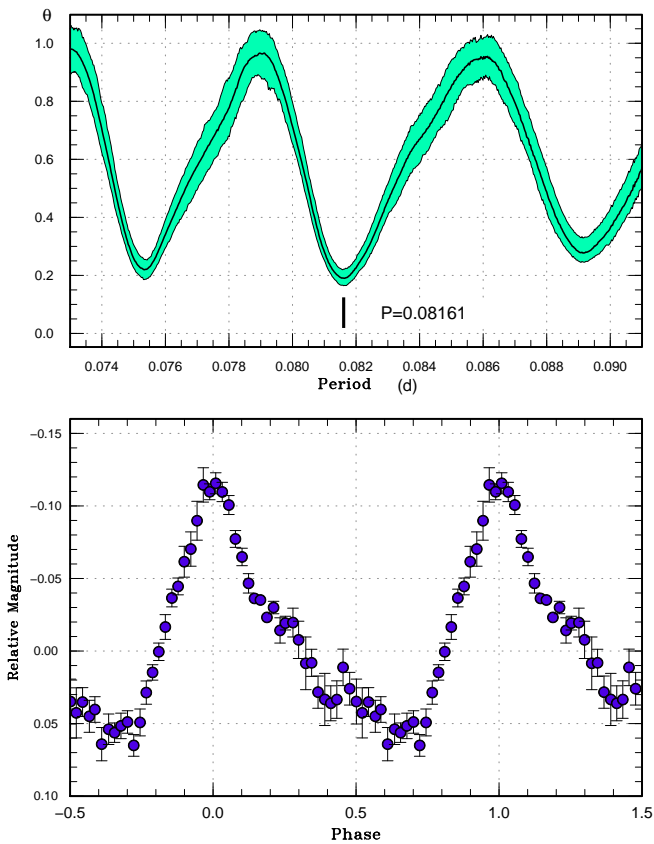}
  \end{center}
  \caption{Superhumps in MASTER J192757 (2017).
     (Upper): PDM analysis.  The alias selection was based on
     the observations on the first night.
     (Lower): Phase-averaged profile.}
  \label{fig:j1927shpdm}
\end{figure}

   This object (hereafter MASTER J192757) is a transient detected
at an unfiltered CCD magnitude of 15.2 on 2014 February 13
\citep{bal14j1927atel6024}.
The 2017 superoutburst was detected at $V$=14.36 on
2017 April 4 by the ASAS-SN team.  The ASAS-SN data
recorded $V$=13.83 on 2017 April 1.
Subsequent observations detected superhumps
(vsnet-alert 20885; figure \ref{fig:j1927shpdm}).
The times of superhump maxima are listed in
e-table \ref{tab:j1927oc2017}.  Although there remained
some ambiguity in the one-day alias, the other aliases
appear to be excluded by using the observations
on the first night.  The best superhump period
by the PDM method was 0.08161(5)~d.

   There were two likely superoutbursts in the past
ASAS-SN data: 2014 March 18 ($V$=13.8) and
2015 June 21 ($V$=14.1).  Since the object is sufficiently
bright, future observations will clarify more detailed
development of superhumps and outburst statistics.

\begin{table}
\caption{Superhump maxima of MASTER J192757 (2017)}\label{tab:j1927oc2017}
\begin{center}
\begin{tabular}{rp{55pt}p{40pt}r@{.}lr}
\hline
\multicolumn{1}{c}{$E$} & \multicolumn{1}{c}{max\commenta} & \multicolumn{1}{c}{error} & \multicolumn{2}{c}{$O-C$\commentb} & \multicolumn{1}{c}{$N$\commentc} \\
\hline
0 & 57849.5593 & 0.0003 & $-$0&0014 & 80 \\
1 & 57849.6437 & 0.0008 & 0&0015 & 30 \\
12 & 57850.5382 & 0.0004 & $-$0&0001 & 85 \\
\hline
  \multicolumn{6}{l}{\commenta BJD$-$2400000.} \\
  \multicolumn{6}{l}{\commentb Against max $= 24849.5607 + 0.081471 E$.} \\
  \multicolumn{6}{l}{\commentc Number of points used to determine the maximum.} \\
\end{tabular}
\end{center}
\end{table}

\subsection{MASTER OT J200904.69$+$825153.6}\label{obj:j2009}

\begin{figure}
  \begin{center}
    \FigureFile(85mm,110mm){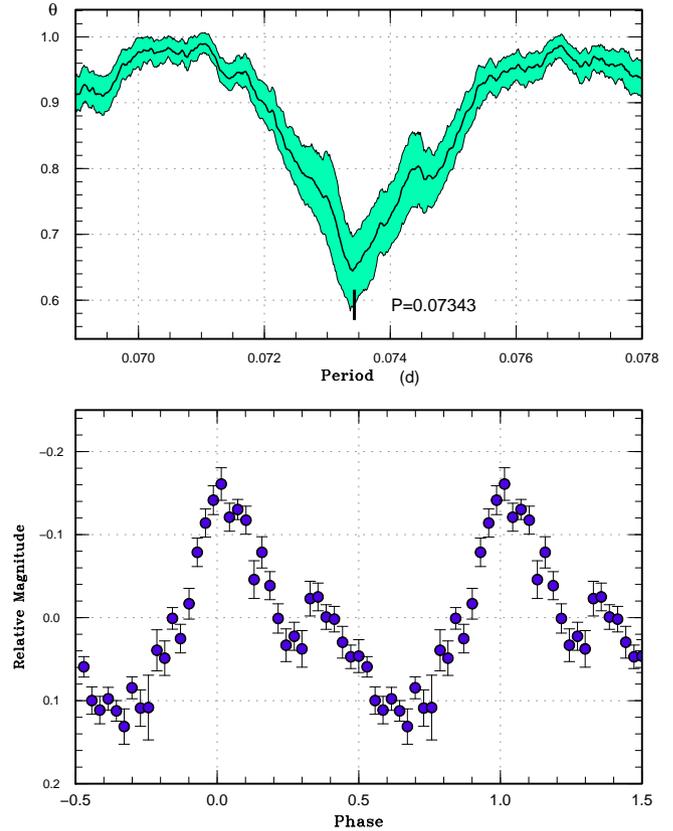}
  \end{center}
  \caption{Superhumps in MASTER J200904 (2017).
     (Upper): PDM analysis.
     (Lower): Phase-averaged profile.}
  \label{fig:j2009shpdm}
\end{figure}

   This object (hereafter MASTER J200904) is a transient detected
at an unfiltered CCD magnitude of 15.6 on 2014 March 10
\citep{bal14j2009atel5974}.  The 2017 outburst was detected
by the ASAS-SN team at $V$=15.70 on 2017 June 2.
The large outburst amplitude \citep{bal14j2009atel5974}
suspected an SU UMa-type dwarf nova.
Subsequent observations detected large-amplitude superhumps
(vsnet-alert 21089, 21093; e-figure \ref{fig:j2009shpdm}).
The times of superhump maxima are listed in
e-table \ref{tab:j2009oc2017}.

\begin{table}
\caption{Superhump maxima of MASTER J200904 (2017)}\label{tab:j2009oc2017}
\begin{center}
\begin{tabular}{rp{55pt}p{40pt}r@{.}lr}
\hline
\multicolumn{1}{c}{$E$} & \multicolumn{1}{c}{max\commenta} & \multicolumn{1}{c}{error} & \multicolumn{2}{c}{$O-C$\commentb} & \multicolumn{1}{c}{$N$\commentc} \\
\hline
0 & 57909.0440 & 0.0038 & $-$0&0025 & 46 \\
1 & 57909.1194 & 0.0008 & $-$0&0004 & 81 \\
2 & 57909.1943 & 0.0007 & 0&0010 & 82 \\
3 & 57909.2593 & 0.0030 & $-$0&0074 & 32 \\
4 & 57909.3378 & 0.0015 & $-$0&0022 & 36 \\
5 & 57909.4139 & 0.0010 & 0&0004 & 78 \\
6 & 57909.4873 & 0.0007 & 0&0004 & 82 \\
7 & 57909.5638 & 0.0009 & 0&0035 & 51 \\
14 & 57910.0791 & 0.0018 & 0&0050 & 73 \\
15 & 57910.1482 & 0.0017 & 0&0006 & 84 \\
16 & 57910.2206 & 0.0014 & $-$0&0004 & 75 \\
19 & 57910.4455 & 0.0012 & 0&0043 & 70 \\
20 & 57910.5157 & 0.0010 & 0&0010 & 70 \\
55 & 57913.0799 & 0.0037 & $-$0&0041 & 65 \\
69 & 57914.1124 & 0.0029 & 0&0007 & 40 \\
\hline
  \multicolumn{6}{l}{\commenta BJD$-$2400000.} \\
  \multicolumn{6}{l}{\commentb Against max $= 2457909.0464 + 0.073409 E$.} \\
  \multicolumn{6}{l}{\commentc Number of points used to determine the maximum.} \\
\end{tabular}
\end{center}
\end{table}

\subsection{MASTER OT J205110.36$+$044842.2}\label{obj:j2051}

\begin{figure}
  \begin{center}
    \FigureFile(85mm,110mm){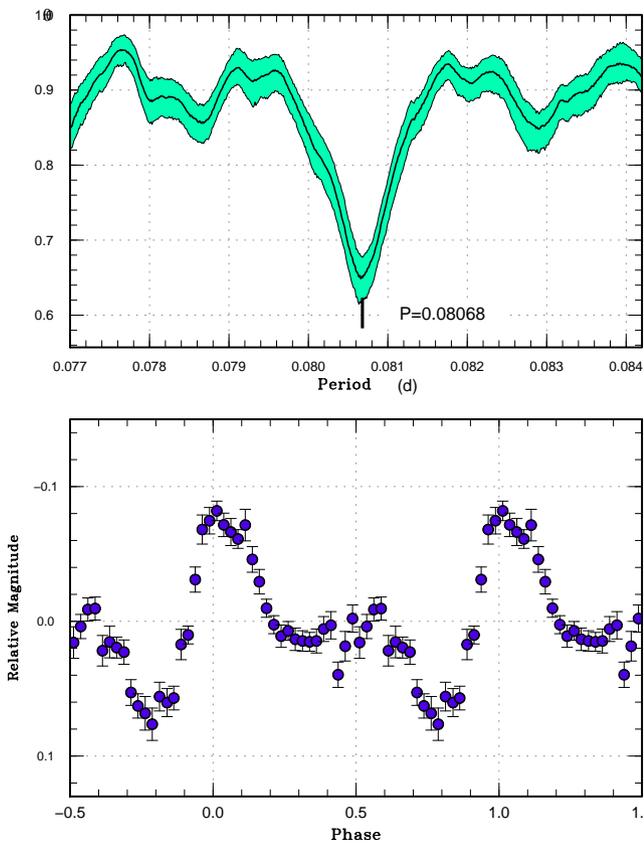}
  \end{center}
  \caption{Superhumps in MASTER J205110 (2017).
     (Upper): PDM analysis.
     (Lower): Phase-averaged profile.}
  \label{fig:j2051shpdm}
\end{figure}

   This object (hereafter MASTER J205110)
was detected as a quasar or a dwarf nova at an unfiltered
CCD magnitude of 15.0 on 2017 September 28 by the MASTER 
network \citep{shu17j2051atel10790}.
Two past outbursts (2001 June and 2002 August) were
detected by D. Denisenko in past images and Denisenko
suggested it to be an SU UMa-type dwarf nova
based on the past outbursts, ROSAT identification
and the SDSS colors (vsnet-alert 21483).
Subsequent observations detected superhumps
(vsnet-alert 21486, 21487, 21488; e-figure \ref{fig:j2051shpdm}).
The times of superhump maxima are listed in
e-table \ref{tab:j2051oc2017}.

\begin{table}
\caption{Superhump maxima of MASTER J205110 (2017)}\label{tab:j2051oc2017}
\begin{center}
\begin{tabular}{rp{55pt}p{40pt}r@{.}lr}
\hline
\multicolumn{1}{c}{$E$} & \multicolumn{1}{c}{max\commenta} & \multicolumn{1}{c}{error} & \multicolumn{2}{c}{$O-C$\commentb} & \multicolumn{1}{c}{$N$\commentc} \\
\hline
0 & 58025.5946 & 0.0005 & 0&0029 & 262 \\
1 & 58025.6720 & 0.0005 & $-$0&0004 & 265 \\
18 & 58027.0415 & 0.0014 & $-$0&0030 & 132 \\
19 & 58027.1245 & 0.0007 & $-$0&0007 & 196 \\
22 & 58027.3685 & 0.0015 & 0&0011 & 88 \\
34 & 58028.3343 & 0.0015 & $-$0&0015 & 79 \\
59 & 58030.3552 & 0.0037 & 0&0016 & 85 \\
\hline
  \multicolumn{6}{l}{\commenta BJD$-$2400000.} \\
  \multicolumn{6}{l}{\commentb Against max $= 2458025.5917 + 0.080710 E$.} \\
  \multicolumn{6}{l}{\commentc Number of points used to determine the maximum.} \\
\end{tabular}
\end{center}
\end{table}

\subsection{MASTER OT J212624.16$+$253827.2}\label{obj:j2126}

   This object (hereafter MASTER J212624)
was detected as a transient at an unfiltered
CCD magnitude of 14.1 on 2013 June 26 by the MASTER network
\citep{den13j2126atel5111}.  The SU UMa-type nature was
confirmed during the 2013 superoutburst.
The object attracted attention since the superhump
period indicates an SU UMa-type dwarf nova
in the period gap see \citet{Pdot5} and \citet{Pdot8}
for more information.

\begin{figure}
  \begin{center}
    \FigureFile(85mm,70mm){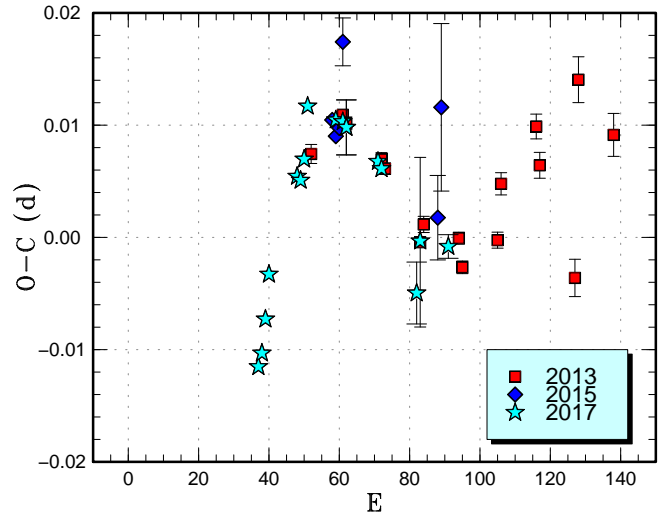}
  \end{center}
  \caption{Comparison of $O-C$ diagrams of MASTER J212624 between different
  superoutbursts.  A period of 0.09135~d was used to draw this figure.
  Approximate cycle counts ($E$) after the start of the superoutburst
  were used.  Since the start of the 2013 superoutburst was not
  well constrained, we shifted the $O-C$ diagram to best fit
  the best-recorded 2017 one.}
  \label{fig:j2126comp}
\end{figure}

   The 2017 superoutburst was detected by the ASAS-SN
team at $V$=14.90 on 2017 August 25.  The object
further brightened to $V$=14.2 in 0.85~d.
The initial observations on 2017 August 26--27 did not
show large-amplitude superhumps.
They became apparent on the next night and grew further
(vsnet-alert 21374, 21377).  This was the first time
in this object growing superhumps were recorded.
The times of superhump maxima are listed in
e-table \ref{tab:j2126oc2017}.  Stage A was impressively
long.  In determining periods, we neglected 32 $\le E \le$ 35,
which were likely a transition phase between stages A and B.
A comparison of the $O-C$ diagrams suggests that we observed
stage A and early half of stage B (e-figure \ref{fig:j2126comp}.
We if could have continued observations, we could have detected
a strongly positive $P_{\rm dot}$ as in the 2013 superoutburst.

   This object adds a new example of long developing time
of superhumps in long-$P_{\rm orb}$ systems
(see subsection \ref{obj:j0809}).

   This object showed relatively regular superoutburst
(e-table \ref{tab:j2126out}).  The supercycle was 345(9)~d.

\begin{table}
\caption{Superhump maxima of MASTER J212624 (2017)}\label{tab:j2126oc2017}
\begin{center}
\begin{tabular}{rp{55pt}p{40pt}r@{.}lr}
\hline
\multicolumn{1}{c}{$E$} & \multicolumn{1}{c}{max\commenta} & \multicolumn{1}{c}{error} & \multicolumn{2}{c}{$O-C$\commentb} & \multicolumn{1}{c}{$N$\commentc} \\
\hline
0 & 57992.3487 & 0.0036 & $-$0&0338 & 93 \\
2 & 57992.5291 & 0.0040 & $-$0&0377 & 62 \\
21 & 57994.3239 & 0.0003 & 0&0068 & 128 \\
22 & 57994.4164 & 0.0002 & 0&0072 & 355 \\
23 & 57994.5108 & 0.0002 & 0&0094 & 323 \\
24 & 57994.6061 & 0.0004 & 0&0127 & 148 \\
32 & 57995.3457 & 0.0002 & 0&0152 & 224 \\
33 & 57995.4367 & 0.0002 & 0&0141 & 332 \\
34 & 57995.5299 & 0.0002 & 0&0152 & 299 \\
35 & 57995.6260 & 0.0006 & 0&0191 & 54 \\
43 & 57996.3555 & 0.0008 & 0&0117 & 74 \\
45 & 57996.5381 & 0.0002 & 0&0100 & 236 \\
46 & 57996.6289 & 0.0024 & 0&0087 & 43 \\
46 & 57996.6289 & 0.0024 & 0&0087 & 43 \\
55 & 57997.4480 & 0.0002 & $-$0&0013 & 309 \\
56 & 57997.5387 & 0.0003 & $-$0&0027 & 227 \\
66 & 57998.4412 & 0.0028 & $-$0&0215 & 65 \\
67 & 57998.5371 & 0.0008 & $-$0&0176 & 131 \\
75 & 57999.2675 & 0.0011 & $-$0&0243 & 97 \\
\hline
  \multicolumn{6}{l}{\commenta BJD$-$2400000.} \\
  \multicolumn{6}{l}{\commentb Against max $= 2457992.3825 + 0.092123 E$.} \\
  \multicolumn{6}{l}{\commentc Number of points used to determine the maximum.} \\
\end{tabular}
\end{center}
\end{table}

\begin{table}
\caption{List of superoutbursts of MASTER J212624 in the ASAS-SN data}\label{tab:j2126out}
\begin{center}
\begin{tabular}{ccccc}
\hline
Year & Month & Day & max\commenta & $V$ mag \\
\hline
2013 & 11 & 16 & 56613 & 14.9 \\
2014 &  9 & 10 & 56911 & 14.5 \\
2015 &  8 & 26 & 57261 & 14.2 \\
2016 &  7 & 31 & 57601 & 14.4 \\
2017 &  8 & 25 & 57991 & 14.2 \\
\hline
  \multicolumn{5}{l}{\commenta JD$-$2400000.} \\
\end{tabular}
\end{center}
\end{table}

\subsection{OT J182142.8$+$212154}\label{obj:j1821}

   This object (hereafter OT J182142) was discovered by K. Itagaki
at an unfiltered CCD magnitude of 14.9 on 2010 April 24
(vsnet-alert 11952).  Subsequent observations confirmed
the SU UMa-type classification \citep{Pdot2}.

   The 2017 superoutburst was detected by the ASAS-SN team
at $V$=14.82 on 2017 May 31.  Subsequent observations
confirmed superhumps (vsnet-alert 21083, 21099).
The times of superhump maxima are listed in
e-table \ref{tab:j1821oc2017}.

   The object faded by 1.4 mag between 2017 June 3 and 5,
and it was likely the termination of the superoutburst.
It was likely that the superoutburst was not detected
sufficiently early and there was a chance that we only
observed stage C superhumps (this may have been also
the case for the 2010 observations).

   This field has been covered by the ASAS-SN team since
2014 March 18.  There was only one previous outburst
(type unknown) in the record at $V$=15.8 on
2016 February 15.

\begin{table}
\caption{Superhump maxima of OT J182142 (2017)}\label{tab:j1821oc2017}
\begin{center}
\begin{tabular}{rp{55pt}p{40pt}r@{.}lr}
\hline
\multicolumn{1}{c}{$E$} & \multicolumn{1}{c}{max\commenta} & \multicolumn{1}{c}{error} & \multicolumn{2}{c}{$O-C$\commentb} & \multicolumn{1}{c}{$N$\commentc} \\
\hline
0 & 57907.1707 & 0.0090 & $-$0&0067 & 30 \\
1 & 57907.2600 & 0.0010 & 0&0005 & 79 \\
2 & 57907.3477 & 0.0016 & 0&0060 & 37 \\
13 & 57908.2475 & 0.0026 & 0&0023 & 45 \\
15 & 57908.4107 & 0.0007 & 0&0012 & 44 \\
16 & 57908.4888 & 0.0008 & $-$0&0028 & 40 \\
39 & 57910.3813 & 0.0016 & 0&0004 & 44 \\
40 & 57910.4622 & 0.0026 & $-$0&0008 & 41 \\
\hline
  \multicolumn{6}{l}{\commenta BJD$-$2400000.} \\
  \multicolumn{6}{l}{\commentb Against max $= 2457907.1774 + 0.082140 E$.} \\
  \multicolumn{6}{l}{\commentc Number of points used to determine the maximum.} \\
\end{tabular}
\end{center}
\end{table}

\subsection{OT J204222.3$+$271211}\label{obj:j2042}

   This object (PNV J20422233$+$2712111, hereafter OT J204222)
is a transient discovered independently by H. Nishimura,
T. Kojima and  Kaneko at an unfiltered
CCD magnitude of 11.1 on 2017 April 13.\footnote{
$<$http://www.cbat.eps.harvard.edu/unconf/followups/J20422233+2712111.html$>$
}
The object was immediately suspected to be a dwarf nova,
not a classical nova as originally suspected, by
the presence of a faint blue object in SDSS and
a GALEX ultraviolet object.
The large outburst amplitude (nearly 9 mag) suggested
a WZ Sge-type dwarf nova (vsnet-alert 20915, see
also $<$https://www.aavso.org/pnv-j204222332712111-new-transient-111-mag-vulpecula$>$).

\begin{figure}
  \begin{center}
    \FigureFile(85mm,110mm){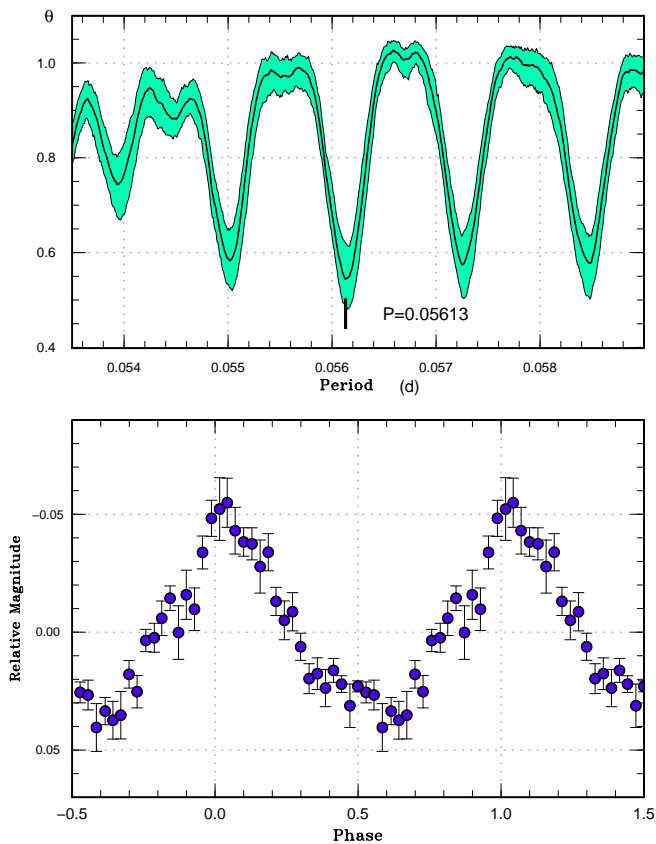}
  \end{center}
  \caption{Superhumps in OT J204222 (2017).
     (Upper): PDM analysis.  The data between BJD 2457865
     and 2457872 were used.  The alias selection was based on
     a long run on 2017 April 25. 
     (Lower): Phase-averaged profile.}
  \label{fig:j2042shpdm}
\end{figure}

   Despite the large outburst amplitude, observations of
this object were rather sparse in the early morning
and expected early superhumps were not detected.
The object was found to show ordinary superhumps 
on 2017 April 22 (vsnet-alert 20941, 20959;
e-figure \ref{fig:j2042shpdm}).  A retrospective examination
suggested that growing superhumps were probably already
present on 2017 April 18 (vsnet-alert 20960), only 5~d
after the outburst detection.
The times of superhump maxima are listed in
e-table \ref{tab:j2042oc2017}.  Due to the lack of
observations, the period of stage A was not determined
and $P_{\rm dot}$ for stage B superhumps was not
well determined (it may be almost zero considering
the large uncertainty).

   In the ASAS-SN data, the object was detected at $V$=12.2
on 2017 April 15 but was still in quiescence or still
very faint on 2017 April 9.  The waiting time of
the emergence of superhumps was thus shorter than 9~d.

\begin{table}
\caption{Superhump maxima of OT J204222 (2017)}\label{tab:j2042oc2017}
\begin{center}
\begin{tabular}{rp{55pt}p{40pt}r@{.}lr}
\hline
\multicolumn{1}{c}{$E$} & \multicolumn{1}{c}{max\commenta} & \multicolumn{1}{c}{error} & \multicolumn{2}{c}{$O-C$\commentb} & \multicolumn{1}{c}{$N$\commentc} \\
\hline
0 & 57861.8675 & 0.0010 & $-$0&0046 & 417 \\
65 & 57865.5307 & 0.0015 & 0&0058 & 61 \\
66 & 57865.5821 & 0.0003 & 0&0010 & 102 \\
119 & 57868.5587 & 0.0003 & $-$0&0009 & 81 \\
120 & 57868.6163 & 0.0005 & 0&0006 & 47 \\
156 & 57870.6483 & 0.0015 & 0&0095 & 25 \\
165 & 57871.1388 & 0.0029 & $-$0&0058 & 25 \\
166 & 57871.1956 & 0.0006 & $-$0&0052 & 61 \\
167 & 57871.2567 & 0.0012 & $-$0&0004 & 16 \\
\hline
  \multicolumn{6}{l}{\commenta BJD$-$2400000.} \\
  \multicolumn{6}{l}{\commentb Against max $= 2457861.8721 + 0.056197 E$.} \\
  \multicolumn{6}{l}{\commentc Number of points used to determine the maximum.} \\
\end{tabular}
\end{center}
\end{table}

\subsection{PNV J20205397$+$2508145}\label{obj:j2020}

\begin{figure}
  \begin{center}
    \FigureFile(85mm,110mm){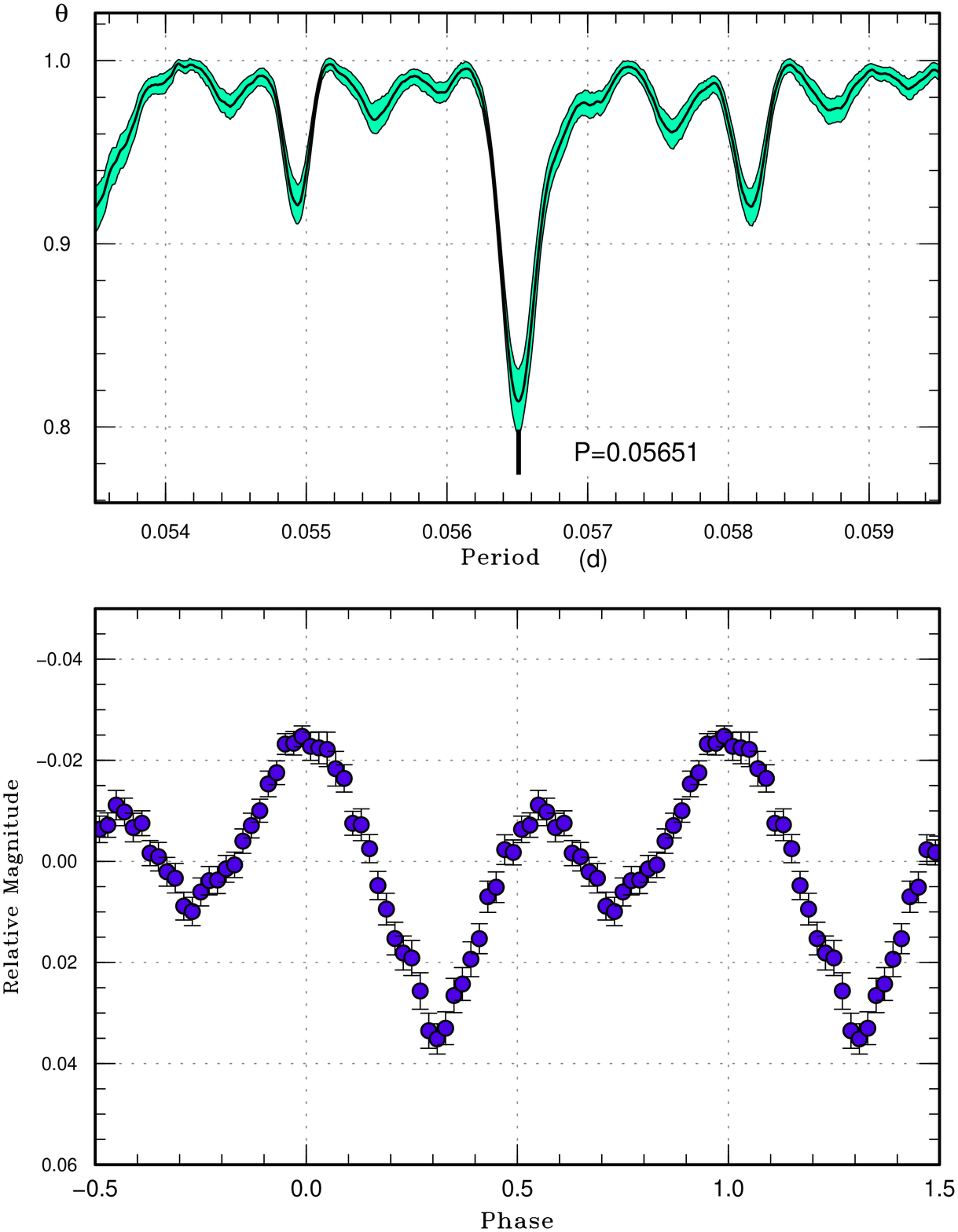}
  \end{center}
  \caption{Early superhumps in PNV J202053 (2017).
     (Upper): PDM analysis.
     (Lower): Phase-averaged profile.}
  \label{fig:j2020eshpdm}
\end{figure}

\begin{figure}
  \begin{center}
    \FigureFile(85mm,110mm){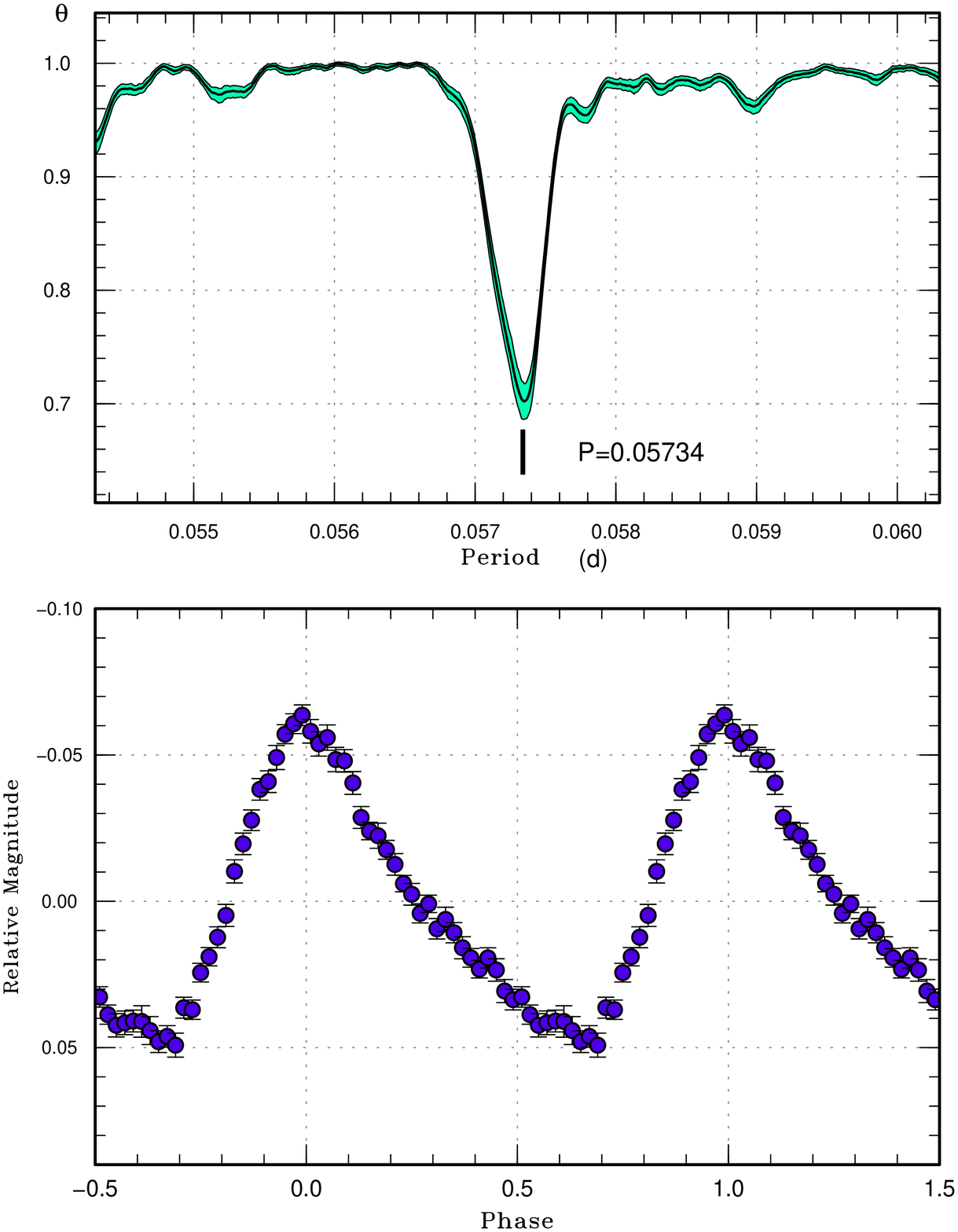}
  \end{center}
  \caption{Ordinary superhumps in PNV J202053 (2017).
     (Upper): PDM analysis.
     (Lower): Phase-averaged profile.}
  \label{fig:j2020shpdm}
\end{figure}

\begin{figure*}
  \begin{center}
    \FigureFile(160mm,200mm){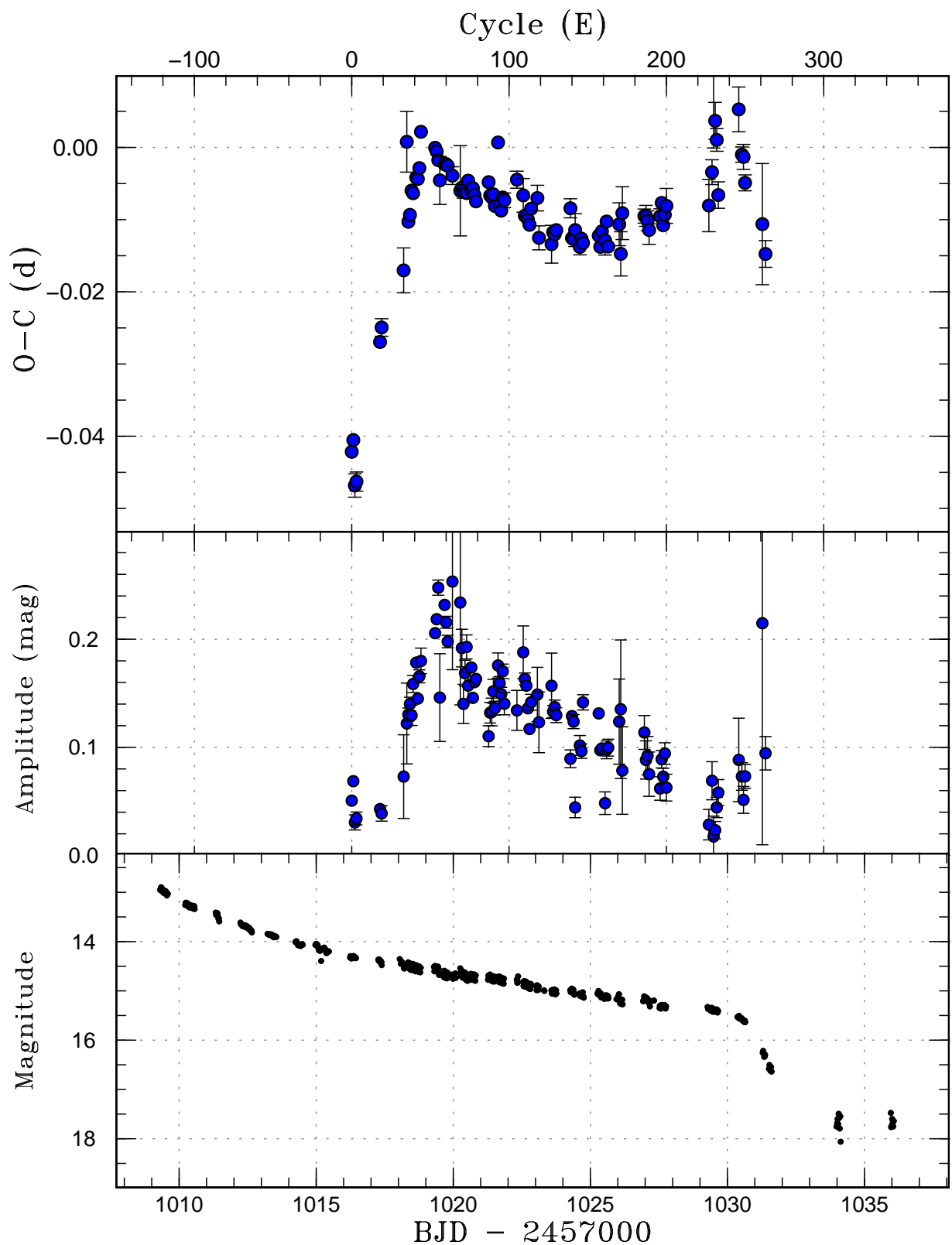}
  \end{center}
  \caption{$O-C$ diagram of superhumps in PNV J202053 (2017).
     (Upper:) $O-C$ diagram.
     We used a period of 0.05739~d for calculating the $O-C$ residuals.
     (Middle:) Amplitudes of superhumps.
     (Lower:) Light curve.  The data were binned to 0.019~d.
  }
  \label{fig:j2020humpall}
\end{figure*}

   This object (hereafter PNV J202053) was detected as
a transient by T. Kojima at an unfiltered CCD magnitude of
12.3 on 2017 September 12.\footnote{
$<$http://www.cbat.eps.harvard.edu/unconf/followups/J20205397$+$2508145.html$>$.
}  The outburst was recorded in the ASAS-SN data
at $V$=12.9 on 2017 September 13.  The object was not
in outburst on 2017 September 8.
A blue quiescent counterpart with $g$=20.41 was identified
by Brian Skiff and astrometry by Andrea Mantero
(cf. vsnet-alert 21424, 21426).  Subsequent photometry
detected early superhumps (vsnet-alert 21427, 21431, 21436;
e-figure \ref{fig:j2020eshpdm})
and the object was identified as a WZ Sge-type dwarf nova.
The object started to show ordinary superhumps
(e-figure \ref{fig:j2020shpdm})
on 2017 September 19 (see vsnet-alert 21455, 21460 and
analysis in this subsection).
The times of maxima of ordinary superhumps are listed
in e-table \ref{tab:j2020oc2017}.
Stages A and B were very clearly recorded and there was
possibly a transition to stage C during the final 
decline from the superoutburst plateau
(e-figure \ref{fig:j2020humpall}).

   The period of early superhump by the PDM method was
0.056509(5)~d (e-figure \ref{fig:j2020eshpdm}).
The value of $\epsilon^*$ for stage A superhumps
was 0.0329(11), which corresponds to $q$=0.090(3).
This value appears to be typical for a WZ Sge-type
dwarf nova.  There was no post-superoutburst rebrightening.

\begin{table*}
\caption{Superhump maxima of PNV J202053 (2017)}\label{tab:j2020oc2017}
\begin{center}
\begin{tabular}{rp{55pt}p{40pt}r@{.}lrrp{55pt}p{40pt}r@{.}lr}
\hline
\multicolumn{1}{c}{$E$} & \multicolumn{1}{c}{max\commenta} & \multicolumn{1}{c}{error} & \multicolumn{2}{c}{$O-C$\commentb} & \multicolumn{1}{c}{$N$\commentc} & \multicolumn{1}{c}{$E$} & \multicolumn{1}{c}{max\commenta} & \multicolumn{1}{c}{error} & \multicolumn{2}{c}{$O-C$\commentb} & \multicolumn{1}{c}{$N$\commentc} \\
\hline
0 & 58016.2541 & 0.0005 & $-$0&0280 & 114 & 111 & 58022.6571 & 0.0002 & 0&0001 & 226 \\
1 & 58016.3131 & 0.0005 & $-$0&0265 & 119 & 112 & 58022.7139 & 0.0002 & $-$0&0005 & 180 \\
2 & 58016.3642 & 0.0016 & $-$0&0328 & 72 & 113 & 58022.7706 & 0.0002 & $-$0&0013 & 164 \\
3 & 58016.4221 & 0.0013 & $-$0&0323 & 74 & 114 & 58022.8302 & 0.0004 & 0&0009 & 50 \\
18 & 58017.3023 & 0.0008 & $-$0&0136 & 121 & 118 & 58023.0612 & 0.0018 & 0&0022 & 48 \\
19 & 58017.3617 & 0.0012 & $-$0&0116 & 129 & 119 & 58023.1131 & 0.0017 & $-$0&0034 & 63 \\
33 & 58018.1731 & 0.0031 & $-$0&0043 & 51 & 127 & 58023.5714 & 0.0026 & $-$0&0046 & 24 \\
35 & 58018.3057 & 0.0042 & 0&0134 & 25 & 128 & 58023.6304 & 0.0003 & $-$0&0029 & 60 \\
36 & 58018.3520 & 0.0003 & 0&0023 & 138 & 129 & 58023.6875 & 0.0004 & $-$0&0033 & 58 \\
37 & 58018.4103 & 0.0003 & 0&0032 & 147 & 130 & 58023.7455 & 0.0004 & $-$0&0027 & 52 \\
38 & 58018.4711 & 0.0006 & 0&0066 & 88 & 139 & 58024.2650 & 0.0013 & $-$0&0001 & 62 \\
39 & 58018.5281 & 0.0004 & 0&0062 & 74 & 140 & 58024.3183 & 0.0003 & $-$0&0043 & 117 \\
41 & 58018.6450 & 0.0002 & 0&0082 & 74 & 141 & 58024.3755 & 0.0004 & $-$0&0044 & 113 \\
42 & 58018.7022 & 0.0002 & 0&0080 & 156 & 142 & 58024.4342 & 0.0023 & $-$0&0032 & 72 \\
43 & 58018.7611 & 0.0003 & 0&0095 & 154 & 145 & 58024.6040 & 0.0011 & $-$0&0057 & 34 \\
44 & 58018.8236 & 0.0007 & 0&0145 & 34 & 146 & 58024.6625 & 0.0005 & $-$0&0046 & 60 \\
53 & 58019.3379 & 0.0001 & 0&0119 & 290 & 147 & 58024.7193 & 0.0004 & $-$0&0052 & 61 \\
54 & 58019.3947 & 0.0001 & 0&0113 & 245 & 157 & 58025.2942 & 0.0002 & $-$0&0046 & 116 \\
55 & 58019.4509 & 0.0002 & 0&0101 & 154 & 158 & 58025.3501 & 0.0004 & $-$0&0062 & 117 \\
56 & 58019.5055 & 0.0033 & 0&0072 & 31 & 159 & 58025.4096 & 0.0006 & $-$0&0041 & 71 \\
59 & 58019.6799 & 0.0002 & 0&0093 & 90 & 161 & 58025.5231 & 0.0020 & $-$0&0055 & 52 \\
60 & 58019.7373 & 0.0002 & 0&0093 & 89 & 162 & 58025.5831 & 0.0006 & $-$0&0029 & 60 \\
61 & 58019.7945 & 0.0002 & 0&0090 & 77 & 163 & 58025.6371 & 0.0006 & $-$0&0064 & 59 \\
64 & 58019.9653 & 0.0012 & 0&0076 & 24 & 170 & 58026.0419 & 0.0030 & $-$0&0036 & 51 \\
69 & 58020.2501 & 0.0063 & 0&0052 & 18 & 171 & 58026.0952 & 0.0031 & $-$0&0077 & 52 \\
70 & 58020.3080 & 0.0007 & 0&0056 & 78 & 172 & 58026.1582 & 0.0037 & $-$0&0021 & 67 \\
71 & 58020.3649 & 0.0010 & 0&0051 & 146 & 186 & 58026.9613 & 0.0010 & $-$0&0031 & 51 \\
72 & 58020.4225 & 0.0005 & 0&0053 & 73 & 187 & 58027.0187 & 0.0015 & $-$0&0031 & 56 \\
73 & 58020.4794 & 0.0004 & 0&0048 & 73 & 188 & 58027.0754 & 0.0014 & $-$0&0039 & 55 \\
74 & 58020.5385 & 0.0006 & 0&0064 & 59 & 189 & 58027.1315 & 0.0020 & $-$0&0052 & 56 \\
76 & 58020.6520 & 0.0002 & 0&0051 & 150 & 196 & 58027.5351 & 0.0010 & $-$0&0036 & 47 \\
77 & 58020.7096 & 0.0002 & 0&0053 & 158 & 197 & 58027.5944 & 0.0006 & $-$0&0018 & 56 \\
78 & 58020.7660 & 0.0002 & 0&0043 & 160 & 198 & 58027.6487 & 0.0007 & $-$0&0049 & 60 \\
79 & 58020.8226 & 0.0004 & 0&0034 & 51 & 199 & 58027.7075 & 0.0008 & $-$0&0035 & 60 \\
87 & 58021.2844 & 0.0006 & 0&0057 & 128 & 200 & 58027.7661 & 0.0024 & $-$0&0023 & 37 \\
88 & 58021.3399 & 0.0005 & 0&0038 & 314 & 227 & 58029.3157 & 0.0036 & $-$0&0034 & 32 \\
89 & 58021.3970 & 0.0007 & 0&0035 & 298 & 229 & 58029.4351 & 0.0017 & 0&0012 & 48 \\
90 & 58021.4548 & 0.0005 & 0&0039 & 167 & 230 & 58029.5082 & 0.0124 & 0&0168 & 36 \\
91 & 58021.5106 & 0.0003 & 0&0022 & 111 & 231 & 58029.5570 & 0.0026 & 0&0082 & 58 \\
93 & 58021.6342 & 0.0007 & 0&0109 & 39 & 232 & 58029.6118 & 0.0016 & 0&0055 & 58 \\
94 & 58021.6830 & 0.0003 & 0&0023 & 57 & 233 & 58029.6615 & 0.0018 & $-$0&0022 & 36 \\
95 & 58021.7395 & 0.0003 & 0&0014 & 60 & 246 & 58030.4194 & 0.0031 & 0&0092 & 128 \\
96 & 58021.7987 & 0.0002 & 0&0032 & 53 & 248 & 58030.5280 & 0.0011 & 0&0028 & 46 \\
97 & 58021.8557 & 0.0010 & 0&0028 & 37 & 249 & 58030.5850 & 0.0017 & 0&0025 & 60 \\
105 & 58022.3177 & 0.0012 & 0&0053 & 54 & 250 & 58030.6388 & 0.0011 & $-$0&0012 & 55 \\
109 & 58022.5451 & 0.0023 & 0&0029 & 29 & 261 & 58031.2644 & 0.0084 & $-$0&0074 & 11 \\
110 & 58022.5997 & 0.0002 & 0&0001 & 71 & 263 & 58031.3750 & 0.0019 & $-$0&0116 & 39 \\
\hline
  \multicolumn{12}{l}{\commenta BJD$-$2400000.} \\
  \multicolumn{12}{l}{\commentb Against max $= 2458016.2821 + 0.057432 E$.} \\
  \multicolumn{12}{l}{\commentc Number of points used to determine the maximum.} \\
\end{tabular}
\end{center}
\end{table*}

\subsection{SDSS J152857.86$+$034911.7}\label{obj:j1528}

\begin{figure}
  \begin{center}
    \FigureFile(85mm,110mm){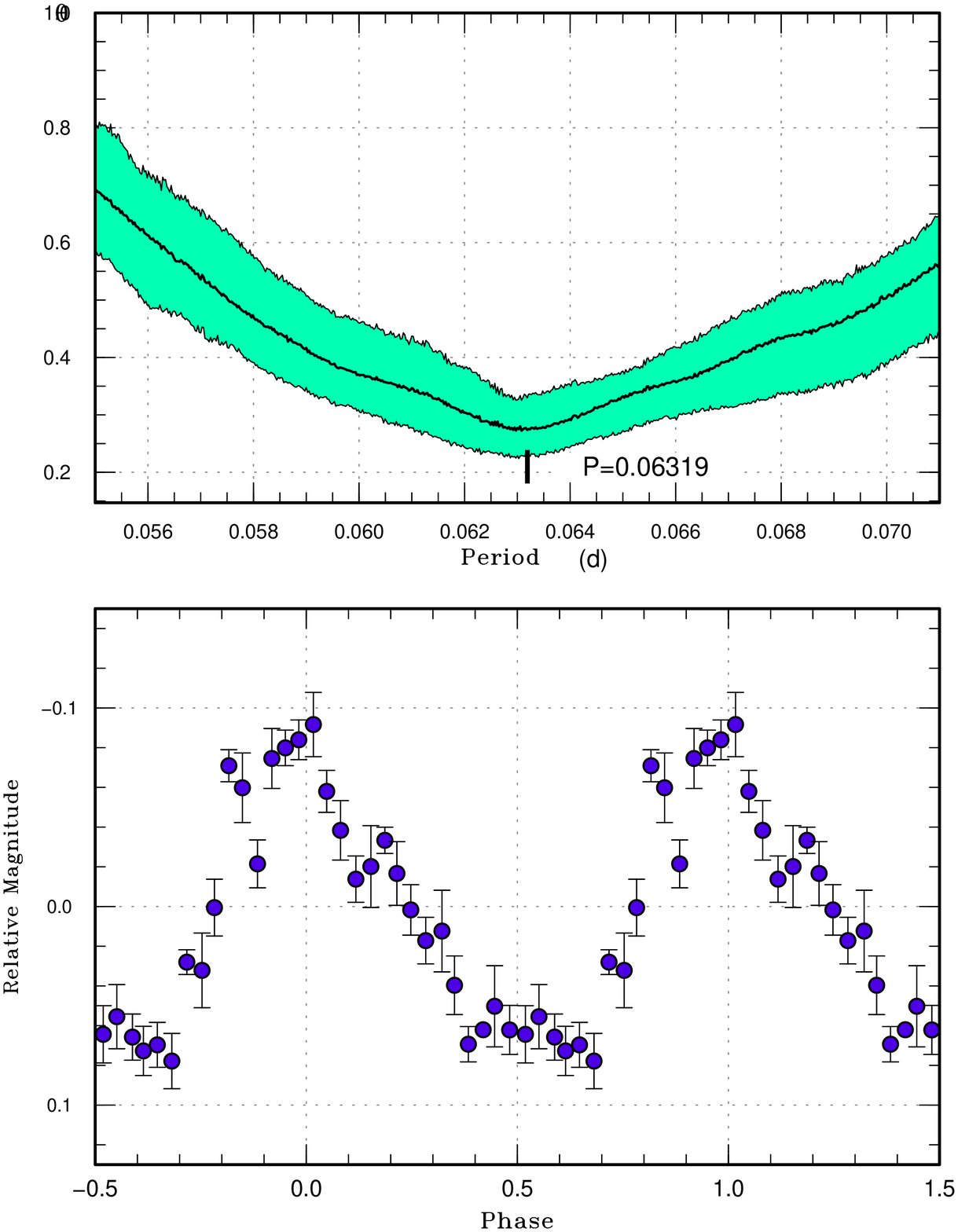}
  \end{center}
  \caption{Superhumps in SDSS J152857 (2017).
     (Upper): PDM analysis.
     (Lower): Phase-averaged profile.}
  \label{fig:j1528shpdm}
\end{figure}

   This object (hereafter SDSS J152857) was selected
as a CV during the course of the SDSS \citep{szk03SDSSCV2}.
Some outbursts were recorded (\cite{wil10j1924};
\cite{dra14CRTSCVs}).
No secure orbital variations were detected by
\citet{wou12SDSSCRTSCVs} and it was concluded that
the object has a very long orbital period or it is
of low inclination.
\citet{kat12DNSDSS} estimated an orbital period of
0.082(3)~d from SDSS colors.
The 2017 outburst was detected by the ASAS-SN team
at $V$=15.62 on 2017 May 16.  Observations on 2017 May 21--22
detected superhumps (vsnet-alert 21056;
e-figure \ref{fig:j1528shpdm};
e-table \ref{tab:j1528oc2017}).

\begin{table}
\caption{Superhump maxima of SDSS J152857 (2017)}\label{tab:j1528oc2017}
\begin{center}
\begin{tabular}{rp{55pt}p{40pt}r@{.}lr}
\hline
\multicolumn{1}{c}{$E$} & \multicolumn{1}{c}{max\commenta} & \multicolumn{1}{c}{error} & \multicolumn{2}{c}{$O-C$\commentb} & \multicolumn{1}{c}{$N$\commentc} \\
\hline
0 & 57895.3966 & 0.0011 & 0&0003 & 41 \\
1 & 57895.4598 & 0.0011 & 0&0000 & 38 \\
2 & 57895.5223 & 0.0007 & $-$0&0008 & 64 \\
3 & 57895.5871 & 0.0028 & 0&0005 & 25 \\
\hline
  \multicolumn{6}{l}{\commenta BJD$-$2400000.} \\
  \multicolumn{6}{l}{\commentb Against max $= 2457895.3963 + 0.063404 E$.} \\
  \multicolumn{6}{l}{\commentc Number of points used to determine the maximum.} \\
\end{tabular}
\end{center}
\end{table}

\subsection{SDSS J153015.04$+$094946.3}\label{obj:j1530}

   This object (hereafter SDSS J153015) was originally
selected as a CV by the SDSS \citep{szk09SDSSCV7}.
The SU UMa-type nature was confirmed during the 2017
March superoutburst \citep{Pdot9}.  The supercycle
was estimated to be very short [84.7(1.2)~d].

   The 2017 June superoutburst was detected by
the ASAS-SN team at $V$=15.86 on 2017 June 13.
Subsequent observations detected superhumps
(vsnet-alert 21121).  The times of superhump maxima
are listed in e-table \ref{tab:j1530oc2017b}.

\begin{table}
\caption{Superhump maxima of SDSS J153015 (2017b)}\label{tab:j1530oc2017b}
\begin{center}
\begin{tabular}{rp{55pt}p{40pt}r@{.}lr}
\hline
\multicolumn{1}{c}{$E$} & \multicolumn{1}{c}{max\commenta} & \multicolumn{1}{c}{error} & \multicolumn{2}{c}{$O-C$\commentb} & \multicolumn{1}{c}{$N$\commentc} \\
\hline
0 & 57919.0487 & 0.0013 & 0&0011 & 86 \\
1 & 57919.1257 & 0.0020 & 0&0029 & 83 \\
4 & 57919.3495 & 0.0017 & 0&0008 & 67 \\
5 & 57919.4249 & 0.0008 & 0&0009 & 86 \\
13 & 57920.0182 & 0.0031 & $-$0&0083 & 63 \\
18 & 57920.4043 & 0.0011 & 0&0012 & 88 \\
26 & 57921.0077 & 0.0013 & 0&0021 & 83 \\
27 & 57921.0717 & 0.0022 & $-$0&0092 & 83 \\
31 & 57921.3873 & 0.0016 & 0&0052 & 25 \\
32 & 57921.4608 & 0.0040 & 0&0034 & 27 \\
\hline
  \multicolumn{6}{l}{\commenta BJD$-$2400000.} \\
  \multicolumn{6}{l}{\commentb Against max $= 2457919.0475 + 0.075310 E$.} \\
  \multicolumn{6}{l}{\commentc Number of points used to determine the maximum.} \\
\end{tabular}
\end{center}
\end{table}

\subsection{SDSS J204817.85$-$061044.8}\label{obj:j2048}

   This object (hereafter SDSS J204817) is a CV selected
during the course of the SDSS \citep{szk03SDSSCV2}.
The object was identified as an SU UMa-type dwarf nova
during the 2009 superoutburst.  \citet{wou10CVperiod}
observed the object in quiescence and obtained
an orbital period of 0.060597(2) d.
\citet{Pdot2} reported a superhump period of 0.06166(2)~d
based on the orbital period by \citet{wou10CVperiod}.

\begin{figure}
  \begin{center}
    \FigureFile(85mm,110mm){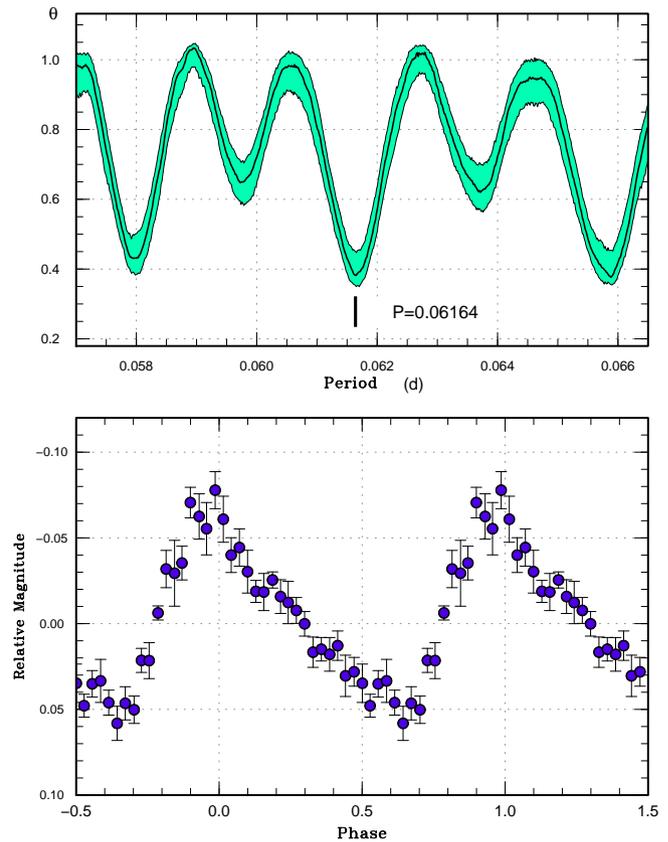}
  \end{center}
  \caption{Superhumps in SDSS J204817 (2009).
     (Upper): PDM analysis.
     (Lower): Phase-averaged profile.}
  \label{fig:j2048shpdm}
\end{figure}

   The 2017 superoutburst was detected by the ASAS-SN
team at $V$=14.55 on 2017 May 30.  Subsequent observations
detected superhumps (vsnet-alert 21087).  Although we
obtained data on two nights, the data on the first
night was of less quality and the apparent hump was not
expressed by the 2009 period, we only used the data
on the second night.  The times of superhump maxima
are listed in e-table \ref{tab:j2048oc2017}.
The quality of the data during the 2017 superoutburst
was poorer and we could not give meaningful
constraint on the superhump period.
We instead provide an updated period analysis
of the 2009 data using a modern (LOWESS) detrending method.
The updated result favors the alias corresponding
to the orbital period (e-figure \ref{fig:j2048shpdm}).

   In the ASAS-SN data, there was a superoutburst
(maximum $V$=14.9 on 2015 October 19)
preceded by a precursor outburst on 2015 October 15
($V$=15.3).  The behavior appears to be typical for
an SU UMa-type dwarf nova.

\begin{table}
\caption{Superhump maxima of SDSS J204817 (2017)}\label{tab:j2048oc2017}
\begin{center}
\begin{tabular}{rp{55pt}p{40pt}r@{.}lr}
\hline
\multicolumn{1}{c}{$E$} & \multicolumn{1}{c}{max\commenta} & \multicolumn{1}{c}{error} & \multicolumn{2}{c}{$O-C$\commentb} & \multicolumn{1}{c}{$N$\commentc} \\
\hline
0 & 57908.4832 & 0.0100 & 0&0006 & 20 \\
1 & 57908.5505 & 0.0013 & $-$0&0011 & 30 \\
2 & 57908.6212 & 0.0031 & 0&0006 & 12 \\
\hline
  \multicolumn{6}{l}{\commenta BJD$-$2400000.} \\
  \multicolumn{6}{l}{\commentb Against max $= 2457908.4826 + 0.069017 E$.} \\
  \multicolumn{6}{l}{\commentc Number of points used to determine the maximum.} \\
\end{tabular}
\end{center}
\end{table}

\subsection{TCP J00332502$-$3518565}\label{obj:j0033}

\begin{figure}
  \begin{center}
    \FigureFile(85mm,110mm){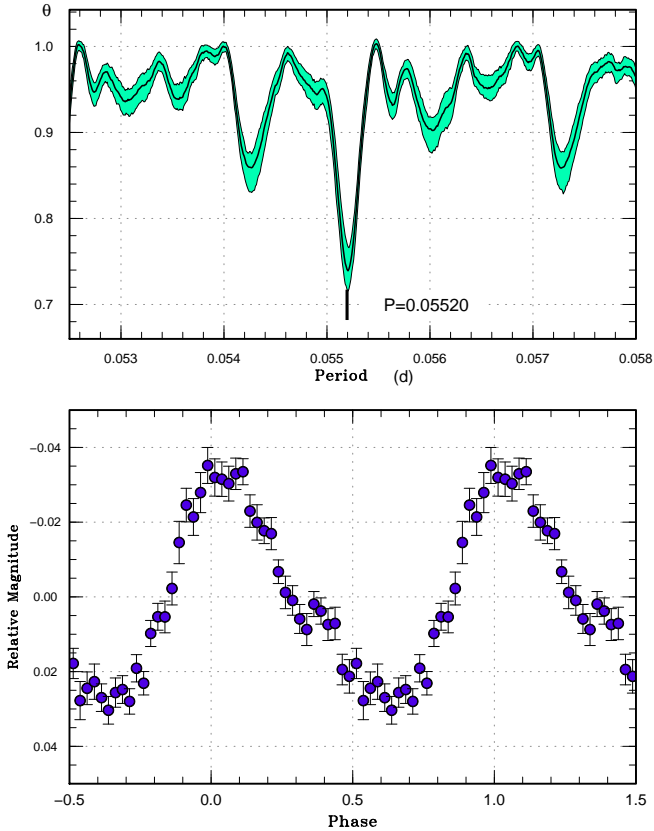}
  \end{center}
  \caption{Ordinary superhumps in TCP J003325 (2017).
     (Upper): PDM analysis.
     (Lower): Phase-averaged profile.}
  \label{fig:j0033shpdm}
\end{figure}

\begin{figure}
  \begin{center}
    \FigureFile(85mm,110mm){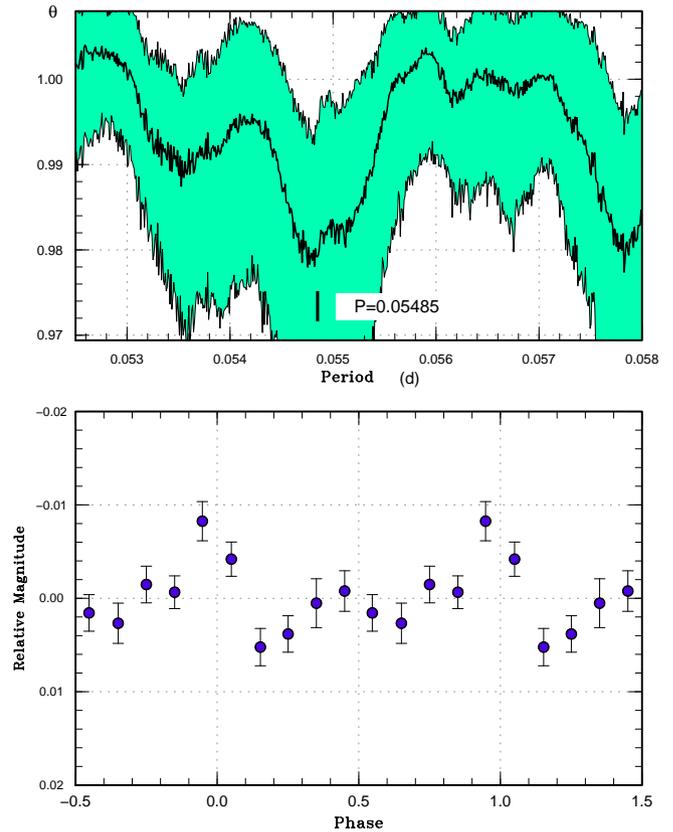}
  \end{center}
  \caption{Possible early superhumps in TCP J003325 (2017).
     (Upper): PDM analysis.
     (Lower): Phase-averaged profile.}
  \label{fig:j0033eshpdm}
\end{figure}

   This object (hereafter TCP J003325) was discovered
by Shigehisa Fujikawa at an unfiltered CCD magnitude
of 12.8--13.3 on 2017 August 5.\footnote{
$<$http://www.cbat.eps.harvard.edu/unconf/followups/J00332502-3518565.html$>$.
}
The object was suspected to be a WZ Sge-type dwarf nova
(vsnet-alert 21320).  There was a 2003 outburst in
the ASAS-3 data (D. Denisenko, vsnet-alert 21321).
The object was also recorded by the ASAS-SN team at $V$=14.7
(rising) in 2017 August 5 and $V$=12.7 on 2017 August 7.
Well-developed superhumps were observed since 2017 August 18
(vsnet-alert 21357; e-figure \ref{fig:j0033shpdm}).
The times of superhump maxima are listed in
e-table \ref{tab:j0033oc2017}.
We retrospectively identified that ordinary superhumps emerged on
2017 August 13 (0 $\le E \le$ 2) based on the flattening
of the outburst light curve and low superhump amplitudes.
Although these superhumps were presumably stage A,
we could not determine the period due to a 4-d gap
in the observation after 2017 August 13.
There were possible very low-amplitude early superhumps
(e-figure \ref{fig:j0033eshpdm}) with a period of
0.05484(16)~d.

   The relatively early (8~d after the outburst rise) appearance
and the presence of the 2003 outburst suggests that
this object is not an extreme WZ Sge-type dwarf nova.

\begin{table}
\caption{Superhump maxima of TCP J003325 (2017)}\label{tab:j0033oc2017}
\begin{center}
\begin{tabular}{rp{55pt}p{40pt}r@{.}lr}
\hline
\multicolumn{1}{c}{$E$} & \multicolumn{1}{c}{max\commenta} & \multicolumn{1}{c}{error} & \multicolumn{2}{c}{$O-C$\commentb} & \multicolumn{1}{c}{$N$\commentc} \\
\hline
0 & 57979.4129 & 0.0018 & $-$0&0003 & 127 \\
1 & 57979.4705 & 0.0023 & 0&0022 & 126 \\
2 & 57979.5338 & 0.0023 & 0&0103 & 127 \\
91 & 57984.4321 & 0.0002 & $-$0&0016 & 127 \\
92 & 57984.4871 & 0.0002 & $-$0&0019 & 124 \\
109 & 57985.4236 & 0.0002 & $-$0&0032 & 127 \\
110 & 57985.4787 & 0.0003 & $-$0&0033 & 98 \\
163 & 57988.4006 & 0.0004 & $-$0&0054 & 107 \\
165 & 57988.5114 & 0.0003 & $-$0&0050 & 127 \\
166 & 57988.5648 & 0.0002 & $-$0&0068 & 123 \\
220 & 57991.5505 & 0.0009 & $-$0&0003 & 127 \\
222 & 57991.6606 & 0.0007 & $-$0&0006 & 109 \\
255 & 57993.4897 & 0.0016 & 0&0080 & 127 \\
256 & 57993.5447 & 0.0013 & 0&0078 & 127 \\
\hline
  \multicolumn{6}{l}{\commenta BJD$-$2400000.} \\
  \multicolumn{6}{l}{\commentb Against max $= 2457979.4132 + 0.055171 E$.} \\
  \multicolumn{6}{l}{\commentc Number of points used to determine the maximum.} \\
\end{tabular}
\end{center}
\end{table}

\subsection{TCP J20100517$+$1303006}\label{obj:j2010}

\begin{figure}
  \begin{center}
    \FigureFile(85mm,110mm){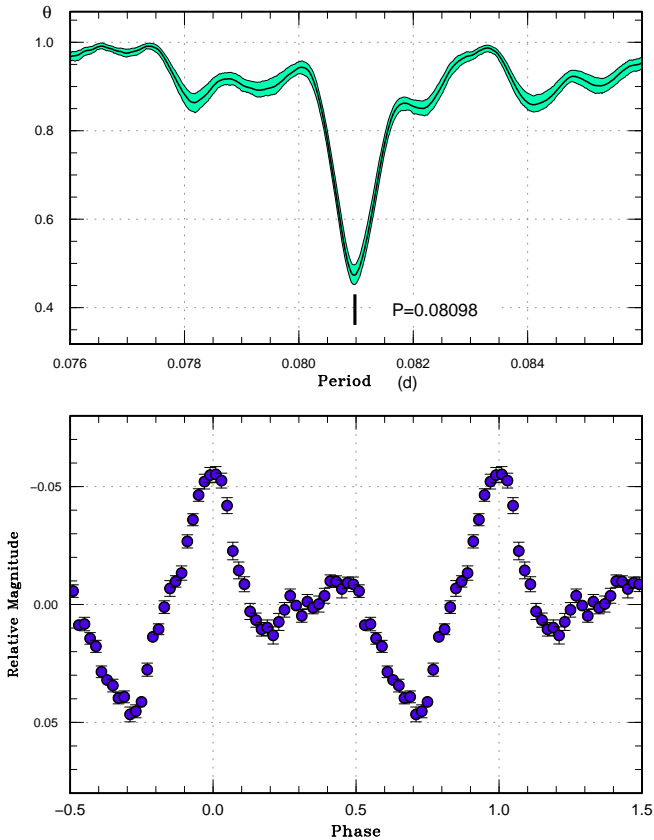}
  \end{center}
  \caption{Superhumps in TCP J201005 (2017).
     (Upper): PDM analysis.  The data before BJD 2457931
     were used.
     (Lower): Phase-averaged profile.}
  \label{fig:j2010shpdm}
\end{figure}

   This object (hereafter TCP J201005) was discovered as
a transient at an unfiltered CCD magnitude of 12.6
on 2017 June 19 by Tadashi Kojima.\footnote{
$<$http://www.cbat.eps.harvard.edu/unconf/followups/J20100517+1303006.html$>$.
}
There is an ROSAT X-ray source 1RXS J201006.4$+$130259
(vsnet-alert 21146) and the object was considered to be
a dwarf nova.  Although initial observations suggested
the presence of early superhumps (vsnet-alert 21149, 21153),
The object was later found to be an ordinary SU UMa-type
dwarf nova (vsnet-alert 21156; the initial claim of
early superhumps was due to the relatively strong
secondary maximum of superhumps; e-figure \ref{fig:j2010shpdm}).  
The object entered the rapidly fading stage on 2017 June 28
(vsnet-alert 21174).

   The times of superhump maxima during the superoutburst
and early post-superoutburst phase are listed in
e-table \ref{tab:j2010oc2017}.  Although there were observations
on BJD 2457931, the hump profile became complex with strong
secondary humps and it was difficult to measure superhump
maxima.  The $O-C$ values for post-superoutburst maxima
are strongly negative, indicating that they were traditional
late superhumps ($\sim$0.5 phase jump around the termination
of the superoutburst).
The superhump stage was unknown.  Since the profile was
already doubly humped at the time of initial observations
(see also e-figure \ref{fig:j2010shpdm}), the observations
were not early enough to detect typical stage B superhumps
The observed superhumps may have been already stage C ones.

   The object was recorded to undergo frequent outbursts
(cf. vsnet-alert 21192, 21197, 21203) and they were once
confused as rebrightenings.  They were shown to be
normal outbursts occurring every 5--10~d (vsnet-alert 21235).
Superoutbursts recorded in the ASAS-SN data
are summarized in e-table \ref{tab:j2010out}.
The coverage by the ASAS-SN observations of this field
was not good enough as in other objects and true maxima
were not necessarily detected.  The data, however, were
sufficient to estimate the supercycle of 103.5(1.3)~d.
The short supercycle suggests a high mass-transfer rate
and it is consistent with the appearance of
traditional late superhumps.

\begin{table}
\caption{Superhump maxima of TCP J201005 (2017)}\label{tab:j2010oc2017}
\begin{center}
\begin{tabular}{rp{55pt}p{40pt}r@{.}lr}
\hline
\multicolumn{1}{c}{$E$} & \multicolumn{1}{c}{max\commenta} & \multicolumn{1}{c}{error} & \multicolumn{2}{c}{$O-C$\commentb} & \multicolumn{1}{c}{$N$\commentc} \\
\hline
0 & 57927.3706 & 0.0096 & $-$0&0061 & 42 \\
1 & 57927.4535 & 0.0010 & $-$0&0039 & 180 \\
2 & 57927.5285 & 0.0018 & $-$0&0095 & 123 \\
9 & 57928.0947 & 0.0039 & $-$0&0078 & 48 \\
10 & 57928.1808 & 0.0008 & $-$0&0023 & 76 \\
11 & 57928.2610 & 0.0012 & $-$0&0028 & 59 \\
13 & 57928.4230 & 0.0003 & $-$0&0021 & 276 \\
14 & 57928.5024 & 0.0004 & $-$0&0034 & 304 \\
15 & 57928.5929 & 0.0011 & 0&0065 & 41 \\
17 & 57928.7515 & 0.0008 & 0&0038 & 87 \\
18 & 57928.8280 & 0.0005 & $-$0&0003 & 110 \\
19 & 57928.9089 & 0.0007 & $-$0&0001 & 57 \\
26 & 57929.4755 & 0.0011 & 0&0020 & 42 \\
30 & 57929.7951 & 0.0015 & $-$0&0010 & 44 \\
31 & 57929.8812 & 0.0007 & 0&0045 & 142 \\
38 & 57930.4484 & 0.0007 & 0&0071 & 189 \\
39 & 57930.5309 & 0.0011 & 0&0090 & 169 \\
40 & 57930.6122 & 0.0016 & 0&0096 & 77 \\
42 & 57930.7740 & 0.0011 & 0&0102 & 185 \\
43 & 57930.8514 & 0.0009 & 0&0069 & 188 \\
44 & 57930.9385 & 0.0014 & 0&0134 & 80 \\
67 & 57932.7677 & 0.0020 & $-$0&0123 & 28 \\
68 & 57932.8509 & 0.0018 & $-$0&0097 & 22 \\
79 & 57933.7416 & 0.0024 & $-$0&0061 & 25 \\
80 & 57933.8229 & 0.0023 & $-$0&0055 & 23 \\
\hline
  \multicolumn{6}{l}{\commenta BJD$-$2400000.} \\
  \multicolumn{6}{l}{\commentb Against max $= 2457927.3767 + 0.080646 E$.} \\
  \multicolumn{6}{l}{\commentc Number of points used to determine the maximum.} \\
\end{tabular}
\end{center}
\end{table}

\begin{table}
\caption{List of superoutbursts of TCP J201005 in the ASAS-SN data}\label{tab:j2010out}
\begin{center}
\begin{tabular}{ccccc}
\hline
Year & Month & Day & max\commenta & $V$ mag \\
\hline
2015 &  6 & 10 & 57184 & 13.4\commentb \\
2015 &  9 & 24 & 57290 & 13.7\commentc \\
2016 &  4 & 10 & 57489 & 13.5 \\
2016 &  7 & 23 & 57593 & 13.7 \\
2016 & 11 & 14 & 57707 & 13.8 \\
\hline
  \multicolumn{5}{l}{\commenta JD$-$2400000.} \\
  \multicolumn{5}{l}{\commentb Following a precursor outburst.} \\
  \multicolumn{5}{l}{\commentc Only late phase was recorded.} \\
\end{tabular}
\end{center}
\end{table}

\end{document}